\documentclass[10pt,journal,compsoc]{IEEEtran}
\usepackage{url}
\usepackage{graphicx}
\usepackage{xcolor}
\usepackage{balance}  
\usepackage{multirow}
\usepackage{threeparttable}
\usepackage{amsmath}
\usepackage{framed}
\usepackage{enumitem}
\usepackage[linesnumbered,ruled,vlined]{algorithm2e}
\usepackage{algorithmicx}  
\usepackage{algpseudocode} 
\usepackage{subfigure}
\usepackage{amsfonts}

\ifCLASSOPTIONcompsoc
  \usepackage[nocompress]{cite}
\else
  \usepackage{cite}
\fi
\ifCLASSINFOpdf
\else
\fi
\begin{document}
%
\title{Efficient Approximate Nearest Neighbor Search for Multiple Weighted \boldmath{$l_{p\leq2}$} Distance Functions}
%
%
%
%

\author{Huan~Hu~and~Jianzhong~Li
\IEEEcompsocitemizethanks{\IEEEcompsocthanksitem H. Hu and J. Li are with the School of Computer Science and Technology, Harbin Institute of Technology, Harbin 150001, China.\protect\\
E-mail: hit\_huhuan@foxmail.com, lijzh@hit.edu.cn.}
}

\IEEEtitleabstractindextext{%
\begin{abstract}
Nearest neighbor search is fundamental to a wide range of applications.
Since the exact nearest neighbor search suffers from the ``curse of dimensionality'', approximate approaches, such as Locality-Sensitive Hashing (LSH), are widely used to trade a little query accuracy for a much higher query efficiency.
In many scenarios, it is necessary to perform nearest neighbor search under multiple weighted distance functions in high-dimensional spaces.
This paper considers the important problem of supporting efficient approximate nearest neighbor search for multiple weighted distance functions in high-dimensional spaces.
To the best of our knowledge, prior work can only solve the problem for the $l_2$ distance.
However, numerous studies have shown that 
the $l_p$ distance with $p\in(0,2)$ could be more effective than the $l_2$ distance in high-dimensional spaces.
We propose a novel method, WLSH, to address the problem
for the $l_p$ distance for $p\in(0,2]$.
WLSH takes the LSH approach and can theoretically guarantee both the efficiency of processing queries and the accuracy of query results while minimizing the required total number of hash tables.
We conduct extensive experiments on synthetic and real data sets, and the results show that WLSH achieves high performance in terms of query efficiency, query accuracy and space consumption.
\end{abstract}

}

\maketitle

\IEEEdisplaynontitleabstractindextext

%
\IEEEpeerreviewmaketitle

\IEEEraisesectionheading{\section{Introduction}\label{Introduction}}
\IEEEPARstart{N}{earest} neighbor search is fundamental to a wide variety of applications, such as recommendation, classification and pattern recognition.
Let $P$ be a set of data points in a $d$-dimensional space. Given a query point $q$, the nearest neighbor search is to find a point $o^*\in P$ such that $o^*$ is the closest point to $q$.

The exact nearest neighbor search has been well studied for low-dimensional cases
\cite{andoni2009nearest,DBLP:journals/corr/abs-1007-0085,samet2006foundations}. A number of methods, such as kd-tree and R-tree based methods \cite{DBLP:journals/cacm/Bentley75,DBLP:journals/sigmod/CheungF98}, have been proposed to efficiently solve the exact nearest neighbor search problem in low-dimensional spaces.

However, conventional methods become even less efficient than the brute-force linear scan when data dimensionality grows due to the ``curse of dimensionality'' \cite{DBLP:reference/db/Chen09}. 
To cope with this problem, approximate approaches are studied to trade query accuracy for query efficiency.
The motivation is that approximate query results are acceptable in most scenarios.
There are several approximate approaches that can perform nearest neighbor search efficiently in high-dimensional spaces \cite{DBLP:conf/visapp/MujaL09,DBLP:conf/stoc/AilonC06,DBLP:journals/pami/MujaL14,DBLP:journals/corr/abs-1806-09823}.
Locality-Sensitive Hashing (LSH) is a popular one among them because of its high performance in both theory and practice  \cite{DBLP:journals/corr/abs-1806-09823,DBLP:journals/corr/WangSSJ14}. 
It relies on LSH families to create hash tables and then process nearest neighbor queries over the hash tables.
Essentially, an LSH family is a set of hash functions that can
hash closer points into the same bucket with higher probability.
A few LSH schemes have been designed based on LSH families in literature \cite{DBLP:conf/stoc/IndykM98,DBLP:conf/vldb/LvJWCL07,DBLP:journals/tods/TaoYSK10,DBLP:conf/sigmod/GanFFN12,DBLP:journals/vldb/HuangFFNW17}.
E2LSH \cite{DBLP:conf/compgeom/DatarIIM04,DBLP:conf/stoc/IndykM98} and C2LSH \cite{DBLP:conf/sigmod/GanFFN12} are two well-known LSH schemes among them. The former ensures that the time cost of processing a nearest neighbor query is sublinear in data size
while the latter ensures that the number of hash tables required to be created grows logarithmically with data size.

While performing nearest neighbor search, distance functions are often used to compute the distance of two points.
To accurately compute the ``real'' distance of two points in practice,  
weighted distance functions are usually adopted \cite{DBLP:journals/classification/Amorim16,DBLP:conf/www/DebnathGM08,DBLP:journals/eaai/EickRBV06}. Briefly, a weighted distance function associates a specific weight vector with the dimensions of a multidimensional space. Thus, a distance function can potentially derive a set of weighted distance functions.
In many scenarios, it is necessary to support nearest neighbor search 
for multiple weighted distance functions in high-dimensional spaces \cite{DBLP:journals/eaai/EickRBV06,DBLP:conf/iccci/HwangST10,DBLP:journals/apin/Liu13,DBLP:conf/sigir/McAuleyTSH15,DBLP:journals/air/WettschereckAM97}.
For example, a personalized recommender system may recommend products based on users' preferences \cite{DBLP:conf/iccci/HwangST10,DBLP:conf/sigir/McAuleyTSH15}.
In such case, products are represented as
a set, $P$, of high-dimensional points, and users' preferences are represented as a set, $\mathcal{S}$, of high-dimensional weight vectors.
Each weight vector in $\mathcal{S}$ determines a weighted distance function.
When a user whose preference is $\vec{W}\in\mathcal{S}$ has already showed interest in product $o\in P$, the system would recommend the product that is the nearest neighbor of $o$ under $\vec{W}$ to the user. 

Traditional LSH schemes like E2LSH and C2LSH are originally proposed to support nearest neighbor search for a distance function. They can not be directly applied to perform nearest neighbor search under multiple weighted distance functions. The fundamental reason is that two close points under a weighted distance function are possibly far away under another weighted distance function.
To the best of our knowledge, SL-ALSH and S2-ALSH \cite{DBLP:conf/icml/LeiHKT19} are the only existing methods in literature that can achieve this goal.
They each introduce an asymmetric LSH family on top of E2LSH, and
the asymmetric LSH families allow them to process each nearest neighbor query flexibly according to the weight vector attached to the query.

Unfortunately, SL-ALSH and S2-ALSH can only work for the $l_2$ distance due to the limitations of the asymmetric LSH families of them. 
Numerous studies over the last few decades have shown that the $l_1$ distance and the fractional distance, which is the $l_p$ distance with $0<p<1$, could be more effective than the $l_2$ distance in high-dimensional spaces \cite{DBLP:conf/icdt/AggarwalHK01,DBLP:conf/vldb/HinneburgAK00,DBLP:conf/ecir/HowarthR05,Hai2012Content}. 

\begin{table}[t]
	\centering
	\footnotesize
	\caption{Time and Space Complexities of SL-ALSH, S2-ALSH and WLSH}
	\label{comparison/complexity}
	\begin{threeparttable}
	\begin{tabular}{c|c|c} 
		\hline
		\textbf{Method}&\textbf{Time Complexity}&\textbf{Space Complexity}\\ \hline
		SL-ALSH&$O(n^{\rho_{SL}}d\log n)$&$O(nd+n^{1+\rho_{SL}})$\\ \hline
		S2-ALSH&$O(n^{\rho_{S2}}d\log n)$&$O(nd+n^{1+\rho_{S2}})$\\ \hline
		WLSH&$O((n+d)\log n)$&$O(nd+n\log n)$\\ \hline
	\end{tabular}
	\begin{tablenotes}
		\item[*] $0<\rho_{SL},\rho_{S2}<1$
	\end{tablenotes}
	\end{threeparttable}
\end{table}

Clearly, it remains a challenging problem to support efficient approximate nearest neighbor search for multiple weighted distance functions with respect to the $l_p$ distance in high-dimensional spaces for $p\in(0,2)$. 

To solve the challenging problem, we propose a novel method called WLSH in this paper. Actually, WLSH is also applicable for the $l_2$ distance. 
A brief overview of WLSH is as follows.
First, WLSH uses the weighted LSH families that are suitable for weighted distance functions with respect to the $l_p$ distance to create hash tables, where $p\in(0,2]$.
Different weighted distance functions correspond to different weighted LSH families. Thus, multiple different groups of hash tables can be created in WLSH.
Second, WLSH obtains derived weighted LSH families from the weighted LSH families to reuse hash tables.
The derived weighted LSH families allow WLSH to reuse a group of hash tables to process nearest neighbor queries under different weighted distance functions.
Third, WLSH partitions the given weighted distance functions into several disjoint subsets such that nearest neighbor search under each of the subsets can be supported by using a single group of hash tables.
In this way, nearest neighbor search under the given weighted distance functions can be supported.
The optimization goal is to process all nearest neighbor queries efficiently with accuracy guarantees while minimizing the required total number of hash tables. 

\textbf{Contributions.} The main contributions of this paper are summarized as follows.
\begin{itemize}[leftmargin=0.4cm]
	\item[1.]{We propose weighted LSH families and derived weighted LSH families that support not only weighted $l_2$ distance functions but also weighted $l_p$ distance functions for $p\in(0,2)$. Actually, the weighted LSH families and derived weighted LSH families can be obtained for various distance measures.}
	\item[2.]{Based on weighted LSH families and derived weighted LSH families, a novel method, WLSH, is proposed to support efficient approximate nearest neighbor search under multiple weighted distance functions with respect to the $l_p$ distance, where $p\in(0,2]$. WLSH offers theoretical guarantees for the query efficiency and query accuracy while minimizing the space consumption.} 
	\item[3.]{To implement the WLSH method, a preprocessing algorithm is designed to create hash tables and a search algorithm is designed to answer nearest neighbor queries over the hash tables. Assuming the number of weighted distance functions is a constant, the time and space complexities of SL-ALSH, S2-ALSH and WLSH are shown in Table \ref{comparison/complexity}, where $n$ is data size, $d$ is data dimensionality, and $\rho_{SL}$ and $\rho_{S2}$ are two real numbers computed by Equations \ref{rho/SL} and \ref{rho/S2} in Appendix \ref{complexity/SL and S2}. As it can be seen, though WLSH is applicable for more than the $l_2$ distance, the space complexity of WLSH is lower than those of SL-ALSH and S2-ALSH, and the time complexity of WLSH is comparable to those of SL-ALSH and S2-ALSH when $d$ is large.}
	\item[4.]{Reasonable trade-offs among the query efficiency, query accuracy and space consumption of WLSH are investigated in order to achieve better practical performance.}
	\item[5.]{Extensive experiments are conducted on both synthetic and real data sets. The results show that WLSH is effective and efficient, and it performs better than SL-ALSH and S2-ALSH in many cases.}
\end{itemize}

The rest of the paper is organized as follows.
Section \ref{Preliminaries} provides the background knowledge and defines the problem we address.
Section \ref{LSH families} introduces the weighted LSH families and derived weighted LSH families. 
The method of WLSH is presented in Section \ref{wlsh}.
The experimental results are reported in Section \ref{Experimental Evaluation}.
We review related work in Section \ref{Related Work} and
conclude the paper in Section \ref{Conclusion}.
\section{Preliminaries}
\label{Preliminaries}
Sections \ref{(R,c)-NN and c-NN}, \ref{Locality Sensitive Hashing Functions} and \ref{LSH schemes} review the LSH approach in the context of approximate nearest neighbor search under a distance function. Section \ref{Problem Description} defines our problem.
\subsection{\boldmath$c$-NN Search and $\left(R,c\right)$-NN Search}
\label{(R,c)-NN and c-NN}
Assume $\mathcal{X}$ is a $d$-dimensional space and $D\left(\cdot,\cdot\right)$ is a distance function defined on $\mathcal{X}$. Let $B(x,R)=\{y\mid D(x,y)\leq R\}$.
In a typical LSH scheme, approximate nearest neighbor search is formalized as 
the \textit{$c$-approximate nearest neighbor ($c$-NN for short) search} problem as follows.
\newtheorem{definition}{Definition}
\begin{definition}
	\label{$c$-NN search}
	(\textsc{$c$-NN Search} \cite{DBLP:journals/corr/WangSSJ14}). Given a $d$-dimensional space $\mathcal{X}$, a distance function $D\left(\cdot,\cdot\right)$, an approximation ratio $c>1$, a set of data points $P\subset \mathcal{X}$ and a user-specified query point $q\in \mathcal{X}$, the $c$-NN search is to return a point $o\in P$ such that $D(q,o)\leq c\times D(q,o^*)$, where $o^*$ is the exact nearest neighbor of $q$ in $P$.
\end{definition}

As will be shown in Section \ref{LSH schemes}, an LSH scheme solves the $c$-NN search problem by reducing it to the \textit{$c$-approximate $R$-near neighbor (($R,c$)-NN for short) search} problem defined as follows.

\begin{definition}
	\label{($R,c$)-NN search}
	(\textsc{($R,c$)-NN Search} \cite{DBLP:journals/corr/WangSSJ14}). Given a $d$-dimensional space $\mathcal{X}$, a distance function $D\left(\cdot,\cdot\right)$, a search radius $R>0$, an approximation ratio $c>1$, a set of data points $P\subset \mathcal{X}$ and a user-specified query point $q\in \mathcal{X}$, the ($R,c$)-NN search is to return a point in $B(q,cR)\cap P$ if $B(q,R)\cap P\neq\emptyset$.
\end{definition}

Essentially, the ($R,c$)-NN search problem is a decision version of the $c$-NN search problem \cite{DBLP:conf/sigmod/GanFFN12}. For convenience, the result of the $c$-NN search is called a $c$-NN, and the result of the ($R,c$)-NN search is called a ($R,c$)-NN.
\subsection{LSH Family}
\label{Locality Sensitive Hashing Functions}
Formally, an \textit{LSH family} is defined as follows.
\begin{definition}
	\label{LSH family}
	(\textsc{LSH Family} \cite{DBLP:journals/corr/WangSSJ14}). An LSH family $\mathcal{H}=\{h:\mathcal{X}\rightarrow U\}$ is called $(R,cR,P_1,P_2)$-sensitive if for any $x,y\in \mathcal{X}$
	\begin{enumerate}
		\item[(1).] if $y\in B(x,R)$, then Pr$[h(x)=h(y)]\geq P_1$;
		\item[(2).] if $y\notin B(x,cR)$, then Pr$[h(x)=h(y)]\leq P_2$.
	\end{enumerate}
\end{definition}

In order for an LSH family to be useful, $P_1$ has to be greater than $P_2$.
Usually, a \textit{collision probability function} $P(\cdot)$ can be obtained for an LSH family such that $P_1=P(R)$ and $P_2=P(cR)$.

We focus on the $l_p$ distance ($p\in(0,2]$) in this paper. The distance function with respect to the $l_p$ distance is 
\begin{equation}
D(x,y)=\sqrt[p]{\sum_{i=1}^{d}\left|x_{i}-y_{i}\right|^p}
\end{equation}
where $\vec{x}=\left(x_{1},x_{2},\ldots,x_{d}\right)\footnote{\text{In this paper, $\vec{x}$ denotes the vector representation of $x\in\mathcal{X}$.}}\in\mathcal{X}=\mathbb{R}^d$ and $\vec{y}=\left(y_{1},y_{2},\ldots,y_{d}\right)\in\mathcal{X}=\mathbb{R}^d$.
A classical LSH family for the $l_p$ distance is $\mathcal{H}_{\vec{a},b}=\{h_{\vec{a},b}:\mathcal{X}\rightarrow U\}$, where 
\begin{equation}
\label{lp/e2lsh}
h_{\vec{a},b}(x)=\lfloor\frac{\vec{a}\cdot\vec{x}+b}{w}\rfloor,
\end{equation}
$\vec{a}$ is a $d$-dimensional vector where each entry is chosen independently from the $p$-stable distribution \cite{Borak2005,DBLP:conf/compgeom/DatarIIM04} (e.g., the 2-stable distribution is the standard normal distribution), $w$ is a user-specified constant, and $b$ is a real number chosen uniformly at random from $[0,w]$. The collision probability function is $P_{l_p}(r)=\int_{0}^{w}\frac{1}{r}F_p(\frac{t}{r})(1-\frac{t}{w})dt$, where $F_p(\cdot)$ is the PDF of the \textit{absolute value} of the $p$-stable distribution \cite{DBLP:conf/compgeom/DatarIIM04}.
\subsection{E2LSH and C2LSH}
\label{LSH schemes}
In the following, we review two well-known LSH schemes, E2LSH \cite{DBLP:conf/stoc/IndykM98,DBLP:journals/cacm/AndoniI08} and C2LSH \cite{DBLP:conf/sigmod/GanFFN12}. 
The preprocessing algorithms and search algorithms for the ($R,c$)-NN search and $c$-NN search are presented for each scheme.

\subsubsection{E2LSH}
We focus on the ($R,c$)-NN search first and then reduce the $c$-NN search to the ($R,c$)-NN search.

\textbf{($R,c$)-NN Search.} The preprocessing algorithm and search algorithm for the ($R,c$)-NN search are as follows.

\textit{Preprocessing algorithm}. 
Given a $(R,cR,P_1,P_2)$-sensitive LSH family $\mathcal{H}$, a compound hash function is of the form: $g=(h_{1},h_{2},\ldots,h_{m})$, where $h_1,h_2,\ldots,h_m$ are LSH functions chosen independently and uniformly at random from $\mathcal{H}$.
The preprocessing algorithm first generates $L$ compound hash functions $g_1, g_2,\ldots, g_L$ independently. Then, it creates $L$ hash tables according to the compound hash functions. As a result, each data point $o\in P$ is hashed into bucket $g_i(o)$ of the $i$th hash table for $1\leq i\leq L$.

\textit{Search algorithm}.
Given a query point $q\in\mathcal{X}$, the search algorithm takes the data points in buckets $g_1(q),g_2(q),\ldots,g_L(q)$ as candidates and computes their distances to $q$ to check if they are ($R,c$)-NNs of $q$. It stops immediately after a ($R,c$)-NN of $q$ is found or $tL$ candidates are checked, where $t$ is a constant.

\textit{Parameter settings}.  
Let $n$ be the number of points in $P$. When we set $m=\log_{1/P_2} n$ and $L=n^{\rho}$ where $\rho=\frac{\ln 1/P_1}{\ln 1/P_2}<1$, the search algorithm above can find a ($R,c$)-NN of $q$ in $P$ with a constant probability if $B(q,R)\cap P\neq\emptyset$ \cite{DBLP:conf/compgeom/DatarIIM04,DBLP:conf/vldb/GionisIM99}.

\textbf{$c$-NN Search.} 
Let $r_{min}$ and $r_{max}$ be the smallest and the largest possible distances between two points respectively.
The $c$-NN search can be performed by issuing a series of ($R,c$)-NN search with increasing radii $R=r_{min},cr_{min},\ldots,c^{\lceil\log_cr_{max}/r_{min}\rceil}r_{min}$ \cite{DBLP:conf/compgeom/DatarIIM04,DBLP:conf/vldb/GionisIM99}.
Thus, the preprocessing algorithm for the $c$-NN search simply calls the preprocessing algorithm for the ($R,c$)-NN search at radii $R=r_{min},cr_{min},\ldots,c^{\lceil\log_cr_{max}/r_{min}\rceil}r_{min}$ respectively. Accordingly, the search algorithm for the $c$-NN search simply calls the search algorithm for the ($R,c$)-NN search at radii $R=r_{min},cr_{min},\ldots,c^{\lceil\log_cr_{max}/r_{min}\rceil}r_{min}$ respectively, and it stops immediately after a $(R,c)$-NN is found or $tL$ candidates are checked at some search radius.

\subsubsection{C2LSH}
\label{C2LSH}
The scheme C2LSH is suitable for the $l_p$ distance.
It relies on two LSH families for the $l_p$ distance,  $\mathcal{H}_{\vec{a},b^*}=\{h_{\vec{a},b^*}:\mathcal{X}\rightarrow U\}$ and $\mathcal{H}_{\vec{a},b^*}^l=\{h_{\vec{a},b^*}^l:\mathcal{X}\rightarrow U\}$, where
\begin{equation}
\label{function1}
h_{\vec{a},b^*}(x)=\lfloor\frac{\vec{a}\cdot\vec{x}+b^*}{w}\rfloor,\quad h_{\vec{a},b^*}^l(x)=\lfloor\frac{h_{\vec{a},b^*}(x)}{l}\rfloor,
\end{equation}
$\vec{a}$ and $w$ are the same as in Equation \ref{lp/e2lsh},
and $b^*$ is a real number chosen uniformly at random from $[0,c^{\lceil\log_cr_{max}/r_{min}\rceil}w]$ \cite{DBLP:journals/tods/TaoYSK10}.
Lemma \ref{lemma 1} in this paper indicates that
$\mathcal{H}_{\vec{a},b^*}$ is $(x,y,P_{l_p}(x),P_{l_p}(y))$-sensitive and $\mathcal{H}_{\vec{a},b^*}^l$ is 
$(xl,yl,P_{l_p}(x),P_{l_p}(y))$-sensitive for any $0<x<y$ and $l\in\{c,c^2,\ldots,c^{\lceil\log_cr_{max}/r_{min}\rceil}\}$.

Buckets defined by an LSH function from $\mathcal{H}_{\vec{a},b^*}$ are called level-1 buckets, and buckets defined by an LSH function from $\mathcal{H}_{\vec{a},b^*}^l$ are called level-$l$ buckets.
From Equation \ref{function1}, 
it is easy to see that a level-$l$ bucket $x$ consists of $l$ consecutive level-1 buckets $x\times l, x \times l+1, \ldots, x\times l+l-1$.
Therefore, the exploration of a level-$l$ bucket 
can be implemented by probing the corresponding $l$ level-1 buckets.
The technique is called \textit{virtual rehashing}.

\textbf{($R,c$)-NN Search.} Assume the search radius is $R=r_{min}l$, where $l\in\{1,c,c^2,\ldots,c^{\lceil\log_cr_{max}/r_{min}\rceil}\}$. The preprocessing algorithm and search algorithm for the ($R,c$)-NN search are described as follows.  

\textit{Preprocessing algorithm}. 
The algorithm first chooses $\beta$ LSH functions $h_{\vec{a},b^*,1},h_{\vec{a},b^*,2},\ldots,h_{\vec{a},b^*,\beta}$ independently and uniformly at random from $\mathcal{H}_{\vec{a},b^*}$. Then,
it creates $\beta$ hash tables according to the LSH functions. As a result, each data point $o\in P$ is hashed into bucket $h_{\vec{a},b^*,i}(o)$ of the $i$th hash table for $1\leq i\leq \beta$.

\textit{Search algorithm}. Some concepts need to be introduced first.
Given a query point $q\in\mathcal{X}$, the \textit{collision number} of a data point $o\in P$ is the number of buckets where $o$ collides with $q$. 
A data point $o\in P$ is called \textit{frequent} if its collision number is greater than or equal to a \textit{collision threshold} $\mu$.
$\gamma$ is the allowable percentage of \textit{false positives} which are frequent points but not ($R,c$)-NNs. 

Let $h_{\vec{a},b^*,i}^l(q)$ be the level-$l$ bucket with respect to $h_{\vec{a},b^*,i}$ and a query point $q\in\mathcal{X}$ for $1\leq i\leq \beta$. The search algorithm 
takes the frequent points in buckets $h_{\vec{a},b^*,1}^l(q), h_{\vec{a},b^*,2}^l(q),\ldots,h_{\vec{a},b^*,\beta}^l(q)$ as candidates and computes their distances to $q$ to check if they are ($R,c$)-NNs of $q$. It stops immediately after a ($R,c$)-NN of $q$ is found or $\gamma n$ candidates are checked.

\textit{Parameter settings}.
The search algorithm above can find a ($R,c$)-NN of $q$ in $P$ if the following two properties hold:
\begin{itemize}
	\item[$\mathcal{P}_1$:] If $o\in B(q,R)\cap P$, then $o$ is frequent, that is, the collision number of $o$ is at least $\mu$.
	\item[$\mathcal{P}_2$:] The total number of false positives is less than $\gamma n$.
\end{itemize}
From \cite{DBLP:conf/sigmod/GanFFN12}, we have Pr$[\mathcal{P}_1\text{ holds}]\geq 1-\epsilon$ and Pr$[\mathcal{P}_2\text{ holds}]> 1/2$ by setting $\beta$ and $\mu$ to:
\begin{equation}
\label{beta}
\beta=\lceil\frac{\ln \frac{1}{\epsilon}}{2(P_{l_p}(x)-P_{l_p}(y))^2}(1+z)^2\rceil
\end{equation}
\begin{equation}
\label{mu}
\mu=\frac{zP_{l_p}(x)+P_{l_p}(y)}{1+z}\beta
\end{equation}
where $z=\sqrt{\frac{\ln\frac{2}{\gamma}}{\ln\frac{1}{\epsilon}}}$, $x=r_{min}$ and $y=cr_{min}$. As a result, a ($R,c$)-NN of $q$ in $P$ can be found with a constant probability if $B(q,R)\cap P\neq\emptyset$.
In practice, $\epsilon$ and $\gamma$ can be set to 0.01 and $100/n$ respectively
\cite{DBLP:conf/sigmod/GanFFN12}.

\textbf{$c$-NN Search.}
Different from E2LSH, the virtual rehashing technique allows C2LSH not to create hash tables at different radii for the $c$-NN search.
Thus, the preprocessing algorithm for the $c$-NN search is the same as the preprocessing algorithm for the ($R,c$)-NN search. The search algorithm for the $c$-NN search calls the search algorithm for the ($R,c$)-NN search at radii $R=r_{min},cr_{min},\ldots,c^{\lceil\log_cr_{max}/r_{min}\rceil}r_{min}$ respectively, and it stops immediately after a $(R,c)$-NN is found or $\gamma n$ candidates are checked at some search radius.
\subsection{Problem Description}
\label{Problem Description}
Generally, for any distance measure, the distance function is of the form $D(x,y)=f(x_1,y_1,x_2,y_2,\ldots,x_d,y_d)$, where $f$ is a function depending on the distance measure, $\vec{x}=(x_1,x_2,\ldots,x_d)\in\mathcal{X}$ and $\vec{y}=(y_1,y_2,\ldots,y_d)\in\mathcal{X}$.
Based on this, a \textit{weighted distance function} is defined as follows.
\begin{definition}
	\label{Dimension Weighting Distance Function}
	(\textsc{Weighted Distance Function}).
	Given a distance function $D(x,y)=f(x_1,y_1,x_2,y_2,\ldots,x_d,y_d)$ and a weight vector $\vec{W}=(w_{1},w_{2},\ldots,w_{d})$ ($w_i>0$ for $1\leq i\leq d$),
	the weighted distance function with respect to them is $D_{\vec{W}}(x,y)=f(w_1x_1,w_1y_1,w_2x_2,w_2y_2,\ldots,w_dx_d,w_dy_d)$.
\end{definition}

Please note that weighted distance functions can take different forms as in  \cite{DBLP:conf/icassp/FanKYWP13,DBLP:conf/cikm/WangZS13,DBLP:conf/sigir/McAuleyTSH15}. Actually, these forms are equivalent to the one in Definition \ref{Dimension Weighting Distance Function}. It can be seen from Definition \ref{Dimension Weighting Distance Function} that a weighted distance function is identified by the associated weight vector for a given distance measure. Thus, we sometimes refer to a weighted distance function by the associated weight vector for ease of discussion.

A weighted distance function can be viewed as a special distance function. Thus, the LSH approach introduced in Sections \ref{(R,c)-NN and c-NN}, \ref{Locality Sensitive Hashing Functions} and \ref{LSH schemes} can be directly extended to support approximate nearest neighbor search for a weighted distance function. 
Let $B_{\vec{W}}(x,R)=\{y\mid D_{\vec{W}}(x,y)\leq R\}$.
Then the $c$-NN search problem is extended to
the \textit{weighted $c$-approximate nearest neighbor ($c$-WNN for short) search} problem in Definition \ref{$c$-WNN search},
and the ($R,c$)-NN search problem is extended to
the \textit{weighted $c$-approximate $R$-near neighbor (($R,c$)-WNN for short) search} problem in Definition \ref{$(R,c)$-WNN search}. 
As it can be seen, the $c$-WNN search problem is a formalization of approximate nearest neighbor search under a weighted distance function. Given a proper LSH family, an LSH scheme, e.g., E2LSH or C2LSH, solves the $c$-WNN search problem by reducing it to the ($R,c$)-WNN search problem.
\begin{definition}
	\label{$c$-WNN search}
	(\textsc{$c$-WNN Search}). Given a $d$-dimensional space $\mathcal{X}$, a distance function $D\left(\cdot,\cdot\right)$, a $d$-dimensional weight vector $\vec{W}$, an approximation ratio $c>1$, a set of data points $P\subset \mathcal{X}$ and a user-specified query point $q\in \mathcal{X}$, the $c$-WNN search is to return a point $o\in P$ such that $D_{\vec{W}}(q,o)\leq c\times D_{\vec{W}}(q,o^*)$, where $o^*$ is the exact nearest neighbor of $q$ in $P$ under $\vec{W}$. 
\end{definition}
\begin{definition}
	\label{$(R,c)$-WNN search}
	(\textsc{($R,c$)-WNN Search}). Given a $d$-dimensional space $\mathcal{X}$, a distance function $D\left(\cdot,\cdot\right)$, a $d$-dimensional weight vector $\vec{W}$, a search radius $R>0$, an approximation ratio $c>1$, a set of data points $P\subset \mathcal{X}$ and a user-specified query point $q\in \mathcal{X}$, the ($R,c$)-WNN search is to return a point in $B_{\vec{W}}(q,cR)\cap P$ if $B_{\vec{W}}(q,R)\cap P\neq\emptyset$.
\end{definition}

Formally, the problem studied in this paper is as follows.

\textbf{Problem Statement.}

\textbf{Input:} A $d$-dimensional space $\mathcal{X}$, a distance function $D\left(\cdot,\cdot\right)$, a set of $d$-dimensional weight vectors $\mathcal{S}$, an approximation ratio $c>1$, a result cardinality $k\geq 1$, a set of data points $P\subset \mathcal{X}$, a user-specified query weight vector $\vec{W_i}\in\mathcal{S}$, and a user-specified query point $q\in \mathcal{X}$.

\textbf{Output:} $kANNs=\{o_1,o_2,\ldots,o_k\}\subset P$ that satisfies $D_{\vec{W_i}}(q,o_j)\leq c\times D_{\vec{W_i}}(q,o_j^*)$ for $1\leq j\leq k$, where the points $o_1^*,o_2^*,\ldots,o_k^*$ are the exact $k$-nearest neighbors of $q$ in $P$ under $\vec{W_i}$.

If $\left|\mathcal{S}\right|$, the cardinality of the input weight vector set $\mathcal{S}$, is equal to 1, then the problem is called the \textit{weighted $c$-approximate $k$-nearest neighbor (($c,k$)-WNN for short) search} problem. The ($c,k$)-WNN search problem is a formalization of approximate $k$-nearest neighbor search under a weighted distance function. It can be solved by repeating the preprocessing algorithm for the $c$-WNN search and modifying the search algorithm for the $c$-WNN search to stop it after $k$ ($R,c$)-WNNs are found or $k$ more candidates than allowed are checked at some search radius \cite{DBLP:conf/sigmod/GanFFN12}. 

It is obvious that our problem is to support the ($c,k$)-WNN search for each weight vector in $\mathcal{S}$.
A naive method for solving our problem is applying the above solution to the ($c,k$)-WNN search problem for each weight vector in $\mathcal{S}$. 
As a result of the naive method, the required total number of hash tables is linear in $\left|\mathcal{S}\right|$.
When $\left|\mathcal{S}\right|$ is not small, the space overhead could be hard to tolerate in practice. 
Therefore, we would like to propose a new method which is more space-efficient than the naive method while ensuring that the time efficiency is still high. We concentrate on the $l_p$ distance.

For convenience, the ($c,k$)-WNN search under a weight vector $\vec{W_i}$ is defined as the ($c,k$)-WNN search where the weighted distance function used is $D_{\vec{W_i}}(\cdot,\cdot)$, and the ($c,k$)-WNN search under a weight vector set $\mathcal{S}$ is defined as the ($c,k$)-WNN search where the weighted distance function used is in the range $\{D_{\vec{W_i}}(\cdot,\cdot)\mid\vec{W_i}\in\mathcal{S}\}$.
The result of the ($c,k$)-WNN search under a weight vector $\vec{W_i}$ is called a ($c,k$)-WNN under $\vec{W_i}$. 
Similar definitions are also introduced for the ($R,c$)-WNN search and the $c$-WNN search. 

Since the techniques for the $c$-WNN search can almost cover the techniques for the ($c,k$)-WNN search, we will limit ourselves to the $c$-WNN search before introducing the method of WLSH. The frequently-used notations in the paper are summarized in Table \ref{notations}.

\begin{table}[htbp]
	\centering
	\footnotesize
	\caption{Frequently-Used Notations}
	\resizebox{0.485\textwidth}{!}{
	\begin{tabular}{l|l}
		\hline
		\textbf{Notation}&\textbf{Description}\\ \hline
		$\mathcal{X}$&A $d$-dimensional space\\ \hline
		$P$&A set of data points in $\mathcal{X}$\\ \hline
		$n$&The cardinality of $P$\\ \hline
		$q$&A query point\\ \hline
		$\vec{W}$&A weight vector\\ \hline
		$D_{\vec{W}}\left(\cdot,\cdot\right)$&The weighted distance function identified by $\vec{W}$\\ \hline
		$B_{\vec{W}}(x,R)$&The ball centered at $x$ with radius $R$ under $\vec{W}$\\ \hline
		\multirow{2}{0.185\columnwidth}{$r_{max}^{\vec{W}}$ ($r_{min}^{\vec{W}}$)}&The largest (smallest) possible distance\\&between two points under $\vec{W}$\\ \hline
		$\mu_{\vec{W}}$&The collision threshold for $\vec{W}$\\ \hline
		$\gamma$&The allowable percentage of false positives\\ \hline
		$\beta_{\vec{W}}$&The required number of hash tables for $\vec{W}$\\ \hline
		$\mathcal{S}$&A weight vector set\\ \hline
		$c$&Approximation ratio\\ \hline
		PDF (PMF)&Abbreviation of probability density (mass) function\\ \hline
	\end{tabular}}
	\label{notations}
\end{table}
\section{LSH Families for Weighted Distance Functions}
\label{LSH families}
In this section, we introduce two kinds of LSH families for weighted distance functions, namely, weighted LSH families and derived weighted LSH families.
\subsection{Weighted LSH Family}
\label{weighted LSH family}
Traditional LSH families are designed to hash closer points under distance functions into the same buckets with higher probabilities, they are not suitable for weighted distance functions. We propose weighted LSH families to cope with this problem. Essentially, weighted LSH families generalize traditional LSH families to weighted distance functions.

Given a weight vector $\vec{W}$, we have $D_{\vec{W}}(o,q)=D(\vec{W}\circ\vec{o},\vec{W}\circ\vec{q})$ from Definition \ref{Dimension Weighting Distance Function}, where $o\in P$, $q\in\mathcal{X}$, and $\circ$ is an element-wise product.
This suggests that searching for a $c$-WNN of $q$ in $P$ under $\vec{W}$ is equivalent to searching for a $c$-NN of $\vec{W}\circ \vec{q}$ in $\vec{W}\circ P$, where $\vec{W}\circ P=\{\vec{W}\circ\vec{o}\mid o\in P\}$. Therefore, given a traditional LSH family $\mathcal{H}=\{h:\mathcal{X}\rightarrow U\}$, 
we can directly obtain a weighted LSH family $\mathcal{H}_{\vec{W}}=\{h:\vec{W}\circ \mathcal{X}\rightarrow U\}$, where $\vec{W}\circ\mathcal{X}=\{\vec{W}\circ\vec{x}\mid x\in \mathcal{X}\}$. 
For convenience, 
define $h_{\vec{W}}(x)=h(\vec{W}\circ \vec{x})$ for $x\in \mathcal{X}$ such that $\mathcal{H}_{\vec{W}}=\{h_{\vec{W}}:\mathcal{X}\rightarrow U\}$.

Next, we determine $P_{1,\vec{W}}$ and $P_{2,\vec{W}}$ such that $\mathcal{H}_{\vec{W}}$ is $(R,cR,P_{1,\vec{W}},P_{2,\vec{W}})$-sensitive.
It is easy to see that $\mathcal{H}_{\vec{W}}$ involves the same set of random variables as $\mathcal{H}$.
Thus, the first step is to set the PDF (or PMF) of each involved random variable for $\mathcal{H}_{\vec{W}}$ to ensure that closer points under $\vec{W}$ can be hashed into the same bucket with higher probability.
The second step is to get
a collision probability function $P_{\vec{W}}(\cdot)$ such that $P_{1,\vec{W}}=P_{\vec{W}}(R)$ and $P_{2,\vec{W}}=P_{\vec{W}}(cR)$, which is purely a mathematical problem.

The weighted LSH family obtained from $\mathcal{H}_{\vec{a},b}=\{h_{\vec{a},b}:\mathcal{X}\rightarrow U\}$ in Section \ref{Locality Sensitive Hashing Functions} is $\mathcal{H}_{\vec{a},b,\vec{W}}=\{h_{\vec{a},b,\vec{W}}:\mathcal{X}\rightarrow U\}$, where 
\begin{equation}
\label{weighted lp/e2lsh}
h_{\vec{a},b,\vec{W}}(x)=\lfloor\frac{\vec{a}\cdot\left(\vec{W}\circ \vec{x}\right)+b}{w}\rfloor,
\end{equation}
and the PDFs of $\vec{a}$ and $b$ are the same as for $\mathcal{H}_{\vec{a},b}$. We can easily know from \cite{DBLP:conf/compgeom/DatarIIM04} that the collision probability function for $\mathcal{H}_{\vec{a},b,\vec{W}}$ is $P_{l_p,\vec{W}}(r)=P_{l_p}(r)=\int_{0}^{w}\frac{1}{r}F_p(\frac{t}{r})(1-\frac{t}{w})dt$. In addition, the weighted LSH families obtained from $\mathcal{H}_{\vec{a},b^*}=\{h_{\vec{a},b^*}:\mathcal{X}\rightarrow U\}$ and $\mathcal{H}_{\vec{a},b^*}^l=\{h_{\vec{a},b^*}^l:\mathcal{X}\rightarrow U\}$ in Section \ref{C2LSH} are $\mathcal{H}_{\vec{a},b^*,\vec{W}}=\{h_{\vec{a},b^*,\vec{W}}:\mathcal{X}\rightarrow U\}$ and $\mathcal{H}_{\vec{a},b^*,\vec{W}}^l=\{h_{\vec{a},b^*,\vec{W}}^l:\mathcal{X}\rightarrow U\}$ respectively, where
\begin{equation}
\label{weighted function1}
\begin{split}
h_{\vec{a},b^*,\vec{W}}(x)=&\lfloor\frac{\vec{a}\cdot\left(\vec{W}\circ \vec{x}\right)+b^*}{w}\rfloor,\\
h_{\vec{a},b^*,\vec{W}}^l(x)=&\lfloor\frac{h_{\vec{a},b^*,\vec{W}}(x)}{l}\rfloor,
\end{split}
\end{equation}
the PDF of $\vec{a}$ is the same as for $\mathcal{H}_{\vec{a},b^*}$, and the PDF of $b^*$ is set to $p(b^*)=1/(c^{\lceil\log_cr_{max}^{\vec{W}}/r_{min}^{\vec{W}}\rceil}w)$ for $b^*\in[0,c^{\lceil\log_cr_{max}^{\vec{W}}/r_{min}^{\vec{W}}\rceil}w]$ 
($r_{min}^{\vec{W}}$ and $r_{max}^{\vec{W}}$ are the smallest and the largest possible distances between two points under $\vec{W}$ respectively). More about
$\mathcal{H}_{\vec{a},b^*,\vec{W}}$ and $\mathcal{H}_{\vec{a},b^*,\vec{W}}^l$ will be introduced in Section \ref{wlsh}. Due to space limitations, we present weighted LSH families for the Hamming distance \cite{DBLP:conf/vldb/GionisIM99} and angular distance \cite{DBLP:conf/stoc/Charikar02} in Appendix \ref{lsh families and weighted lsh families}.
\subsection{Derived Weighted LSH Family}
\label{derived weighted LSH family}
Given a weighted LSH family $\mathcal{H}_{\vec{W}}=\{h_{\vec{W}}:\mathcal{X}\rightarrow U\}$ and a weight vector $\vec{W'}$, 
the derived weighted LSH family with respect to them is denoted by $\mathcal{H}_{\vec{W}\rightarrow\vec{W'}}=\{h_{\vec{W}\rightarrow\vec{W'}}:\mathcal{X}\rightarrow U\}$. 
Suppose a group of hash tables have been created according to weighted LSH functions from $\mathcal{H}_{\vec{W}}$ to support the $c$-WNN search under $\vec{W}$. 
The goal of $\mathcal{H}_{\vec{W}\rightarrow\vec{W'}}$ is to reuse the hash tables for the $c$-WNN search under $\vec{W'}$ (assume $\vec{W}\neq\vec{W'}$).
First, we let $h_{\vec{W}\rightarrow\vec{W'}}(x)=h_{\vec{W}}(x)$ for $x\in \mathcal{X}$. That is, $\mathcal{H}_{\vec{W}\rightarrow\vec{W'}}$ hashes points in the same way as $\mathcal{H}_{\vec{W}}$.
The next step is to determine $P_{1,\vec{W}\rightarrow\vec{W'}}$ and $P_{2,\vec{W}\rightarrow\vec{W'}}$ such that $\mathcal{H}_{\vec{W}\rightarrow\vec{W'}}$ is $(R,cR,P_{1,\vec{W}\rightarrow\vec{W'}},P_{2,\vec{W}\rightarrow\vec{W'}})$-sensitive. 
If $P_{1,\vec{W}\rightarrow\vec{W'}}>P_{2,\vec{W}\rightarrow\vec{W'}}$, then $\mathcal{H}_{\vec{W}\rightarrow\vec{W'}}$ is an effective LSH family under $\vec{W'}$.
In order to determine $P_{1,\vec{W}\rightarrow\vec{W'}}$ and $P_{2,\vec{W}\rightarrow\vec{W'}}$, 
we make the following assumption. 
\newtheorem{assumption}{Assumption}
\begin{assumption}
	\label{assumption1}
	The collision probability $P_{\vec{W}}(r)$ is inversely proportional to $r$.
\end{assumption}
\noindent In fact, the assumption is reasonable and it holds for all the weighted LSH families in the paper. According to Definition \ref{LSH family}, $\mathcal{H}_{\vec{W}\rightarrow\vec{W'}}$ is $(R,cR,P_{1,\vec{W}\rightarrow\vec{W'}},P_{2,\vec{W}\rightarrow\vec{W'}})$-sensitive if and only if
(1) any point in $B_{\vec{W'}}(q,R)$
collides with $q$ with probability at least $P_{1,\vec{W}\rightarrow\vec{W'}}$ and (2) any point not in $B_{\vec{W'}}(q,cR)$ collides with $q$ with probability at most $P_{2,\vec{W}\rightarrow\vec{W'}}$. The two conditions are formulated as:
\begin{equation}
\label{tt1}
\begin{split}
P_{1,\vec{W}\rightarrow\vec{W'}}&\leq\min\limits_{o\in B_{\vec{W'}}(q,R)}\text{Pr}[h_{\vec{W}\rightarrow\vec{W'}}(o)=h_{\vec{W}\rightarrow\vec{W'}}(q)]\\
&=\min\limits_{o\in B_{\vec{W'}}(q,R)}\text{Pr}[h_{\vec{W}}(o)=h_{\vec{W}}(q)]\\
P_{2,\vec{W}\rightarrow\vec{W'}}&\geq\max\limits_{o\notin B_{\vec{W'}}(q,cR)}\text{Pr}[h_{\vec{W}\rightarrow\vec{W'}}(o)=h_{\vec{W}\rightarrow\vec{W'}}(q)]\\
&=\max\limits_{o\notin B_{\vec{W'}}(q,cR)}\text{Pr}[h_{\vec{W}}(o)=h_{\vec{W}}(q)]
\end{split}
\end{equation}
Define $R^{\uparrow}$ as an upper bound of the largest possible distance between $q$ and $o\in B_{\vec{W'}}(q,R)$ under $\vec{W}$,
and define $(cR)^{\downarrow}$ as a lower bound of the smallest possible distance between $q$ and $o\notin B_{\vec{W'}}(q,cR)$ under $\vec{W}$. That is,
\begin{equation}
\label{R_up}
\begin{split}
R^{\uparrow}\geq&\max\limits_{o\in B_{\vec{W'}}(q,R)}D_{\vec{W}}(o,q)\\
(cR)^{\downarrow}\leq&\min\limits_{o\notin B_{\vec{W'}}(q,cR)}D_{\vec{W}}(o,q)
\end{split}
\end{equation}
Combining Equation \ref{R_up} and Assumption \ref{assumption1}, we obtain:
\begin{equation}
\label{upper}
\begin{split}
P_{\vec{W}}(R^{\uparrow})\leq&\min\limits_{o\in B_{\vec{W'}}(q,R)}\text{Pr}[h_{\vec{W}}(o)=h_{\vec{W}}(q)]\\
P_{\vec{W}}((cR)^{\downarrow})\geq&\max\limits_{o\notin B_{\vec{W'}}(q,cR)}\text{Pr}[h_{\vec{W}}(o)=h_{\vec{W}}(q)]
\end{split}
\end{equation}
Further, combining Equations \ref{tt1} and \ref{upper}, we can safely set
$P_{1,\vec{W}\rightarrow\vec{W'}}=P_{\vec{W}}(R^{\uparrow})$ and $P_{2,\vec{W}\rightarrow\vec{W'}}= P_{\vec{W}}((cR)^{\downarrow})$. As a result, $\mathcal{H}_{\vec{W}\rightarrow\vec{W'}}$ is
$(R,cR,P_{\vec{W}}(R^{\uparrow}),P_{\vec{W}}((cR)^{\downarrow}))$-sensitive.

In order for $\mathcal{H}_{\vec{W}\rightarrow\vec{W'}}$ to be useful, $R^{\uparrow}<(cR)^{\downarrow}$ should hold so that $P_{\vec{W}}(R^{\uparrow})>P_{\vec{W}}((cR)^{\downarrow})$. Actually, $R^{\uparrow}<(cR)^{\downarrow}$ may not hold even when $R^{\uparrow}$ and $(cR)^{\downarrow}$ are both tight bounds. 
In the following Theorem \ref{R_bound}, we give $R^{\uparrow}$s and $(cR)^{\downarrow}$s for the $l_p$ distance, Hamming distance and angular distance. The proof can be found in Appendix \ref{proof of theorem 1}.
\newtheorem{theorem}{Theorem}
\begin{theorem}
	\label{R_bound}
	Given any $R>0$ and any $\mathcal{H}_{\vec{W}\rightarrow\vec{W'}}$ where $\vec{W}=(w_{1},w_{2},\ldots,w_{d})$ and $\vec{W'}=(w_{1}',w_{2}',\ldots,w_{d}')$, we have:
	\begin{enumerate}
		\item[(1).]{For the $l_p$ distance, $R^{\uparrow}=R\max_{1\leq i\leq d}(w_i/w_i')$ and $(cR)^{\downarrow}=cR\min_{1\leq i\leq d}(w_i/w_i')$.}
		\item[(2).]{For the Hamming distance, $R^{\uparrow}=R\max_{1\leq i\leq d}(w_i/w_i')$ and $(cR)^{\downarrow}=cR\min_{1\leq i\leq d}(w_i/w_i')$.}
		\item[(3).]{For the angular distance, $R^{\uparrow}=\arccos\left(\max\left(-1,X\right)\right)$ and $(cR)^{\downarrow}=\arccos\left(\min\left(1,Y\right)\right)$, where $X=\cos(R)+(N-M)/M$, $Y=M\cos(cR)/N+(M-N)/N$, $M=\max_{1\leq i\leq d}(w_i^2/w_i'^2)$, and $N=\min_{1\leq i\leq d}(w_i^2/w_i'^2)$.}
	\end{enumerate}
\end{theorem} 
\section{WLSH}
\label{wlsh}
Our problem is to efficiently support the ($c,k$)-WNN search under a given weight vector set $\mathcal{S}$ for the $l_p$ distance.
Recall the overview of WLSH in Section \ref{Introduction}.
WLSH addresses the problem by partitioning $\mathcal{S}$ into several disjoint subsets and then supporting the ($c,k$)-WNN search under each of the subsets respectively. The partition method can ensure that the ($c,k$)-WNN search under each obtained subset of $\mathcal{S}$ requires only a single group of hash tables.
In Section \ref{challenge1}, we introduce how WLSH supports the ($c,k$)-WNN search under a subset of $\mathcal{S}$ by creating a group of hash tables, and in Section \ref{challenge2}, we introduce how WLSH partitions $\mathcal{S}$ into several disjoint subsets appropriately.
Finally, we present a preprocessing algorithm and a search algorithm for the ($c,k$)-WNN search under $\mathcal{S}$.
\subsection{\boldmath ($c,k$)-WNN Search under a Subset of $\mathcal{S}$}
\label{challenge1}
We base WLSH on the LSH scheme of C2LSH \cite{DBLP:conf/sigmod/GanFFN12}. Thus, the weighted LSH families given by Equation \ref{weighted function1} are used for WLSH. We focus on the $c$-WNN search first. Let $\mathcal{S}^{\circ}$ be a subset of $\mathcal{S}$.
Suppose the $c$-WNN search under $\mathcal{S}^{\circ}$ will be supported using a group of hash tables created according to weighted LSH functions from $\mathcal{H}_{\vec{a},b^*,\vec{W^{\circ}}}$, where $\vec{W^{\circ}}=\{w^{\circ}_1,w^{\circ}_2,\ldots,w^{\circ}_d\}$ is some weight vector in $\mathcal{S}$.
We need to determine the required number of hash tables $\beta_{\vec{W_i}}$ and the collision threshold $\mu_{\vec{W_i}}$ for each $\vec{W_i}\in\mathcal{S}^{\circ}$ from Section \ref{C2LSH}.
The number of hash tables required to be created is $\beta_{\mathcal{S}^{\circ}}=\max_{\vec{W_i}\in\mathcal{S}^{\circ}}(\beta_{\vec{W_i}})$. 
To obtain the parameters $\mathcal{B}_{\mathcal{S}^{\circ}}=\{(\beta_{\vec{W_i}},\mu_{\vec{W_i}})\mid\vec{W_i}\in\mathcal{S}^{\circ}\}\bigcup\{\beta_{\mathcal{S}^{\circ}}\}$, the following Lemma \ref{lemma 1} is necessary.
\newtheorem{lemma}{Lemma}
\begin{lemma}
	\label{lemma 1}
	Given any two real numbers $x$ and $y$ satisfying $0<x<y$, we have: 
	\begin{enumerate}
		\item[(1).]{$\mathcal{H}_{\vec{a},b^*,\vec{W^{\circ}}}$ is $(x,y,P_{l_p,\vec{W^{\circ}}}(x),P_{l_p,\vec{W^{\circ}}}(y))$-sensitive.}
		\item[(2).]{$\mathcal{H}_{\vec{a},b^*,\vec{W^{\circ}}}^l$ is $(xl,yl,P_{l_p,\vec{W^{\circ}}}(x),P_{l_p,\vec{W^{\circ}}}(y))$-sensitive for $l\in\{c,c^2,\ldots,c^{\lceil\log_cr_{max}^{\vec{W^{\circ}}}/r_{min}^{\vec{W^{\circ}}}\rceil}\}$.} 
	\end{enumerate}
\end{lemma}

Let $\mathcal{H}_{\vec{a},b^*,\vec{W^{\circ}}\rightarrow\vec{W_i}}$ be the derived weighted LSH family with respect to $\mathcal{H}_{\vec{a},b^*,\vec{W^{\circ}}}$ and $\vec{W_i}$ for $\vec{W_i}\in\mathcal{S}^{\circ}$, and let $\mathcal{H}_{\vec{a},b^*,\vec{W^{\circ}}\rightarrow\vec{W_i}}^l$ be the derived weighted LSH family with respect to $\mathcal{H}_{\vec{a},b^*,\vec{W^{\circ}}}^l$ and $\vec{W_i}$ for $\vec{W_i}\in\mathcal{S}^{\circ}$. Based on Theorem \ref{R_bound}(1) and Lemma \ref{lemma 1}, we can obtain the following Theorem \ref{theorem 2}.
\begin{theorem}
	\label{theorem 2}
	Given any $\vec{W_i}=\{w_{i1},w_{i2},\ldots,w_{id}\}\in\mathcal{S}^{\circ}$ and any two real numbers $x$ and $y$ satisfying $0<x^{\uparrow}<y^{\downarrow}$, where $x^{\uparrow}=x\max_{1\leq j\leq d}(w^{\circ}_{j}/w_{ij})$ and $y^{\downarrow}=y\min_{1\leq j\leq d}(w^{\circ}_{j}/w_{ij})$, we have: 
	\begin{enumerate}
		\item[(1).]{$\mathcal{H}_{\vec{a},b^*,\vec{W^{\circ}}\rightarrow\vec{W_i}}$ is $(x,y,P_{l_p,\vec{W^{\circ}}}(x^{\uparrow}),P_{l_p,\vec{W^{\circ}}}(y^{\downarrow}))$-sensitive.}
		\item[(2).]{$\mathcal{H}_{\vec{a},b^*,\vec{W^{\circ}}\rightarrow\vec{W_i}}^l$ is $(xl,yl,P_{l_p,\vec{W^{\circ}}}(x^{\uparrow}),P_{l_p,\vec{W^{\circ}}}(y^{\downarrow}))$-sensitive for $l\in\{c,c^2,\ldots,c^{\lceil\log_cr_{max}^{\vec{W_i}}/r_{min}^{\vec{W_i}}\rceil}\}$.}
	\end{enumerate}
\end{theorem}

The proofs of Lemma \ref{lemma 1} and Theorem \ref{theorem 2} are shown in Appendices \ref{proof of lemma 1} and \ref{proof of theorem 2} respectively. To guarantee the correctness of Lemma \ref{lemma 1} and Theorem \ref{theorem 2}, the PDF of $b^*$ involved should be set to $p(b^*)=1/(c^{\lceil\log_cr_{max/min}^{\mathcal{S}^{\circ}}\rceil} w)$ for $b^*\in[0,c^{\lceil\log_cr_{max/min}^{\mathcal{S}^{\circ}}\rceil} w]$, where $r_{max/min}^{\mathcal{S}^{\circ}}$ is a real number greater than or equal to $\max_{\vec{W_i}\in\mathcal{S}^{\circ}}(r_{max}^{\vec{W_i}}/r_{min}^{\vec{W_i}})$. In the experiments in Section \ref{Experimental Evaluation}, $w$ is empirically set to $r_{min}^{\vec{W^{\circ}}}$ to achieve good performance. 

The $c$-WNN search under $\vec{W_i}\in\mathcal{S}^{\circ}$ can be performed by issuing a series of ($R,c$)-WNN search under $\vec{W_i}$ with increasing radii $R=r_{min}^{\vec{W_i}},cr_{min}^{\vec{W_i}},\ldots,c^{\lceil\log_cr_{max}^{\vec{W_i}}/r_{min}^{\vec{W_i}}\rceil}r_{min}^{\vec{W_i}}$.
Following C2LSH, a ($R,c$)-WNN of $q$ in $P$ under $\vec{W_i}\in\mathcal{S}^{\circ}$ can be found if the following two properties hold:
\begin{itemize}
	\item[$\mathcal{P}_1'$:] If $o\in B_{\vec{W_i}}(q,R)\cap P$, then $o$ is frequent, that is, the collision number of $o$ is at least $\mu_{\vec{W_i}}$.
	\item[$\mathcal{P}_2'$:] The total number of false positives which are frequent points but not ($R,c$)-WNNs is less than $\gamma n$.
\end{itemize}
Combining Theorem \ref{theorem 2} and Equations \ref{beta} and \ref{mu}, we can easily know that Pr$[\mathcal{P}_1'\text{ holds}]\geq 1-\epsilon$ and Pr$[\mathcal{P}_2'\text{ holds}]> 1/2$ if $\beta_{\vec{W_i}}$ and $\mu_{\vec{W_i}}$ are set to:
\begin{equation}
\label{c2lsh2}
\beta_{\vec{W_i}}=\lceil\frac{\ln \frac{1}{\epsilon}}{2(P_{l_p,\vec{W^{\circ}}}(x^{\uparrow})-P_{l_p,\vec{W^{\circ}}}(y^{\downarrow}))^2}(1+z)^2\rceil
\end{equation}
\begin{equation}
\label{mu2}
\mu_{\vec{W_i}}=\frac{zP_{l_p,\vec{W^{\circ}}}(x^{\uparrow})+P_{l_p,\vec{W^{\circ}}}(y^{\downarrow})}{1+z}\beta_{\vec{W_i}}
\end{equation}
where $z=\sqrt{\frac{\ln\frac{2}{\gamma}}{\ln\frac{1}{\epsilon}}}$, $x=r_{min}^{\vec{W_i}}$, $y=cr_{min}^{\vec{W_i}}$ and $\vec{W_i}\in\mathcal{S}^{\circ}$.
As in \cite{DBLP:conf/sigmod/GanFFN12}, we set $\epsilon=0.01$ and $\gamma=100/n$. Therefore, $\beta_{\vec{W_i}}=O(\ln n)$ for $\vec{W_i}\in\mathcal{S}^{\circ}$. Please note that the partition method introduced in Section \ref{challenge2} will guarantee that $x^{\uparrow}<y^{\downarrow}$ holds for $x=r_{min}^{\vec{W_i}}$, $y=cr_{min}^{\vec{W_i}}$ and $\vec{W_i}\in\mathcal{S}^{\circ}$ so that Theorem \ref{theorem 2} is available to derive Equations \ref{c2lsh2} and \ref{mu2}.

\begin{function}[t]
	\footnotesize
	\caption{ComputeParam($\mathcal{S}^{\circ},\vec{W^{\circ}},c$)}
	\ForEach{$\vec{W_i}\in\mathcal{S}^{\circ}$}
	{Compute $\beta_{\vec{W_i}}$ and $\mu_{\vec{W_i}}$ according to Equations \ref{c2lsh2} and \ref{mu2};}
	$\beta_{\mathcal{S}^{\circ}}\gets\max_{\vec{W_i}\in\mathcal{S}^{\circ}}(\beta_{\vec{W_i}})$;\\
	\Return{$\mathcal{B}_{\mathcal{S}^{\circ}}=\{(\beta_{\vec{W_i}},\mu_{\vec{W_i}})\mid\vec{W_i}\in\mathcal{S}^{\circ}\}\bigcup\{\beta_{\mathcal{S}^{\circ}}\}$};
\end{function}
\begin{function}[t]
	\footnotesize
	\caption{GenerateHF($\mathcal{B}_{\mathcal{S}^{\circ}},\vec{W^{\circ}}$)}
	\For{$i=1\cdots \beta_{\mathcal{S}^{\circ}}$}
	{Choose a function $h_i^{\mathcal{S}^{\circ}}$ uniformly at random from $\mathcal{H}_{\vec{a},b^*,\vec{W^{\circ}}}$;}
	\Return{$H_{\mathcal{S}^{\circ}}=\{h_i^{\mathcal{S}^{\circ}}\mid1\leq i\leq \beta_{\mathcal{S}^{\circ}}\}$};
\end{function}
\begin{function}[t]
	\footnotesize
	\caption{CreateHT($P,\mathcal{B}_{\mathcal{S}^{\circ}},H_{\mathcal{S}^{\circ}}$)}
	\For{$i=1\cdots \beta_{\mathcal{S}^{\circ}}$}
	{Create a hash table $T_i^{\mathcal{S}^{\circ}}$ according to $h_i^{\mathcal{S}^{\circ}}$;}
	\Return{$T_{\mathcal{S}^{\circ}}=\{T_i^{\mathcal{S}^{\circ}}\mid1\leq i\leq\beta_{\mathcal{S}^{\circ}}\}$};
\end{function}
\begin{function}[t]
	\footnotesize
	\caption{SearchHT($P,q,\vec{W_i},k,\mathcal{B}_{\mathcal{S}^{\circ}},H_{\mathcal{S}^{\circ}},T_{\mathcal{S}^{\circ}}$)}
	\tcc{$h_i^{\mathcal{S}^{\circ},l}(q)$ denotes the level-$l$ bucket with respect to $h_i^{\mathcal{S}^{\circ}}$ and $q$.}
	Obtain $\beta_{\vec{W_i}}$ and $\mu_{\vec{W_i}}$ from $\mathcal{B}_{\mathcal{S}^{\circ}}$;\\
	$R\gets r_{min}^{\vec{W_i}}$, $C\gets\emptyset$, $l\gets1$;\\
	\While{$TRUE$}
	{\For{$i= 1\cdots \beta_{\vec{W_i}}$}
		{Count collision for points in bucket $h_i^{\mathcal{S}^{\circ},l}(q)$ of hash table $T_i^{\mathcal{S}^{\circ}}$ and add frequent points to $C$;\\
			\If{$|\{o\mid o\in C\land D_{\vec{W_i}}(o,q)\leq c\times R\}|\geq k$}{\Return{exact $k$-nearest neighbors of $q$ in $C$ under $\vec{W_i}$};}
			\If{$|C|\geq k+\gamma n$}
			{\Return{exact $k$-nearest neighbors of $q$ in $C$ under $\vec{W_i}$};}}
		$R\gets c\times R$, $l\gets c\times l$;}
\end{function}

Now we give the preprocessing algorithm and search algorithm for the ($c,k$)-WNN search under $\mathcal{S}^{\circ}$.

The preprocessing algorithm is straightforward and as follows. 
First, Function \textit{ComputeParam()} is called to compute the parameters $\mathcal{B}_{\mathcal{S}^{\circ}}=\{(\beta_{\vec{W_i}},\mu_{\vec{W_i}})\mid\vec{W_i}\in\mathcal{S}^{\circ}\}\bigcup\{\beta_{\mathcal{S}^{\circ}}\}$. Then, Function \textit{GenerateHF()} is called to obtain $\beta_{\mathcal{S}^{\circ}}$ weighted LSH functions from $\mathcal{H}_{\vec{a},b^*,\vec{W^{\circ}}}$. Finally, Function \textit{CreateHT()} is called to create $\beta_{\mathcal{S}^{\circ}}$ hash tables according to the weighted LSH functions. 

The search algorithm is shown in Function \textit{SearchHT()}.
To find a ($c,k$)-WNN of $q$ in $P$ under $\vec{W_i}\in\mathcal{S}^{\circ}$, $\beta_{\vec{W_i}}$ and $\mu_{\vec{W_i}}$ are retrieved from $\mathcal{B}_{\mathcal{S}^{\circ}}$ first (Line 1). Then, the search radius starts from $r_{min}^{\vec{W_i}}$ (Line 2) and increases
by a factor of $c$ for each iteration of the while loop (Line 10).
Frequent points are identified by doing collision counting (Line 5).
The search algorithm stops after one of the following conditions is satisfied: 
(1) At least $k$ ($R,c$)-WNNs under $\vec{W_i}$ have been found (Lines 6-7);
(2) At least $k+\gamma n$ frequent points have been checked (Lines 8-9).
The two conditions are respectively supported by properties $\mathcal{P}_1'$ and $\mathcal{P}_2'$.
\subsection{\boldmath ($c,k$)-WNN Search under $\mathcal{S}$}
\label{challenge2}
We introduce the partition method first.
Suppose that $\mathcal{S}$ will be partitioned into the set of disjoint subsets $\widetilde{\mathcal{S}}$
and the ($c,k$)-WNN search under $\mathcal{S}_i\in\widetilde{\mathcal{S}}$ will be supported using a group of hash tables created according to weighted LSH functions from $\mathcal{H}_{\vec{a},b^*,\vec{W_{\mathcal{S}_i}}}$, where $\vec{W_{\mathcal{S}_i}}$ is some weight vector in $\mathcal{S}$.
Let $\mathcal{S^*}=\{\vec{W_{\mathcal{S}_i}}\mid\mathcal{S}_i\in\widetilde{\mathcal{S}}\}$.
From Section \ref{challenge1}, we need to determine the parameters $\mathcal{B}_{\mathcal{S}_i}=\{(\beta_{\vec{W_j}},\mu_{\vec{W_j}})\mid\vec{W_j}\in\mathcal{S}_i\}\bigcup\{\beta_{\mathcal{S}_i}\}$ for $\mathcal{S}_i\in\widetilde{\mathcal{S}}$. Let $\mathcal{B}_{\mathcal{S}}=\{\mathcal{B}_{\mathcal{S}_i}\mid\mathcal{S}_i\in\widetilde{\mathcal{S}}\}$. The total number of hash tables required to be created is $\beta_{\mathcal{S}}=\sum_{\mathcal{S}_i\in\widetilde{\mathcal{S}}}\beta_{\mathcal{S}_i}$. 
In order to efficiently perform the ($c,k$)-WNN search under $\mathcal{S}$, we use a pre-specified threshold $\tau$ to limit the required number of hash tables for each weight vector in $\mathcal{S}$.
Moreover, it is desired to minimize $\beta_{\mathcal{S}}$. As a result, the problem of partitioning $\mathcal{S}$ is formulated as the below optimization problem:
\begin{equation}
\begin{aligned}
\label{setcover}
&\min\ \ \beta_{\mathcal{S}}=\sum_{\mathcal{S}_i\in\widetilde{\mathcal{S}}}\beta_{\mathcal{S}_i}\\
&\begin{array}{lll}
\text{s.t.}&\bigcup_{\mathcal{S}_i\in\widetilde{\mathcal{S}}}\mathcal{S}_i=\mathcal{S}&\\
&\mathcal{S}_i\bigcap\mathcal{S}_j=\emptyset,&\mathcal{S}_i,\mathcal{S}_j\in\widetilde{\mathcal{S}}\land i\neq j\\
&\beta_{\mathcal{S}_i}\leq\tau,&\mathcal{S}_i\in\widetilde{\mathcal{S}}\\
\end{array}
\end{aligned}
\end{equation}

\begin{function}[t]
	\footnotesize
	\caption{Partition($\mathcal{S},\tau,c$)}
	$Y\gets\emptyset$;\\
	\ForEach{$\vec{W_i}\in\mathcal{S}$}{
	$A_{\vec{W_i}}\gets\{\beta_{\vec{W_k}}\mid(\vec{W_k}\in\mathcal{S})\land ((r_{min}^{\vec{W_k}})^{\uparrow}<(cr_{min}^{\vec{W_k}})^{\downarrow})\}$;\\
	\If{$A_{\vec{W_i}}\neq\emptyset$}{
	\For{$j=1\cdots\left|A_{\vec{W_i}}\right|$}{ $X_{\vec{W_i}}^j\gets\{\text{top-}j\text{ smallest elements in }A_{\vec{W_i}}\}$;\\ $w_{i,j}\gets\max(X_{\vec{W_i}}^j)$;\\\If{$w_{i,j}\leq\tau$}{ $B_{\vec{W_i}}^j\gets\{\vec{W_k}\mid\beta_{\vec{W_k}}\in X_{\vec{W_i}}^j\}$;\\Add $(B_{\vec{W_i}}^j,w_{i,j})$ to $Y$;}}}}
	$\{\widetilde{Y},\mathcal{S^*}\}\gets$\text{GenerateWSC($Y$)};\\
	$\{\widetilde{\mathcal{S}},\mathcal{B}_{\mathcal{S}}\}\gets$Process($\widetilde{Y},\mathcal{S^*}$);\\
	\Return{$\{\mathcal{S^*},\widetilde{\mathcal{S}},\mathcal{B}_{\mathcal{S}}\}$};
\end{function}

The threshold $\tau$ should be set to ensure the existence of a solution of the above optimization problem.
Consider an extreme case: the ($c,k$)-WNN search under each weight vector in $\mathcal{S}$ is supported using an individual group of hash tables.
Denote $\max_{\vec{W_i}\in\mathcal{S}}(\beta_{\vec{W_i}})$ in this case by $\tau_{min}$.
Then $\tau$ can be set to a value no less than $\tau_{min}$.

We determine
$\mathcal{S^*}$, $\widetilde{\mathcal{S}}$ and $\mathcal{B}_{\mathcal{S}}$ by solving the above optimization problem. The steps are shown in Function \textit{Partition()} and explained as follows.

\textbf{Step 1} (Lines 1-9).
For each $\vec{W_i}\in\mathcal{S}$, suppose $\mathcal{H}_{\vec{a},b^*,\vec{W_i}}$ will be used for creating hash tables and then construct all subsets of $\mathcal{S}$ that satisfy the two conditions: (1) The required number of hash tables for the ($c,k$)-WNN search under each of the subsets is less than or equal to $\tau$; (2) No more weight vectors in $\mathcal{S}$ can be added to any of the subsets without increasing the required number of hash tables for the ($c,k$)-WNN search under it.

\textbf{Step 2} (Lines 10-11).
Take each of the subsets of $\mathcal{S}$ obtained in Step 1 as a weighted set with the weight being the required number of hash tables for the ($c,k$)-WNN search under it. 
Further, take all these weighted sets, denoted by $Y$, as the input of the weighted set cover problem \cite{DBLP:conf/icde/AjamiC19}.
The output will be a weighted set cover $\widetilde{Y}\subseteq Y$ that covers 
$\mathcal{S}$ and has a minimal sum of weights. Actually, the reason why we use Condition (2) in Step 1 is that it can dramatically reduce the input size while not increasing the sum of weights of the output weighted set cover. As it can be seen, the minimum required total number of hash tables for the ($c,k$)-WNN search under $\mathcal{S}$ is equal to the sum of weights of $\widetilde{Y}$. 
Once $\widetilde{Y}$ is produced, $\mathcal{S^*}$ can be determined immediately.

The decision version of the weighted set cover problem is NP-complete \cite{DBLP:conf/icde/AjamiC19}. Thus, a number of algorithms have been proposed to return non-optimal weighted set covers \cite{DBLP:journals/ior/CapraraFT99,DBLP:journals/mor/Chvatal79,DBLP:journals/eor/LanDW07}. In the experiments in Section \ref{Experimental Evaluation}, we will adopt the approximate algorithm in \cite{DBLP:journals/mor/Chvatal79}, which can provide an approximation ratio of $O(\ln \left|\mathcal{S}\right|)$.

\textbf{Step 3} (Line 12).
A weight vector may be contained in multiple weighted sets of $\widetilde{Y}$.
Thus, we obtain $\widetilde{\mathcal{S}}$ by deduplicating $\widetilde{Y}$.
Given $\mathcal{S^*}$ and $\widetilde{\mathcal{S}}$, $\mathcal{B}_{\mathcal{S}}$ is determined in the same way as in Section \ref{challenge1}.

\begin{algorithm}[t]
	\caption{$Preprocess(\mathcal{S},P,\tau,c)$}
	\footnotesize
	\label{prerpocess}
	\tcc{ $\mathcal{S^*}=\{\vec{W_{\mathcal{S}_i}}\mid\mathcal{S}_i\in\widetilde{\mathcal{S}}\}$, $\mathcal{B}_{\mathcal{S}}=\{\mathcal{B}_{\mathcal{S}_i}\mid\mathcal{S}_i\in\widetilde{\mathcal{S}}\}$, $H_{\mathcal{S}}=\{H_{\mathcal{S}_i}\mid\mathcal{S}_i\in\widetilde{\mathcal{S}}\}$, $T_{\mathcal{S}}=\{T_{\mathcal{S}_i}\mid\mathcal{S}_i\in\widetilde{\mathcal{S}}\}$.}
	$\{\mathcal{S^*},\widetilde{\mathcal{S}},\mathcal{B}_{\mathcal{S}}\}\gets$Partition($\mathcal{S},\tau,c$);\\
	\ForEach{$\mathcal{S}_i\in\widetilde{\mathcal{S}}$}{
		$H_{\mathcal{S}_i}\gets$GenerateHF($\mathcal{B}_{\mathcal{S}_i},\vec{W_{\mathcal{S}_i}}$);\\
		$T_{\mathcal{S}_i}\gets$CreateHT($P,\mathcal{B}_{\mathcal{S}_i},H_{\mathcal{S}_i}$);}
	\Return{$\{\widetilde{\mathcal{S}},\mathcal{B}_{\mathcal{S}},H_{\mathcal{S}},T_{\mathcal{S}}\}$};
\end{algorithm}
\begin{algorithm}[t]
	\footnotesize
	\caption{$Search(P,q,\vec{W_i},k,\widetilde{\mathcal{S}},\mathcal{B}_{{\mathcal{S}}},H_{\mathcal{S}},T_{\mathcal{S}})$}
	\label{search}
	Find $\mathcal{S}_j\in\widetilde{\mathcal{S}}$ where $\vec{W_i}$ is contained;\\
	Locate the positions of $\mathcal{B}_{{\mathcal{S}_j}}\in\mathcal{B}_{{\mathcal{S}}}$, $H_{\mathcal{S}_j}\in H_{\mathcal{S}}$ and $T_{\mathcal{S}_j}\in T_{\mathcal{S}}$;\\
	$kANNs\gets$SearchHT($P,q,\vec{W_i},k,\mathcal{B}_{{\mathcal{S}_j}},H_{\mathcal{S}_j},T_{\mathcal{S}_j}$);\\
	\Return{$kANNs$};
\end{algorithm}

Formally, the preprocessing algorithm for the ($c,k$)-WNN search under $\mathcal{S}$ is presented in Algorithm \ref{prerpocess}. 
First, the algorithm determines $\mathcal{S^*}$, $\widetilde{\mathcal{S}}$ and $\mathcal{B}_{\mathcal{S}}$ by calling Function \textit{Partition()} (Line 1). Then, for each $\mathcal{S}_i\in\widetilde{\mathcal{S}}$, the algorithm generates weighted LSH functions and creates hash tables for the ($c,k$)-WNN search under $\mathcal{S}_i$ by calling Functions \textit{GenerateHF()} and \textit{CreateHT()}, respectively (Lines 2-4).

Algorithm \ref{search} gives the search algorithm for the ($c,k$)-WNN search under $\mathcal{S}$. 
Given a query point $q\in\mathcal{X}$ and a query weight vector $\vec{W_i}\in\mathcal{S}$,
the algorithm first obtains the necessary index information (Lines 1-2) and then retrieves a ($c,k$)-WNN of $q$ in $P$ under $\vec{W_i}$ by calling Function \textit{SearchHT()} (Line 3).

\textbf{Complexity Analysis.} 
Since the threshold $\tau$ can be $O(\log n)$,
the storage cost for a group of hash tables can be $O(n\log n)$. As a result,
the total storage cost for all the hash tables is $O(\left|\widetilde{\mathcal{S}}\right|n\log n)$. 
Besides, the storage cost for the data set $P$ is $O(nd)$, and the storage cost for $\widetilde{\mathcal{S}}$ is $O(\left|\mathcal{S}\right|d)$. The storage costs for $\mathcal{B}_{\mathcal{S}}$ and $H_{\mathcal{S}}$ are negligible.
Therefore, the space complexity of WLSH is $O(\left|\widetilde{\mathcal{S}}\right|n\log n+nd+\left|\mathcal{S}\right|d)$.
Obviously, $\left|\widetilde{\mathcal{S}}\right|$ is no greater than $\left|\mathcal{S}\right|$, and it is dependent on both the weight vector set $\mathcal{S}$ and the approximate (or heuristic) weighted set cover algorithm adopted.

It is easy to know that Lines 1-2 of Algorithm \ref{search} and Line 1 of Function \textit{SearchHT()} can be processed in $O(d)$ time by using a trie-like structure. This will not increase the space complexity of WLSH as the storage cost for the trie-like structure is $O(\left|\mathcal{S}\right|d)$.
The processing time for 
Lines 2-10 of Function \textit{SearchHT()}
consists of three parts: the time of locating the buckets to read, the time of collision counting, and the time of checking candidates. They are respectively $O(d\log n)$, $O(n\log n)$ and $O((k+\gamma n)d)=O(d)$. As a result, the time complexity of WLSH is $O(d)+O(d\log n)+O(n\log n)+O(d)=O(n\log n+d\log n)$.
\subsubsection{Trade-offs}
\label{trade-offs}
For better practical performance, we need to balance the query efficiency, query accuracy and space consumption of WLSH.
Usually, it is expensive to provide a worst-case guarantee on the query accuracy. Thus, we can optionally improve the query efficiency and reduce the space consumption at the sacrifice of the query accuracy guarantee.

Following the idea, we introduce two approaches for optimizing WLSH: \textit{bound relaxation} and \textit{collision threshold reduction}. The goal of the bound relaxation is to reduce the space consumption, while the goal of the collision threshold reduction is to improve the query efficiency. The two approaches can be applied together.

\textbf{Bound Relaxation.} WLSH is based on Theorem \ref{R_bound}(1). As we can see, the bounds $R^{\uparrow}$ and $(cR)^{\downarrow}$ given by Theorem \ref{R_bound}(1) are independent of $P$ and $q$. This implies that they are probably too strict in practice.
We can relax them as follows:
\begin{equation}
R^{\uparrow}=RT^{(v)}
\end{equation}
\begin{equation}
(cR)^{\downarrow}=cRT^{(d+1-v')}
\end{equation}
where $T^{(v)}$ is the $v$th largest number in $T$, $T^{(d+1-v')}$ is the $v'$th smallest number in $T$, $T=\{w_i/w_i'\mid1\leq i\leq d\}$, and $1\leq v\leq d+1-v'\leq d$. 
Obviously, $R^{\uparrow}$ and $(cR)^{\downarrow}$ above are the same as in Theorem \ref{R_bound}(1) when $v=v'=1$. 
If $v$ and $v'$ are set to be greater than 1, then the required number of hash tables for each weight vector tends to be reduced according to Equation \ref{c2lsh2}. As a result, the required total number of hash tables is likely to be reduced.

\textbf{Collision Threshold Reduction.} 
We can improve the query efficiency of WLSH by using a smaller collision threshold.
The intuition is that, the data points that are close to $q$ can be identified as frequent points earlier with a smaller collision threshold.
Actually, C2LSH has already adopted the collision threshold reduction approach to improve its query efficiency \cite{DBLP:conf/sigmod/GanFFN12}.
In the experiments in Section \ref{Experimental Evaluation}, we will use the collision threshold $\widetilde{\mu_{\vec{W_i}}}=X\mu_{\vec{W_i}}$ for the ($c,k$)-WNN search under $\vec{W_i}\in\mathcal{S}$ when collision threshold reduction is applied, where $X=P_{l_p,\vec{W_i}}((c^2r_{min}^{\vec{W_i}})^{\uparrow})/P_{l_p,\vec{W_i}}((r_{min}^{\vec{W_i}})^{\uparrow})<1$ and $\mu_{\vec{W_i}}$ is given by Equation \ref{mu2}. 
The collision threshold $\widetilde{\mu_{\vec{W_i}}}$ is simply extended from the corresponding one in C2LSH.
\section{Experiments}
\label{Experimental Evaluation}
The experiments are targeted to answer four questions:
\begin{enumerate}
	\item[(1).]{How does the space consumption of WLSH vary with various parameters? (Section \ref{space/various paramenters})}
	\item[(2).]{How does the space consumption of WLSH compare with those of SL-ALSH and S2-ALSH? (Section \ref{space/comparison})}
	\item[(3).]{How do the query efficiency and query accuracy of WLSH vary with various parameters? (Section \ref{efficiency/various parameters})}
	\item[(4).]{How do the query efficiency and query accuracy of WLSH compare with those of SL-ALSH and S2-ALSH? (Section \ref{efficiency/comparison})}
\end{enumerate}

We focus on the $l_1$ distance and $l_2$ distance while evaluating the performance of WLSH due to space limitations. Moreover, we compare WLSH with SL-ALSH and S2-ALSH only for the $l_2$ distance since SL-ALSH and S2-ALSH are both designed for the $l_2$ distance.
\subsection{Experimental Setup}
\subsubsection{Data Sets, Weight Vector Sets and Query Sets}
The data sets, weight vector sets and query sets used in the experiments are introduced as follows.

\begin{table}[!t]
	\centering
	\footnotesize
	\caption{Parameter Settings for Synthetic Data Sets}
	\label{synthetic data sets}
	\begin{tabular}{c|c} 
		\hline
		\textbf{Dimensionality \boldmath$d$}&100, 200, \underline{400}, 800, 1.6k\\ \hline
		\textbf{Cardinality \boldmath$n$}&100k, 200k, \underline{400k}, 800k, 1.6m\\ \hline
		\textbf{Value Range}&[0, 10,000]\\ \hline
	\end{tabular}
\end{table}
\begin{table}[t]
	\centering
	\footnotesize
	\caption{Statistics of Real Data Sets}
	\label{real data sets}
	\begin{tabular}{c|c|c|c} 
		\hline
		\textbf{Data Set}&\textbf{Dimensionality \boldmath$d$}&\textbf{Cardinality \boldmath$n$}&\textbf{Value Range}\\ \hline
		\textit{Sift}&128&994,411&[0, 255]\\ \hline
		\textit{Ukbench}&128&1,097,857&[0, 255]\\ \hline
		\textit{Notre}&128&332,618&[0, 255]\\ \hline
		\textit{Sun}&512&79,056&[0, 10,000]\\ \hline
	\end{tabular}
\end{table}
\begin{table}[t]
	\centering
	\footnotesize
	\caption{Parameter Settings for Weight Vector Sets}
	\label{weight vector sets}
	\begin{tabular}{c|c} 
		\hline
		\textbf{Cardinality \boldmath$\left|\mathcal{S}\right|$}&1k, 3k, \underline{5k}, 7k, 9k\\\hline
		\boldmath$\#Subset$&50, 100, \underline{200}, 500, 1k\\\hline
		\boldmath$\#Subrange$&5, 10, \underline{20}, 50, 100\\\hline
	\end{tabular}
\end{table}
\textbf{Data Sets.} 
We use both synthetic and real data sets.
The parameter settings for the synthetic data sets are summarized in Table \ref{synthetic data sets}, where the default dimensionality and cardinality are underlined. 
For each synthetic data set, the values in each dimension are integers uniformly distributed in the range [0, 10,000].
The real data sets\footnote{https://github.com/DBWangGroupUNSW/nns\_benchmark} used are \textit{Sift}, \textit{Ukbench}, 
\textit{Notre} and \textit{Sun}.
The statistics of them are summarized in Table \ref{real data sets}.
For each real data set, the values in each dimension are normalized to be integers 
in a given range shown in Table \ref{real data sets}. More details about them are available in \cite{DBLP:journals/corr/LiZSWZL16}.

\textbf{Weight Vector Sets.} 
The data sets with the same dimensionality share the same collection of weight vector sets. 
For $d$-dimensional data sets, the weight vector sets used are identified by weight vector set cardinality $\left|\mathcal{S}\right|$ and the other two parameters $\#Subset$ and $\#Subrange$. Given $\left|\mathcal{S}\right|$, $\#Subset$ and $\#Subrange$, the corresponding weight vector set is the union of $\#Subset$ equal-size subsets, each of which is generated as follows.
First, the range $\text{[1, 10]}$ is divided into $\#Subrange$ equal-width subranges, denoted by $X$. 
Then, a set of $d$ subranges, denoted by $Y=\{Y_i\mid1\leq i\leq d\}$, are chosen uniformly at random from $X$.
Finally, the values in the $i$th dimension are chosen uniformly at random from the subrange $Y_i$ for $1\leq i\leq d$.
Obviously, a number of different weight vector sets can be obtained by varying $\left|\mathcal{S}\right|$, $\#Subset$ and $\#Subrange$. For example, a set of uniformly random weight vectors from $\text{[1, 10]}^d$ can be obtained by setting $\#Subset=\left|\mathcal{S}\right|$ and $\#Subrange=1$.

Besides the uniformly random weight vector sets, 
we also use some other weight vector sets whose parameter settings are shown in Table \ref{weight vector sets}, where the default values of $\left|\mathcal{S}\right|$, $\#Subset$ and $\#Subrange$ are underlined.

\textbf{Query Sets.} Given a data set and a weight vector set, a query set is obtained in the following way. First, 50 points are chosen uniformly at random from the data set as a query point set (these points are removed from the data set at the same time).
Then, 10 weight vectors are chosen uniformly at random from the weight vector set as a query weight vector set. 
Finally, the Cartesian product of the query point set and the query weight vector set is taken as a query set.
\subsubsection{Evaluation Metrics}
Following previous work \cite{DBLP:conf/sigmod/GanFFN12,DBLP:conf/sigmod/ZhengGTW16}, 
three metrics are adopted in our evaluations.

\textbf{Space Consumption.}
The space consumption is measured by the required total number of hash tables.

\textbf{Query Efficiency.}
Since the ($c,k$)-WNN search is I/O intensive,
the query efficiency is measured by the I/O cost averaged over a query set.
Specifically, the I/O cost consists of the cost for identifying candidates and the cost for checking candidates.

\textbf{Query Accuracy.}
As in \cite{DBLP:journals/tods/TaoYSK10,DBLP:conf/sigmod/GanFFN12,DBLP:conf/sigmod/ZhengGTW16}, the query accuracy is measured by the \textit{average overall ratio},
which is the mean of the \textit{overall ratios} over a query set. For a query whose query point and query weight vector are $q$ and $\vec{W}$ respectively, 
the overall ratio is computed by:
\begin{equation}
\label{overall ratio}
R(q,\vec{W})=\frac{1}{k}\sum_{i=1}^{k}\frac{D_{\vec{W}}(o_i,q)}{D_{\vec{W}}(o_i^*,q)}
\end{equation}
where $o_1,o_2,\ldots,o_k$ are 
the reported approximate $k$-nearest neighbors sorted in non-decreasing order of their distances to $q$,
and $o_1^*,o_2^*,\ldots,o_k^*$ are the exact $k$-nearest neighbors sorted in the same way.
\subsubsection{Settings of Other Parameters}
The block size is set to 4KB.
The result cardinality $k$ is in the range \{10, 100\}. 
We set the threshold $\tau$ to 1,000 for the $l_1$ distance and 500 for the $l_2$ distance. 
Unless otherwise specified, the approximation ratio $c$ is in the range $\{2, 3, 4, 5, 6\}$, and its default value is 3.
For a $d$-dimensional weight vector set, we set the parameters $v=v'=d/4$ when bound relaxation is applied.
\subsection{Space Consumption of WLSH}
\subsubsection{Impacts of Various Parameters}
\label{space/various paramenters}
\begin{table*}[htbp]
	\centering
	\footnotesize
	\caption{Space Consumption of WLSH}
	\resizebox{\textwidth}{!}{
	\begin{threeparttable}
	\begin{tabular}{c|c|c|c|c|c|c|c|c|c|c|c}
		\hline
		\multicolumn{6}{c|}{$l_1$ distance}&\multicolumn{6}{c}{$l_2$ distance}\\ \hline\hline
		\boldmath$d$&100 & 200 & 400 & 800 & 1.6k&\boldmath$d$&100 & 200 & 400 & 800 & 1.6k\\ \hline
		\boldmath$\beta_{\mathcal{S}}$&453,613&1,224,539&2,097,571&\textbf{2,160,000}&\textbf{2,160,000}&\boldmath$\beta_{\mathcal{S}}$&336,180&872,652&1,174,561&\textbf{1,180,000}&\textbf{1,180,000}\\ \hline
		\boldmath$\beta_{\mathcal{S}}^{br}$&106,863&102,799&98,310&94,243&94,344&\boldmath$\beta_{\mathcal{S}}^{br}$&59,829&57,892&54,158&51,537&51,641\\ \hline
		\hline
		\boldmath$n$&100k & 200k & 400k & 800k & 1.6m&\boldmath$n$&100k & 200k & 400k & 800k & 1.6m\\ \hline
		\boldmath$\beta_{\mathcal{S}}$&1,760,436&1,943,748&2,097,571&2,228,029&2,339,999&\boldmath$\beta_{\mathcal{S}}$&1,041,430&1,115,838&1,174,561&1,228,088&1,284,165\\  \hline
		\boldmath$\beta_{\mathcal{S}}^{br}$&89,314&93,846&98,310&102,686&106,546&\boldmath$\beta_{\mathcal{S}}^{br}$&49,346&52,972&54,158&57,197&59,363\\ \hline
		\hline
		\boldmath$c$&2 & 3 & 4 & 5& 6&\boldmath$c$&2 & 3 & 4 & 5& 6\\ \hline
		\boldmath$\beta_{\mathcal{S}}$&\textbf{4,150,000}&2,097,571&251,398&158,190&128,783&\boldmath$\beta_{\mathcal{S}}$&\textbf{2,205,000}&1,174,561&201,161&90,272&74,172\\ \hline
		\boldmath$\beta_{\mathcal{S}}^{br}$&197,636&98,310&72,993&62,101&56,789&\boldmath$\beta_{\mathcal{S}}^{br}$&237,194&54,158&42,092&35,675&31,685\\ \hline
		\hline
		\boldmath$\#Subrange$&5 & 10 & 20 & 50& 100&\boldmath$\#Subrange$&5 & 10 & 20 & 50& 100\\ \hline
		\boldmath$\beta_{\mathcal{S}}$&\textbf{2,160,000}&\textbf{2,160,000}&2,097,571&142,255&108,178&\boldmath$\beta_{\mathcal{S}}$&\textbf{1,180,000}&\textbf{1,180,000}&1,174,561&77,997&60,113\\ \hline
		\boldmath$\beta_{\mathcal{S}}^{br}$&136,804&108,279&98,310&88,273&87,174&\boldmath$\beta_{\mathcal{S}}^{br}$&76,460&59,855&54,158&48,471&47,666\\ \hline
		\hline
		\boldmath$\#Subset$&50 & 100 & 200 & 500& 1k&\boldmath$\#Subset$&50 & 100 & 200 & 500& 1k\\ \hline
		\boldmath$\beta_{\mathcal{S}}$&1,946,151&2,036,838&2,097,571&2,146,358&2,156,237&\boldmath$\beta_{\mathcal{S}}$&1,159,154&1,164,454&1,174,561&1,177,607&1,179,613\\ \hline
		\boldmath$\beta_{\mathcal{S}}^{br}$&31,728&52,781&98,310&231,124&462,875&\boldmath$\beta_{\mathcal{S}}^{br}$&16,613&29,023&54,158&126,052&252,494\\ \hline
		\hline
		\boldmath$\left|\mathcal{S}\right|$&1k & 3k & 5k & 7k& 9k&\boldmath$\left|\mathcal{S}\right|$&1k & 3k & 5k & 7k& 9k\\ \hline
		\boldmath$\beta_{\mathcal{S}}$&430,984&1,286,204&2,097,571&2,940,547&3,741,455&\boldmath$\beta_{\mathcal{S}}$&235,980&706,746&1,174,561&1,645,914&2,111,501\\ \hline
		\boldmath$\beta_{\mathcal{S}}^{br}$&92,361&92,988&98,310&100,863&106,095&\boldmath$\beta_{\mathcal{S}}^{br}$&50,438&50,798&54,158&56,479&59,601\\ \hline
	\end{tabular}
	\begin{tablenotes}
		\item[*] $\beta_{\mathcal{S}}$ is the required total number of hash tables when not using bound relaxation. $\beta_{\mathcal{S}}^{br}$ is the required total number of hash tables when using bound relaxation.
	\end{tablenotes}
	\end{threeparttable}}
	\label{Space WLSH/l1 l2}
\end{table*}

As the space consumption of WLSH is independent of data distribution, we use only synthetic data sets to study the impacts of various parameters on it.
The parameters involved are $d$, $n$, $c$, $\#Subrange$, $\#Subset$ and $\left|\mathcal{S}\right|$. 
We evaluate the impact of each parameter by varying one parameter and setting the other five parameters to the default values. 
The experimental results are presented in Table \ref{Space WLSH/l1 l2}, and the followings are brief explanations of them.

\textbf{Impacts of \boldmath$d$, \boldmath$n$ and \boldmath$c$.}
The required number of hash tables for each weight vector is directly affected by $x^{\uparrow}$ and $y^{\downarrow}$ rather than $d$ from Equation \ref{c2lsh2}. 
However, we observe that, due to our weight vector set generation method, $x^{\uparrow}$ tends to increase with $d$ and $y^{\downarrow}$ tends to decrease with $d$ when bound relaxation is not used, while $x^{\uparrow}$ tends to decrease with $d$ and $y^{\downarrow}$ tends to increase with $d$ when bound relaxation is used.
Therefore, according to Equation \ref{c2lsh2}, the required number of hash tables for each weight vector tends to increase with $d$ when bound relaxation is not used and decrease with $d$ when bound relaxation is used. The required total number of hash tables has the same trend as the required number of hash tables for each weight vector.

The required number of hash tables for each weight vector tends to increase with $n$ and decrease with $c$ according to Equation \ref{c2lsh2}.
Thus, the required total number of hash tables tends to increase with $n$ and decrease with $c$.

\textbf{Impacts of \boldmath$\#Subrange$, \boldmath$\#Subset$ and \boldmath$\left|\mathcal{S}\right|$.}
According to our weight vector set generation method,
a weight vector set is the union of $\#Subset$ equal-size subsets and each of the subsets contains similar weight vectors. As $\#Subrange$ increases, the weight vectors in the same subset become more similar to each other and tend to share more hash tables so as to minimize the required total number of hash tables. Thus, the required total number of hash tables tends to decrease with $\#Subrange$. 
Since the weight vectors in different subsets are quite dissimilar, they can hardly share hash tables. Consequently, the required total number of hash tables tends to increase with $\#Subset$.

The impact of $\left|\mathcal{S}\right|$ depends on $\#Subrange$.
When $\#Subrange$ is sufficiently large, the weight vectors in the same subset can share the same group of hash tables, no matter how big $\left|\mathcal{S}\right|$ is.
In this case, the required total number of hash tables tends to remain relatively stable with the increase of $\left|\mathcal{S}\right|$.
However, when $\#Subrange$ is not so large (e.g., $\#Subrange$ is set to the default value), the weight vectors in the same subset
can not share the same group of hash tables. 
In this case, the required total number of hash tables tends to increase with $\left|\mathcal{S}\right|$.

Additionally, we can draw the following conclusions from Table \ref{Space WLSH/l1 l2}. 
First, the space consumption of WLSH can be significantly reduced by using bound relaxation. 
Actually, without using bound relaxation, WLSH can even degenerate into the naive method mentioned in Section \ref{Problem Description} in a small fraction of cases (marked in bold in Table \ref{Space WLSH/l1 l2}). 
Second, the space consumption of WLSH for the $l_1$ distance is higher than that for the $l_2$ distance in almost all cases. The fundamental reason is that, the required number of hash tables for each weight vector tends to be larger for the $l_1$ distance than for the $l_2$ distance according to Equation \ref{c2lsh2}.

Though the space consumption of WLSH can also be reduced by increasing $c$ from Table \ref{Space WLSH/l1 l2}, 
we experimentally demonstrate the necessity of introducing bound relaxation to WLSH in Appendix \ref{necessity of bound relaxation}.
\subsubsection{Comparison with SL-ALSH and S2-ALSH}
\label{space/comparison}
\begin{table}[t]
	\centering
	\footnotesize
	\caption{Space Consumptions of SL-ALSH and S2-ALSH ($R=1000$)}
	\resizebox{0.485\textwidth}{!}{
	\begin{threeparttable}
	\begin{tabular}{c|c|c|c|c|c}
		\hline
		\multicolumn{6}{c}{$l_2$ distance}\\
		\hline
		\hline
		\boldmath$d$&100 & 200 & 400 & 800 & 1.6k\\ \hline
		\boldmath{$L_{SL}$}&333,626&328,715&322,001&320,141&317,238\\ \hline
		\boldmath{$L_{S2}$}&330,158&325,574&319,349&317,624&314,963\\ \hline
		\hline
		\boldmath$n$&100k & 200k & 400k & 800k & 1.6m\\ \hline
		\boldmath{$L_{SL}$}&82,405&162,881&322,001&636,530&1,258,699\\ \hline
		\boldmath{$L_{S2}$}&81,793&161,618&319,349&631,015&1,246,850\\ \hline
		\hline
		\boldmath$c$&2 & 3 & 4 & 5& 6\\ \hline
		\boldmath{$L_{SL}$}&387,551&322,001&192,819&76,763&20,945\\ \hline
		\boldmath{$L_{S2}$}&387,251&319,349&189,406&71,811&18,093\\ \hline
		\hline
		\boldmath$\#Subrange$&5 & 10 & 20 & 50& 100\\ \hline
		\boldmath{$L_{SL}$}&322,571&321,165&322,001&321,098&327,221\\ \hline
		\boldmath{$L_{S2}$}&319,832&318,556&319,349&318,396&324,201\\ \hline
		\hline
		\boldmath$\#Subset$&50 & 100 & 200 & 500& 1k\\ \hline
		\boldmath{$L_{SL}$}&324,596&322,799&322,001&323,635&323,740\\ \hline
		\boldmath{$L_{S2}$}&321,727&320,001&319,349&320,775&320,902\\ \hline
		\hline
		\boldmath$\left|\mathcal{S}\right|$&1k & 3k & 5k & 7k& 9k\\ \hline
		\boldmath{$L_{SL}$}&322,793&322,508&322,001&321,816&323,251\\ \hline
		\boldmath{$L_{S2}$}&320,117&319,766&319,349&319,154&320,545\\ \hline
	\end{tabular}
	\begin{tablenotes}
		\item[*] $L_{SL}$ and $L_{S2}$ are the required total numbers of hash tables for SL-ALSH and S2-ALSH, respectively.
	\end{tablenotes}
	\end{threeparttable}}
	\label{Space S2 and SL}
\end{table}

We evaluate the space consumptions of SL-ALSH and S2-ALSH first.
The space complexities of SL-ALSH and S2-ALSH are discussed in Appendix \ref{complexity/SL and S2}.
It is not explicitly stated in \cite{DBLP:conf/icml/LeiHKT19} how to create hash tables at different radii for SL-ALSH and S2-ALSH.
That seems not an easy task.
Here we simply assume that hash tables are created at a single radius $R=1000$ for both SL-ALSH and S2-ALSH (as a consequence, the space consumptions of SL-ALSH and S2-ALSH are underestimated).
Then the required total numbers of hash tables for SL-ALSH and S2-ALSH are computed by $L_{SL}=n^{\rho_{SL}}$ and $L_{S2}=n^{\rho_{S2}}$ respectively, where $\rho_{SL}$ and $\rho_{S2}$ are shown in Appendix \ref{complexity/SL and S2}.
The settings of $d$, $n$, $c$, $\#Subrange$, $\#Subset$ and $\left|\mathcal{S}\right|$ are the same as in Section \ref{space/various paramenters}. 
The results are presented in Table \ref{Space S2 and SL}.

It can be seen from Table \ref{Space S2 and SL} that the required total numbers of hash tables for SL-ALSH and S2-ALSH increase with $n$ and decrease with $c$, and they are not significantly affected by the other four parameters.
We can calculate that $\rho_{SL}$ and $\rho_{S2}$ are both greater than 0.98 (very close to 1) when $c\leq3$ and greater than 0.75 when $c\leq6$. This indicates that the required total numbers of hash tables for SL-ALSH and S2-ALSH can increase very fast with $n$.
In contrast, the required total number of hash tables for WLSH grows logarithmically with $n$.
Combining Section \ref{space/various paramenters} and Table \ref{Space S2 and SL}, we conclude that the space consumptions of SL-ALSH and S2-ALSH are more sensitive to the data set, while the space consumption of WLSH is more sensitive to the weight vector set.
\subsection{Query Efficiency and Query Accuracy of WLSH}
\label{efficiency}
Since bound relaxation is effective in reducing the space consumption of WLSH, we apply it to WLSH in this section.
\subsubsection{Impacts of Various Parameters}
\label{efficiency/various parameters}
\begin{figure*}[t]
	\centering
	\subfigure[I/O cost vs. $d$]{\includegraphics[width=0.246\textwidth]{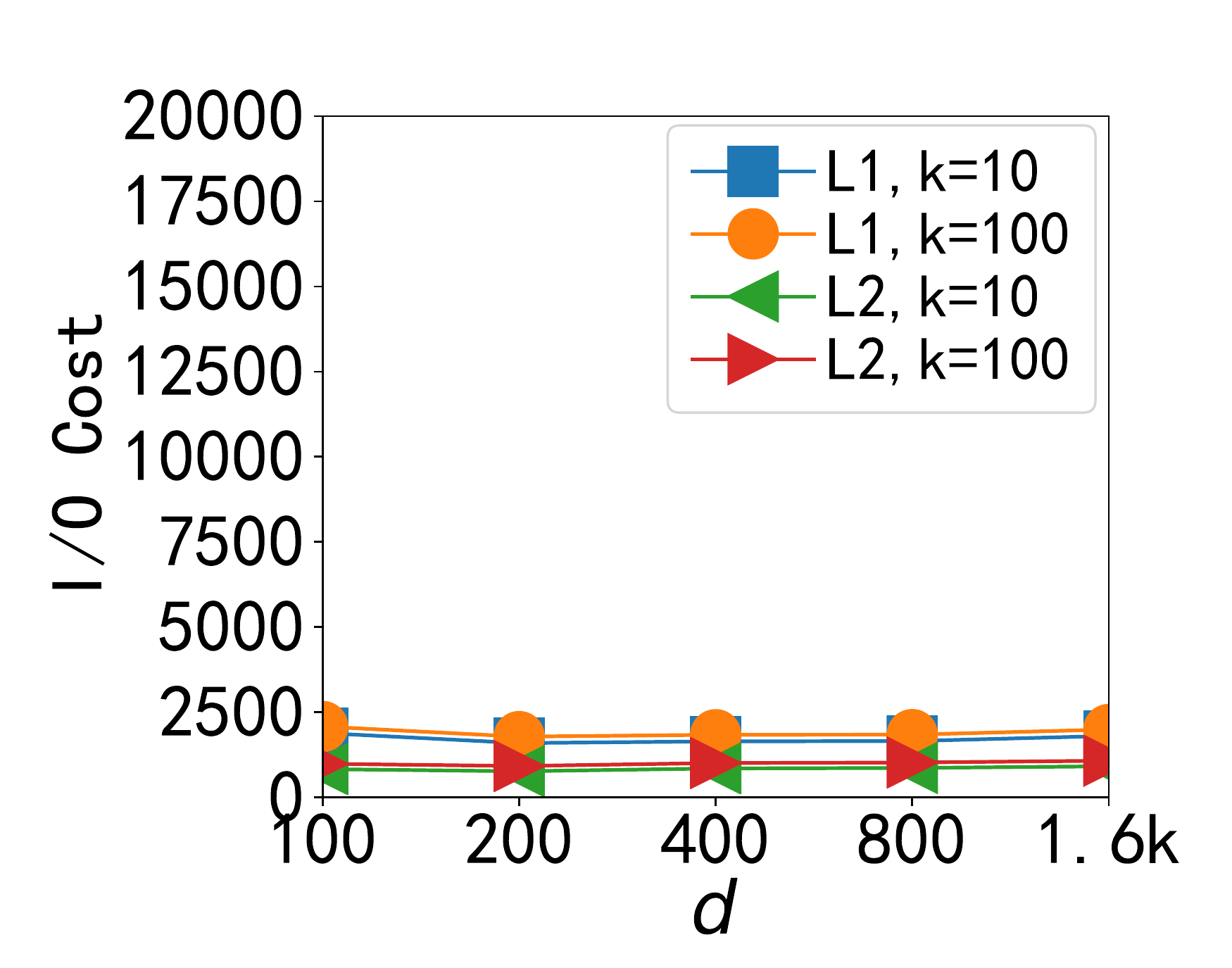}\label{rand/d/IO_useCt=1}}
	\subfigure[I/O cost vs. $n$]{\includegraphics[width=0.246\textwidth]{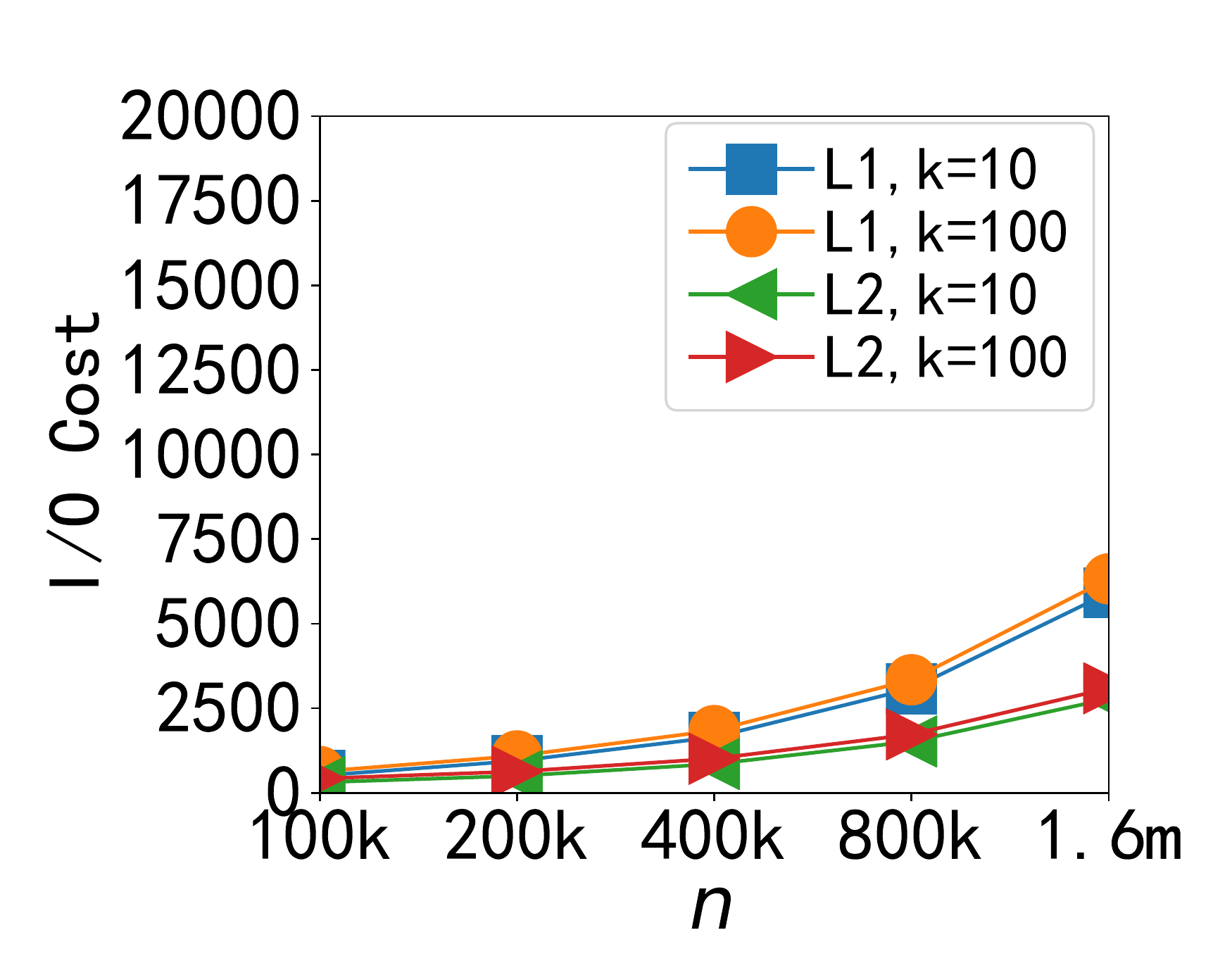}\label{rand/n/IO_useCt=1}}
	\subfigure[I/O cost vs. $c$]{\includegraphics[width=0.246\textwidth]{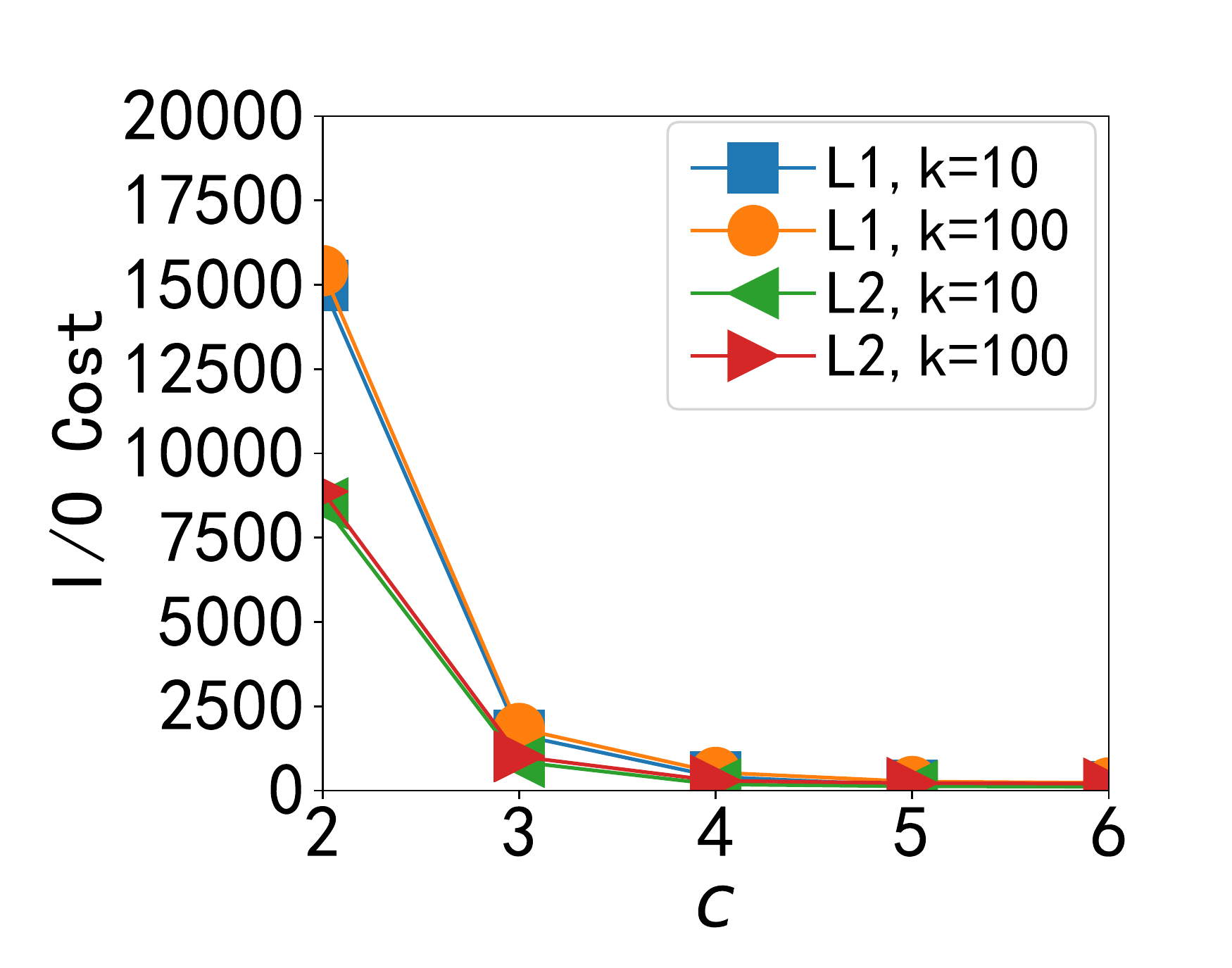}\label{rand/c/IO_useCt=1}}
	\subfigure[I/O cost vs. $\#Subrange$]{\includegraphics[width=0.246\textwidth]{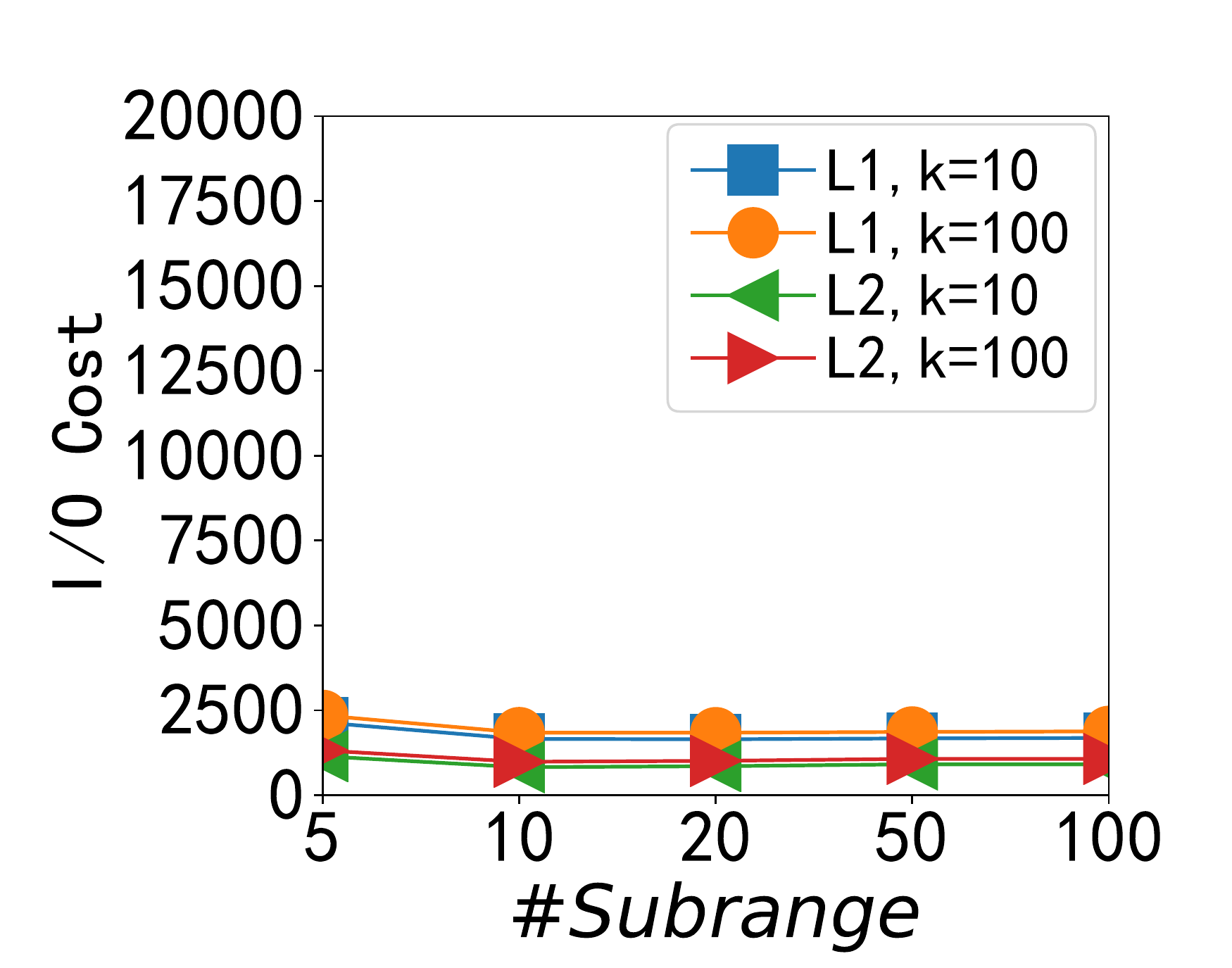}\label{rand/subrange/IO_useCt=1}}
	
	\subfigure[I/O cost vs. $\#Subset$]{\includegraphics[width=0.246\textwidth]{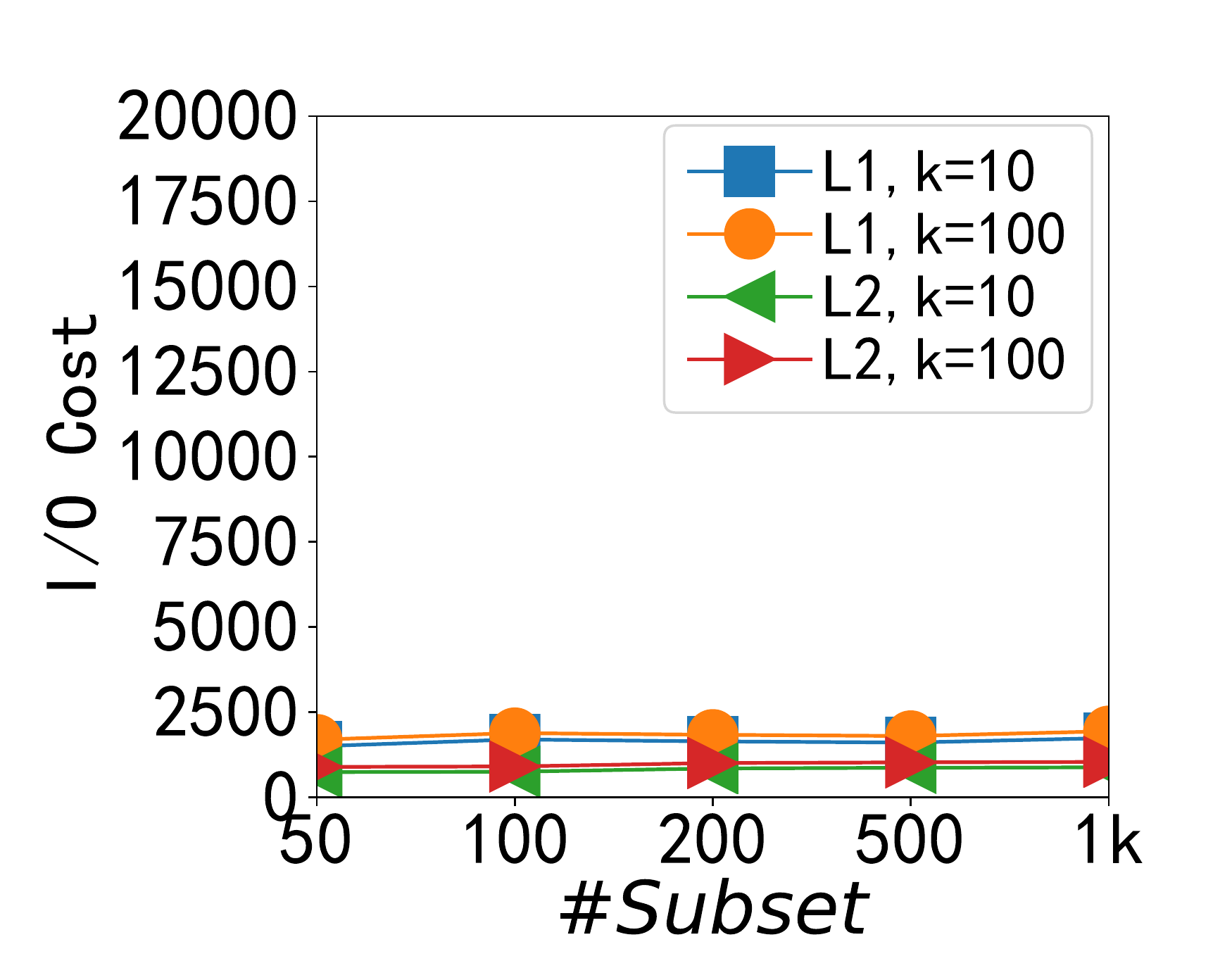}\label{rand/subset/IO_useCt=1}}
	\subfigure[I/O cost vs. $\left|S\right|$]{\includegraphics[width=0.246\textwidth]{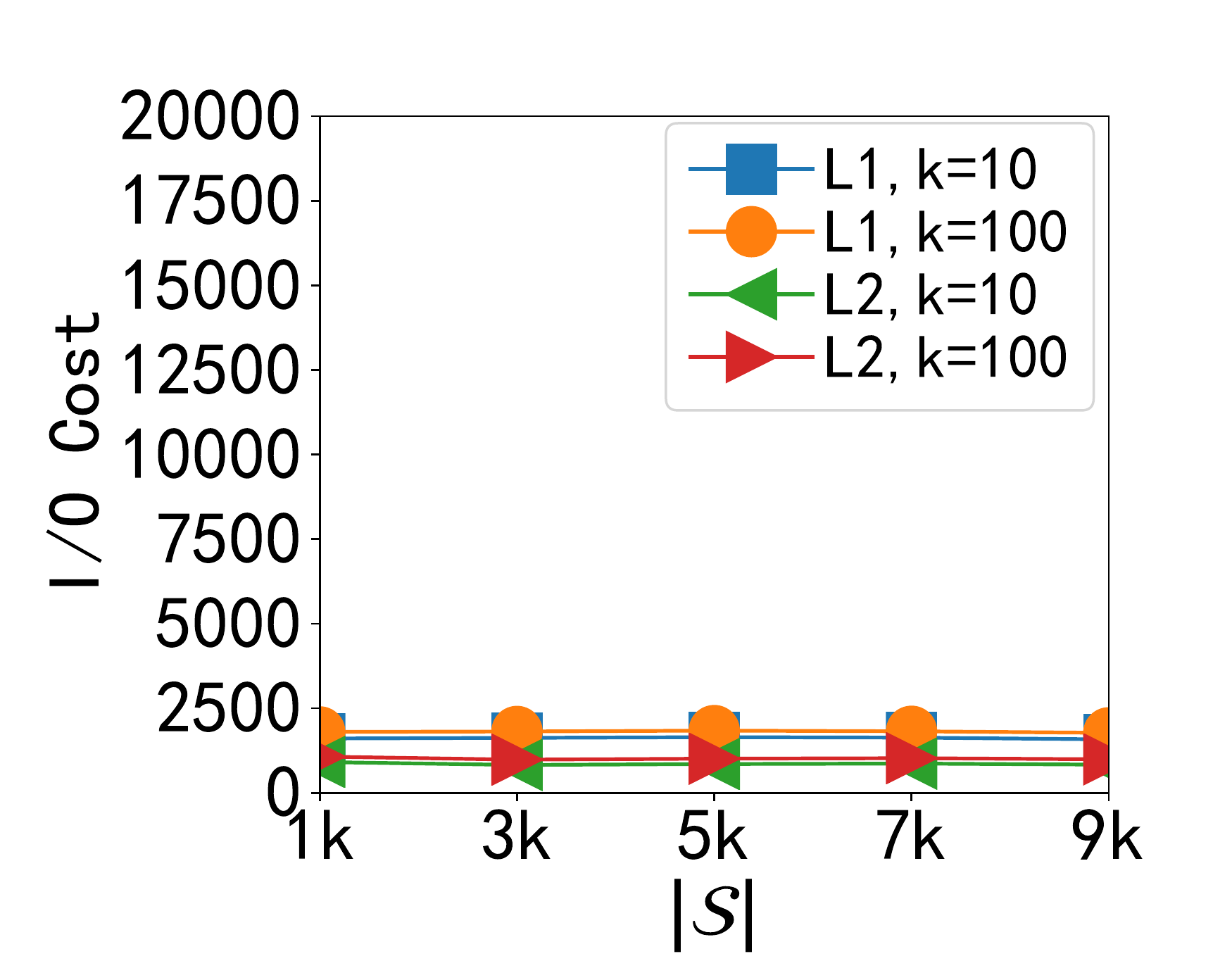}\label{rand/S/IO_useCt=1}}
	\subfigure[Ratio vs. $d$]{\includegraphics[width=0.246\textwidth]{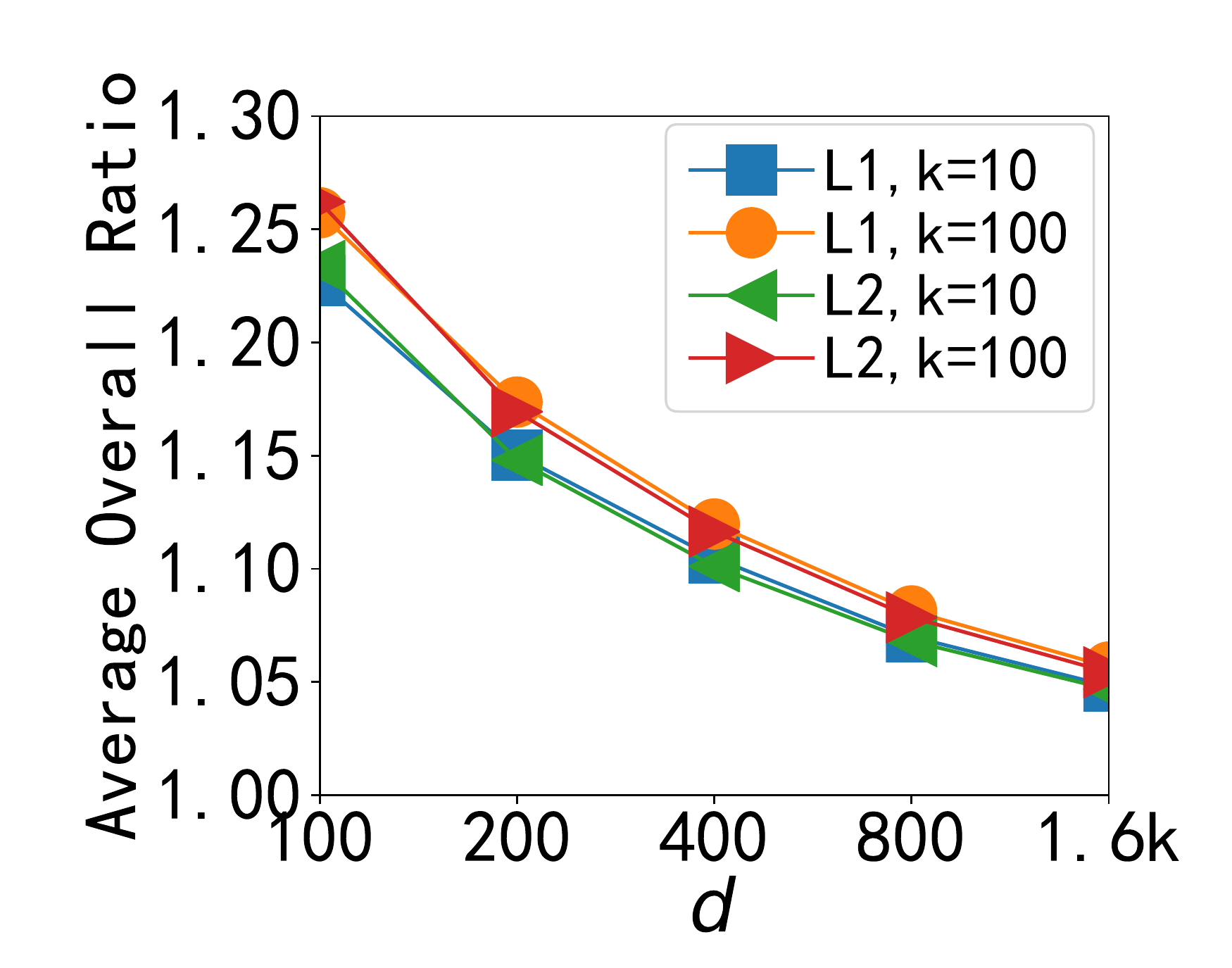}\label{rand/d/ratio_useCt=1}}
	\subfigure[Ratio vs. $n$]{\includegraphics[width=0.246\textwidth]{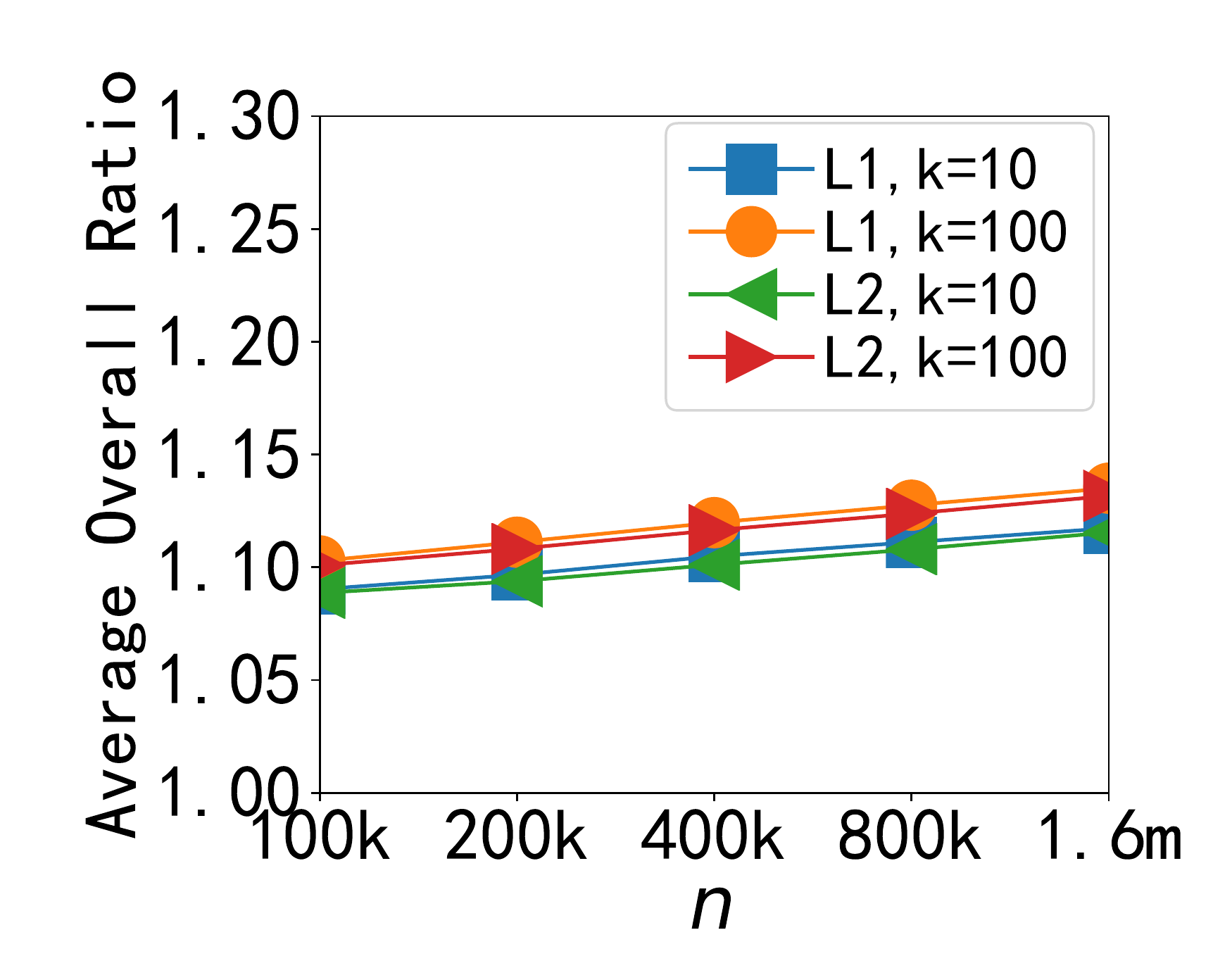}\label{rand/n/ratio_useCt=1}}
	
	\subfigure[Ratio vs. $c$]{\includegraphics[width=0.246\textwidth]{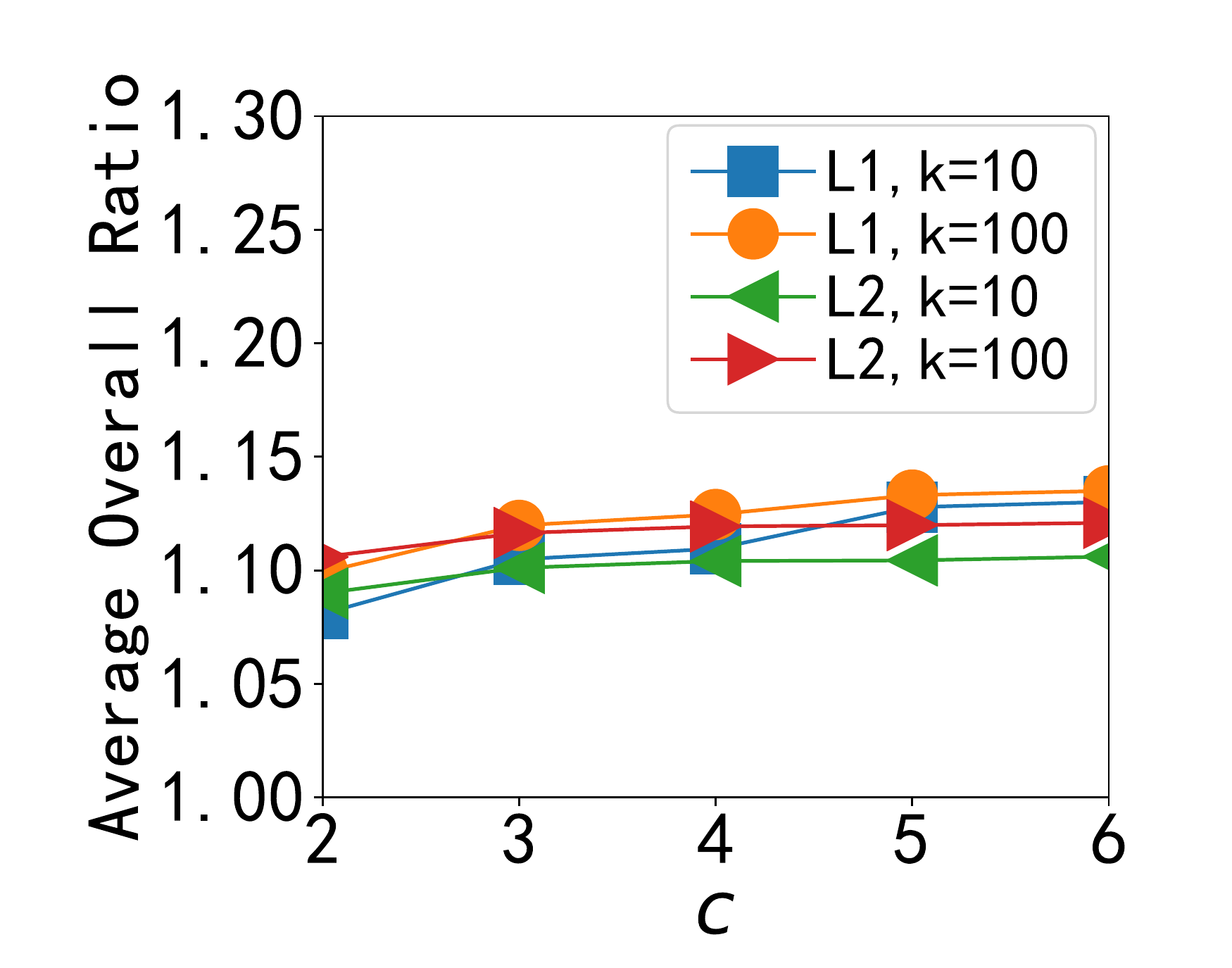}\label{rand/c/ratio_useCt=1}}
	\subfigure[Ratio vs. $\#Subrange$]{\includegraphics[width=0.246\textwidth]{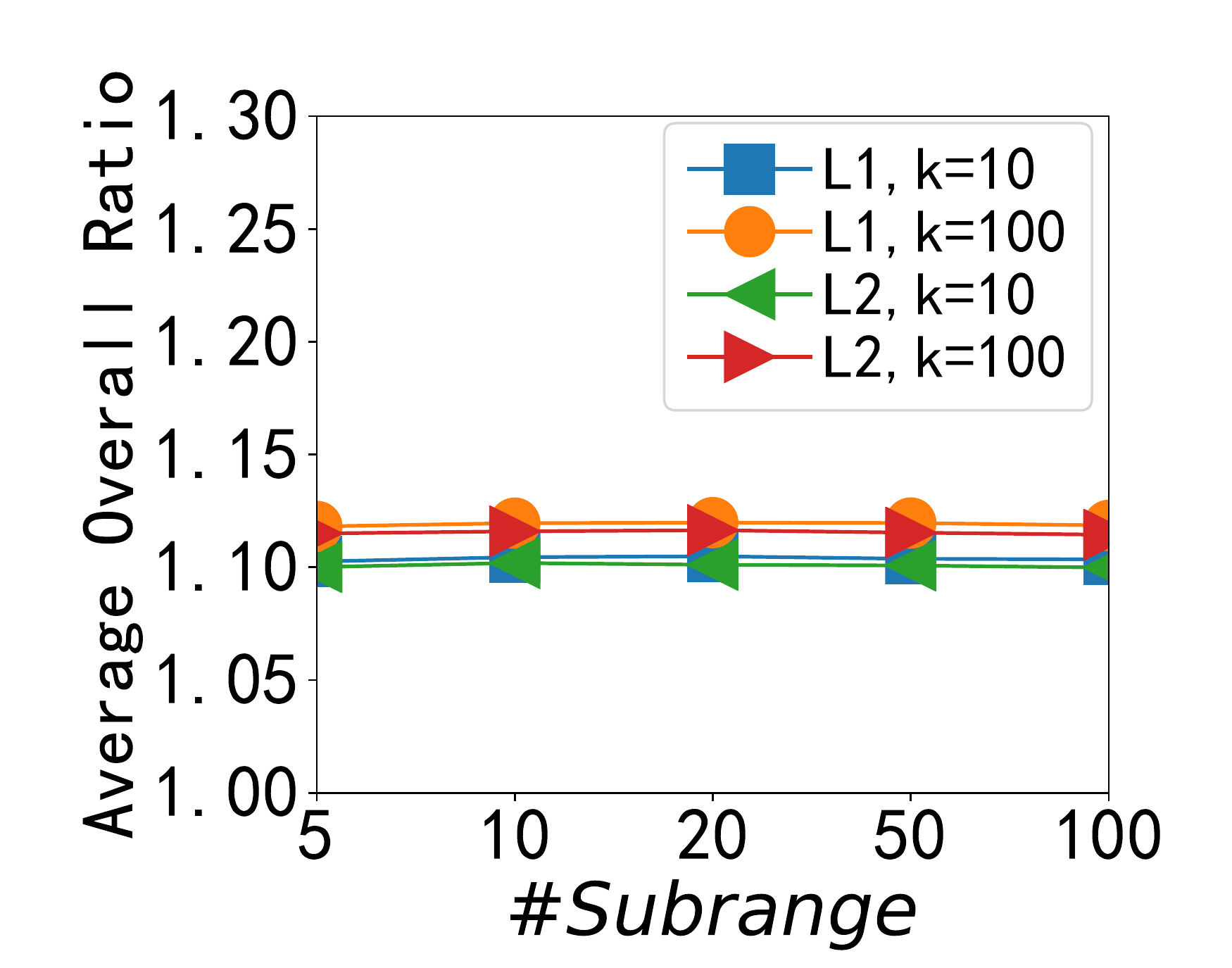}\label{rand/subrange/ratio_useCt=1}}
	\subfigure[Ratio vs. $\#Subset$]{\includegraphics[width=0.246\textwidth]{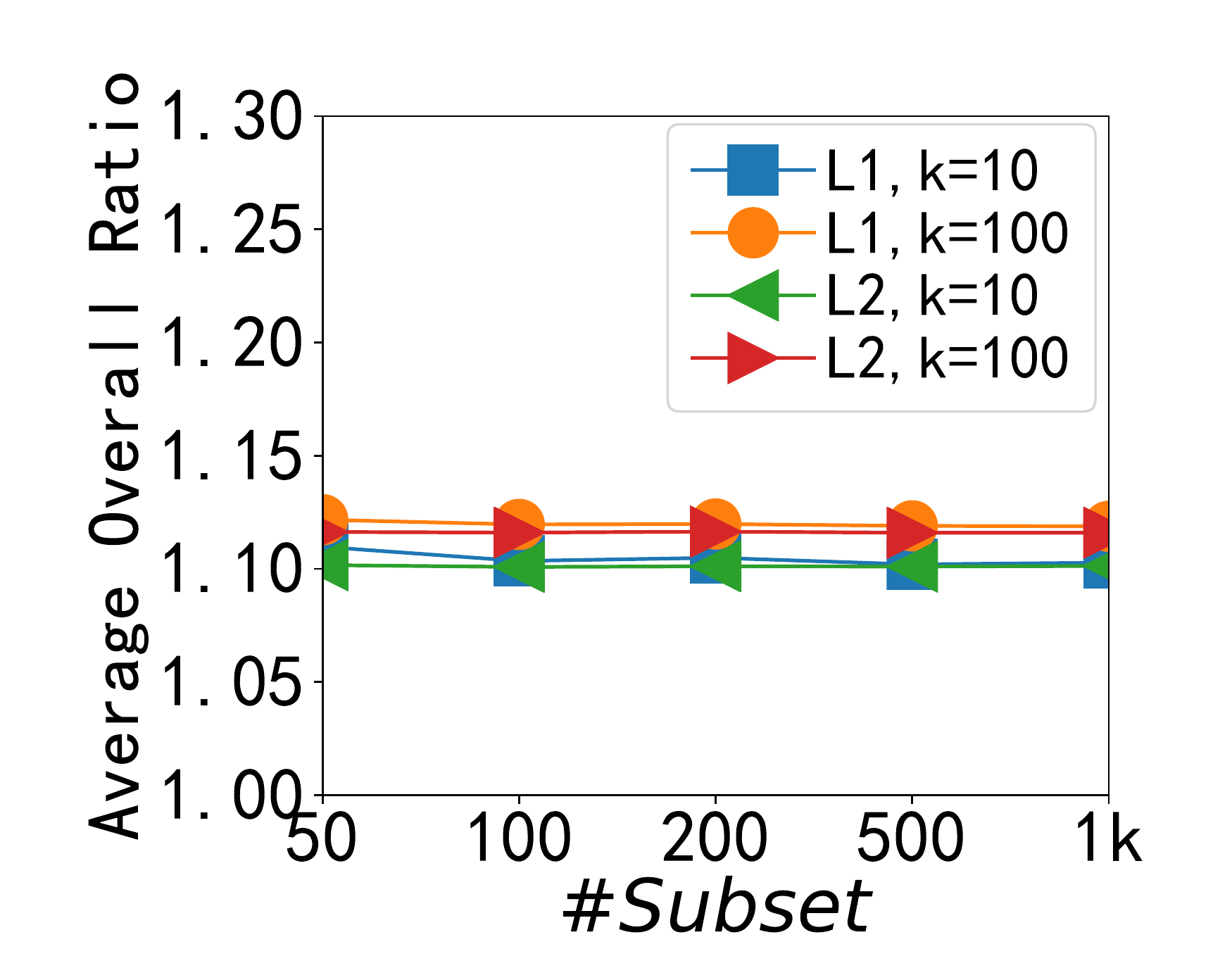}\label{rand/subset/ratio_useCt=1}}
	\subfigure[Ratio vs. $\left|S\right|$]{\includegraphics[width=0.246\textwidth]{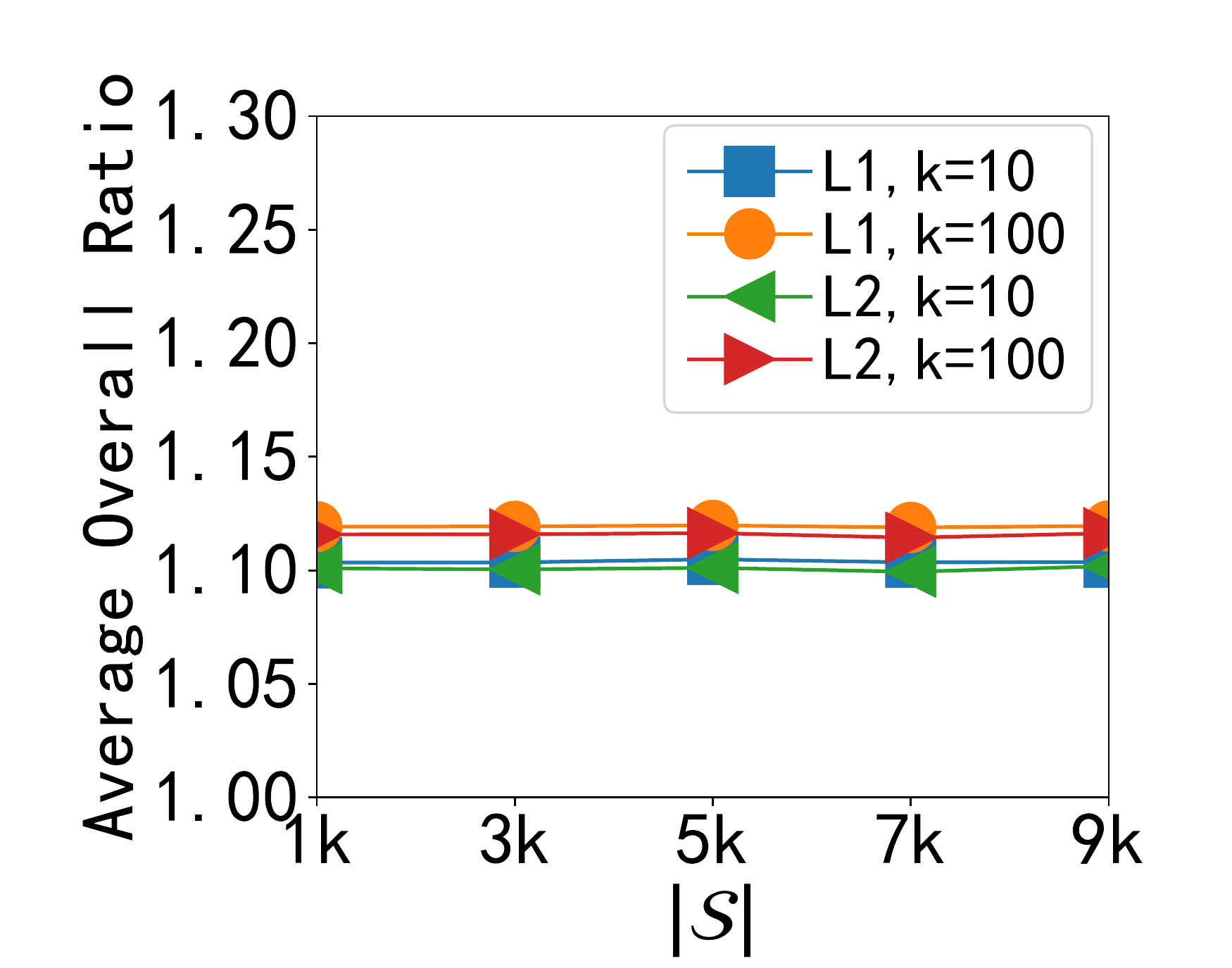}\label{rand/S/ratio_useCt=1}}
	
	\caption{Query efficiency and query accuracy of WLSH on synthetic data when using collision threshold reduction}
	\label{rand/efficiency and accuracy/L1/useCt=1}
\end{figure*}

Both synthetic and real data sets are used to study the impacts of various parameters on the query efficiency and query accuracy of WLSH. 
The evaluation on synthetic data sets involves the parameters $d$, $n$, $c$, $\#Subrange$, $\#Subset$ and $\left|\mathcal{S}\right|$, and the evaluation on real data sets involves the parameters $c$, $\#Subrange$, $\#Subset$ and $\left|\mathcal{S}\right|$. 
As in Section \ref{space/various paramenters}, 
the impact of each parameter is evaluated by varying one parameter and setting the other parameters to the default values. 

The results on synthetic data sets for WLSH using collision threshold reduction are shown in Fig. \ref{rand/efficiency and accuracy/L1/useCt=1}, and the results on real data sets for WLSH using collision threshold reduction are shown in Figs. \ref{real/efficiency and accuracy/L1/useCt=1/k=10}-\ref{real/efficiency and accuracy/L2/useCt=1} in Appendix \ref{WLSH/query efficiency and query accuracy/more experiments}. Additionally, Figs. \ref{rand/efficiency and accuracy/L1}-\ref{rand/efficiency and accuracy/L2} in Appendix \ref{WLSH/query efficiency and query accuracy/more experiments} show the results on synthetic data sets for WLSH without using collision threshold reduction. It can be seen from Fig. \ref{rand/efficiency and accuracy/L1/useCt=1} and Figs. \ref{rand/efficiency and accuracy/L1}-\ref{rand/efficiency and accuracy/L2} in Appendix \ref{WLSH/query efficiency and query accuracy/more experiments} that by adopting collision threshold reduction, 
the query efficiency of WLSH can be improved significantly while the query accuracy of WLSH is still high.
Thus, we apply collision threshold reduction to WLSH in the subsequent experiments. In the following, we analyze the results in Fig. \ref{rand/efficiency and accuracy/L1/useCt=1} and Figs. \ref{real/efficiency and accuracy/L1/useCt=1/k=10}-\ref{real/efficiency and accuracy/L2/useCt=1} in Appendix \ref{WLSH/query efficiency and query accuracy/more experiments}.

In Fig. \ref{rand/efficiency and accuracy/L1/useCt=1}, 
the I/O cost of WLSH increases with $n$ and decreases with $c$, and remains stable as $d$, $\#Subrange$, $\#Subset$ or $\left|\mathcal{S}\right|$ increases. 
Moreover, the average overall ratio of WLSH increases with $n$ or $c$ and decreases with $d$, and remains stable as $\#Subrange$, $\#Subset$ or $\left|\mathcal{S}\right|$ increases.
The impacts of $c$, $\#Subrange$, $\#Subset$ and $\left|\mathcal{S}\right|$ in Figs. \ref{real/efficiency and accuracy/L1/useCt=1/k=10}-\ref{real/efficiency and accuracy/L2/useCt=1} in Appendix \ref{WLSH/query efficiency and query accuracy/more experiments} are similar to those in Fig. \ref{rand/efficiency and accuracy/L1/useCt=1}.
The reason why the I/O cost of WLSH increases with $n$ is that we need to do collision counting for more data points as $n$ increases, and the reason why the average overall ratio of WLSH increases with $n$ and decreases with $d$ is that the distances from a query point to the exact $k$-nearest neighbors tend to decrease with $n$ and increase with $d$ while the distances from a query point to the reported approximate $k$-nearest neighbors tend to be relatively stable as $n$ or $d$ increases. 
Besides, the reason why the I/O cost of WLSH decreases with $c$ and the average overall ratio of WLSH increases with $c$ is that,
as $c$ increases, the collision threshold gets smaller such that the I/O cost for collision counting tends to decrease and the distances from a query point to the identified frequent points tend to increase.

From Fig. \ref{rand/efficiency and accuracy/L1/useCt=1} and Figs. \ref{real/efficiency and accuracy/L1/useCt=1/k=10}-\ref{real/efficiency and accuracy/L2/useCt=1} in Appendix \ref{WLSH/query efficiency and query accuracy/more experiments}, we can also see that the I/O cost of WLSH slightly increases with $k$. 
The reason is that more frequent points need to be identified by doing collision counting as $k$ increases.
In addition, the I/O cost of WLSH for the $l_1$ distance is higher than that for the $l_2$ distance in most cases.
The reason is that the collision threshold tends to be larger for the $l_1$ distance than for the $l_2$ distance such that the I/O cost for collision counting tends to be higher for the $l_1$ distance than for the $l_2$ distance.
\subsubsection{Comparison with SL-ALSH and S2-ALSH}
\label{efficiency/comparison}
\begin{table}[h]
	\centering
	\footnotesize
	\caption{Comparison of Average Overall Ratios of WLSH, SL-ALSH and S2-ALSH Using Uniformly Random Weight Vector Sets ($\left|\mathcal{S}\right|=5\text{k}$)}
	\begin{tabular}{l|c|c|c|c}
		\hline
		\multicolumn{5}{c}{$k=10$}\\\hline\hline
		\textbf{Data Set}&\textit{Sift}&\textit{Ukbench}&\textit{Notre}&\textit{Sun}\\\hline
		\textbf{Ratio (WLSH)}&1.823645&2.078039&2.169291&1.411036\\\hline
		\textbf{Ratio (SL-ALSH)}&1.69267&2.05576&1.68895&1.41198\\\hline
		\textbf{Ratio (S2-ALSH)}&1.82768&2.47739&1.93986&1.696\\\hline
		\hline
		\multicolumn{5}{c}{$k=100$}\\\hline\hline
		\textbf{Data Set}&\textit{Sift}&\textit{Ukbench}&\textit{Notre}&\textit{Sun}\\\hline
		\textbf{Ratio (WLSH)}&1.823147&2.150079&1.864421&1.450632\\\hline
		\textbf{Ratio (SL-ALSH)}&2.04149&2.38945&1.90809&1.59371\\\hline
		\textbf{Ratio (S2-ALSH)}&1.66179&2.28655&1.78331&1.52213\\\hline
	\end{tabular}
	\label{uniform/compare}
\end{table}

In the subsubsection, we compare WLSH, SL-ALSH and S2-ALSH in terms of query efficiency and query accuracy.
For WLSH, SL-ALSH or S2-ALSH, given a query, the average overall ratio is inversely proportional to the I/O cost.
Thus, we only need to let the I/O costs of SL-ALSH and S2-ALSH equal that of WLSH and compare the average overall ratios of them.
The settings of the parameters specific to SL-ALSH and S2-ALSH are shown in Table \ref{parameter settings/SL-ALSH,S2-ALSH} in Appendix \ref{WLSH/query efficiency and query accuracy/more experiments}.

Two groups of experiments are carried out on real data sets.
In the first group of experiments,  
we vary the parameters $c$, $\#Subrange$, $\#Subset$ and $\left|\mathcal{S}\right|$ in the same way as in Section \ref{efficiency/various parameters}. 
Due to space limitations, the results are shown in Figs. \ref{compare/efficiency and accuracy/L2/useCt=1_k=10}-\ref{compare/efficiency and accuracy/L2/useCt=1_k=100} in Appendix \ref{WLSH/query efficiency and query accuracy/more experiments}. In summary, the average overall ratio of WLSH is smaller than those of SL-ALSH and S2-ALSH in 120/160 and 115/160 of cases respectively.
In the second group of experiments, we use uniformly random weight vector sets and set $c$ to 8 so that the space consumptions of WLSH, SL-ALSH and S2-ALSH are similar and moderate. The experimental results are presented in Table \ref{uniform/compare}. In the table, the average overall ratio of WLSH is smaller than those of SL-ALSH and S2-ALSH in 5/8 and 5/8 of cases respectively. 
As a whole, WLSH performs better than SL-ALSH and S2-ALSH in the two groups of experiments.
This implies that the asymmetric LSH families for SL-ALSH and S2-ALSH are less effective than the weighted LSH families and derived weighted LSH families for WLSH. 
Actually, SL-ALSH and S2-ALSH are not so appropriate as WLSH for solving our problem since they create hash tables regardless of the weight vector set due to their asymmetric LSH families.
\section{Related Work}
\label{Related Work}
\subsection{Locality-Sensitive Hashing for Approximate Nearest Neighbor Search}
\label{Near/Nearest Neighbor Search and LSH}
Nearest neighbor search arises in a wide variety of applications,
such as product recommendation \cite{DBLP:journals/apin/Liu13}, 
image retrieval \cite{DBLP:journals/cviu/PengBQ99} and clustering \cite{DBLP:journals/ml/ModhaS03}. 
Locality-Sensitive Hashing (LSH) is a popular approach for approximate nearest neighbor search in high-dimensional spaces  \cite{DBLP:journals/corr/abs-1806-09823,DBLP:journals/corr/WangSSJ14}. Several important LSH schemes are briefly reviewed as follows.

The first LSH scheme is E2LSH proposed by Indyk et al. \cite{DBLP:conf/stoc/IndykM98}. It can answer nearest neighbor queries in sublinear time.
However, it needs to create hash tables at a series of radii, which is space-consuming. 
The multi-probe LSH scheme proposed by Lv et al. reduces the required number of hash tables by checking not only the data points that fall in the same bucket as the query point but also the data points that fall in the nearby buckets \cite{DBLP:conf/vldb/LvJWCL07}. 
LSB-tree/LSB-forest proposed by Tao et al. is the first LSH scheme that avoids creating hash tables at different radii by exploiting virtual rehashing \cite{DBLP:journals/tods/TaoYSK10}.
Since LSB-tree/LSB-forest uses compound hash functions to create hash tables, the number of hash buckets to read increases exponentially with the search radius, which is undesirable.
To overcome the drawback, Gan et al. propose C2LSH where each hash table is created by using a single LSH function \cite{DBLP:conf/sigmod/GanFFN12}.
Combining the virtual rehashing and collision counting techniques, C2LSH is proven to be better than LSB-tree/LSB-forest.
Recently, QALSH/QALSH$^+$ proposed by Huang et al. achieves better performance than C2LSH by introducing
query-aware LSH families and a query-aware bucket partition strategy \cite{DBLP:journals/vldb/HuangFFNW17}.
\subsection{SL-ALSH, S2-ALSH and LazyLSH}
In the following, we cover the prior methods related to our work. They all take the LSH approach.

\textbf{SL-ALSH and S2-ALSH.} 
The problem studied in \cite{DBLP:conf/icml/LeiHKT19} is most similar to ours. It is to support efficient approximate nearest neighbor search for all possible weighted distance functions with respect to the $l_2$ distance. 
SL-ALSH and S2-ALSH \cite{DBLP:conf/icml/LeiHKT19} are proposed to address the problem.
The novelty of SL-ALSH and S2-ALSH is that they each introduce an asymmetric LSH family to create hash tables and process queries. 
Each of the asymmetric LSH families involves
two mappings which can map data points and query points to different hyperspheres respectively.
The mapping for data points is independent of any weighted distance function,
while the mapping for query points is sensitive to the associated weighted distance functions. Since the two asymmetric LSH families target the $l_2$ distance, they are not available for the $l_p$ distance for $p\in(0,2)$.

\textbf{LazyLSH.} The work of \cite{DBLP:conf/sigmod/ZhengGTW16} studies how to support efficient approximate nearest neighbor search under the distance function with respect to the $l_p$ distance, where $p$ is taken as a variable. 
The method proposed for solving this problem is called LazyLSH \cite{DBLP:conf/sigmod/ZhengGTW16}.
Briefly, LazyLSH first creates a group of hash tables using an LSH family originally proposed for the $l_1$ distance, and then processes all incoming queries with various values of $p$ over the hash tables.
The motivation behind LazyLSH is that 
if two points are close under the $l_{p_1}$ distance function, then they are likely to be close under the $l_{p_2}$ distance function, where $p_1\neq p_2$.
The idea of LazyLSH is not available for solving our problem since two close points under a weighted distance function
can be far away under another weighted distance function.

\section{Conclusion}
\label{Conclusion}
In this paper, we consider the important problem of supporting efficient approximate nearest neighbor search for multiple weighted distance functions in high-dimensional spaces.
The WLSH method is proposed to address the problem for the $l_p$ distance with $p\in(0,2]$.
Remarkably, it is the first method that can solve the problem for the $l_p$ distance for $p\in(0,2)$.
Theoretically, WLSH can guarantee the efficiency of processing queries and the accuracy of query results while minimizing the space consumption.
To achieve better practical performance, WLSH can also make reasonable trade-offs among the query efficiency, query accuracy and space consumption.
Extensive experiments are carried out on both synthetic and real data sets, and the results verify the effectiveness and efficiency of WLSH.
However, we observe that the performance of WLSH degrades as $p$ decreases.
A promising direction for future work is to base WLSH on QALSH/QALSH$^+$ \cite{DBLP:journals/vldb/HuangFFNW17} instead of on C2LSH so that the performance of WLSH could be further improved. We believe that this can be done without the need of making many modifications to our main techniques in this paper.


%

\appendices
\section{Time and Space Complexities of SL-ALSH and S2-ALSH}
\label{complexity/SL and S2}
Both the SL-ALSH and S2-ALSH methods \cite{DBLP:conf/icml/LeiHKT19} can be used to address the problem studied in the paper. Here we give the time complexities of SL-ALSH and S2-ALSH for processing queries and the space complexities of SL-ALSH and S2-ALSH for storing hash tables and original data.

For SL-ALSH and S2-ALSH, suppose $o,q\in[0,V]^d\subset\mathbb{R}^d$ for all $o\in P$ and $q$ for $0<V\leq\pi$. Otherwise, we can shift and/or rescale $o\in P$ and $q$ without changing the query results. Further, suppose 
$\Vert\vec{W_i}\Vert_1=1$ for all $\vec{W_i}\in\mathcal{S}$. Otherwise, we can also rescale $\vec{W_i}\in\mathcal{S}$ without changing the query results.
It is not explicitly stated in \cite{DBLP:conf/icml/LeiHKT19} how SL-ALSH and S2-ALSH create hash tables at different radii. Thus, we simply assume that they only create hash tables at a radius $R$ for all $\vec{W_i}\in\mathcal{S}$. 
Let $\eta_{\vec{W_i}}=\sqrt{d}\Vert\vec{W_i}\Vert_2$ for $\vec{W_i}\in\mathcal{S}$.
According to Theorem 3 in \cite{DBLP:conf/icml/LeiHKT19} and Definition \ref{Dimension Weighting Distance Function}, 
the time and space complexities of SL-ALSH are $O(n^{\rho_{SL}}d\log n)$ and $O(nd+n^{1+\rho_{SL}})$ respectively, where
\begin{multline}
\label{rho/SL}
\rho_{SL}=\\\min_{w,V}\left(\max_{\vec{W_i}\in\mathcal{S}}\left(\frac{\ln\left(P_{l_2}\left(\sqrt{2\eta_{\vec{W_i}}-2+R}\right)\right)}{\ln\left(P_{l_2}\left(\sqrt{2\eta_{\vec{W_i}}-2+cR-\frac{1}{12}V^4}\right)\right)}\right)\right)
\end{multline}
According to Theorem 4 in \cite{DBLP:conf/icml/LeiHKT19} and Definition \ref{Dimension Weighting Distance Function}, the time and space complexities of S2-ALSH are $O(n^{\rho_{S2}}d\log n)$ and $O(nd+n^{1+\rho_{S2}})$ respectively, where
\begin{equation}
\label{rho/S2}
\rho_{S2}=\min_{V}\left(\max_{\vec{W_i}\in\mathcal{S}}\left(\frac{\ln\left(1-\frac{1}{\pi}\arccos\left(\frac{1-\frac{1}{2}R}{\eta_{\vec{W_i}}}\right)\right)}{\ln\left(1-\frac{1}{\pi}\arccos\left(\frac{1-\frac{1}{2}cR+\frac{1}{24}V^4}{\eta_{\vec{W_i}}}\right)\right)}\right)\right)
\end{equation}

Please note that in Equation \ref{rho/SL}, $P_{l_2}(r)=\int_{0}^{w}\frac{1}{r}F_2(\frac{t}{r})(1-\frac{t}{w})dt$, where $F_2(\cdot)$ is the PDF of the \textit{absolute value} of the standard normal distribution. It can be verified that $P_{l_2}(r)$ is smaller than 1 and decreases with $r$.
Both SL-ALSH and S2-ALSH require that $cR-\frac{1}{12}V^4>R$ \cite{DBLP:conf/icml/LeiHKT19}. Thus, $0<\rho_{SL},\rho_{S2}<1$.
The required total number of hash tables for SL-ALSH is $n^{\rho_{SL}}$, and the required total number of hash tables for S2-ALSH is $n^{\rho_{S2}}$. 
\section{Traditional LSH Families and Weighted LSH Families for Hamming Distance and Angular Distance}
\label{lsh families and weighted lsh families}
In the following, we give the distance functions with respect to the Hamming distance and angular distance.

\textbf{Hamming Distance:} Let $\vec{x}=\left(x_{1},x_{2},\ldots,x_{d}\right)\in\{0,1\}^d$ and $\vec{y}=\left(y_{1},y_{2},\ldots,y_{d}\right)\in\{0,1\}^d$. The distance function with respect to the Hamming distance is
\begin{equation}
D(x,y)=\sum_{i=1}^{d}\left|x_{i}-y_{i}\right|
\end{equation}

\textbf{Angular Distance:} Let $\vec{x}=\left(x_{1},x_{2},\ldots,x_{d}\right)\in\mathbb{R}^d$ and $\vec{y}=\left(y_{1},y_{2},\ldots,y_{d}\right)\in\mathbb{R}^d$. The distance function with respect to the angular distance is
\begin{equation}
D(x,y)=\arccos\left(\frac{\vec{x}\cdot\vec{y}}{\Vert\vec{x}\Vert_2\times\Vert\vec{y}\Vert_2}\right)
\end{equation}

In Table \ref{distance functions}, we present a classical traditional LSH family for the Hamming distance and a classical traditional LSH family for the angular distance.
The corresponding weighted LSH families are shown in Table \ref{distance functions-weighted}.
\begin{table*}
	\centering
	\caption{Traditional LSH Families for Hamming Distance and Angular Distance}
	\begin{threeparttable}
		\begin{tabular}{|p{0.39\columnwidth}|p{0.5\columnwidth}|p{0.6\columnwidth}|p{0.35\columnwidth}|} 
			\hline
			\multirow{1}{0.39\columnwidth}{\textit{Distance Measure}}&\multirow{1}{0.53\columnwidth}{$h(x)$}&\textit{PMF (or PDF)}&\multirow{1}{0.35\columnwidth}{\textit{$P(\cdot)$}}\\ \hline\hline
			
			\multirow{1}{0.39\columnwidth}{Hamming distance \cite{DBLP:conf/vldb/GionisIM99}}&\multirow{1}{0.5\columnwidth}{$h_k(x)=x_k$, $k\in\{1,2,\ldots,d\}$}&$p(k)=\frac{1}{d}$ for $k\in\{1,2,\ldots,d\}$.&\multirow{1}{0.35\columnwidth}{$P_{H}(r)=1-\frac{r}{d}$}\\ \hline
			
			\multirow{3}{0.39\columnwidth}{Angular distance \cite{DBLP:conf/stoc/Charikar02}}&\multirow{3}{0.5\columnwidth}{$h_{\vec{u}}(x)=\text{sign}\left(\vec{u}\cdot\vec{x}\right)$, $\vec{u}=(u_1,u_2,\ldots,u_d)\in\mathbb{R}^d$}&$p(\vec{u})=\prod_{i=1}^{d}f_2(u_i)$, where $f_2$ is the PDF of the 2-stable distribution (i.e., the standard normal distribution).&\multirow{3}{0.38\columnwidth}{$P_{\theta}(r)=1-\frac{r}{\pi}$}\\ 
			\hline
		\end{tabular}
		\begin{tablenotes}
			\item[*] $\vec{x}=(x_1,x_2,\ldots,x_d)$
		\end{tablenotes}
	\end{threeparttable}
	\label{distance functions}
\end{table*}
\begin{table*}
	\centering
	\caption{Weighted LSH Families for Hamming Distance and Angular Distance}
	\begin{threeparttable}
		\begin{tabular}{|p{0.39\columnwidth}|p{0.5\columnwidth}|p{0.6\columnwidth}|p{0.35\columnwidth}|} 
			\hline
			\multirow{1}{0.39\columnwidth}{\textit{Distance Measure}}&\multirow{1}{0.53\columnwidth}{$h_{\vec{W}}(x)$}&\textit{PMF (or PDF)}&\multirow{1}{0.35\columnwidth}{\textit{$P_{\vec{W}}(\cdot)$}}\\ \hline\hline
			
			\multirow{1}{0.39\columnwidth}{Hamming distance \cite{DBLP:conf/vldb/GionisIM99}}&$h_{k,\vec{W}}(x)=w_kx_k, k\in\{1,2,\ldots,d\}$&$p(k)=\frac{w_k}{\sum_{j=1}^{d}w_j}$ for $k\in\{1,2,\ldots,d\}$.&\multirow{1}{0.38\columnwidth}{$P_{H,\vec{W}}(r)=1-\frac{r}{\sum_{i=1}^{d}w_i}$}\\ \hline
			
			\multirow{3}{0.39\columnwidth}{Angular distance \cite{DBLP:conf/stoc/Charikar02}}&\multirow{3}{0.53\columnwidth}{$h_{\vec{u},\vec{W}}(x)=\text{sign}(\vec{u}\cdot(\vec{W}\circ \vec{x}))$, $\vec{u}=(u_1,u_2,\ldots,u_d)\in\mathbb{R}^d$}&$p(\vec{u})=\prod_{i=1}^{d}f_2(u_i)$, where $f_2$ is the PDF of the 2-stable distribution (i.e., the standard normal distribution).&\multirow{3}{0.35\columnwidth}{$P_{\theta,\vec{W}}(r)=1-\frac{r}{\pi}$}\\ 
			\hline
		\end{tabular}
		\begin{tablenotes}
			\item[*] $\vec{x}=(x_1,x_2,\ldots,x_d)$ and $\vec{W}=(w_{1},w_{2},\ldots,w_{d})$
		\end{tablenotes}
	\end{threeparttable}
	\label{distance functions-weighted}
\end{table*}
\section{Proof of Theorem \ref{R_bound}}
\label{proof of theorem 1}
\begin{IEEEproof}
	\mbox{}\par
	(1). Let $\vec{x}=\left(x_{1},x_{2},\ldots,x_{d}\right)\in\mathbb{R}^d$, $\vec{y}=\left(y_{1},y_{2},\ldots,y_{d}\right)\in\mathbb{R}^d$, $\frac{w_s}{w_s'}=\max_{1\leq i\leq d}(\frac{w_i}{w_i'})$ and $\frac{w_t}{w_t'}=\min_{1\leq i\leq d}(\frac{w_i}{w_i'})$. 
	For the $l_p$ distance, we have $D_{\vec{W}}(x,y)=\sqrt[p]{\sum_{i=1}^{d}\left|w_ix_{i}-w_iy_{i}\right|^p}$ and $D_{\vec{W'}}(x,y)=\sqrt[p]{\sum_{i=1}^{d}\left|w_i'x_{i}-w_i'y_{i}\right|^p}$.
	
	First, we need to prove $R^{\uparrow}=R\frac{w_s}{w_s'}\geq\max(D_{\vec{W}}(x,y))$ under the constraint of $D_{\vec{W'}}(x,y)\leq R$. 
	It is equivalent to prove $R^p(\frac{w_s}{w_s'})^p\geq\max(\sum_{i=1}^{d}w_i^p\left|x_{i}-y_{i}\right|^p)$ under the constraint of $R^p\geq\sum_{i=1}^{d}(w_i')^p\left|x_{i}-y_{i}\right|^p$. 
	As $R^p(\frac{w_s}{w_s'})^p\geq\sum_{i=1}^{d}(w_i')^p(\frac{w_s}{w_s'})^p\left|x_{i}-y_{i}\right|^p$ can be obtained from $R^p\geq\sum_{i=1}^{d}(w_i')^p\left|x_{i}-y_{i}\right|^p$,
	it is sufficient to prove $\sum_{i=1}^{d}(w_i')^p(\frac{w_s}{w_s'})^p\left|x_{i}-y_{i}\right|^p\geq\max(\sum_{i=1}^{d}w_i^p\left|x_{i}-y_{i}\right|^p)$. 
	Actually, it always holds because 
	$(w_i')^p(\frac{w_s}{w_s'})^p\left|x_{i}-y_{i}\right|^p\geq (w_i')^p(\frac{w_i}{w_i'})^p\left|x_{i}-y_{i}\right|^p=w_i^p\left|x_{i}-y_{i}\right|^p$ for any $x_i,y_i\in\mathbb{R}$ and $1\leq i\leq d$.
	
	Next, we need to prove $(cR)^{\downarrow}=cR\frac{w_t}{w_t'}\leq\min(D_{\vec{W}}(x,y))$ under the constraint of $D_{\vec{W'}}(x,y)\geq cR$. Similarly, we can know that it is sufficient to prove $\sum_{i=1}^{d}(w_i')^p(\frac{w_t}{w_t'})^p\left|x_{i}-y_{i}\right|^p\leq\min(\sum_{i=1}^{d}w_i^p\left|x_{i}-y_{i}\right|^p)$, and it always holds because $(w_i')^p(\frac{w_t}{w_t'})^p\left|x_{i}-y_{i}\right|^p\leq (w_i')^p(\frac{w_i}{w_i'})^p\left|x_{i}-y_{i}\right|^p=w_i^p\left|x_{i}-y_{i}\right|^p$ for any $x_i,y_i\in\mathbb{R}$ and $1\leq i\leq d$.
	
	(2). The proof is almost the same as the one above and thus omitted.
	
	(3). Let $\vec{x}=\left(x_{1},x_{2},\ldots,x_{d}\right)\in\mathbb{R}^d$, $\vec{y}=\left(y_{1},y_{2},\ldots,y_{d}\right)\in\mathbb{R}^d$, $S=\Vert \vec{W}\circ\vec{x}\Vert_2\times\Vert \vec{W}\circ\vec{y}\Vert_2$, $S'=\Vert \vec{W'}\circ\vec{x}\Vert_2\times\Vert \vec{W'}\circ\vec{y}\Vert_2$, $T=\left(\vec{W}\circ\vec{x}\right)\cdot\left(\vec{W}\circ\vec{y}\right)$, $T'=\left(\vec{W'}\circ\vec{x}\right)\cdot\left(\vec{W'}\circ\vec{y}\right)$, 
	$M=\max_{1\leq i\leq d}\left(\frac{w_i^2}{w_i'^2}\right)$ and $N=\min_{1\leq i\leq d}\left(\frac{w_i^2}{w_i'^2}\right)$. 
	For the angular distance, we have $D_{\vec{W}}(x,y)=\arccos\left(\frac{T}{S}\right)$ and $D_{\vec{W'}}(x,y)=\arccos\left(\frac{T'}{S'}\right)$.
	Let $U=\{i\mid x_{i}y_{i}>0\}$. Construct $\vec{x'}=\left(x_{1}',x_{2}',\ldots,x_{d}'\right)$ and $\vec{y'}=\left(y_{1}',y_{2}',\ldots,y_{d}'\right)$, where $x_i'=x_i$ for $i\in U$ and $x_i'=0$ for $i\notin U$.
	Let $U'=\{i\mid x_{i}y_{i}<0\}$. Construct $\vec{x''}=\left(x_{1}'',x_{2}'',\ldots,x_{d}''\right)$ and $\vec{y''}=\left(y_{1}'',y_{2}'',\ldots,y_{d}''\right)$, where $x_i''=x_i$ for $i\in U'$ and $x_i''=0$ for $i\notin U'$.
	
	First, we need to prove $R^{\uparrow}=\arccos\left(\max\left(-1,\cos\left(R\right)+\frac{N-M}{M}\right)\right)\geq\max(D_{\vec{W}}(x,y))$ under the constraint of $D_{\vec{W'}}(x,y)\leq R$. After a simple calculation, we have $S\leq\max_{1\leq i,j\leq d}\left(\frac{w_iw_j}{w_i'w_j'}\right)S'=MS'$ and $T\geq\min_{i\in U}\left(\frac{w_i^2}{w_i'^2}\right)\times\left(\sum_{i\in U}w_i'^2x_{i}y_{i}\right)+\max_{i\notin U}\left(\frac{w_i^2}{w_i'^2}\right)\times\left(\sum_{i\notin U}w_i'^2x_{i}y_{i}\right)\geq N\times\left(\sum_{i\in U}w_i'^2x_{i}y_{i}\right)+M\times\left(\sum_{i\notin U}w_i'^2x_{i}y_{i}\right)$. 
	Further, as $T'=\sum_{i\in U}w_i'^2x_{i}y_{i}+\sum_{i\notin U}w_i'^2x_{i}y_{i}$ and
	$R\geq D_{\vec{W'}}(x,y)=\arccos\left(\frac{T'}{S'}\right)$, we have
	$T\geq M\times\left(S'\cos\left(R\right)\right)+\left(N-M\right)\times\left(\sum_{i\in U}w_i'^2x_{i}y_{i}\right)$. 
	Therefore, $\frac{T}{S}\geq\cos\left(R\right)+\frac{\left(N-M\right)\times\left(\sum_{i\in U}w_i'^2x_{i}y_{i}\right)}{MS'}\geq\cos\left(R\right)+\frac{N-M}{M}$, given that
	$\frac{\left(\sum_{i\in U}w_i'^2x_{i}y_{i}\right)}{S'}\leq\frac{\left(\vec{W'}\circ\vec{x'}\right)\cdot\left(\vec{W'}\circ\vec{y'}\right)}{\Vert \vec{W'}\circ\vec{x'}\Vert_2\times\Vert \vec{W'}\circ\vec{y'}\Vert_2}\leq1$ and $N-M\leq0$. 
	Obviously, $\frac{T}{S}\geq -1$. 
	Thus, $\arccos\left(\frac{T}{S}\right)\leq\arccos\left(\max\left(-1,\cos\left(R\right)+\frac{N-M}{M}\right)\right)$ for any $\vec{x},\vec{y}\in\mathbb{R}^d$. That is, $R^{\uparrow}\geq\max(D_{\vec{W}}(x,y))$.
	
	Next, we need to prove $(cR)^{\downarrow}=\arccos\left(\min\left(1,\frac{M\cos\left(cR\right)}{N}+\frac{M-N}{N}\right)\right)\leq\min(D_{\vec{W}}(x,y))$ under the constraint of $D_{\vec{W'}}(x,y)\geq cR$.
	After a simple calculation, we have $S\geq\min_{1\leq i,j\leq d}\left(\frac{w_iw_j}{w_i'w_j'}\right)S'=NS'$ and $T\leq\min_{i\in U'}\left(\frac{w_i^2}{w_i'^2}\right)\times\left(\sum_{i\in U'}w_i'^2x_{i}y_{i}\right)+\max_{i\notin U'}\left(\frac{w_i^2}{w_i'^2}\right)\times\left(\sum_{i\notin U'}w_i'^2x_{i}y_{i}\right)\leq N\times\left(\sum_{i\in U'}w_i'^2x_{i}y_{i}\right)+M\times\left(\sum_{i\notin U'}w_i'^2x_{i}y_{i}\right)$.
	Further, as $T'=\sum_{i\in U'}w_i'^2x_{i}y_{i}+\sum_{i\notin U'}w_i'^2x_{i}y_{i}$ and $cR\leq D_{\vec{W'}}(x,y)=\arccos\left(\frac{T'}{S'}\right)$, we have
	$T\leq M\times\left(S'\cos\left(cR\right)\right)+\left(N-M\right)\times\left(\sum_{i\in U'}w_i'^2x_{i}y_{i}\right)$. 
	Therefore, $\frac{T}{S}\leq\frac{M\cos\left(cR\right)}{N}+\frac{\left(N-M\right)\times\left(\sum_{i\in U'}w_i'^2x_{i}y_{i}\right)}{NS'}\leq\frac{M\cos\left(cR\right)}{N}+\frac{M-N}{N}$, given that $\frac{\left(\sum_{i\in U'}w_i'^2x_{i}y_{i}\right)}{S'}\geq\frac{\left(\vec{W'}\circ\vec{x''}\right)\cdot\left(\vec{W'}\circ\vec{y''}\right)}{\Vert \vec{W'}\circ\vec{x''}\Vert_2\times\Vert \vec{W'}\circ\vec{y''}\Vert_2}\geq-1$ and $N-M\leq0$.
	Obviously, $\frac{T}{S}\leq 1$. Thus, $\arccos\left(\frac{T}{S}\right)\geq\arccos\left(\min\left(1,\frac{M\cos\left(cR\right)}{N}+\frac{M-N}{N}\right)\right)$ for any $\vec{x},\vec{y}\in\mathbb{R}^d$. That is, $(cR)^{\downarrow}\leq\min(D_{\vec{W}}(x,y))$.
\end{IEEEproof}
\section{Proof of Lemma \ref{lemma 1}}
\label{proof of lemma 1}
Please note that $\mathcal{X}=\mathbb{R}^d$ for the $l_p$ distance in the paper. Given any integer $i\geq1$, define a traditional LSH family $\mathcal{H}_{\vec{a},b}'=\{h_{\vec{a},b}':\mathcal{X}\rightarrow U\}$, where 
\begin{equation}
h_{\vec{a},b}'(x)=\lfloor\frac{\vec{a}\cdot\vec{x}+bi}{w}\rfloor,
\end{equation}
and $\vec{a}$, $b$ and $w$ are the same as in Equation \ref{lp/e2lsh}.
The proof of Lemma \ref{lemma 1} is based on the proof of Observation 1 in \cite{DBLP:journals/tods/TaoYSK10}. In brief, the proof of Observation 1 in \cite{DBLP:journals/tods/TaoYSK10} shows that the probability of any two points in $\mathcal{X}$ colliding under an LSH function from $\mathcal{H}_{\vec{a},b}'$ is not affected by $i$.

\begin{IEEEproof}
	\mbox{}\par
	(1). The weighted LSH family obtained from $\mathcal{H}_{\vec{a},b}'$ is $\mathcal{H}_{\vec{a},b,\vec{W}}'=\{h_{\vec{a},b,\vec{W}}':\mathcal{X}\rightarrow U\}$, where 
	\begin{equation}
	h_{\vec{a},b,\vec{W}}'(x)=\lfloor\frac{\vec{a}\cdot\left(\vec{W}\circ\vec{x}\right)+bi}{w}\rfloor,
	\end{equation}
	and $\vec{a}$, $b$ and $w$ are the same as in Equation \ref{lp/e2lsh}. According to the proof of Observation 1 in \cite{DBLP:journals/tods/TaoYSK10}, the probability of any two points in $\vec{W}\circ\mathcal{X}$ colliding under an LSH function from $\mathcal{H}_{\vec{a},b}'$ is not affected by $i$ since $\vec{W}\circ\mathcal{X}=\mathcal{X}$, where $\vec{W}\circ\mathcal{X}=\{\vec{W}\circ\vec{x}\mid x\in \mathcal{X}\}$. 
	Therefore, the probability of any two points in $\mathcal{X}$ colliding under an LSH function from $\mathcal{H}_{\vec{a},b,\vec{W}}'$ is not affected by $i$. Based on this, we can have that $\mathcal{H}_{\vec{a},b,\vec{W}}'$ is $(x,y,P_{l_p,\vec{W}}(x),P_{l_p,\vec{W}}(y))$-sensitive for any $0<x<y$, given the facts that $\mathcal{H}_{\vec{a},b,\vec{W}}$ introduced in Section \ref{weighted LSH family} is $(x,y,P_{l_p,\vec{W}}(x),P_{l_p,\vec{W}}(y))$-sensitive for any $0<x<y$ and $\mathcal{H}_{\vec{a},b,\vec{W}}'$ is the same as $\mathcal{H}_{\vec{a},b,\vec{W}}$ when $i=1$.
	By letting $i=c^{\lceil\log_cr_{max/min}^{\mathcal{S}^{\circ}}\rceil}$ and $\vec{W}=\vec{W^{\circ}}$ for $\mathcal{H}_{\vec{a},b,\vec{W}}'$, we obtain that $\mathcal{H}_{\vec{a},b^*,\vec{W^{\circ}}}$ is $(x,y,P_{l_p,\vec{W^{\circ}}}(x),P_{l_p,\vec{W^{\circ}}}(y))$-sensitive for any $0<x<y$.
	
	(2). Let $l\in\{c,c^2,\ldots,c^{\lceil\log_cr_{max/min}^{\mathcal{S}^{\circ}}\rceil}\}$. From Equation \ref{weighted function1}, $\mathcal{H}_{\vec{a},b^*,\vec{W^{\circ}}}^l=\{h_{\vec{a},b^*,\vec{W^{\circ}}}^l:\mathcal{X}\rightarrow U\}$, where $h_{\vec{a},b^*,\vec{W^{\circ}}}^l(x)=\lfloor\frac{\lfloor\frac{\vec{a}\cdot\left(\vec{W^{\circ}}\circ\vec{x}\right)+b^*}{w}\rfloor}{l}\rfloor=\lfloor\frac{\vec{a}\cdot\left(\vec{W^{\circ}}\circ\vec{x}\right)+b^*}{wl}\rfloor$.
	Create a space $\mathcal{X'}$ by dividing all coordinates of $\mathcal{X}$ by $l$. It is easy to see that the distance between two points in space $\mathcal{X}$ under $\vec{W^{\circ}}$ is $l$ times the distance between their converted points in space $\mathcal{X'}$ under $\vec{W^{\circ}}$.
	Define $\mathcal{H}_{\vec{a},b^*,\vec{W^{\circ}}}'=\{h_{\vec{a},b^*,\vec{W^{\circ}}}':\mathcal{X'}\rightarrow U\}$, where
	\begin{equation}
	h_{\vec{a},b^*,\vec{W^{\circ}}}'(x')=\lfloor\frac{\vec{a}\cdot\left(\vec{W^{\circ}}\circ\vec{x'}\right)+\left(\frac{b^*}{f}\right)\times\left(\frac{f}{l}\right)}{w}\rfloor,
	\end{equation}
	$f=c^{\lceil\log_cr_{max/min}^{\mathcal{S}^{\circ}}\rceil}$, and $\vec{a}$, $b^*$ and $w$ are the same as in $\mathcal{H}_{\vec{a},b^*,\vec{W^{\circ}}}^l$. As it can be seen, $\frac{b^*}{f}$ is a real number chosen uniformly at random from $[0,w]$, and $\frac{f}{l}$ is an integer greater than or equal to 1. 
	Thus, since $\mathcal{H}_{\vec{a},b,\vec{W}}'$ in the proof of Lemma \ref{lemma 1}(1) is $(x,y,P_{l_p,\vec{W}}(x),P_{l_p,\vec{W}}(y))$-sensitive for any $0<x<y$, 
	we have that $\mathcal{H}_{\vec{a},b^*,\vec{W^{\circ}}}'$ is $(x,y,P_{l_p,\vec{W^{\circ}}}(x),P_{l_p,\vec{W^{\circ}}}(y))$-sensitive for any $0<x<y$.
	Let $o$ be any point in space $\mathcal{X}$ and $o'$ be the corresponding point in space $\mathcal{X'}$. It is easy to know that $h_{\vec{a},b^*,\vec{W^{\circ}}}^l(o)=h_{\vec{a},b^*,\vec{W^{\circ}}}'(o')$. 
	This means that the probability of any two points in space $\mathcal{X}$ colliding under an LSH function from $\mathcal{H}_{\vec{a},b^*,\vec{W^{\circ}}}^l$ is the same as the probability of the corresponding two points in space $\mathcal{X'}$ colliding under an LSH function from $\mathcal{H}_{\vec{a},b^*,\vec{W^{\circ}}}'$.
	As a result, $\mathcal{H}_{\vec{a},b^*,\vec{W^{\circ}}}^l$ is $(xl,yl,P_{l_p,\vec{W^{\circ}}}(x),P_{l_p,\vec{W^{\circ}}}(y))$-sensitive for any $0<x<y$ and $l\in\{c,c^2,\ldots,c^{\lceil\log_cr_{max/min}^{\mathcal{S}^{\circ}}\rceil}\}$. Since $r_{max/min}^{\mathcal{S}^{\circ}}\geq r_{max}^{\vec{W^{\circ}}}/r_{min}^{\vec{W^{\circ}}}$, Lemma \ref{lemma 1}(2) holds.
\end{IEEEproof}
\section{Proof of Theorem \ref{theorem 2}}
\label{proof of theorem 2}
\begin{IEEEproof}
	\mbox{}\par
	(1). Theorem \ref{theorem 2}(1) can be directly obtained from Section \ref{derived weighted LSH family} and Lemma \ref{lemma 1}(1) (It is easy to know that $0<x^{\uparrow}<y^{\downarrow}$ implies $0<x<y$ such that Lemma \ref{lemma 1}(1) is available).
	
	(2).
	Define $P_{l_p,\vec{W^{\circ}},l}(r)=P_{l_p,\vec{W^{\circ}}}(\frac{r}{l})$ for $r>0$. 
	According to the proof of Lemma \ref{lemma 1}(2), $\mathcal{H}_{\vec{a},b^*,\vec{W^{\circ}}}^l$ is $(xl,yl,P_{l_p,\vec{W^{\circ}},l}(xl),P_{l_p,\vec{W^{\circ}},l}(yl))$-sensitive for any $0<x<y$ and $l\in\{c,c^2,\ldots,c^{\lceil\log_cr_{max/min}^{\mathcal{S}^{\circ}}\rceil}\}$.
	Thus, $\mathcal{H}_{\vec{a},b^*,\vec{W^{\circ}}\rightarrow\vec{W_i}}^l$ is 
	$(xl,yl,P_{l_p,\vec{W^{\circ}},l}((xl)^{\uparrow}),P_{l_p,\vec{W^{\circ}},l}((yl)^{\downarrow}))$-sensitive for $l\in\{c,c^2,\ldots,c^{\lceil\log_cr_{max/min}^{\mathcal{S}^{\circ}}\rceil}\}$ from Section \ref{derived weighted LSH family}.
	It is easy to know that
	$(xl)^{\uparrow}=x^{\uparrow}l$ and $(yl)^{\downarrow}=y^{\downarrow}l$ from Theorem \ref{R_bound}(1).
	As a result,
	$\mathcal{H}_{\vec{a},b^*,\vec{W^{\circ}}\rightarrow\vec{W_i}}^l$ is 
	$(xl,yl,P_{l_p,\vec{W^{\circ}}}(x^{\uparrow}),P_{l_p,\vec{W^{\circ}}}(y^{\downarrow}))$-sensitive for $l\in\{c,c^2,\ldots,c^{\lceil\log_cr_{max/min}^{\mathcal{S}^{\circ}}\rceil}\}$. Since $r_{max/min}^{\mathcal{S}^{\circ}}\geq r_{max}^{\vec{W_i}}/r_{min}^{\vec{W_i}}$, Theorem \ref{theorem 2}(2) holds.
\end{IEEEproof}
\section{More Experimental Results}
\subsection{Necessity of Bound Relaxation}
\label{necessity of bound relaxation}
\begin{table}[h]
	\centering
	\footnotesize
	\caption{Evaluation of Bound Relaxation ($\left|\mathcal{S}\right|=5\text{k}$)}
		\begin{threeparttable}
		\begin{tabular}{c|c|c|c|c|c}
			\hline
			\multicolumn{6}{c}{$l_1$ distance}\\\hline\hline
			\multicolumn{6}{c}{\textit{Sift}}\\\hline
			\boldmath$c$&$5$&$7$&$9$&$11$&$13$\\\hline
			\boldmath$\beta_{\mathcal{S}}$&1,500,000&1,280,000&1,180,000&1,120,000&1,080,000\\\hline
			\boldmath$\beta_{\mathcal{S}}^{br}$&451,165&15,651&4,965&2,562&2,254\\\hline
			\multicolumn{6}{c}{\textit{Ukbench}}\\\hline
			\boldmath$\beta_{\mathcal{S}}$&1,510,000&1,290,000&1,185,000&1,125,000&1,085,000\\\hline
			\boldmath$\beta_{\mathcal{S}}^{br}$&467,427&16,450&5,123&2,577&2,825\\\hline
			\multicolumn{6}{c}{\textit{Notre}}\\\hline
			\boldmath$\beta_{\mathcal{S}}$&1,400,000&1,195,000&1,100,000&1,045,000&1,010,000\\\hline
			\boldmath$\beta_{\mathcal{S}}^{br}$&304,462&11,230&4,134&2,995&2,103\\\hline
			\multicolumn{6}{c}{\textit{Sun}}\\\hline
			\boldmath$\beta_{\mathcal{S}}$&1,265,000&1,080,000&995,000&945,000&910,000\\\hline
			\boldmath$\beta_{\mathcal{S}}^{br}$&1,262,616&26,658&2,754&2,252&1,969\\\hline
			\hline
			\multicolumn{6}{c}{$l_2$ distance}\\\hline\hline
			\multicolumn{6}{c}{\textit{Sift}}\\\hline
			\boldmath$c$&$5$&$7$&$9$&$11$&$13$\\\hline
			\boldmath$\beta_{\mathcal{S}}$&840,000&720,000&665,000&635,000&615,000\\\hline
			\boldmath$\beta_{\mathcal{S}}^{br}$&394,355&17,729&3,212&1,761&1,591\\\hline
			\multicolumn{6}{c}{\textit{Ukbench}}\\\hline
			\boldmath$\beta_{\mathcal{S}}$&845,000&725,000&670,000&640,000&620,000\\\hline
			\boldmath$\beta_{\mathcal{S}}^{br}$&405,239&18,307&3,712&1,767&1,602\\\hline
			\multicolumn{6}{c}{\textit{Notre}}\\\hline
			\boldmath$\beta_{\mathcal{S}}$&785,000&675,000&625,000&595,000&575,000\\\hline
			\boldmath$\beta_{\mathcal{S}}^{br}$&278,013&11,813&2,859&1,639&1,487\\\hline
			\multicolumn{6}{c}{\textit{Sun}}\\\hline
			\boldmath$\beta_{\mathcal{S}}$&710,000&610,000&565,000&535,000&520,000\\\hline
			\boldmath$\beta_{\mathcal{S}}^{br}$&710,000&95,039&3,644&1,336&1,177\\\hline
	\end{tabular}
	\begin{tablenotes}
		\item[*] $\beta_{\mathcal{S}}$ is the required total number of hash tables when not using bound relaxation. $\beta_{\mathcal{S}}^{br}$ is the required total number of hash tables when using bound relaxation.
	\end{tablenotes}
	\end{threeparttable}
	\label{effectiveness of bound relaxation}
\end{table}
The experimental results in Table \ref{Space WLSH/l1 l2} indicate that the space consumption of WLSH can be reduced by setting the approximation ratio $c$ to a larger value. One may doubt the necessity of introducing bound relaxation to WLSH. Here we demonstrate it by using uniformly random weight vector sets and varying $c$ in the range \{5, 7, 9, 11, 13\}. The experiments are conducted on real data sets, and the results are shown in Table \ref{effectiveness of bound relaxation}. As it can be seen from the table, when bound relaxation is not used, the required total number of hash tables for WLSH decreases slower with the increase of $c$. Worse still, it is still very large when $c=13$. This implies that it is limited to reduce the space consumption of WLSH only by increasing $c$.
In contrast, when bound relaxation is used, the required total number of hash tables for WLSH is acceptable if $c\geq 7$. 
Thus, it can be concluded that the approach of bound relaxation is necessary for reducing the space consumption of WLSH.
\subsection{Query Efficiency and Query Accuracy of WLSH}
\label{WLSH/query efficiency and query accuracy/more experiments}
\begin{figure*}[t]
	\centering
	\subfigure[I/O cost vs. $d$]{\includegraphics[width=0.246\textwidth]{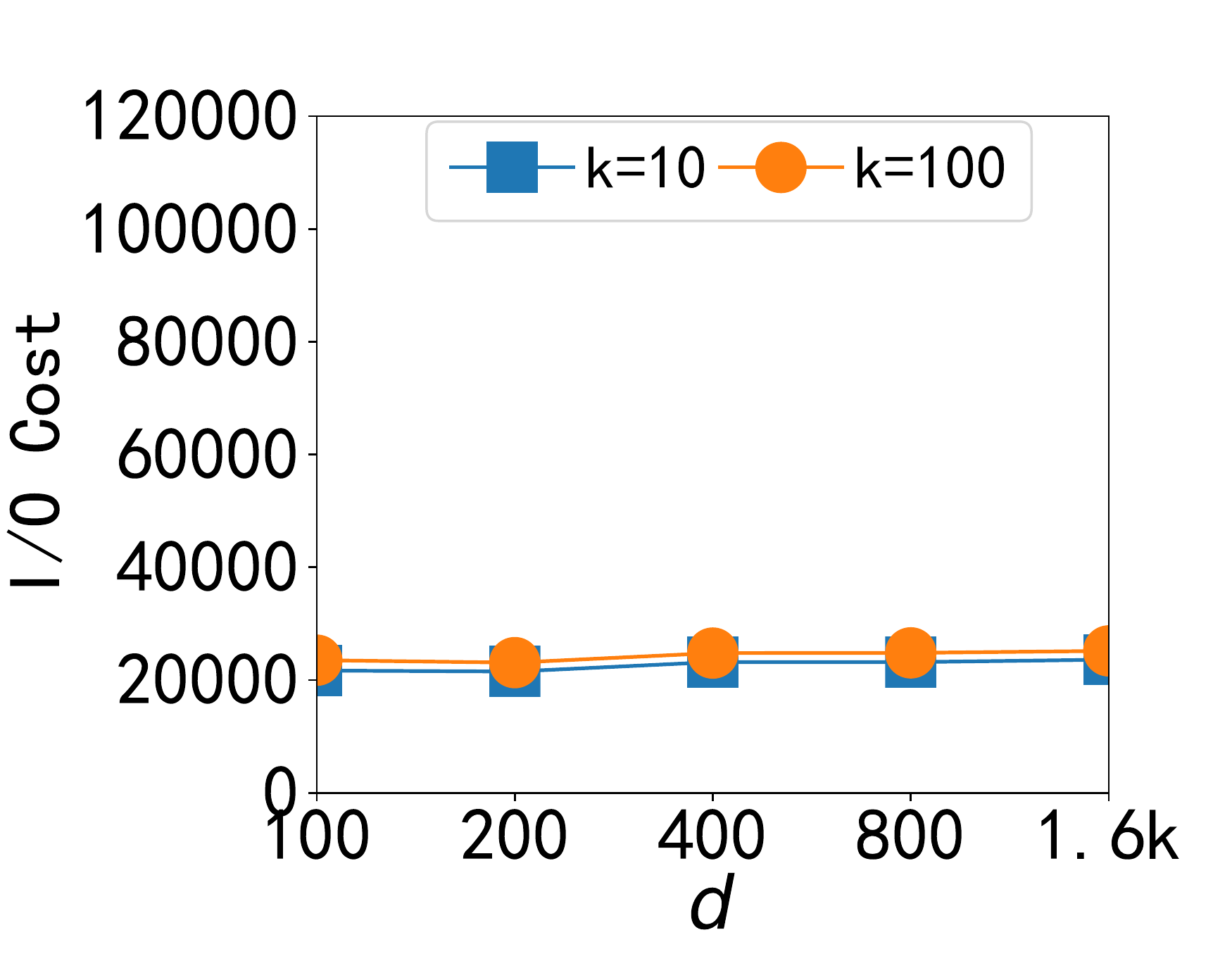}\label{rand/d/IO_useCt=0_L1}}
	\subfigure[I/O cost vs. $n$]{\includegraphics[width=0.246\textwidth]{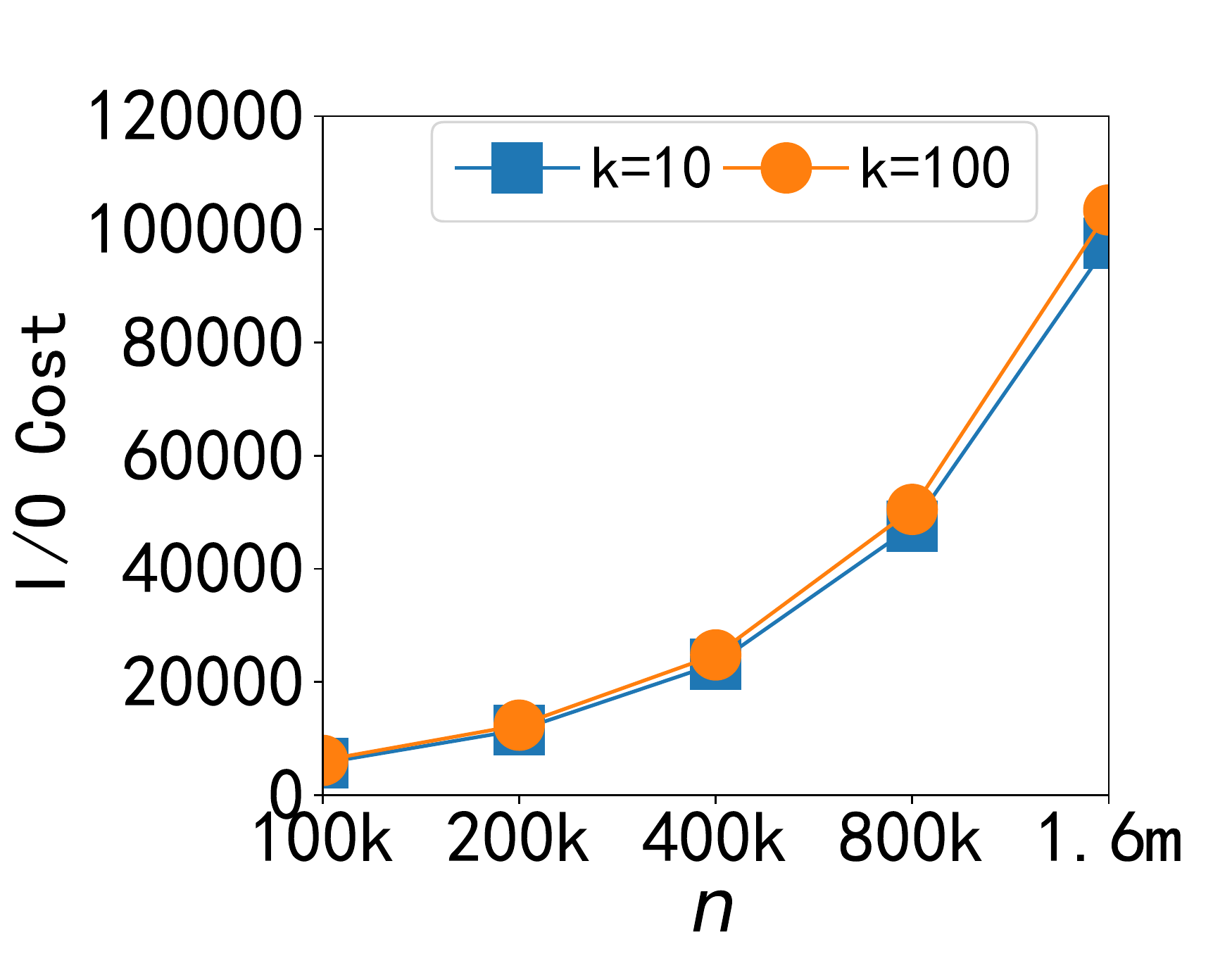}\label{rand/n/IO_useCt=0_L1}}
	\subfigure[I/O cost vs. $c$]{\includegraphics[width=0.246\textwidth]{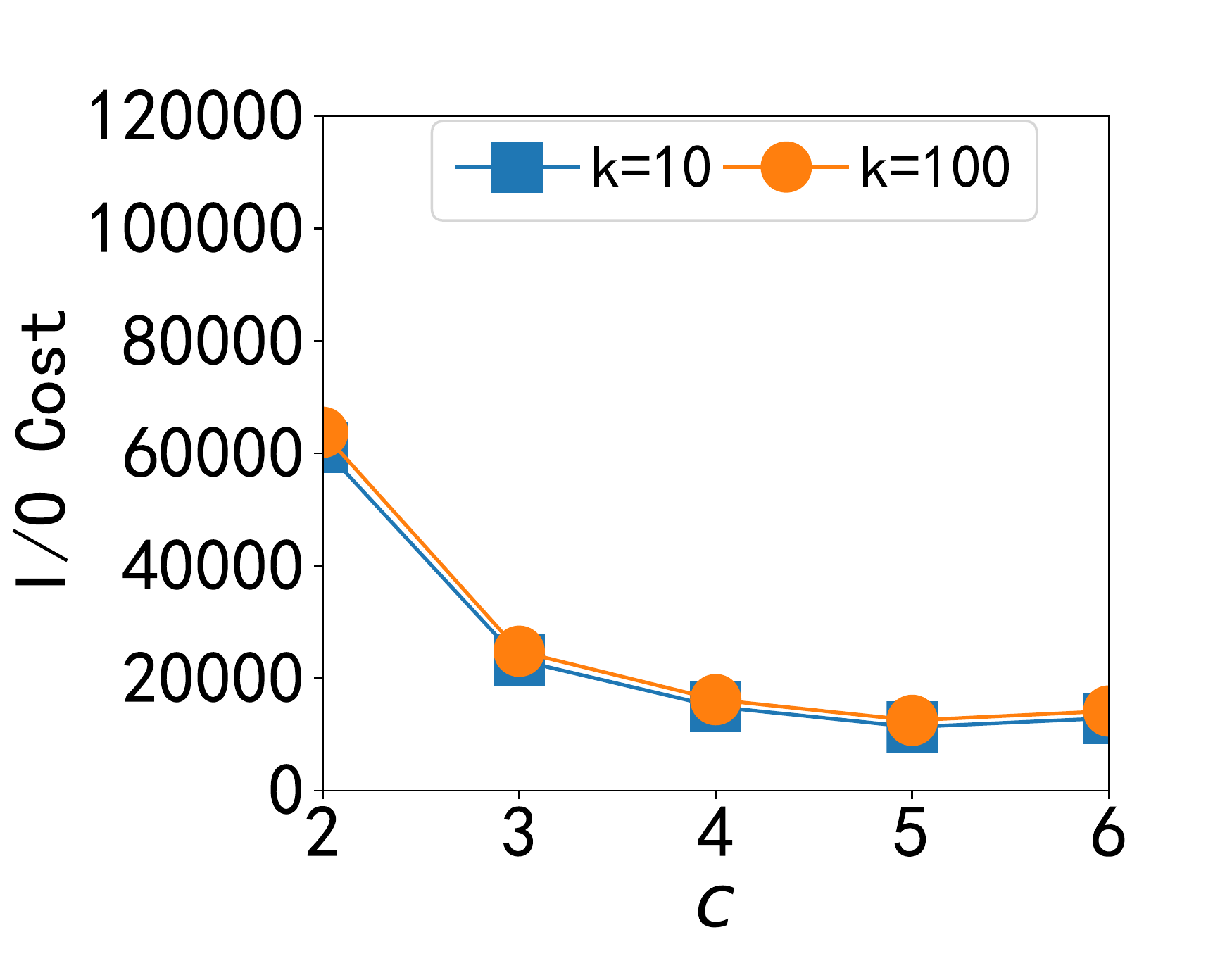}\label{rand/c/IO_useCt=0_L1}}
	\subfigure[I/O cost vs. $\#Subrange$]{\includegraphics[width=0.246\textwidth]{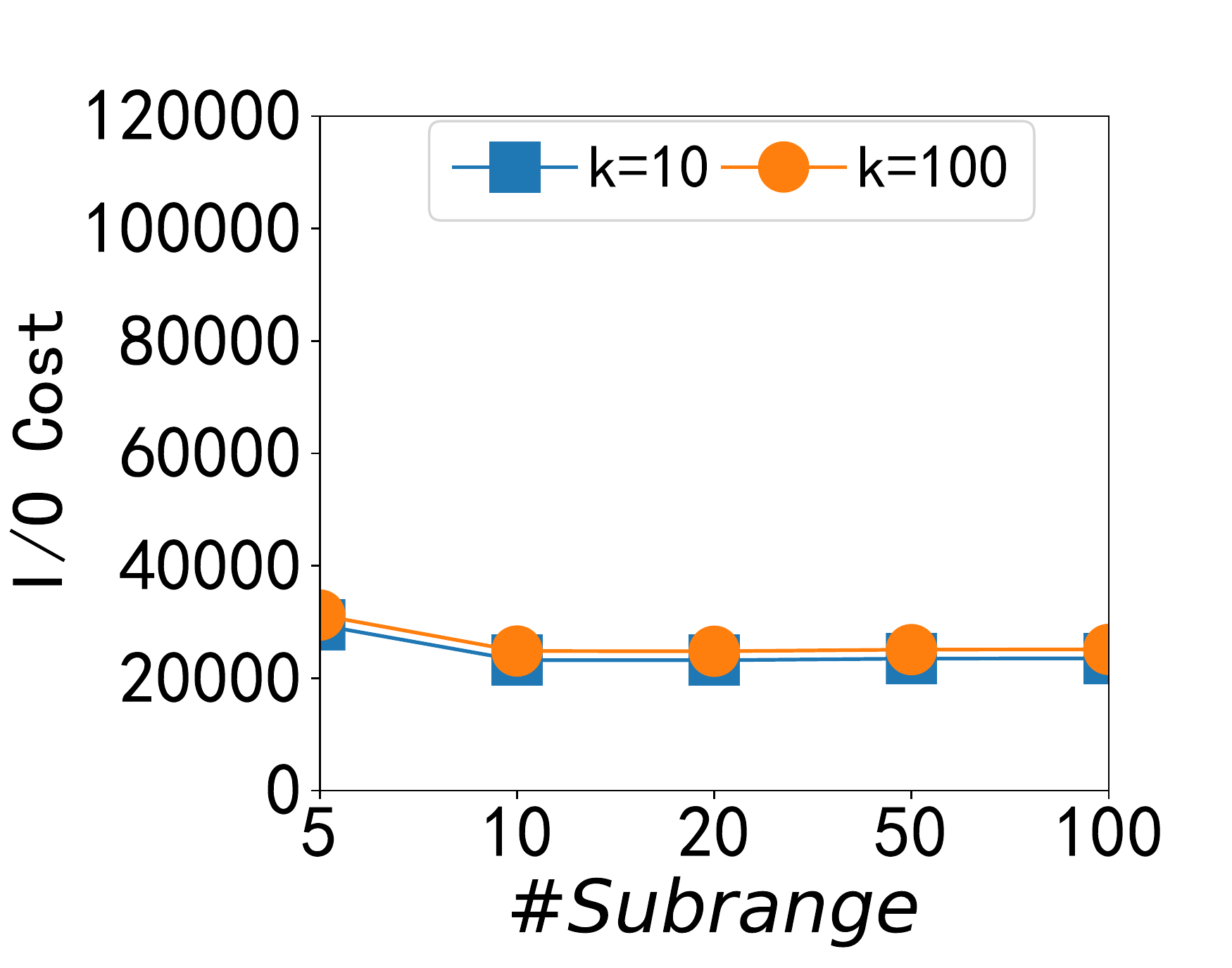}\label{rand/subrange/IO_useCt=0_L1}}
	
	\subfigure[I/O cost vs. $\#Subset$]{\includegraphics[width=0.246\textwidth]{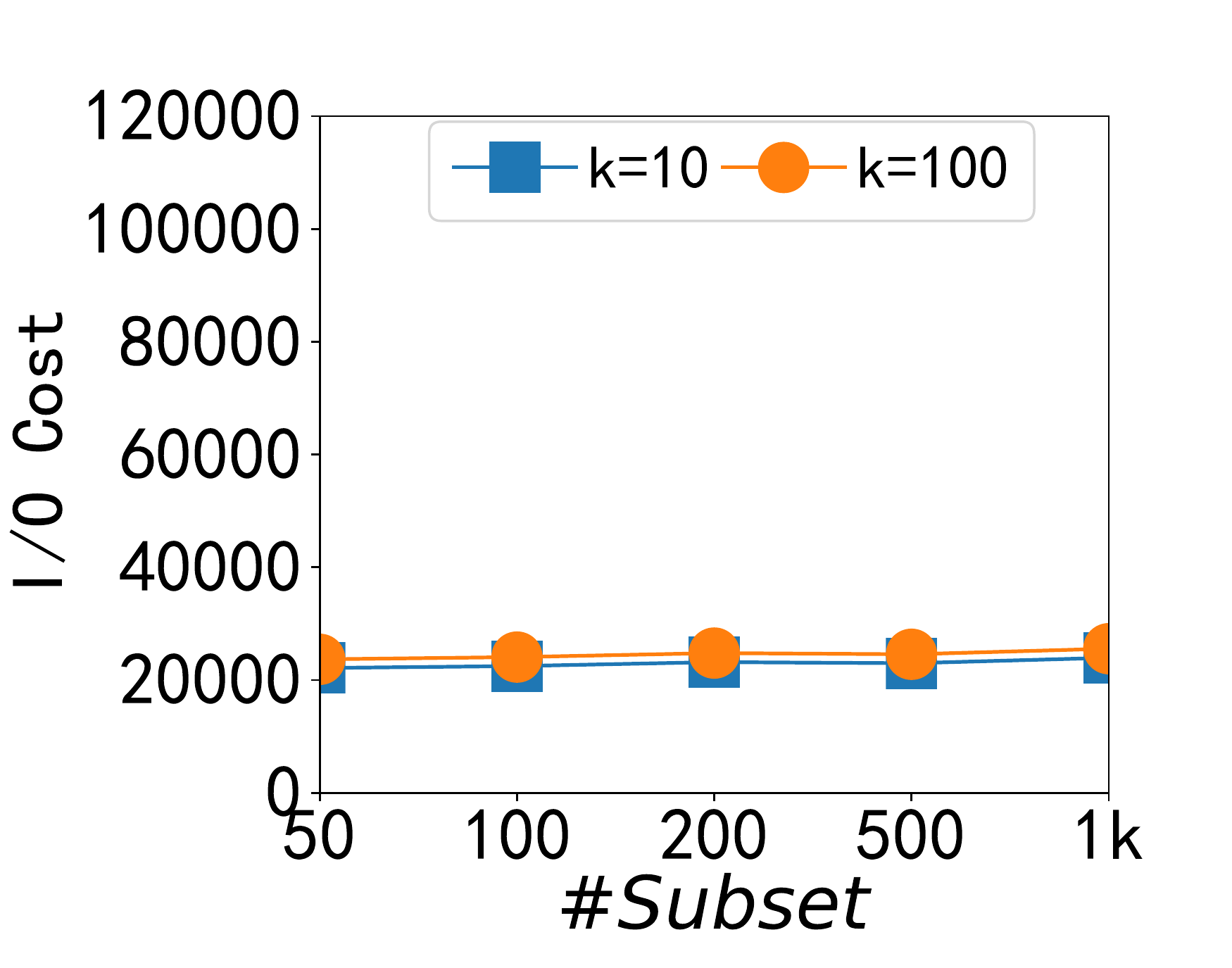}\label{rand/subset/IO_useCt=0_L1}}
	\subfigure[I/O cost vs. $\left|S\right|$]{\includegraphics[width=0.246\textwidth]{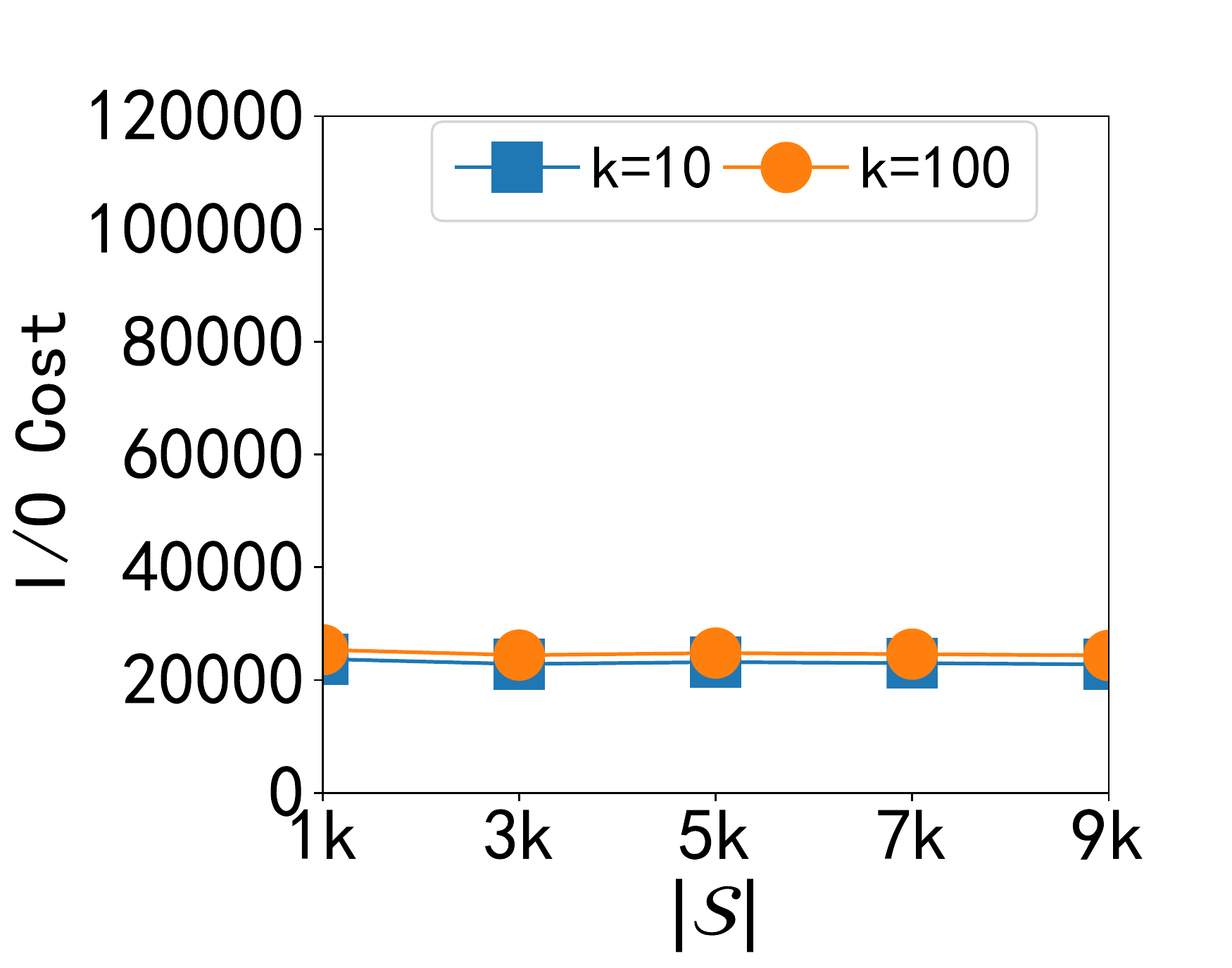}\label{rand/S/IO_useCt=0_L1}}
	\subfigure[Ratio vs. $d$]{\includegraphics[width=0.246\textwidth]{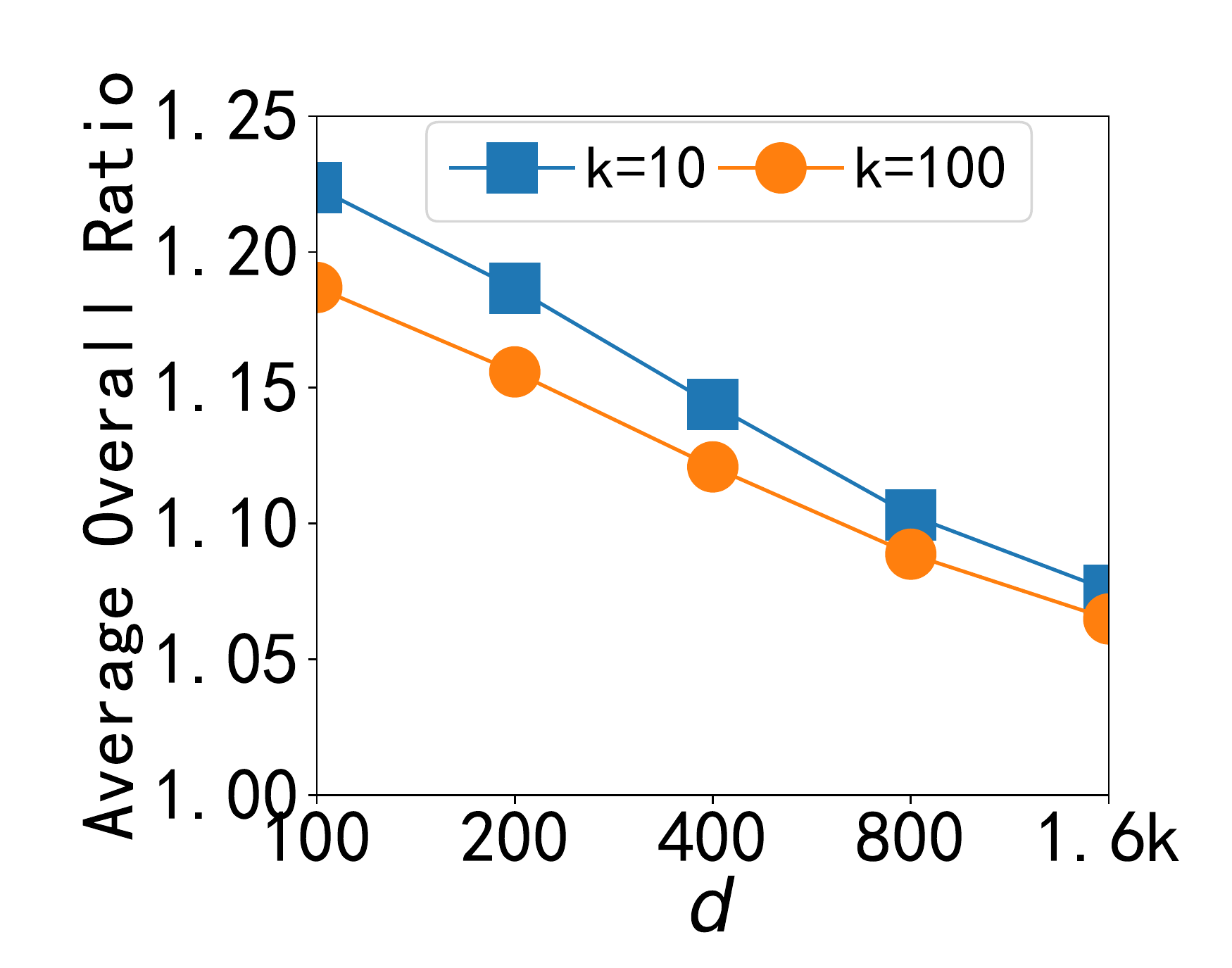}\label{rand/d/ratio_useCt=0_L1}}
	\subfigure[Ratio vs. $n$]{\includegraphics[width=0.246\textwidth]{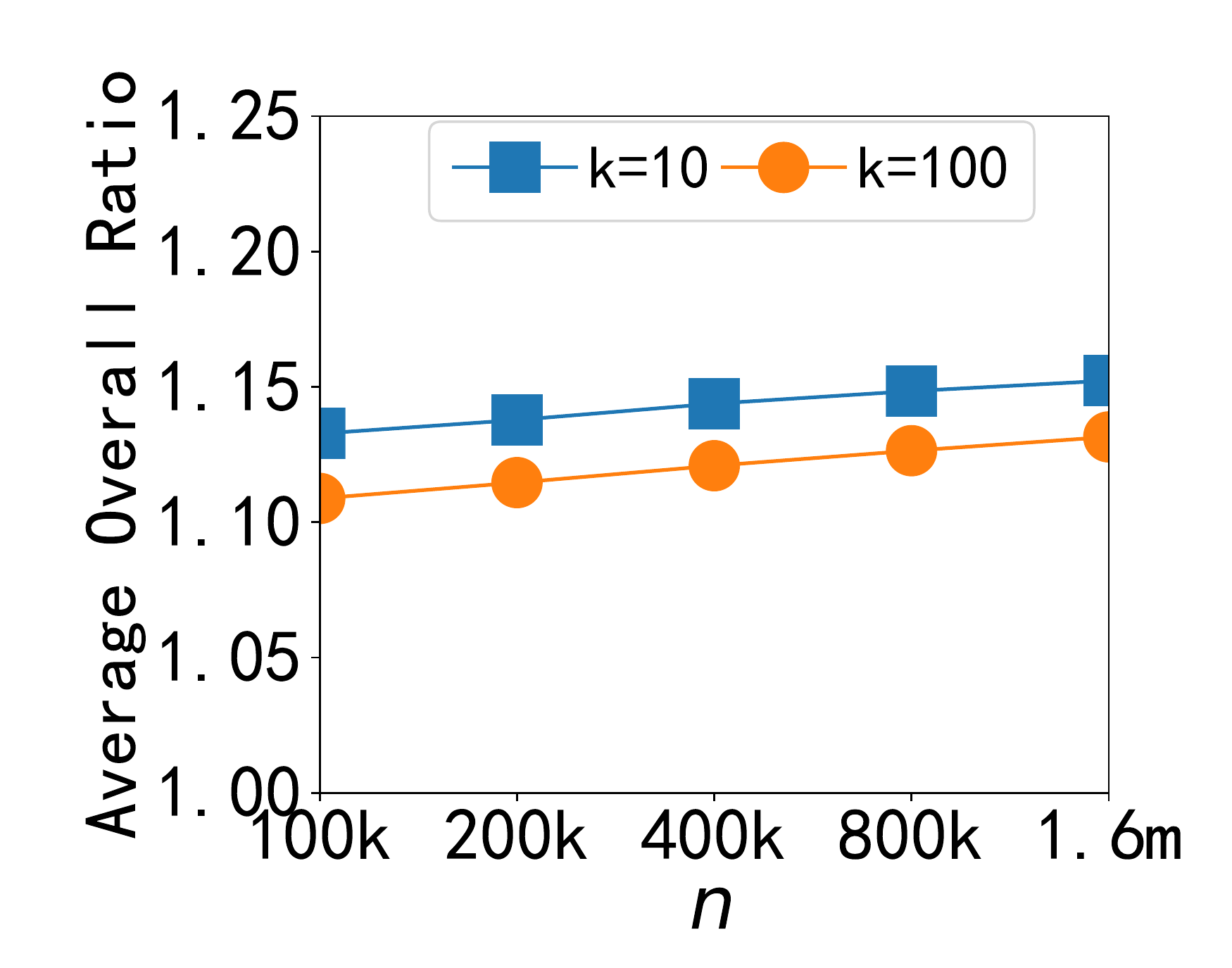}\label{rand/n/ratio_useCt=0_L1}}
	
	\subfigure[Ratio vs. $c$]{\includegraphics[width=0.246\textwidth]{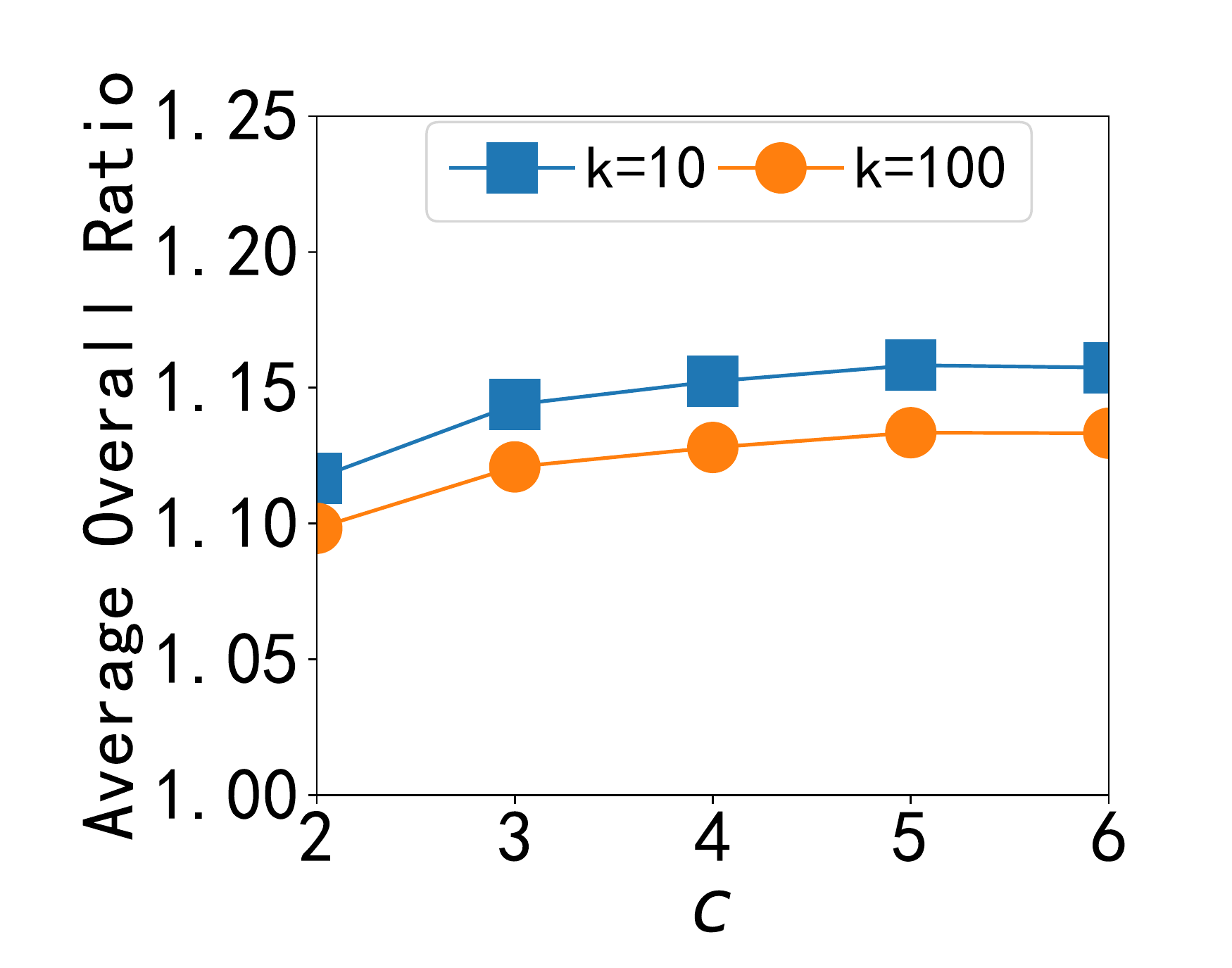}\label{rand/c/ratio_useCt=0_L1}}
	\subfigure[Ratio vs. $\#Subrange$]{\includegraphics[width=0.246\textwidth]{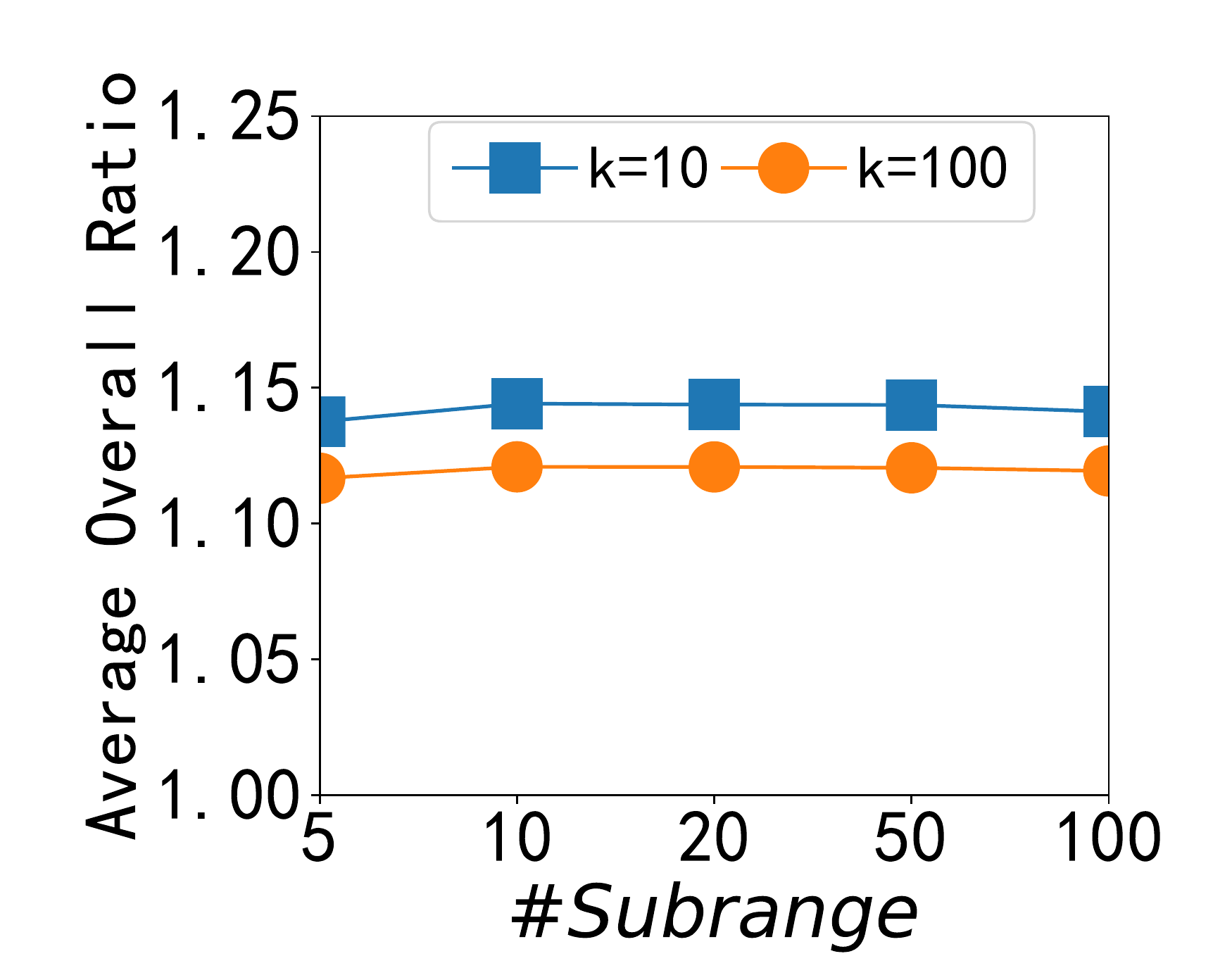}\label{rand/subrange/ratio_useCt=0_L1}}
	\subfigure[Ratio vs. $\#Subset$]{\includegraphics[width=0.246\textwidth]{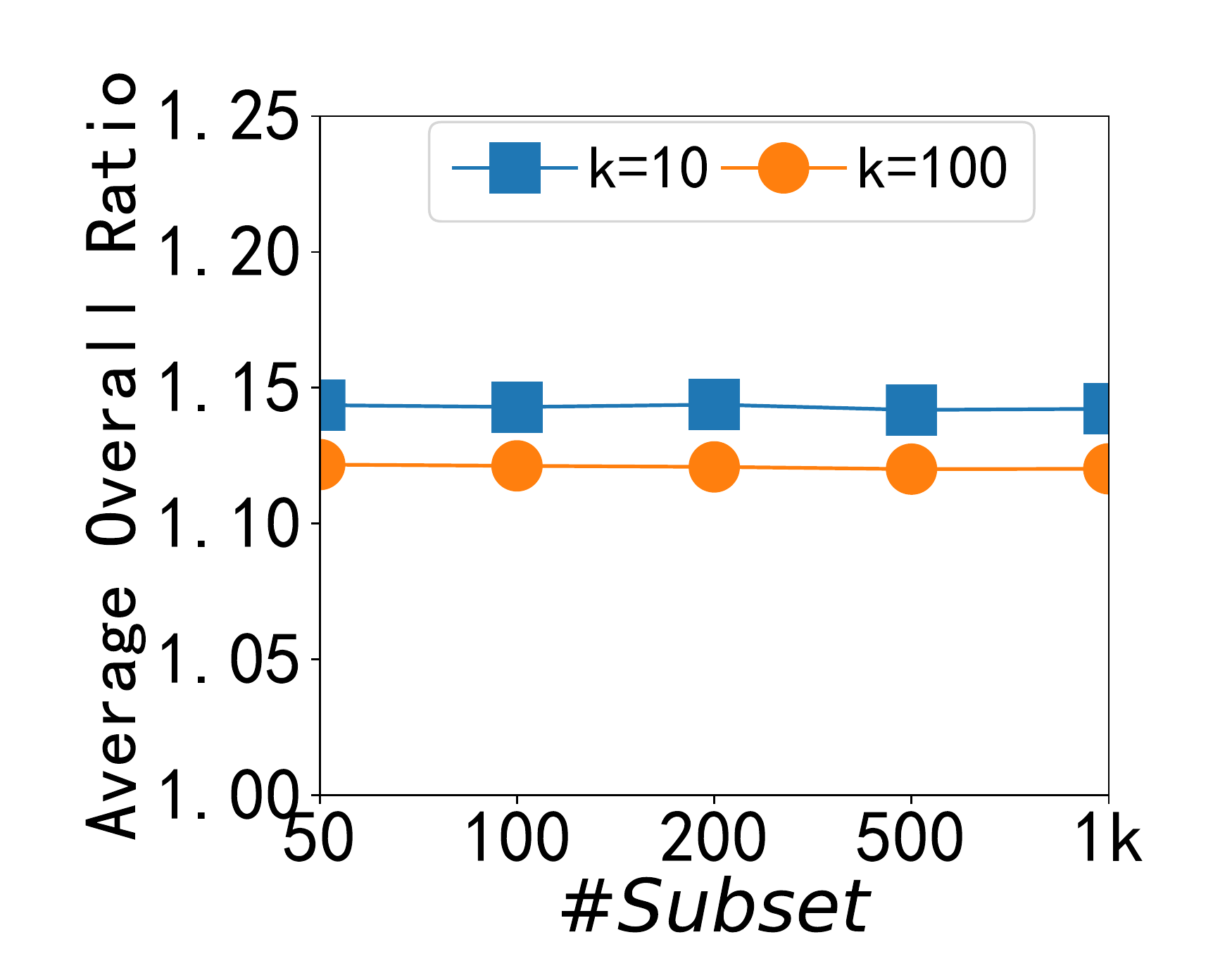}\label{rand/subset/ratio_useCt=0_L1}}
	\subfigure[Ratio vs. $\left|S\right|$]{\includegraphics[width=0.246\textwidth]{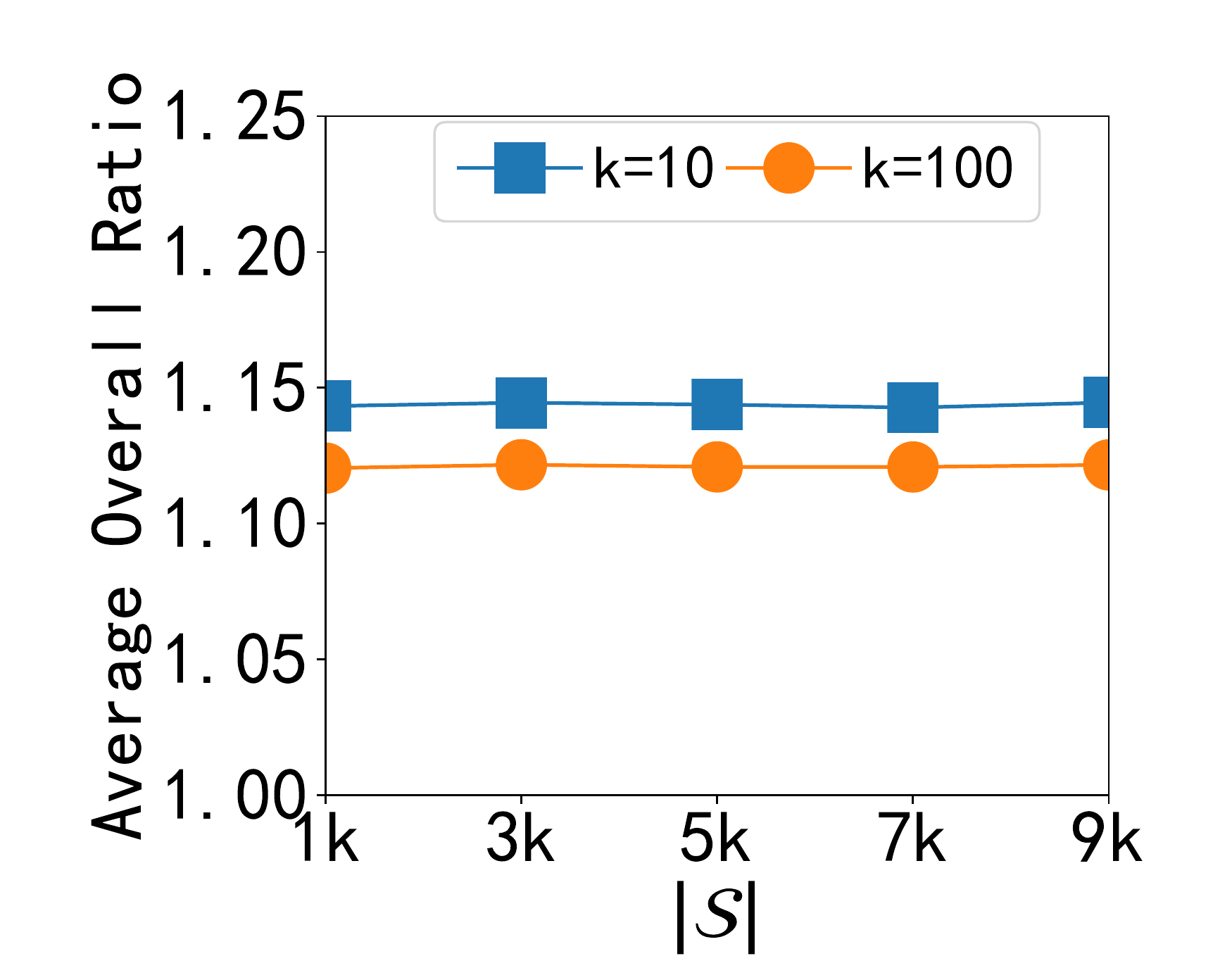}\label{rand/S/ratio_useCt=0_L1}}
	
	\caption{Query efficiency and query accuracy of WLSH on synthetic data when not using collision threshold reduction, $l_1$ distance}
	\label{rand/efficiency and accuracy/L1}
\end{figure*}

\begin{figure*}[t]
	\centering
	\subfigure[I/O cost vs. $d$]{\includegraphics[width=0.246\textwidth]{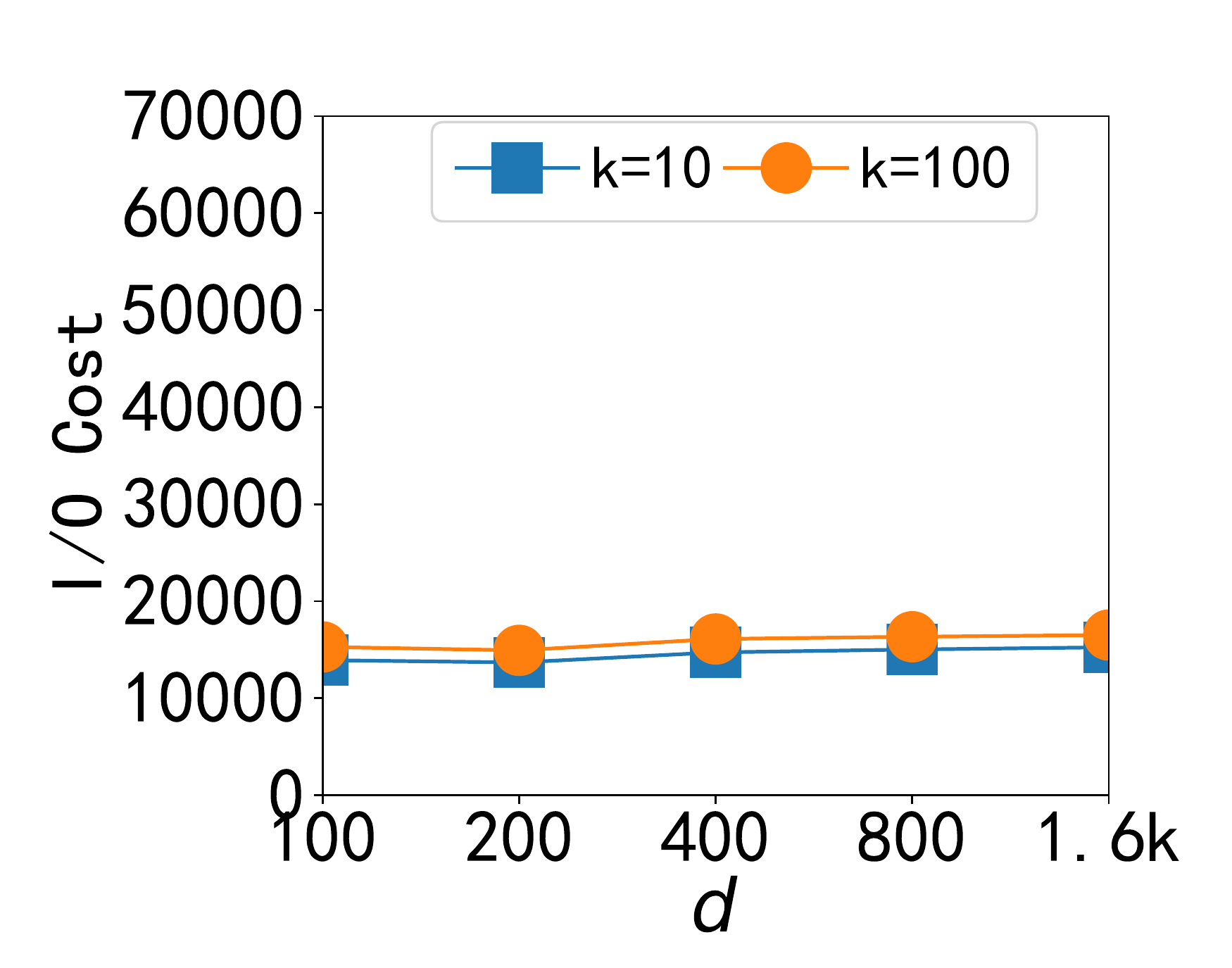}\label{rand/d/IO_useCt=0_L2}}
	\subfigure[I/O cost vs. $n$]{\includegraphics[width=0.246\textwidth]{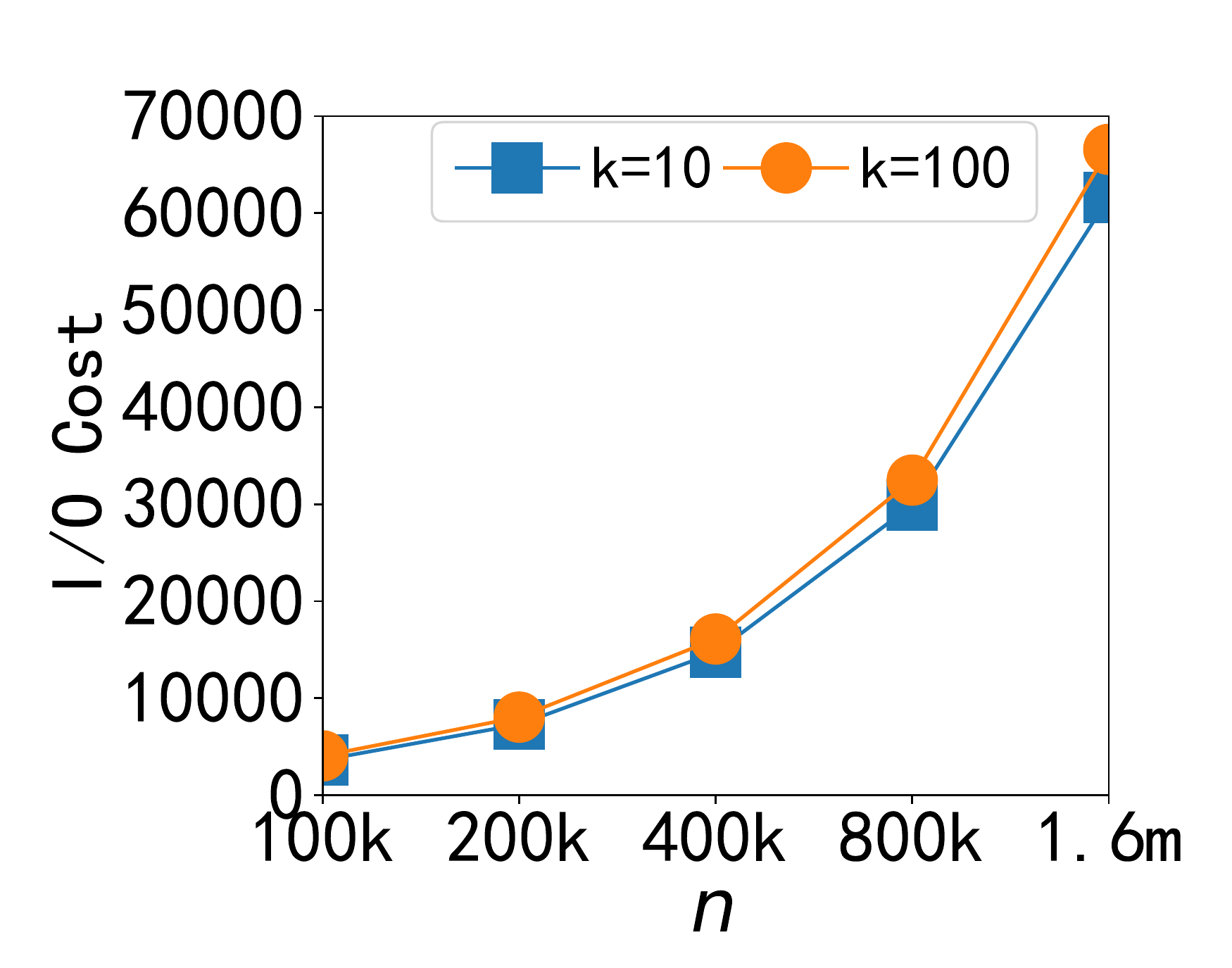}\label{rand/n/IO_useCt=0_L2}}
	\subfigure[I/O cost vs. $c$]{\includegraphics[width=0.246\textwidth]{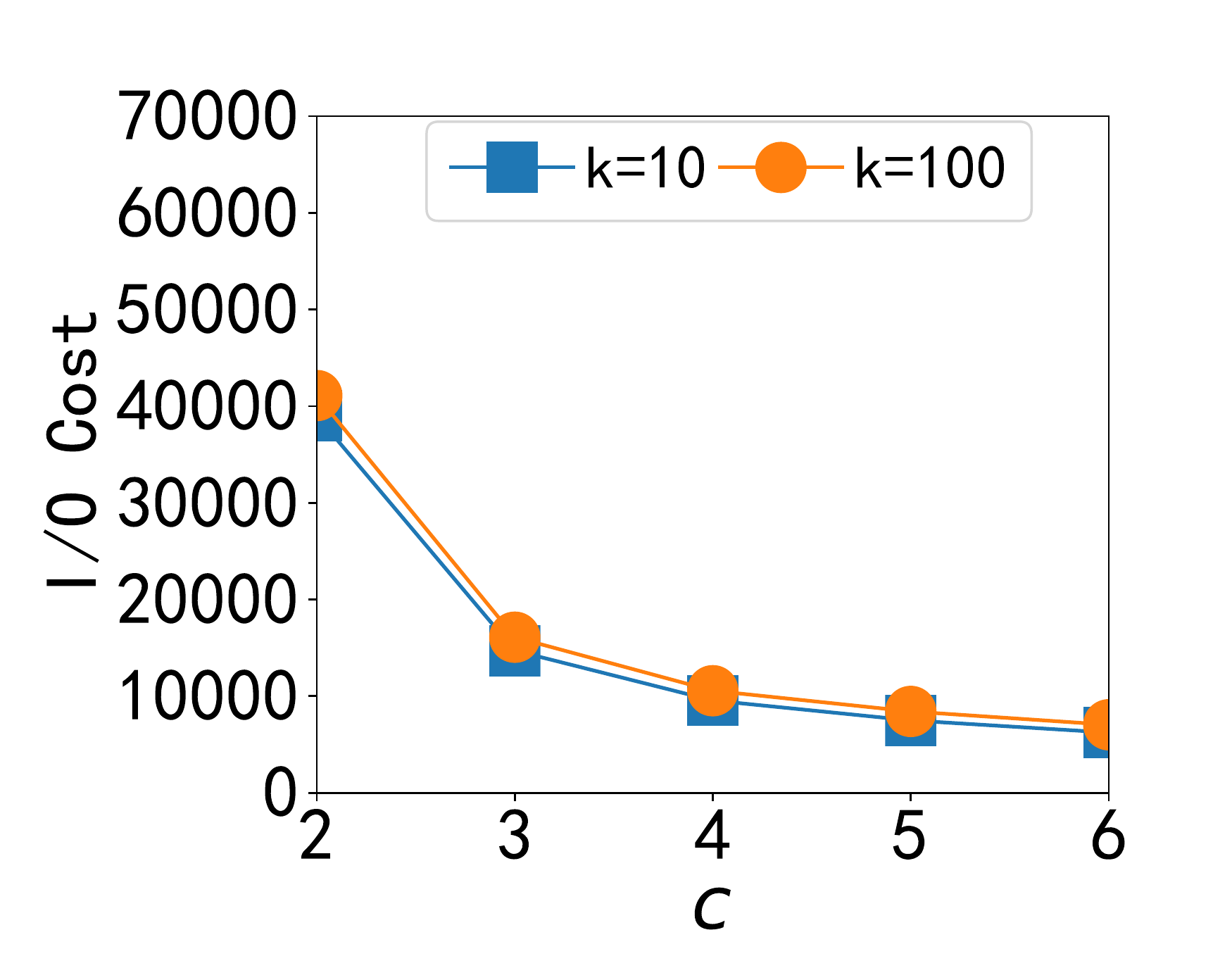}\label{rand/c/IO_useCt=0_L2}}
	\subfigure[I/O cost vs. $\#Subrange$]{\includegraphics[width=0.246\textwidth]{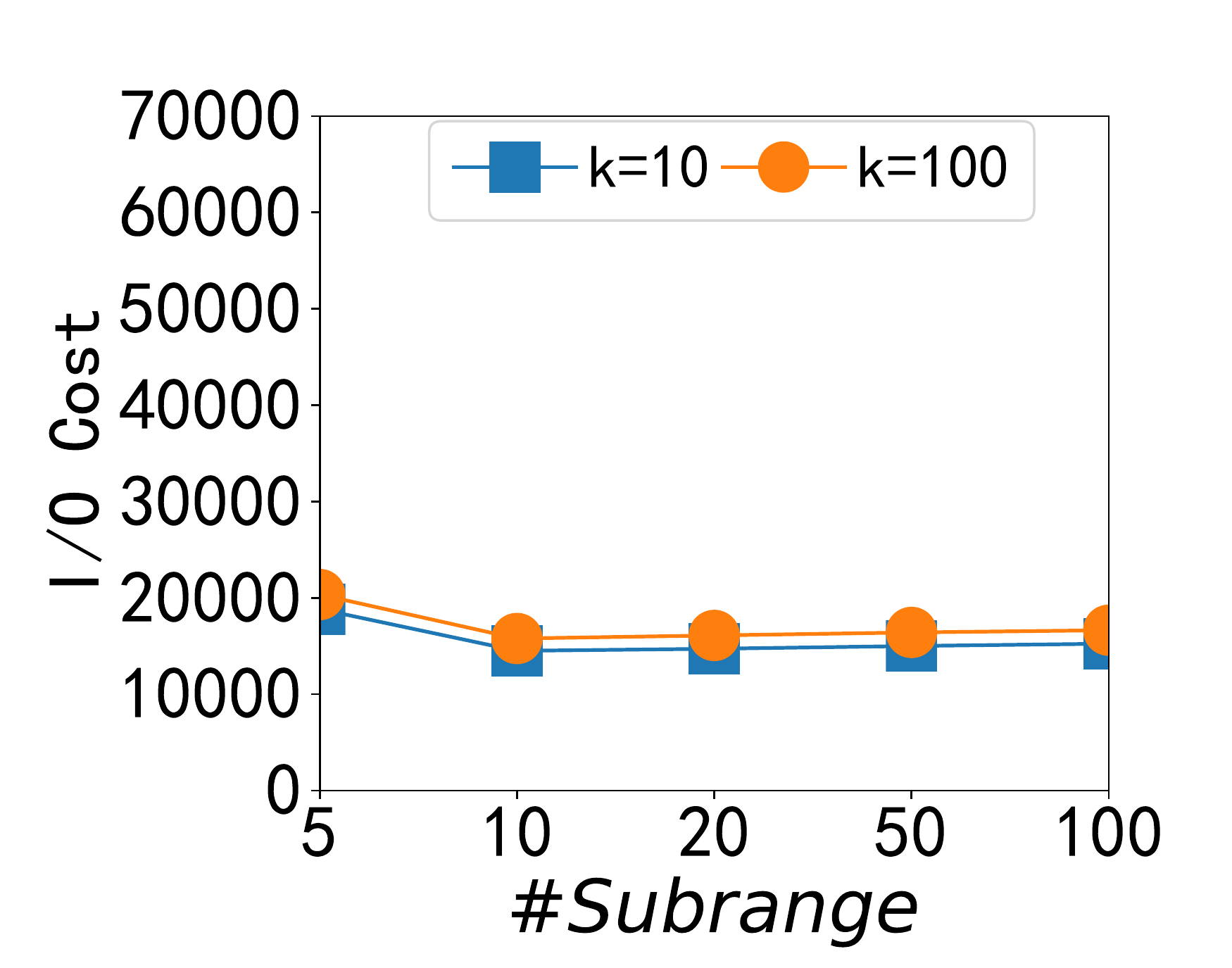}\label{rand/subrange/IO_useCt=0_L2}}
	
	\subfigure[I/O cost vs. $\#Subset$]{\includegraphics[width=0.246\textwidth]{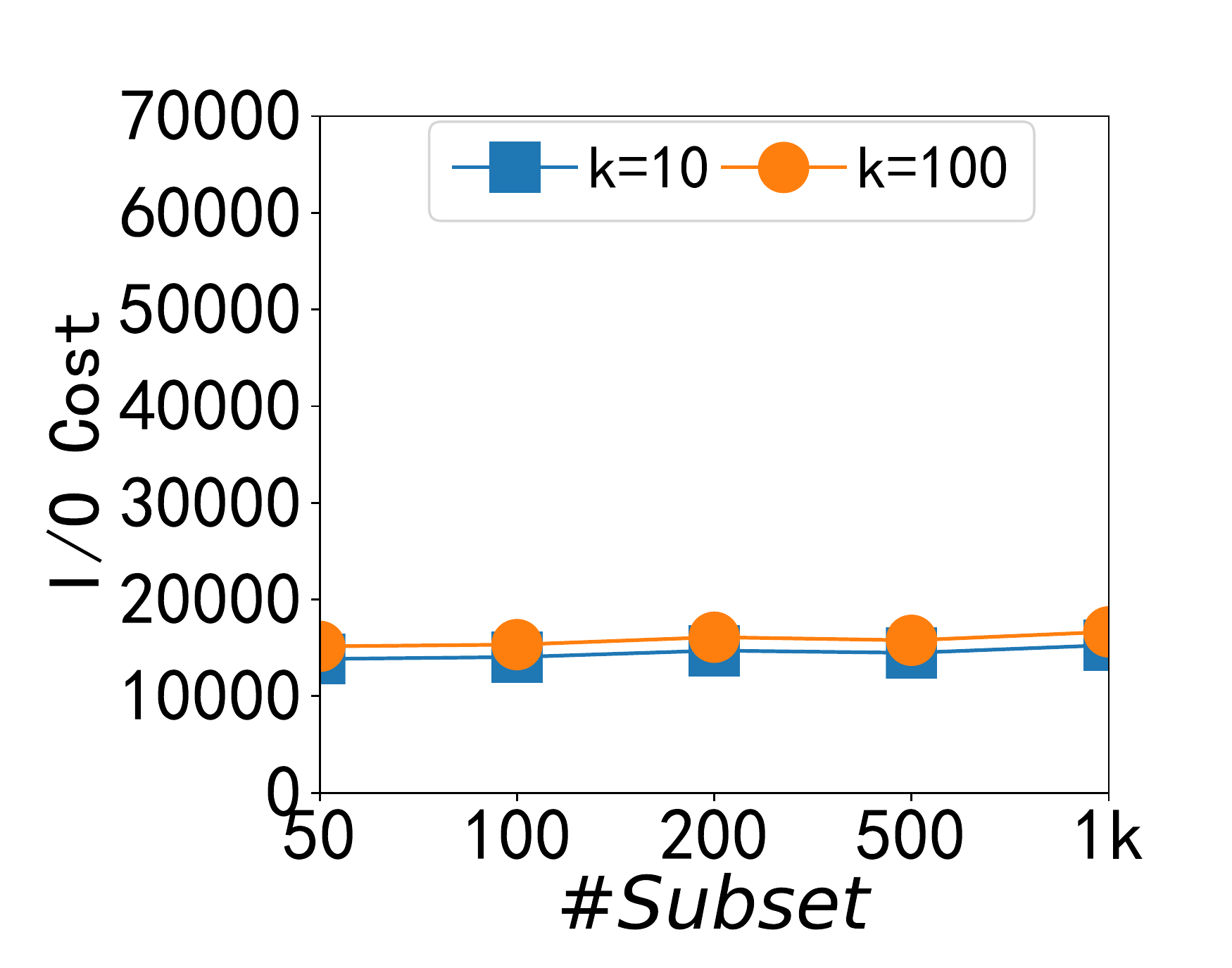}\label{rand/subset/IO_useCt=0_L2}}
	\subfigure[I/O cost vs. $\left|S\right|$]{\includegraphics[width=0.246\textwidth]{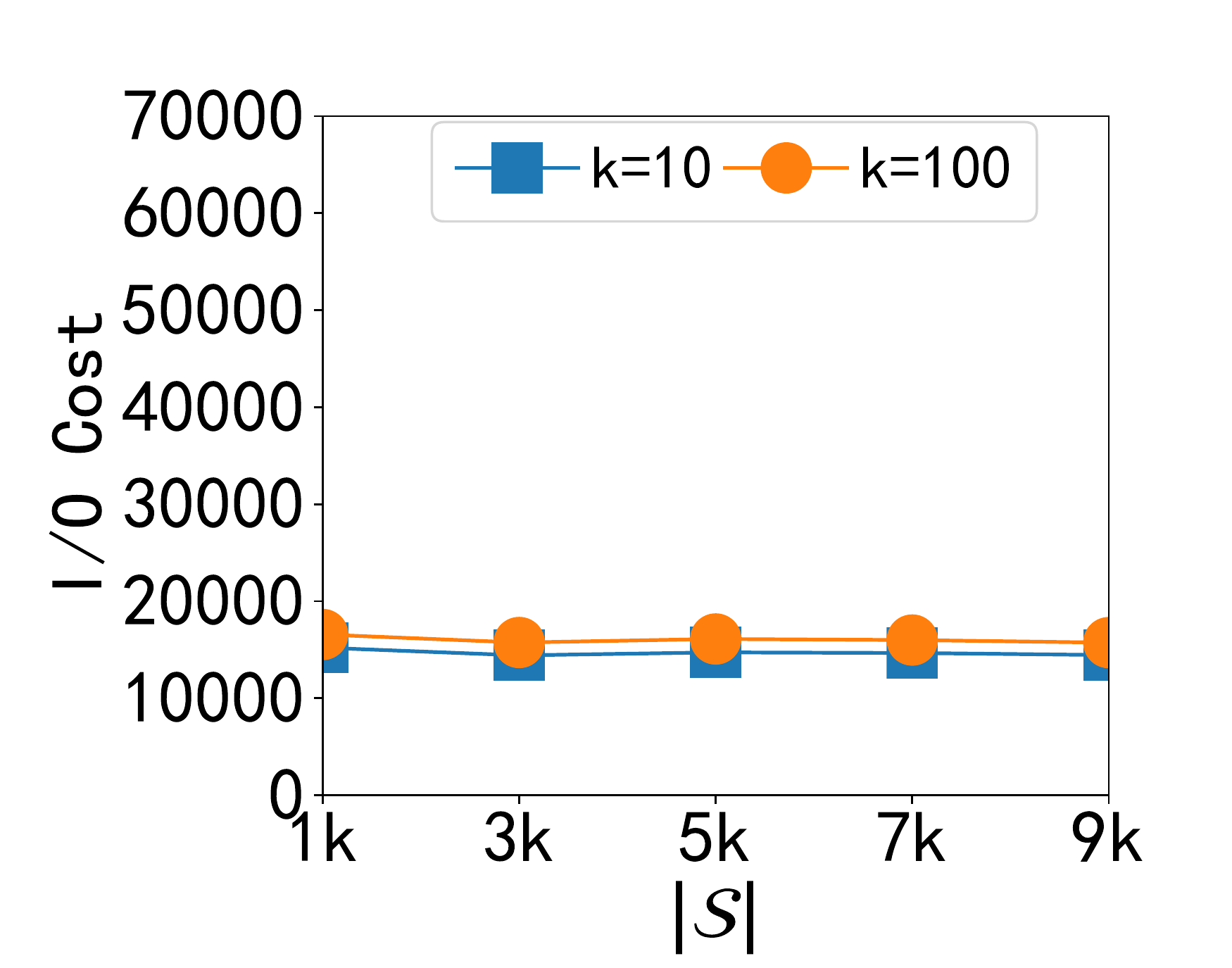}\label{rand/S/IO_useCt=0_L2}}
	\subfigure[Ratio vs. $d$]{\includegraphics[width=0.246\textwidth]{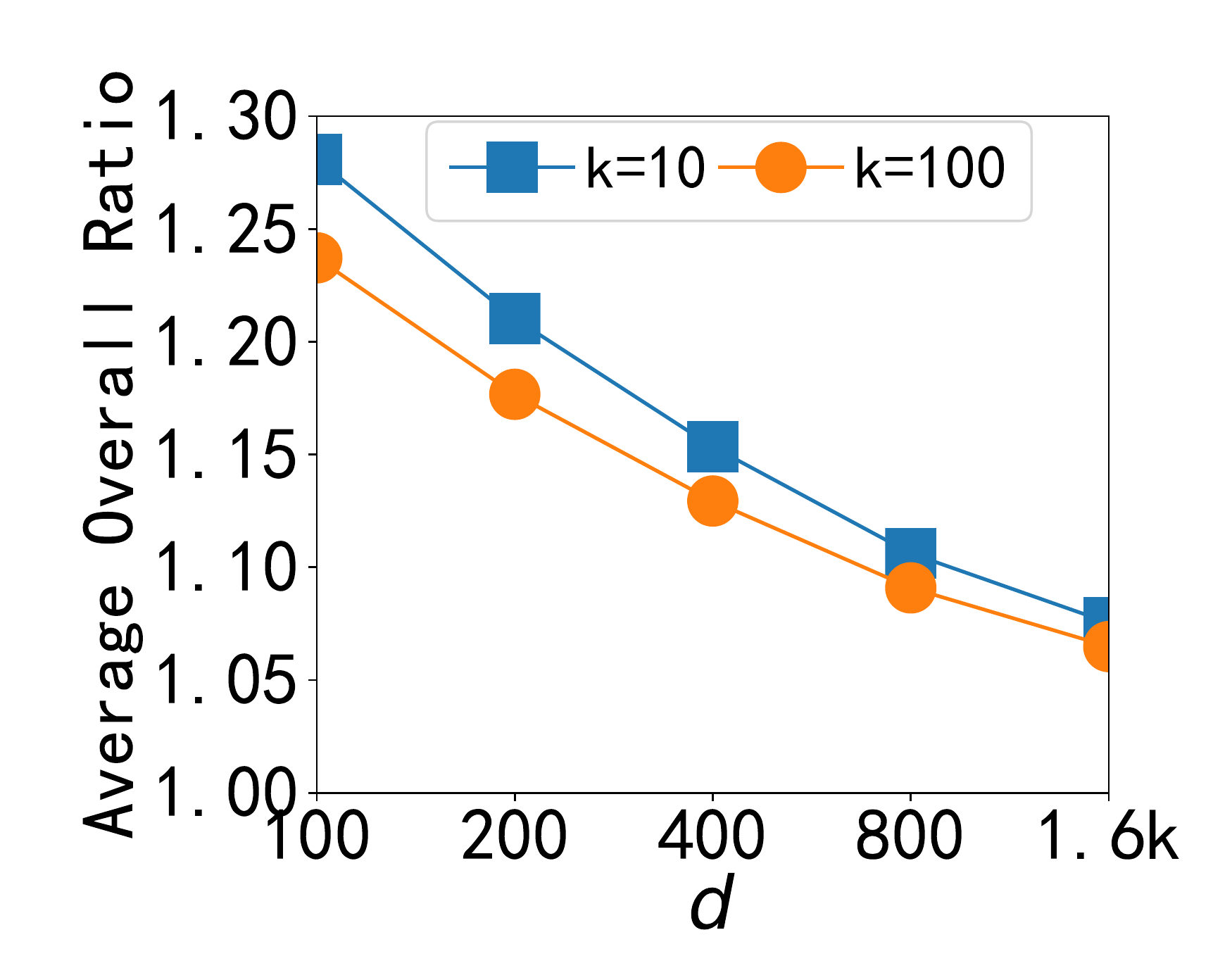}\label{rand/d/ratio_useCt=0_L2}}
	\subfigure[Ratio vs. $n$]{\includegraphics[width=0.246\textwidth]{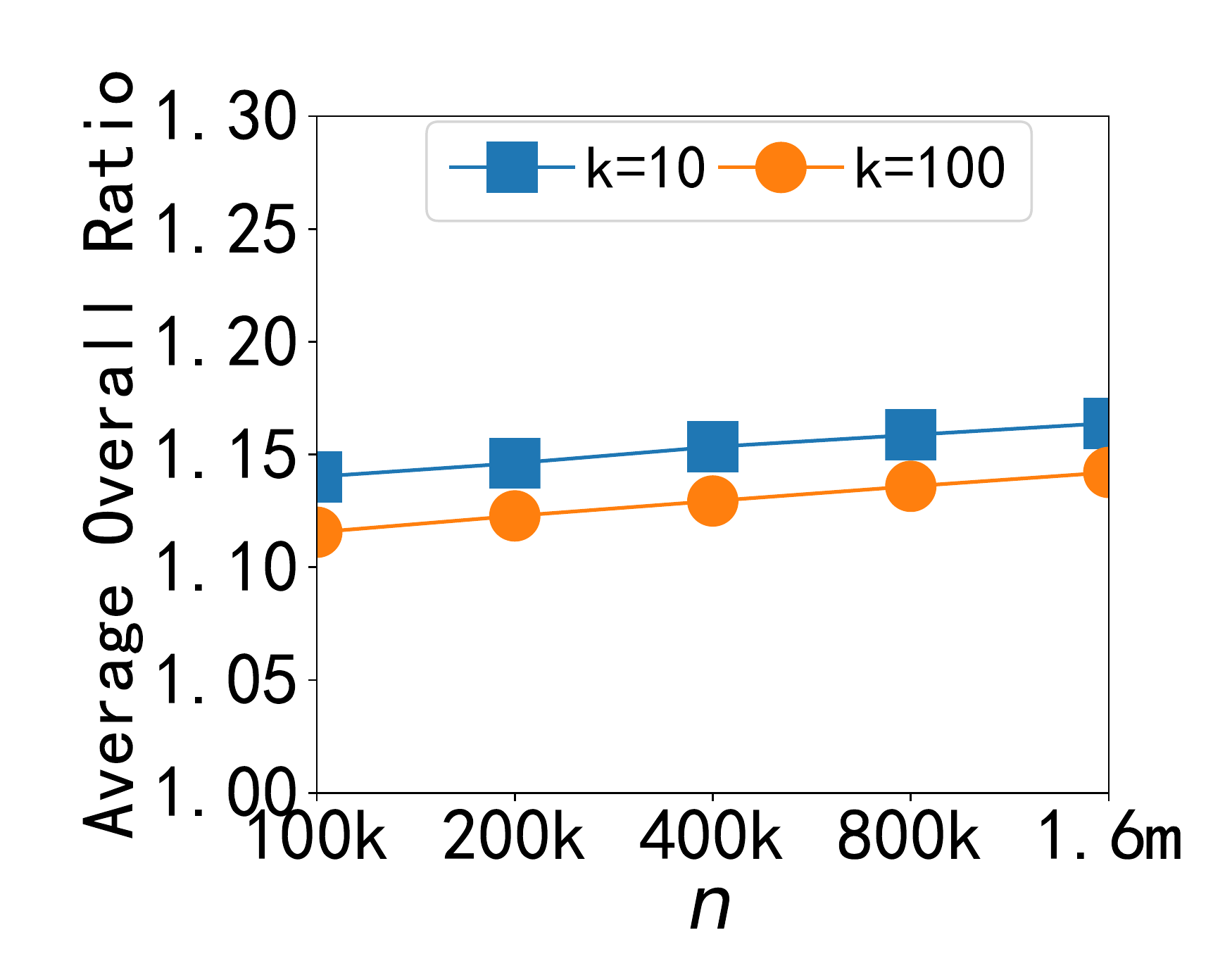}\label{rand/n/ratio_useCt=0_L2}}
	
	\subfigure[Ratio vs. $c$]{\includegraphics[width=0.246\textwidth]{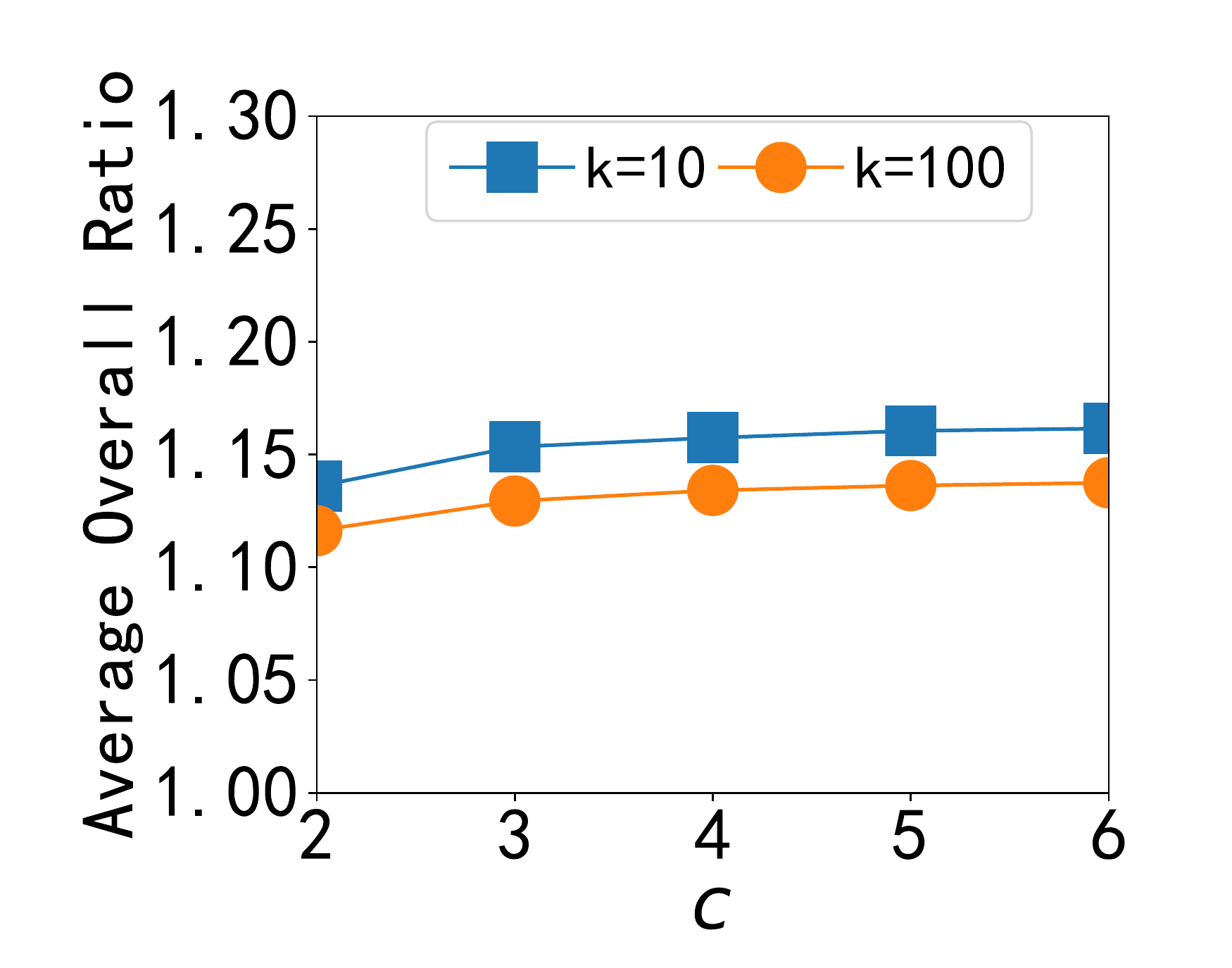}\label{rand/c/ratio_useCt=0_L2}}
	\subfigure[Ratio vs. $\#Subrange$]{\includegraphics[width=0.246\textwidth]{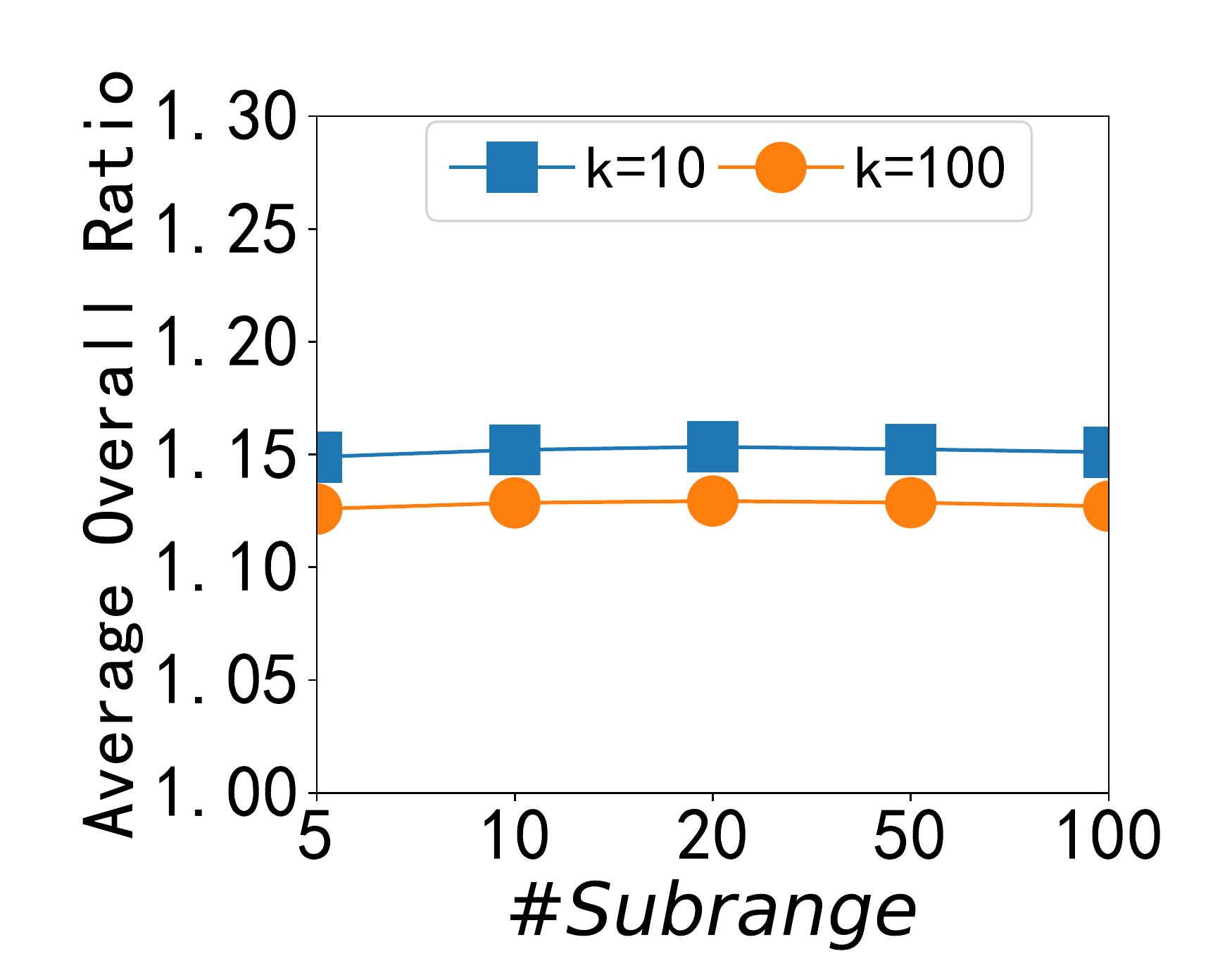}\label{rand/subrange/ratio_useCt=0_L2}}
	\subfigure[Ratio vs. $\#Subset$]{\includegraphics[width=0.246\textwidth]{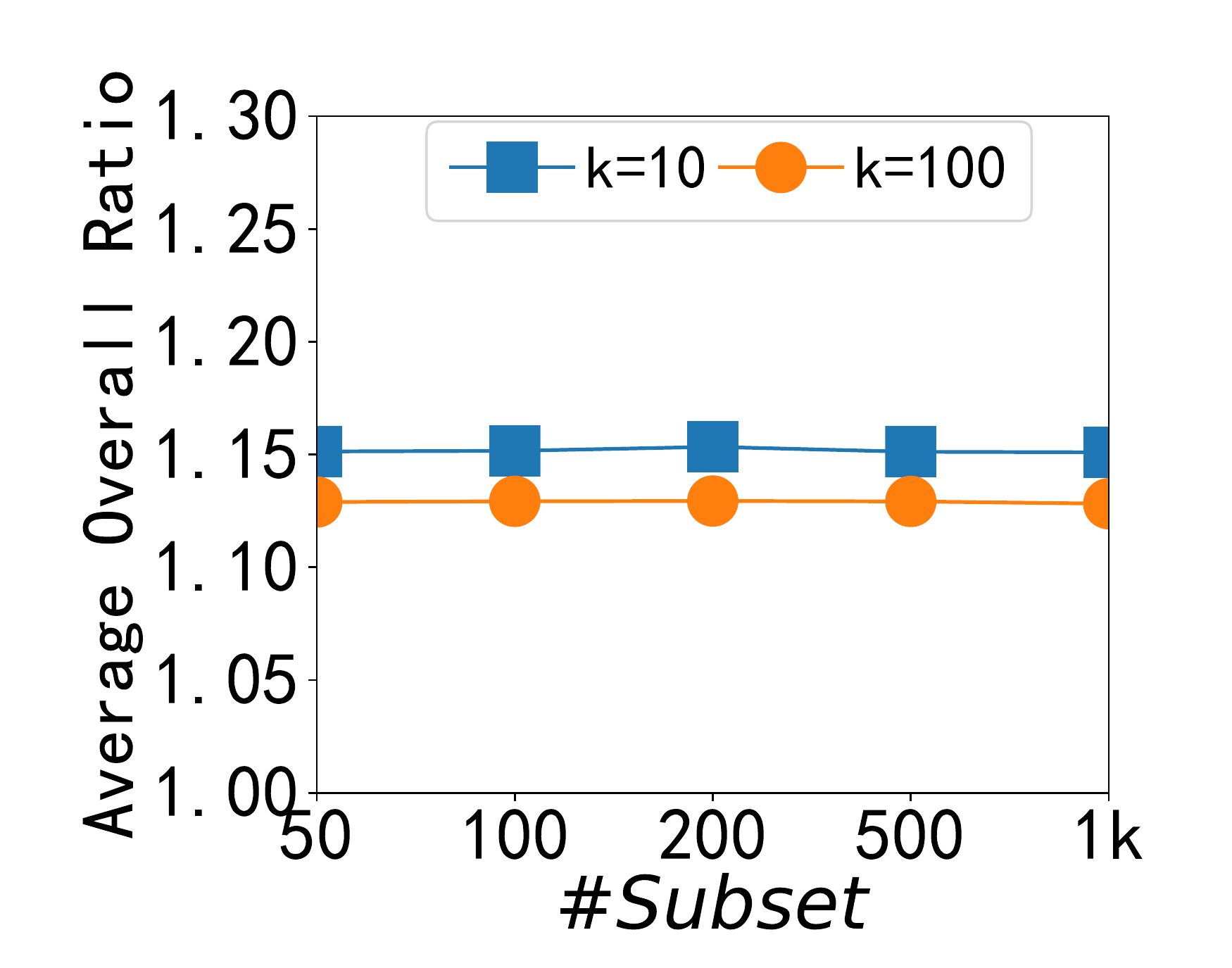}\label{rand/subset/ratio_useCt=0_L2}}
	\subfigure[Ratio vs. $\left|S\right|$]{\includegraphics[width=0.246\textwidth]{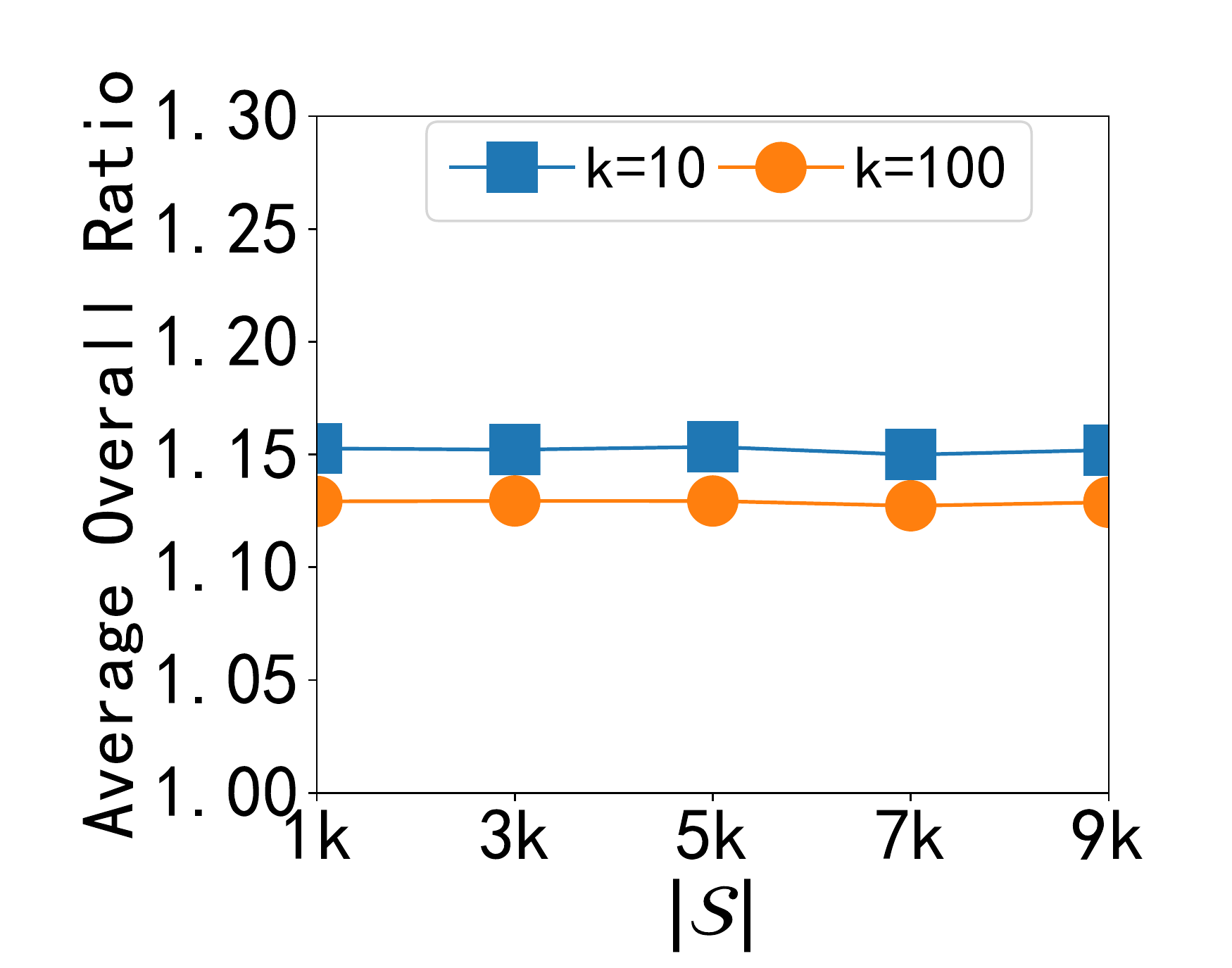}\label{rand/S/ratio_useCt=0_L2}}
	
	\caption{Query efficiency and query accuracy of WLSH on synthetic data when not using collision threshold reduction, $l_2$ distance}
	\label{rand/efficiency and accuracy/L2}
\end{figure*}

\begin{figure}[t]
	\centering
	\subfigure[I/O cost vs. $c$]{\includegraphics[width=0.24\textwidth]{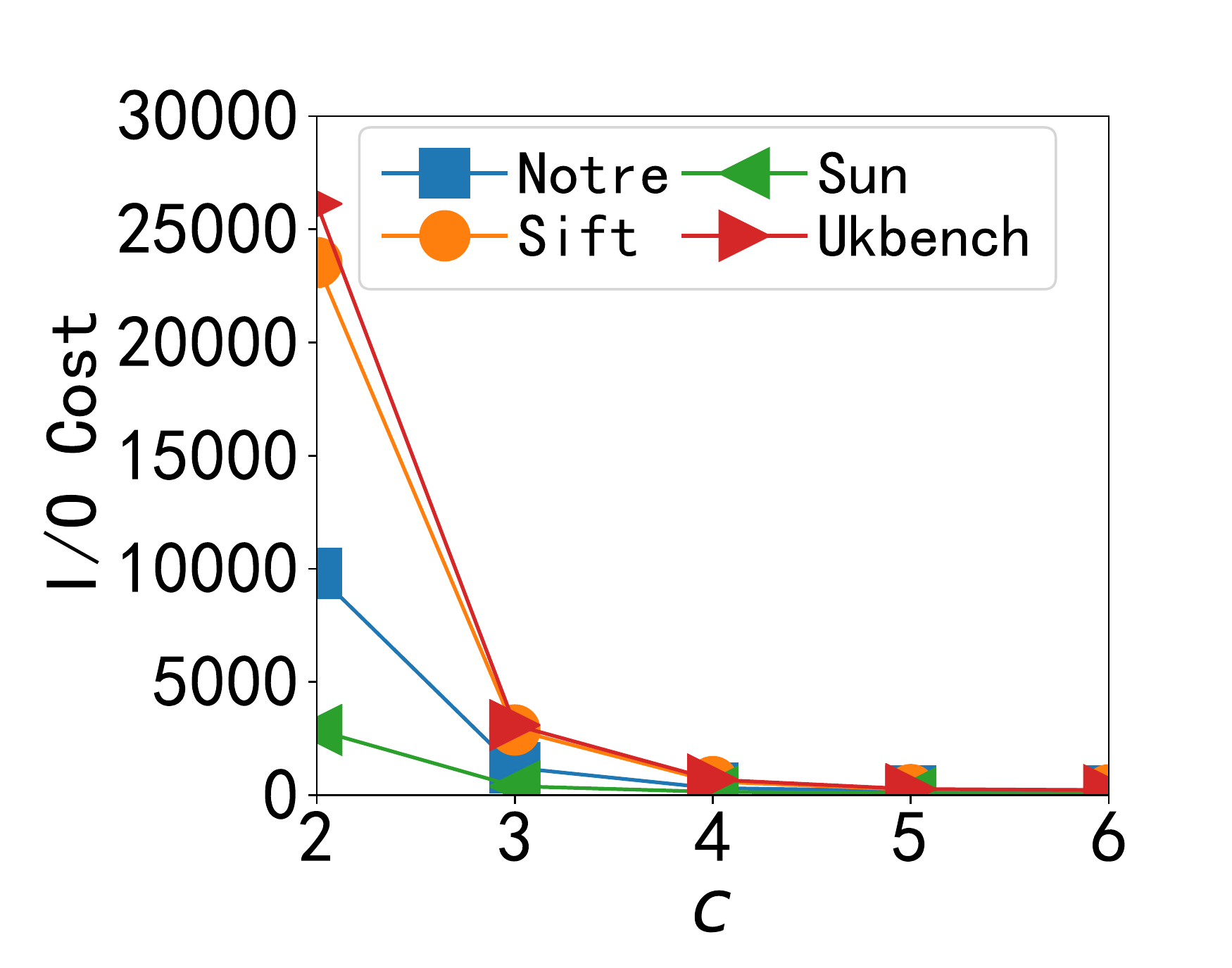}\label{real/c/IO_useCt=1_L1/k=10}}
	\subfigure[Ratio vs. $c$]{\includegraphics[width=0.24\textwidth]{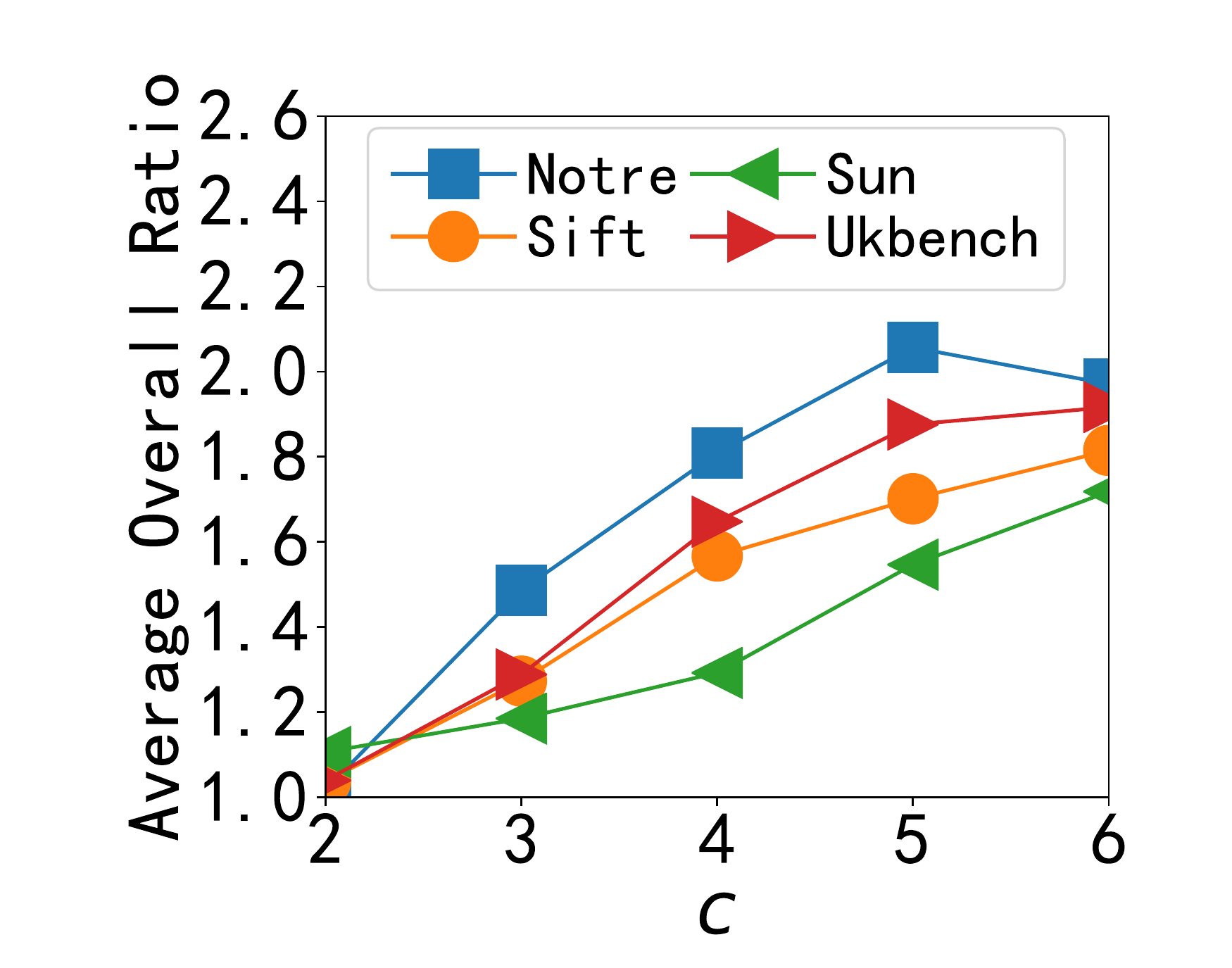}\label{real/c/ratio_useCt=1_L1/k=10}}
	
	\subfigure[I/O cost vs. $\#Subrange$]{\includegraphics[width=0.24\textwidth]{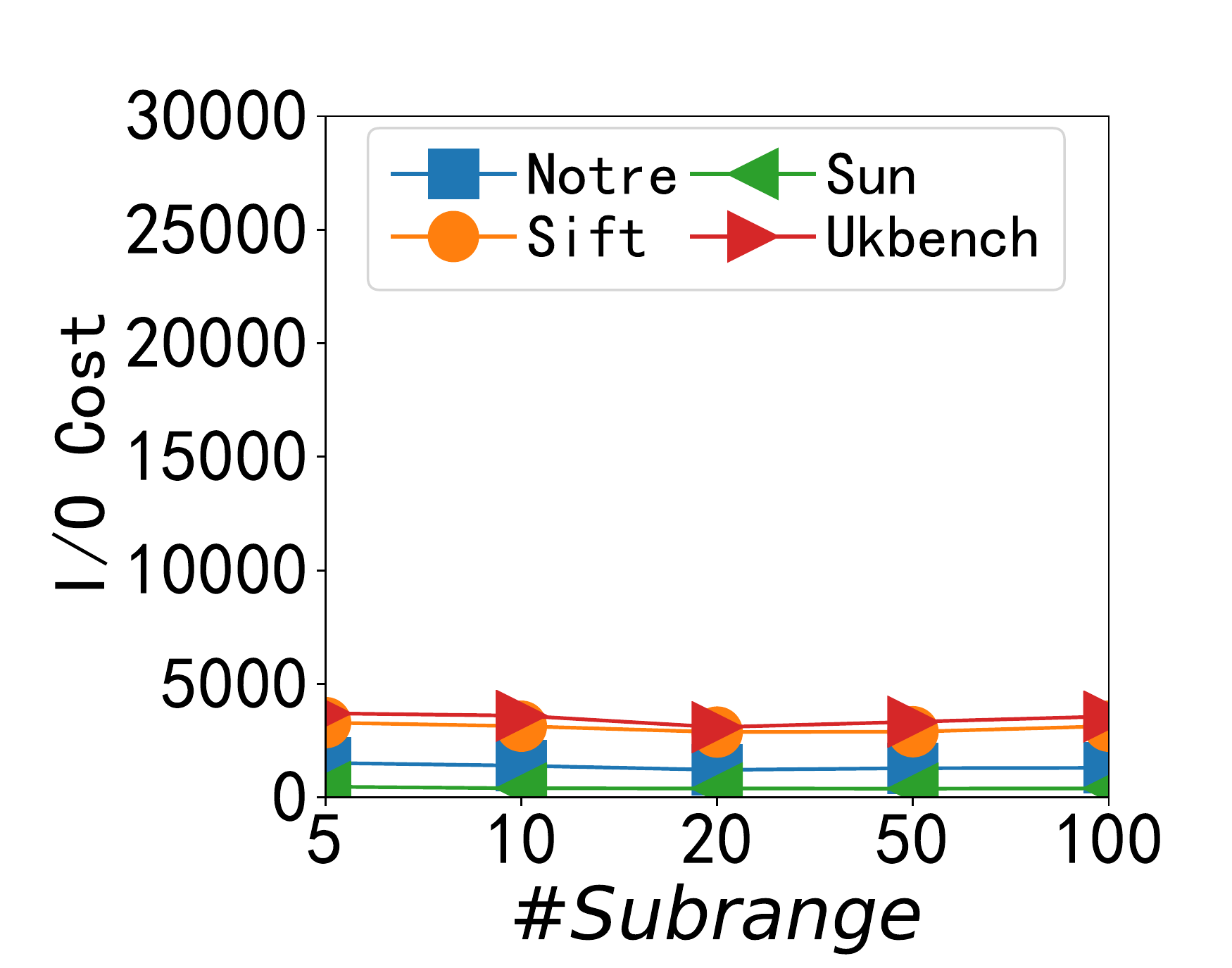}\label{real/subrange/IO_useCt=1_L1/k=10}}
	\subfigure[Ratio vs. $\#Subrange$]{\includegraphics[width=0.24\textwidth]{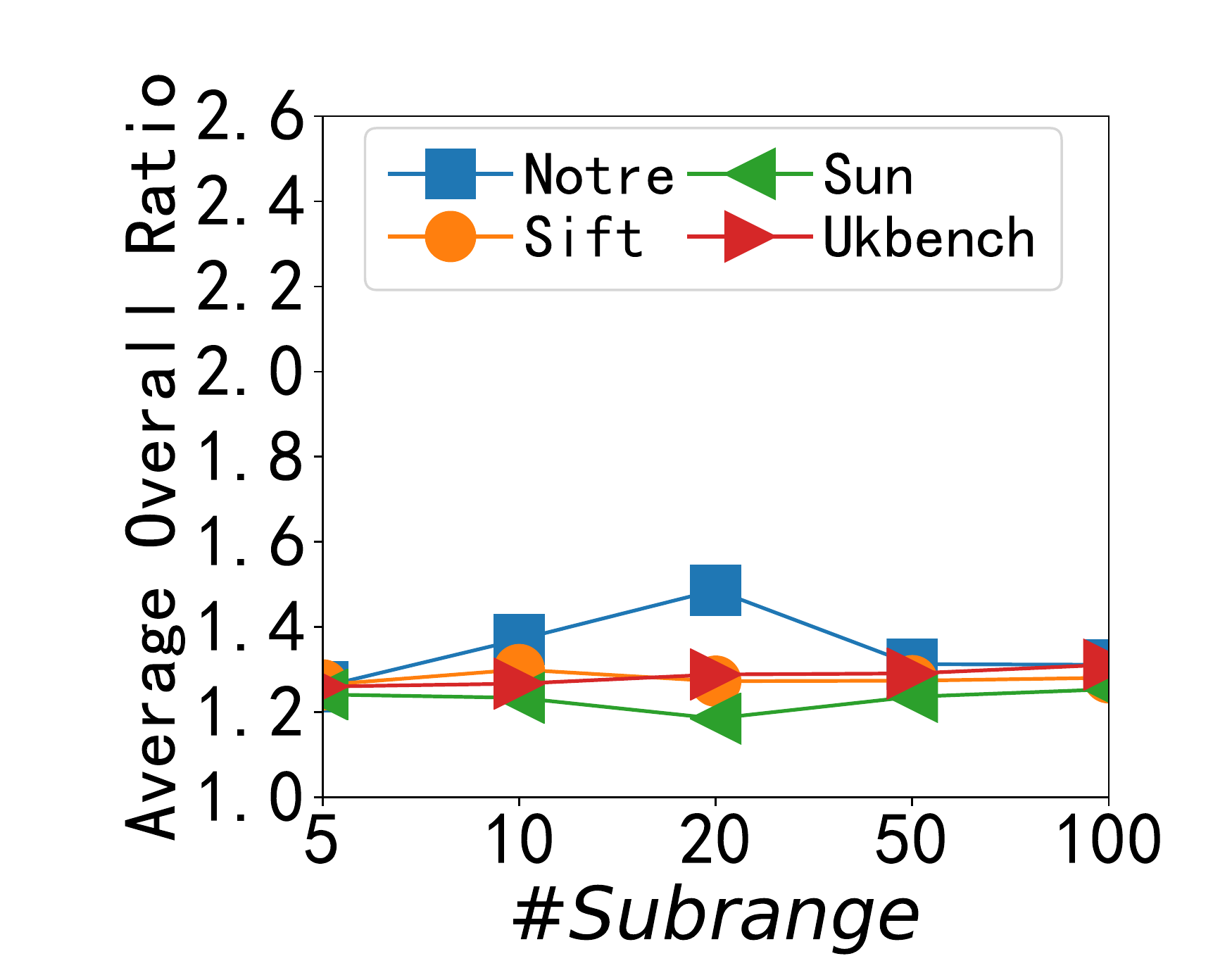}\label{real/subrange/ratio_useCt=1_L1/k=10}}
	
	\subfigure[I/O cost vs. $\#Subset$]{\includegraphics[width=0.24\textwidth]{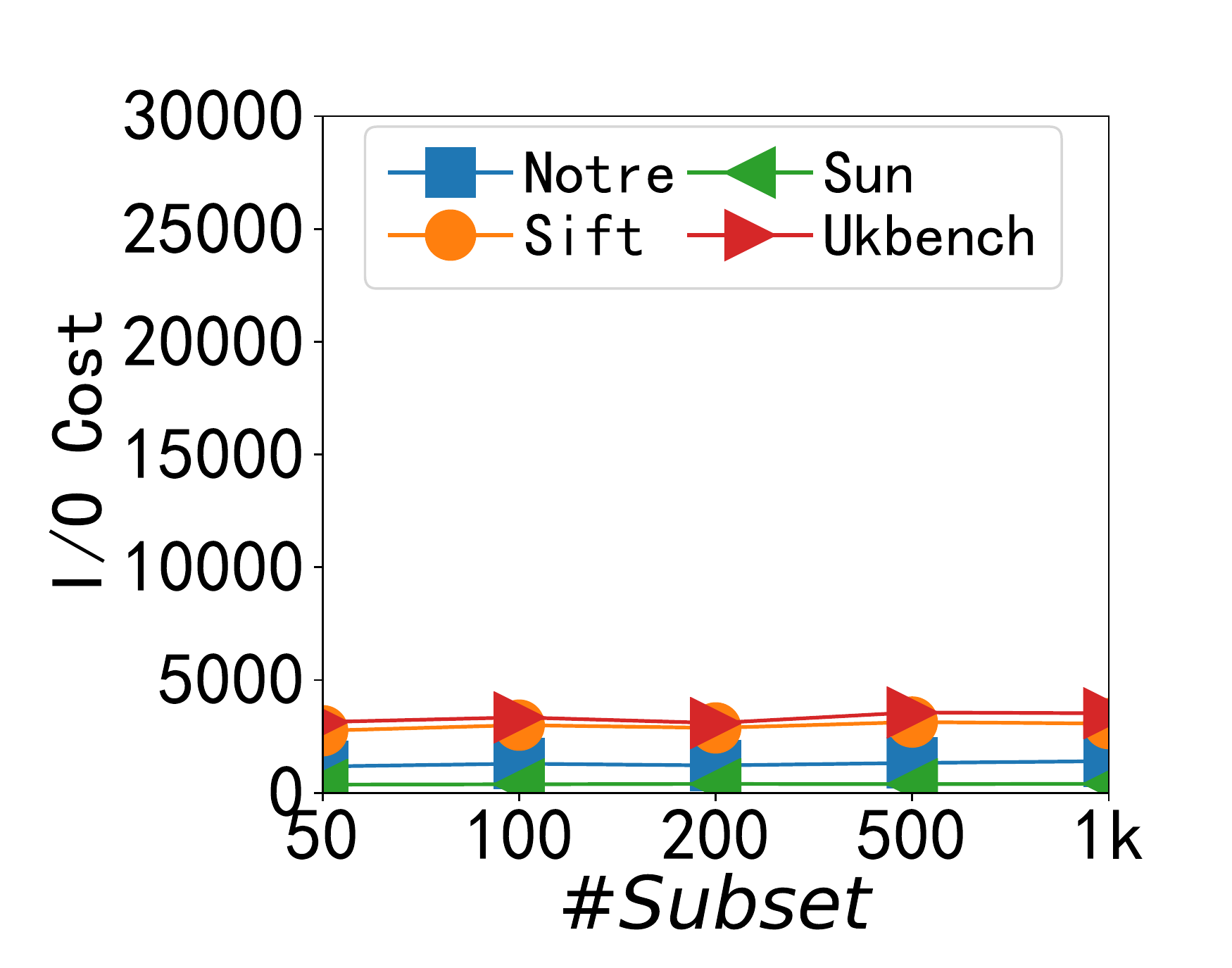}\label{real/subset/IO_useCt=1_L1/k=10}}
	\subfigure[Ratio vs. $\#Subset$]{\includegraphics[width=0.24\textwidth]{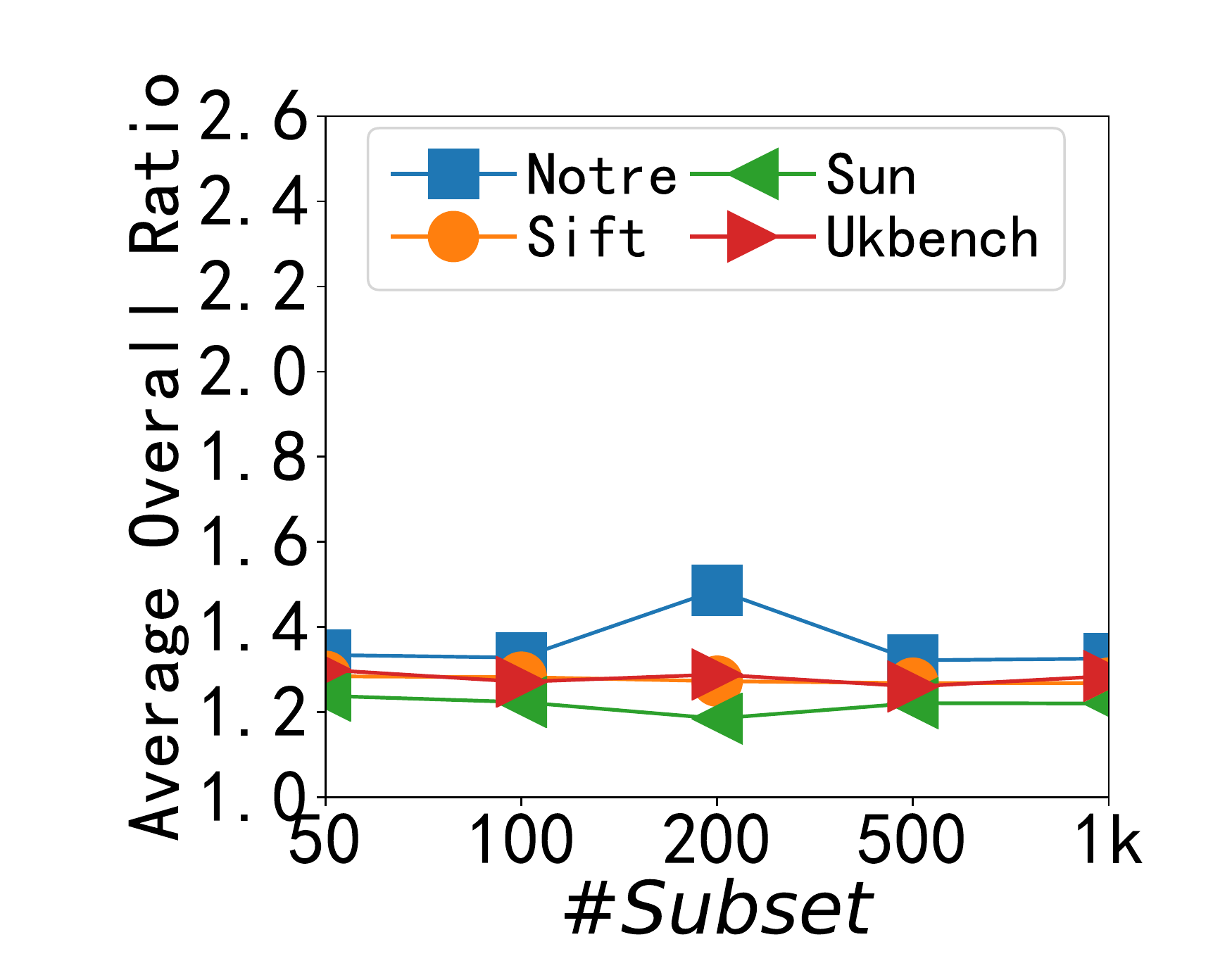}\label{real/subset/ratio_useCt=1_L1/k=10}}
	
	\subfigure[I/O cost vs. $\left|S\right|$]{\includegraphics[width=0.24\textwidth]{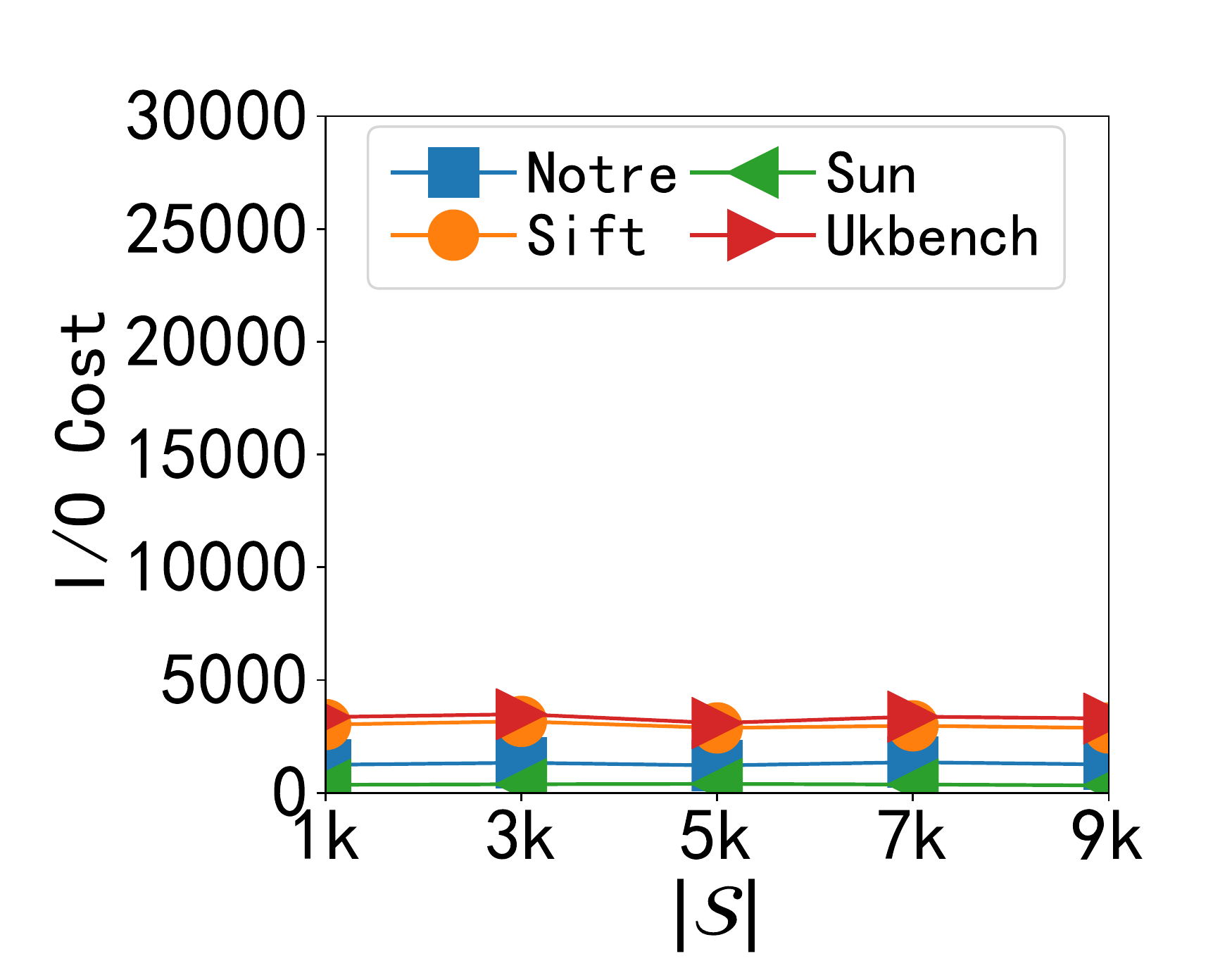}\label{real/S/IO_useCt=1_L1/k=10}}
	\subfigure[Ratio vs. $\left|S\right|$]{\includegraphics[width=0.24\textwidth]{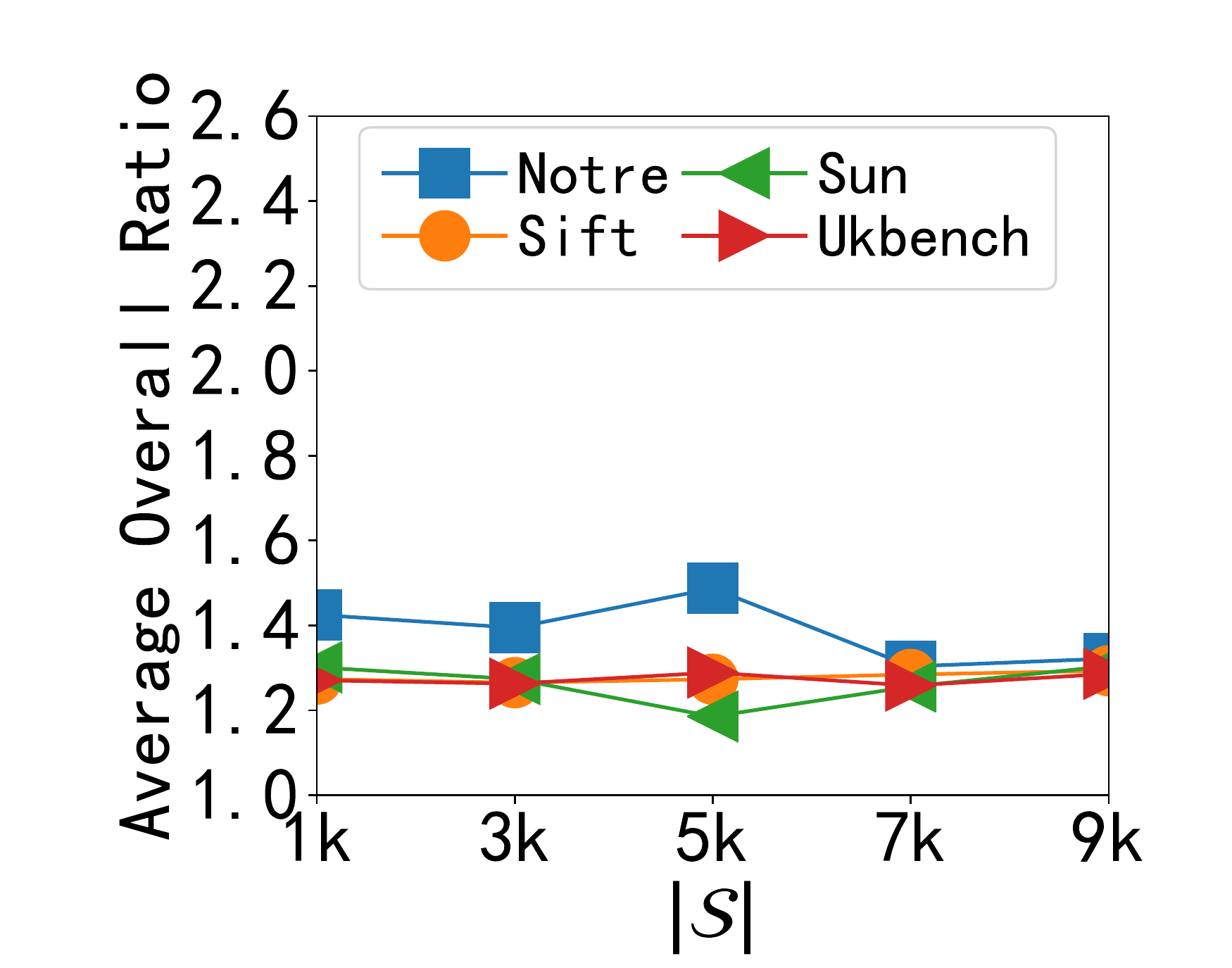}\label{real/S/ratio_useCt=1_L1/k=10}}
	
	\caption{Query efficiency and query accuracy of WLSH on real data when using collision threshold reduction, $l_1$ distance, $k=10$}
	\label{real/efficiency and accuracy/L1/useCt=1/k=10}
\end{figure}

\begin{figure}[t]
	\centering
	\subfigure[I/O cost vs. $c$]{\includegraphics[width=0.24\textwidth]{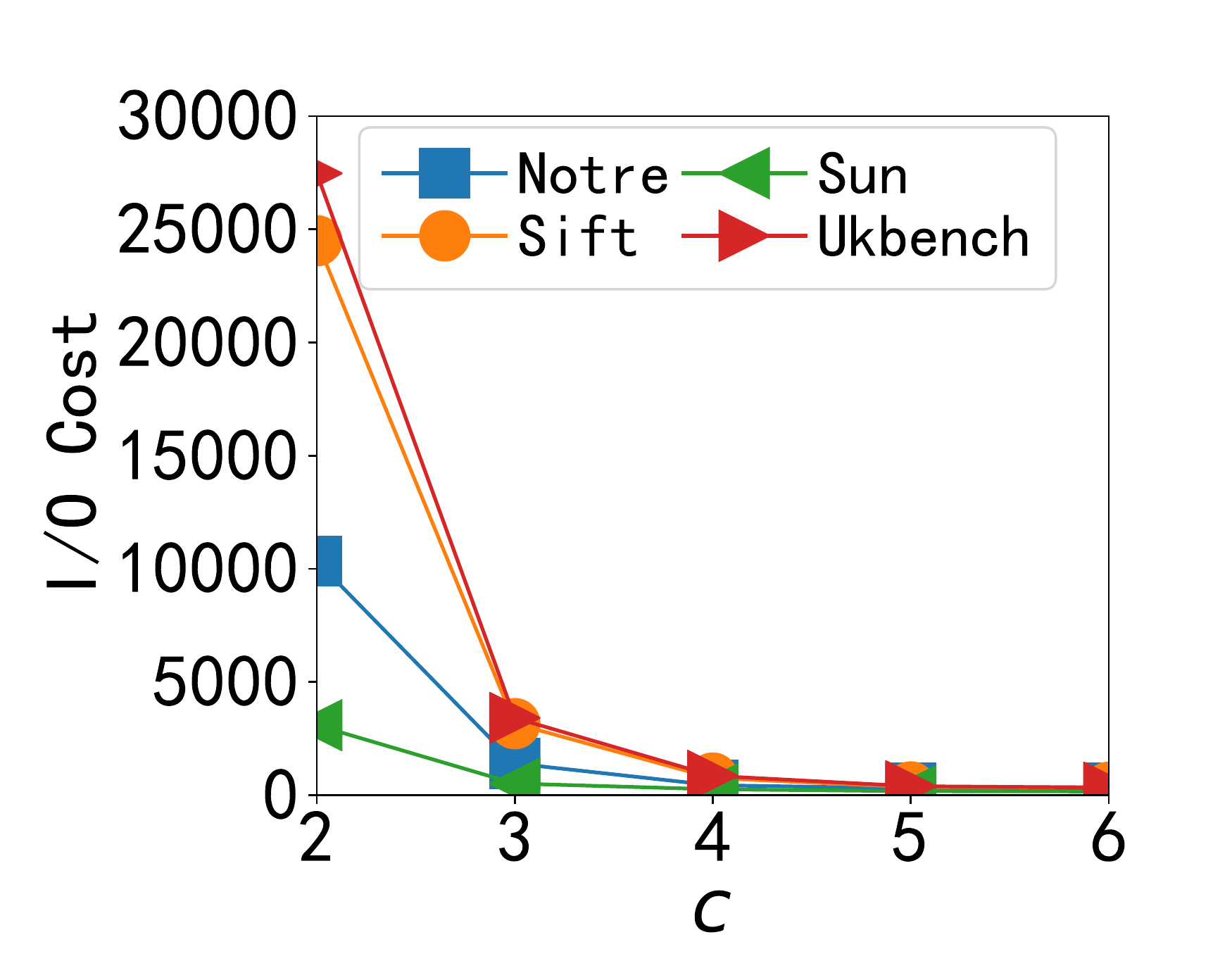}\label{real/c/IO_useCt=1_L1}}
	\subfigure[Ratio vs. $c$]{\includegraphics[width=0.24\textwidth]{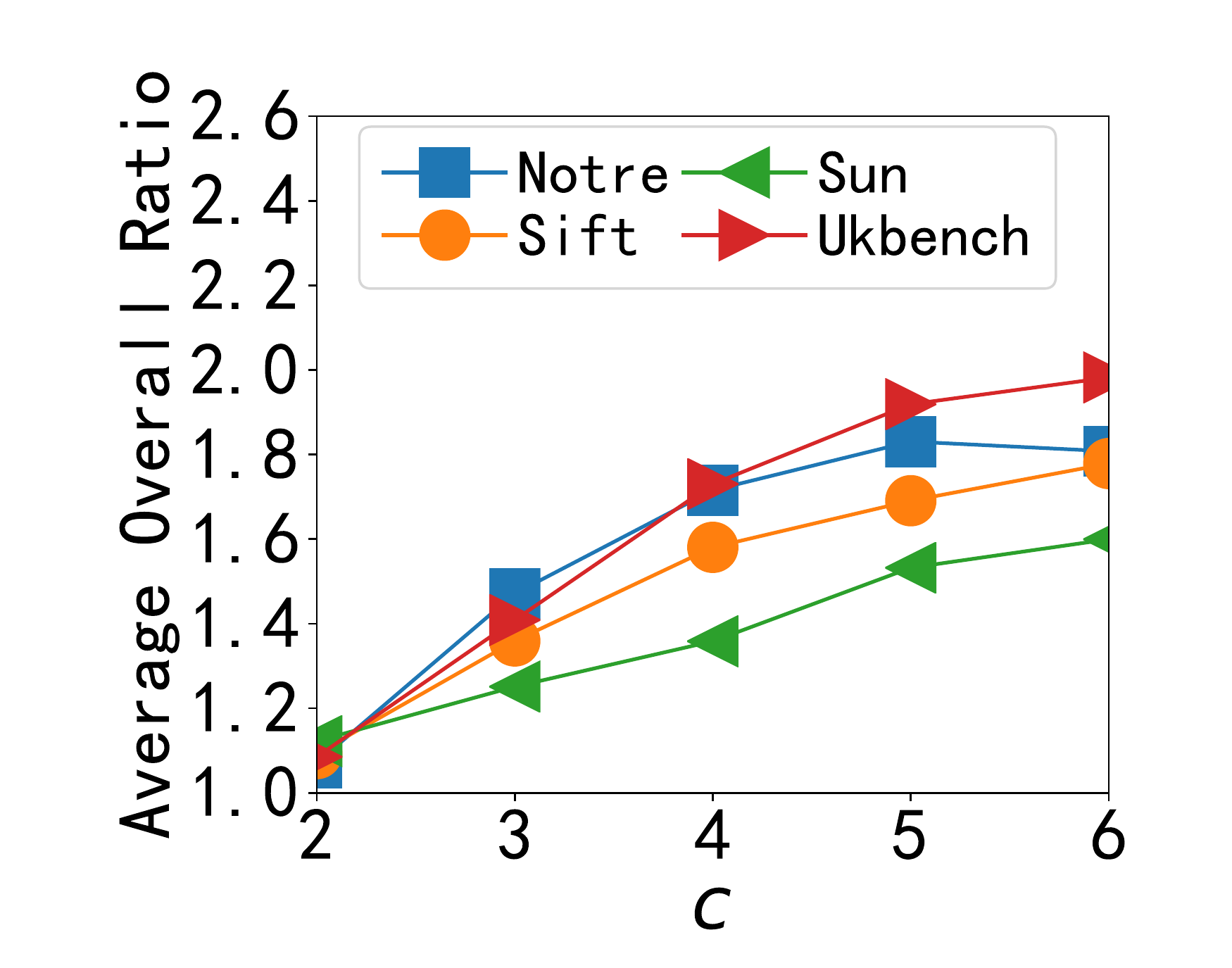}\label{real/c/ratio_useCt=1_L1}}
	
	\subfigure[I/O cost vs. $\#Subrange$]{\includegraphics[width=0.24\textwidth]{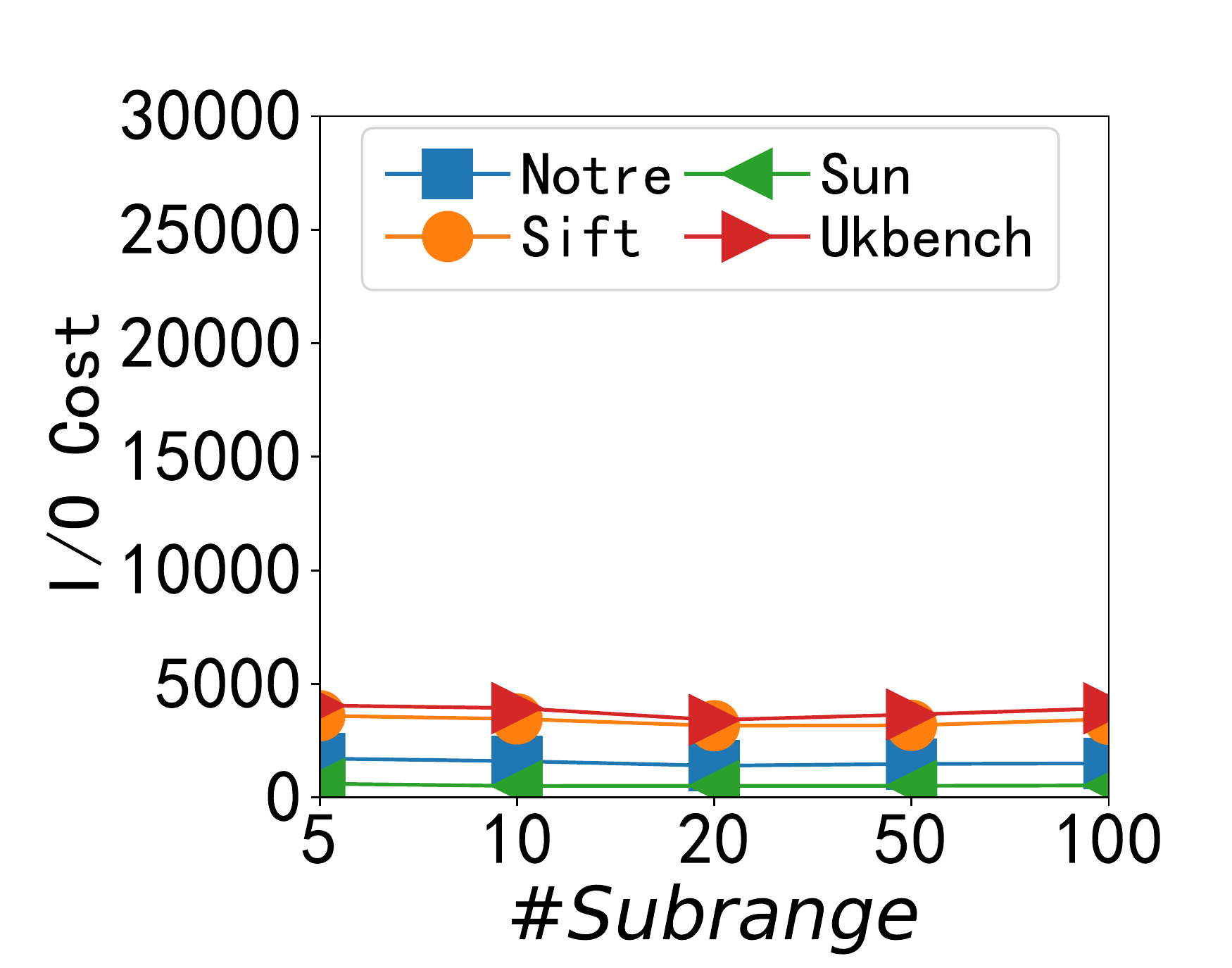}\label{real/subrange/IO_useCt=1_L1}}
	\subfigure[Ratio vs. $\#Subrange$]{\includegraphics[width=0.24\textwidth]{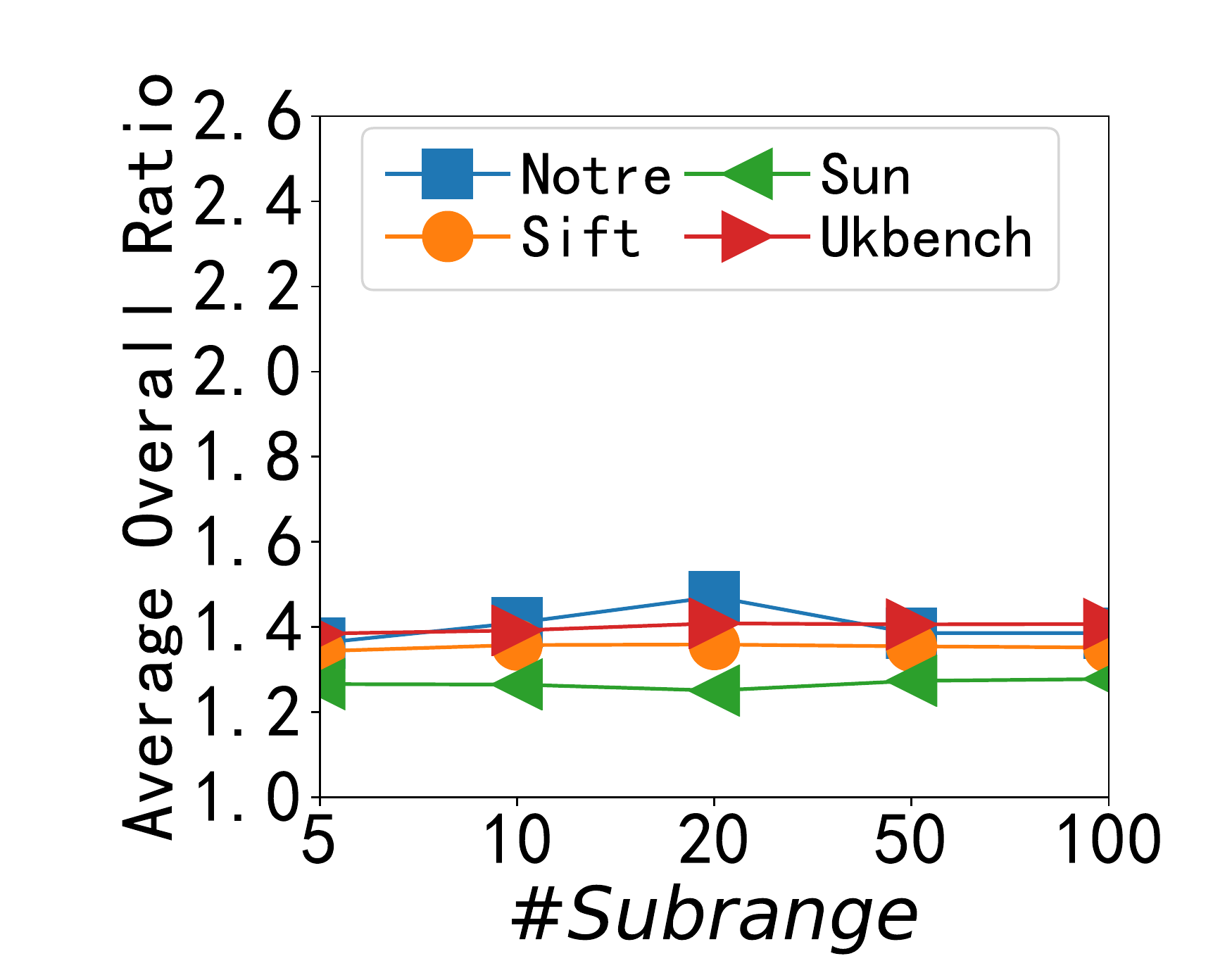}\label{real/subrange/ratio_useCt=1_L1}}
	
	\subfigure[I/O cost vs. $\#Subset$]{\includegraphics[width=0.24\textwidth]{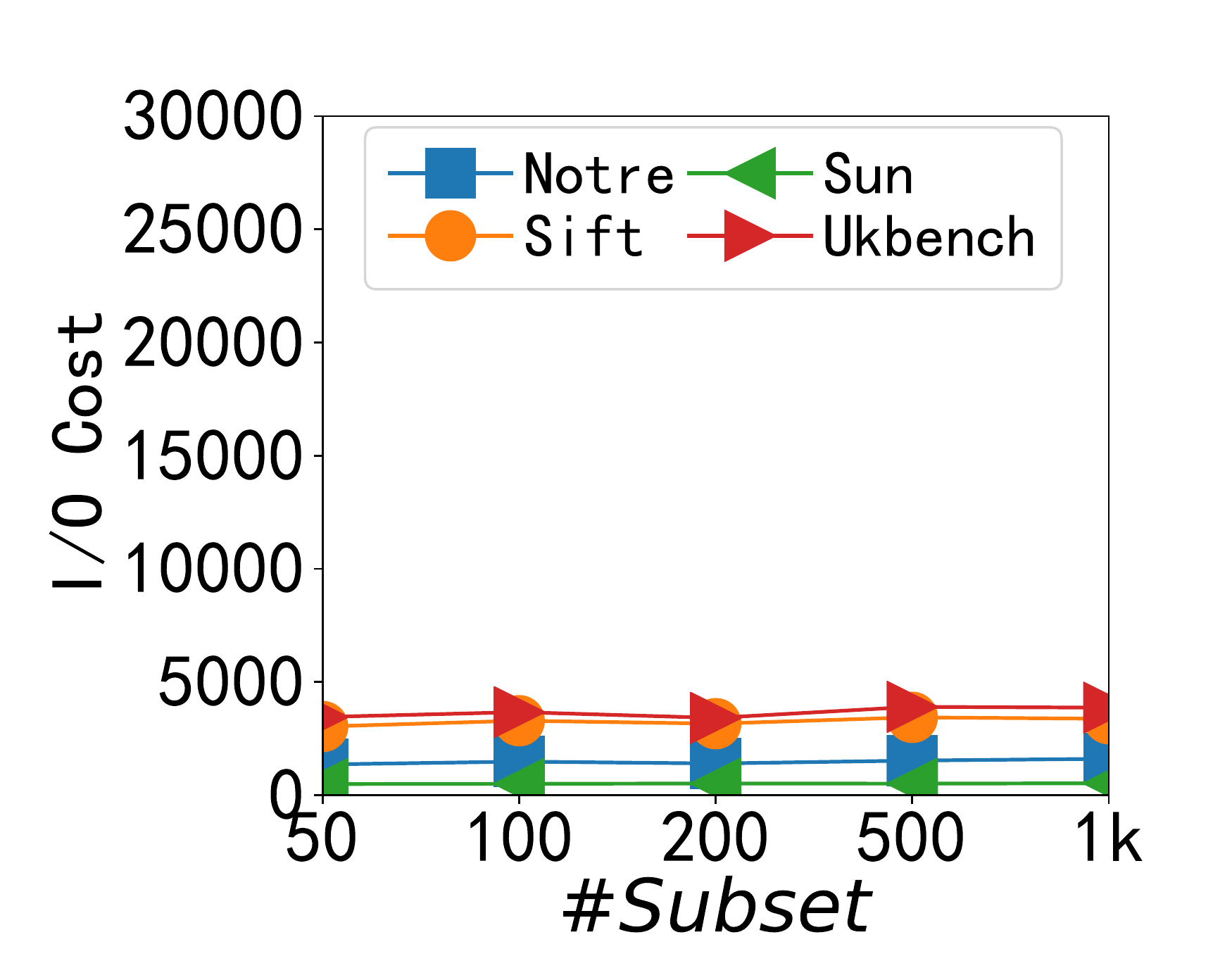}\label{real/subset/IO_useCt=1_L1}}
	\subfigure[Ratio vs. $\#Subset$]{\includegraphics[width=0.24\textwidth]{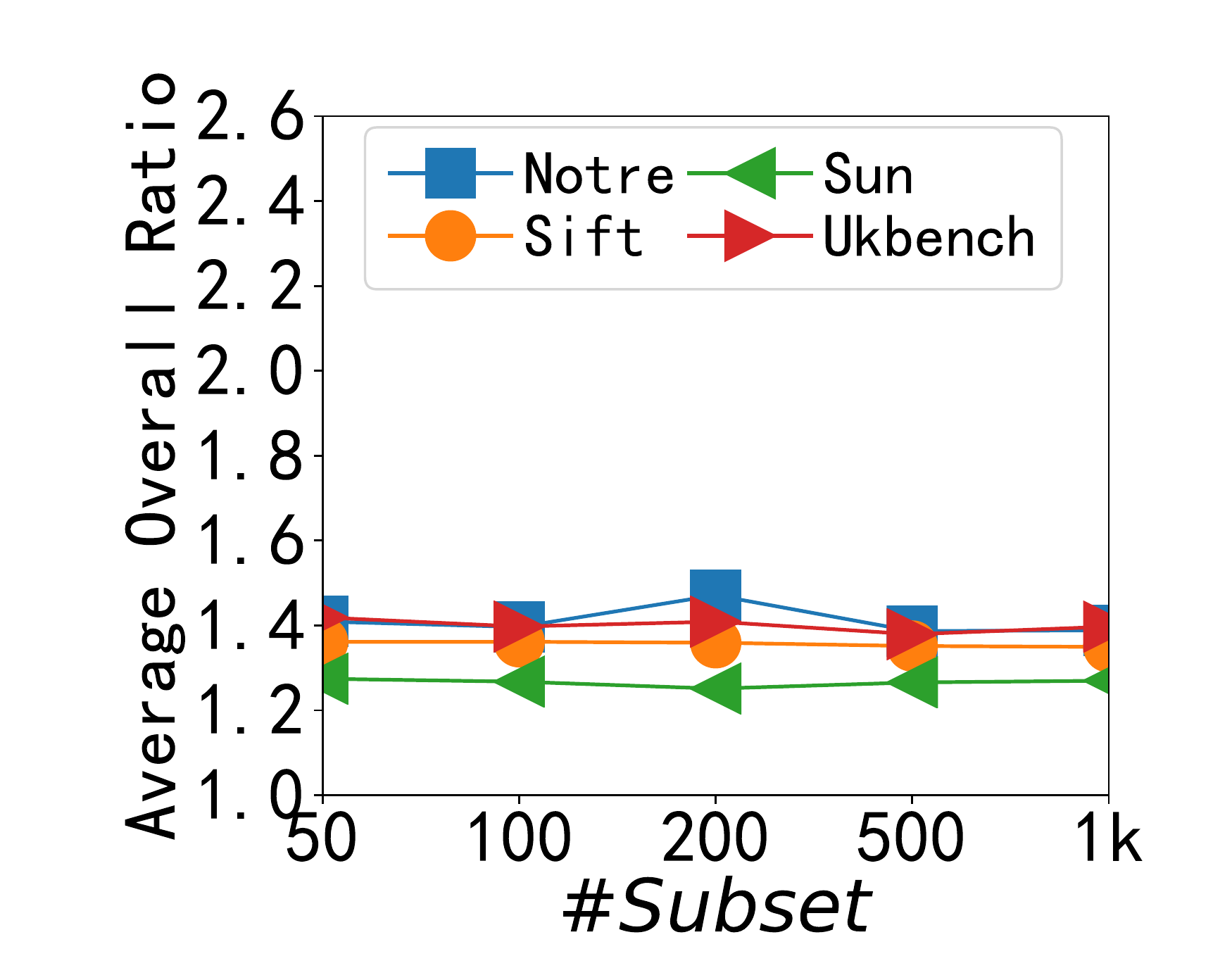}\label{real/subset/ratio_useCt=1_L1}}
	
	\subfigure[I/O cost vs. $\left|S\right|$]{\includegraphics[width=0.24\textwidth]{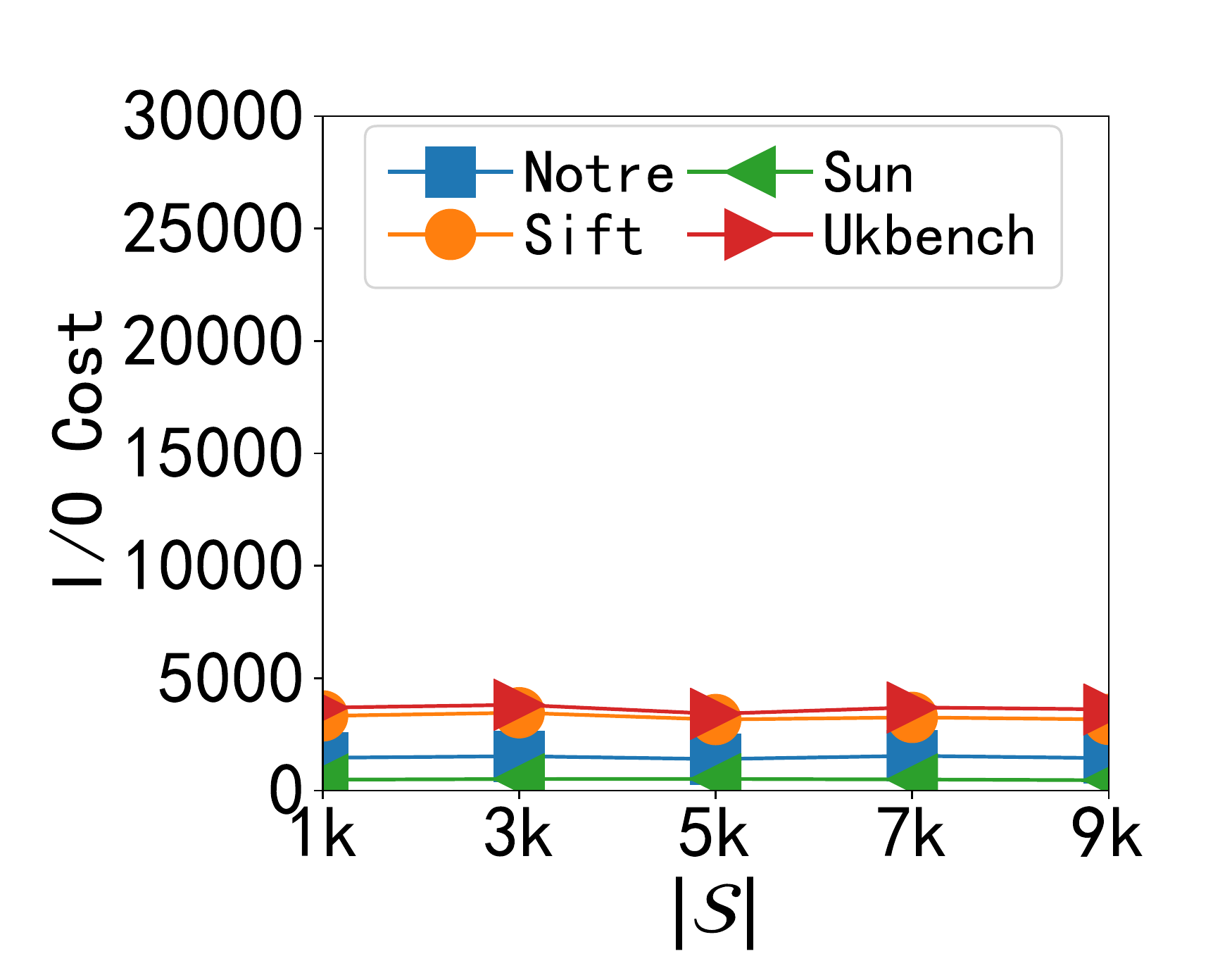}\label{real/S/IO_useCt=1_L1}}
	\subfigure[Ratio vs. $\left|S\right|$]{\includegraphics[width=0.24\textwidth]{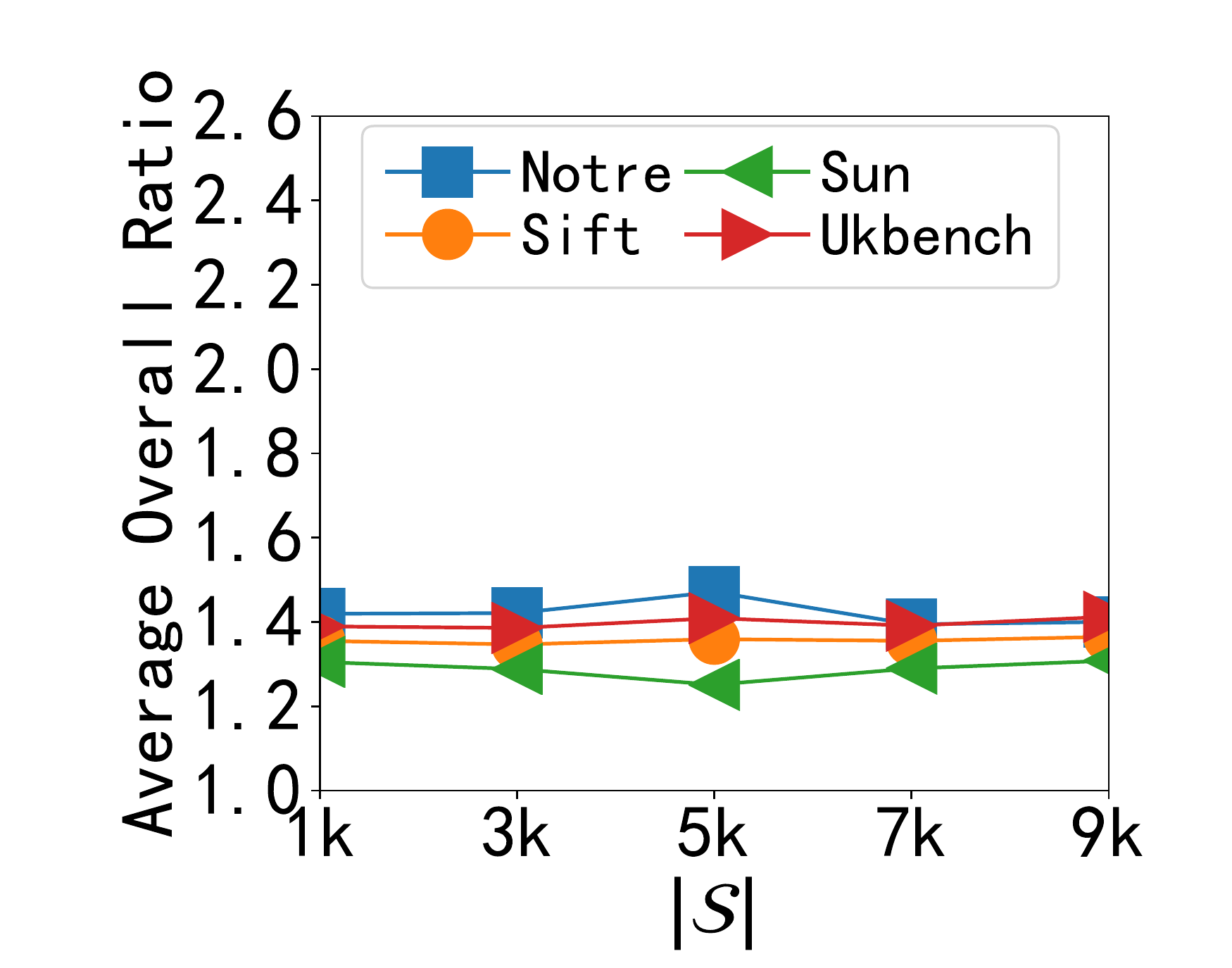}\label{real/S/ratio_useCt=1_L1}}
	
	\caption{Query efficiency and query accuracy of WLSH on real data when using collision threshold reduction, $l_1$ distance, $k=100$}
	\label{real/efficiency and accuracy/L1/useCt=1}
\end{figure}

\begin{figure}[t]
	\centering
	\subfigure[I/O cost vs. $c$]{\includegraphics[width=0.24\textwidth]{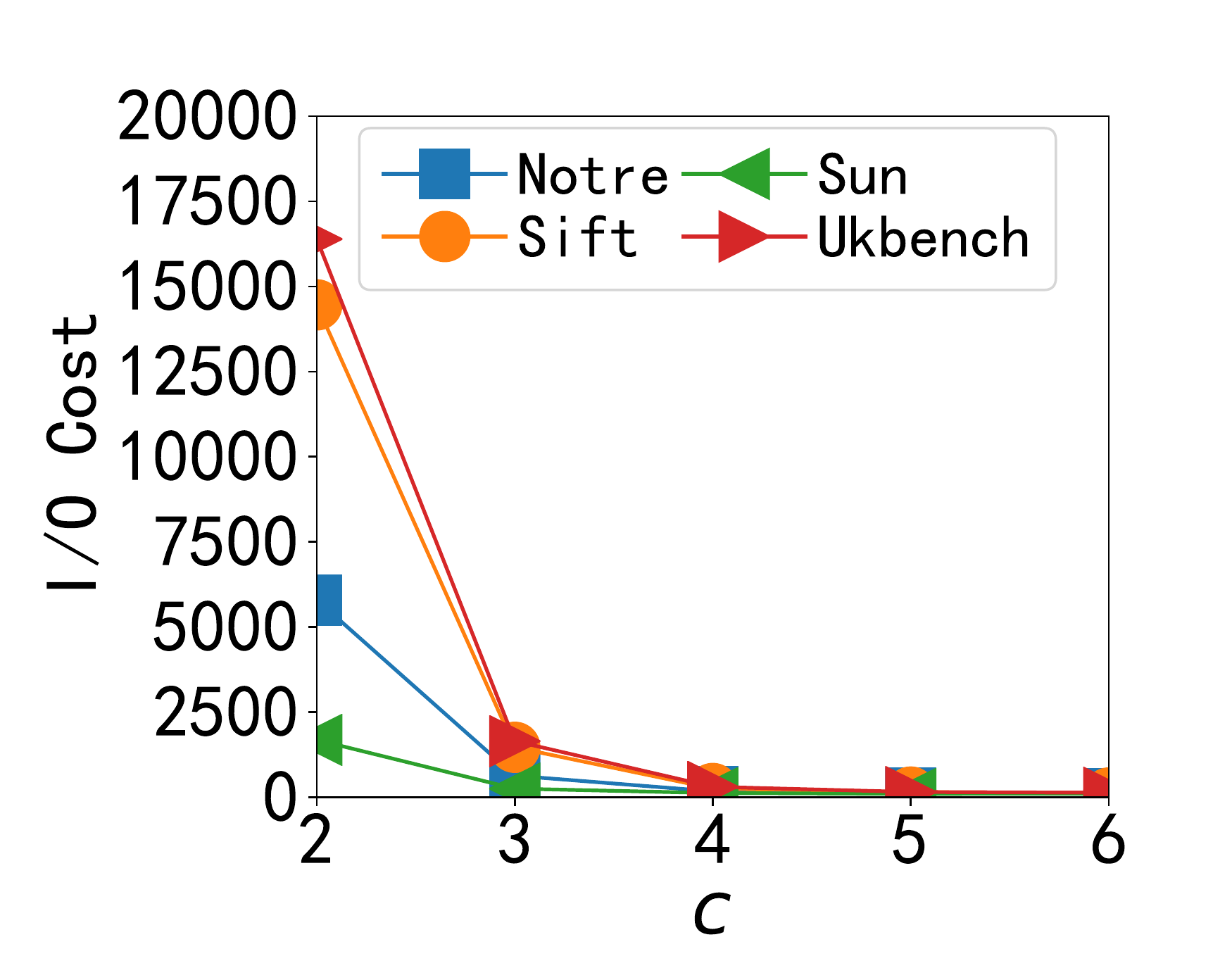}\label{real/c/IO_useCt=1_L2/k=10}}
	\subfigure[Ratio vs. $c$]{\includegraphics[width=0.24\textwidth]{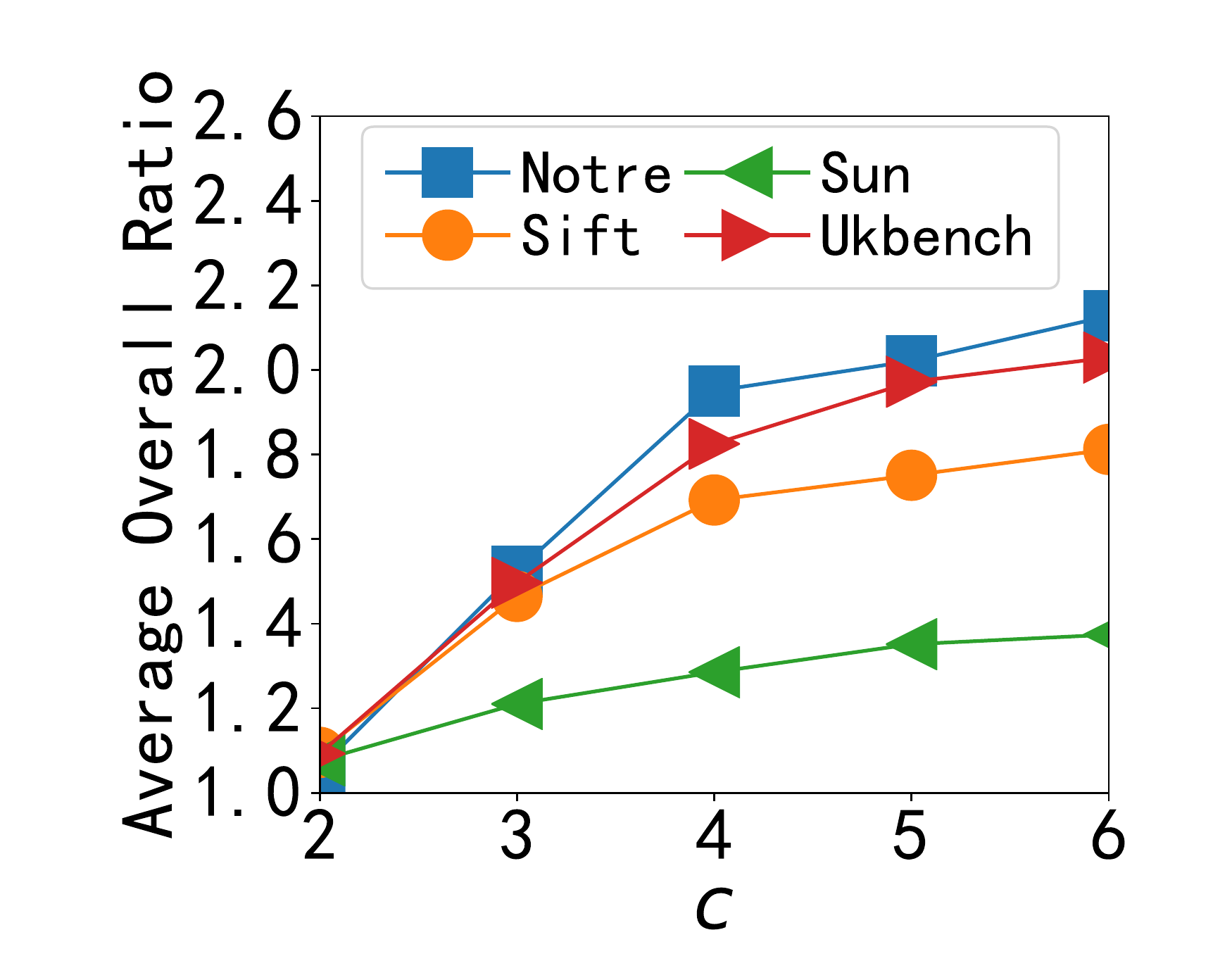}\label{real/c/ratio_useCt=1_L2/k=10}}
	
	\subfigure[I/O cost vs. $\#Subrange$]{\includegraphics[width=0.24\textwidth]{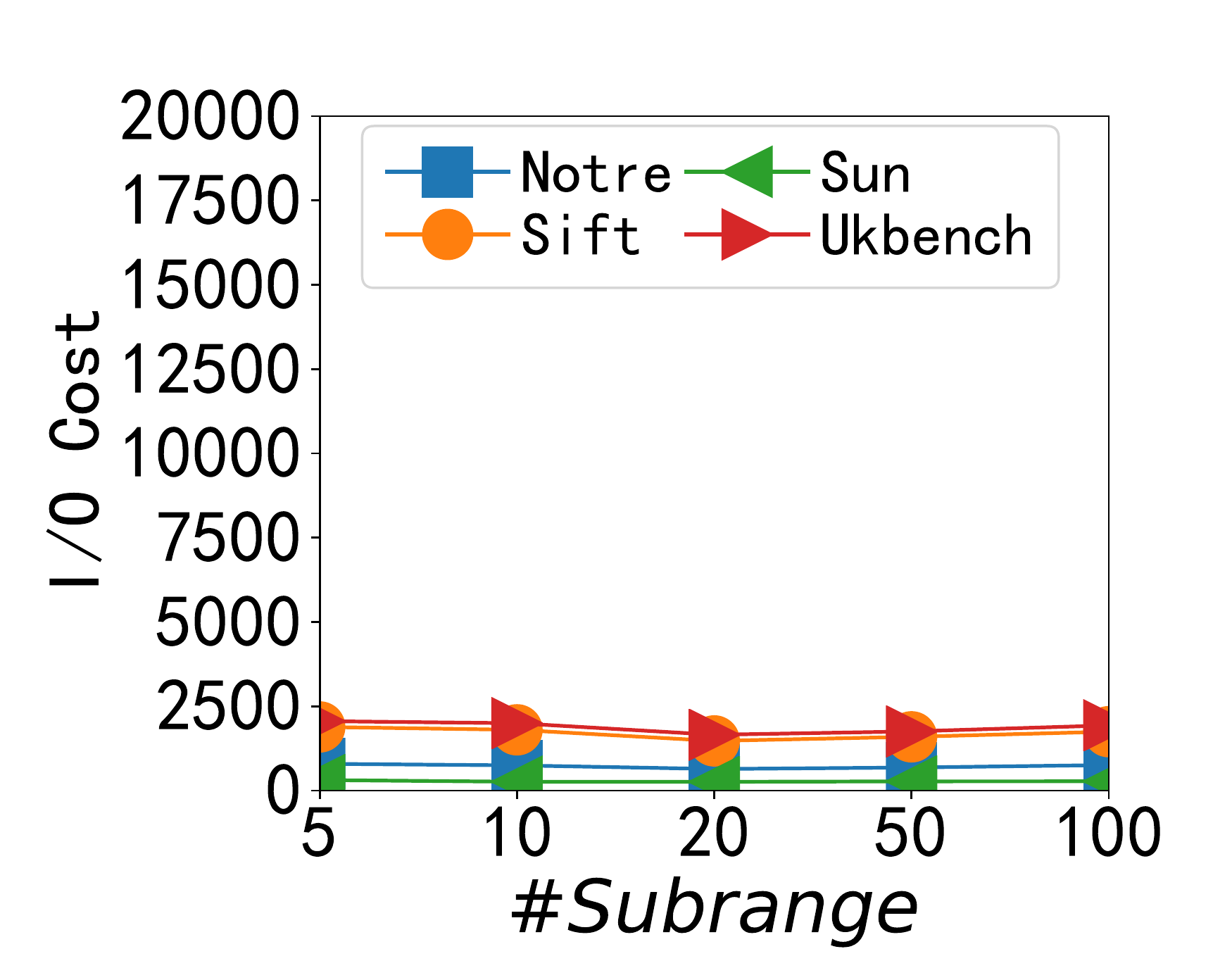}\label{real/subrange/IO_useCt=1_L2/k=10}}
	\subfigure[Ratio vs. $\#Subrange$]{\includegraphics[width=0.24\textwidth]{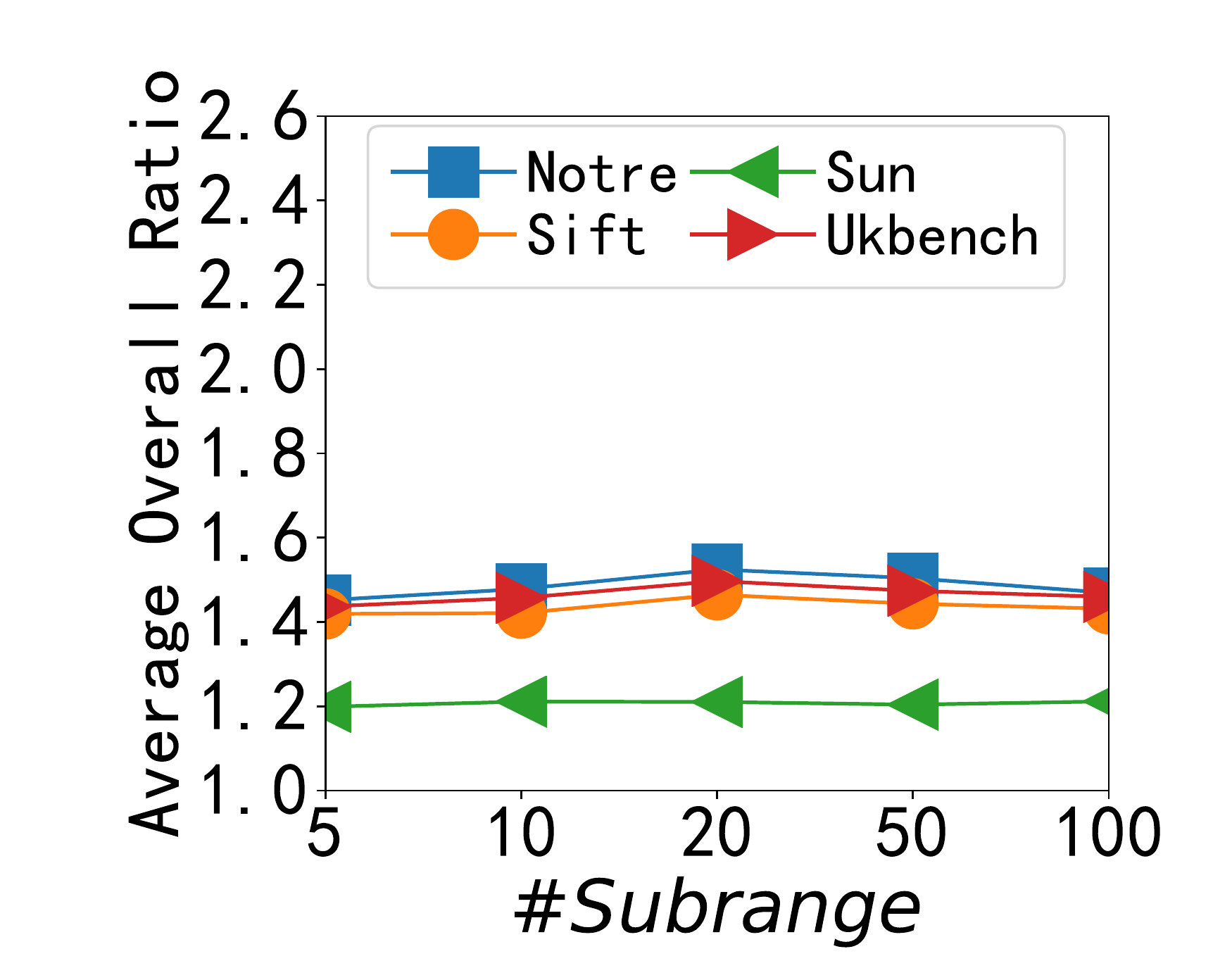}\label{real/subrange/ratio_useCt=1_L2/k=10}}
	
	\subfigure[I/O cost vs. $\#Subset$]{\includegraphics[width=0.24\textwidth]{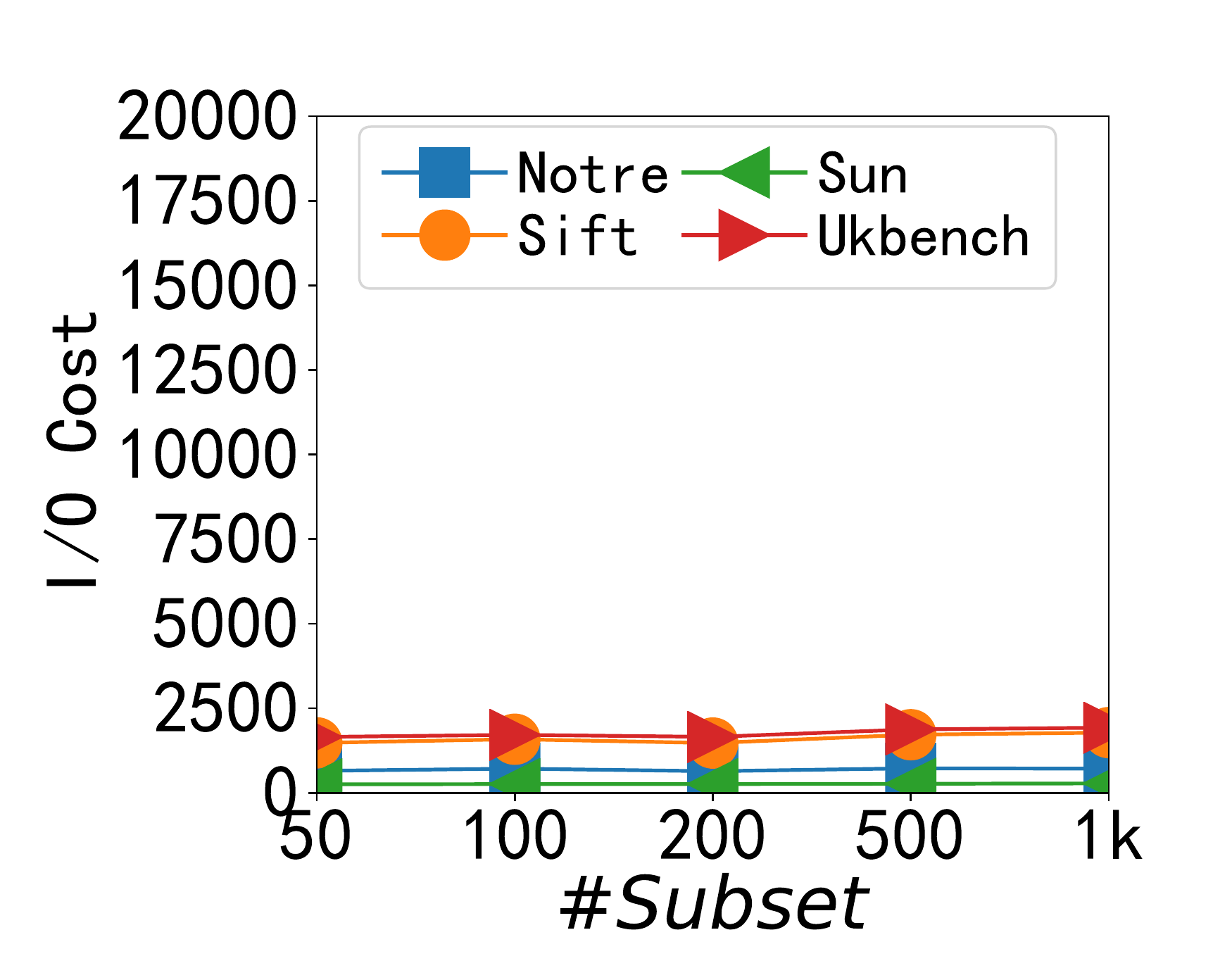}\label{real/subset/IO_useCt=1_L2/k=10}}
	\subfigure[Ratio vs. $\#Subset$]{\includegraphics[width=0.24\textwidth]{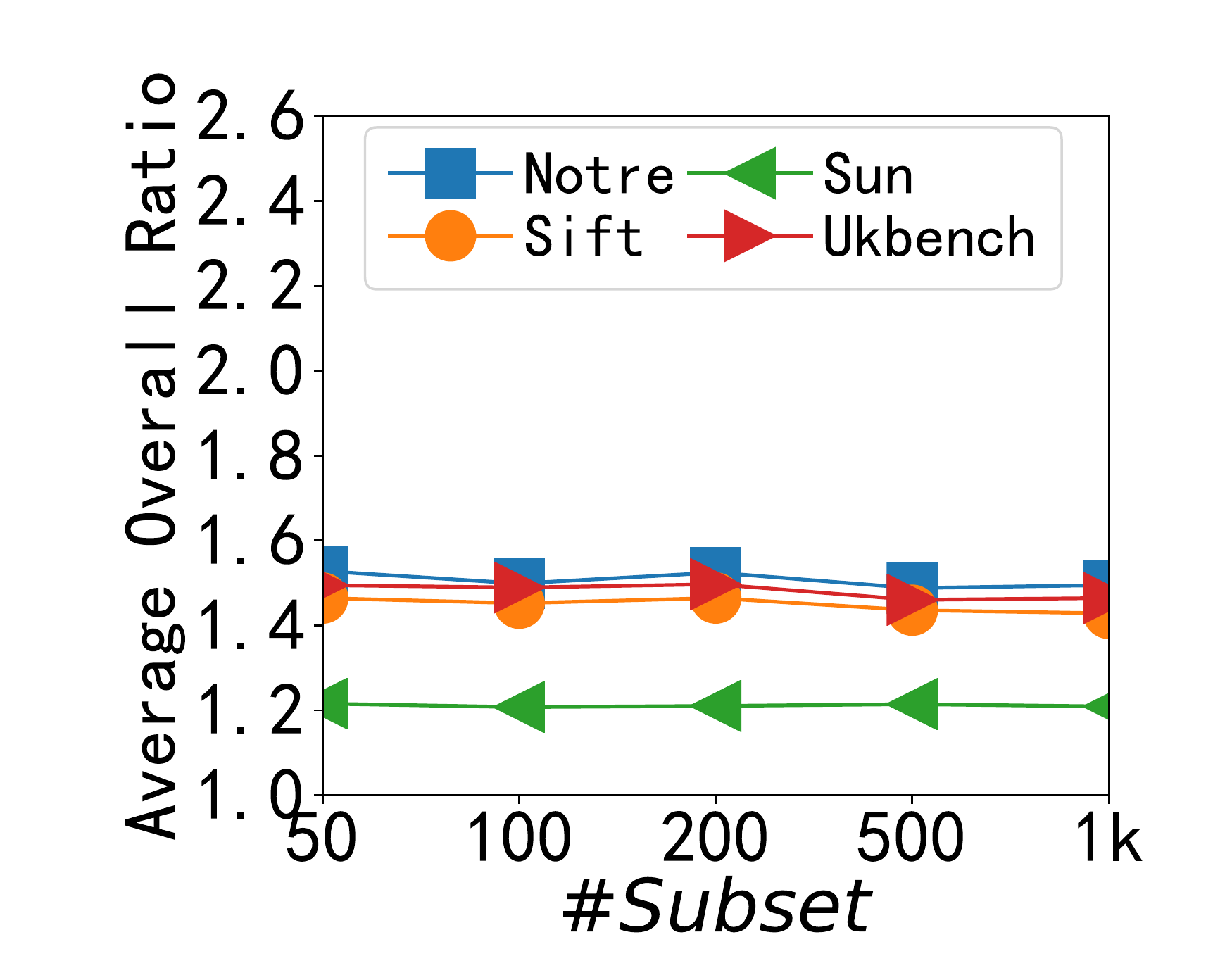}\label{real/subset/ratio_useCt=1_L2/k=10}}
	
	\subfigure[I/O cost vs. $\left|S\right|$]{\includegraphics[width=0.24\textwidth]{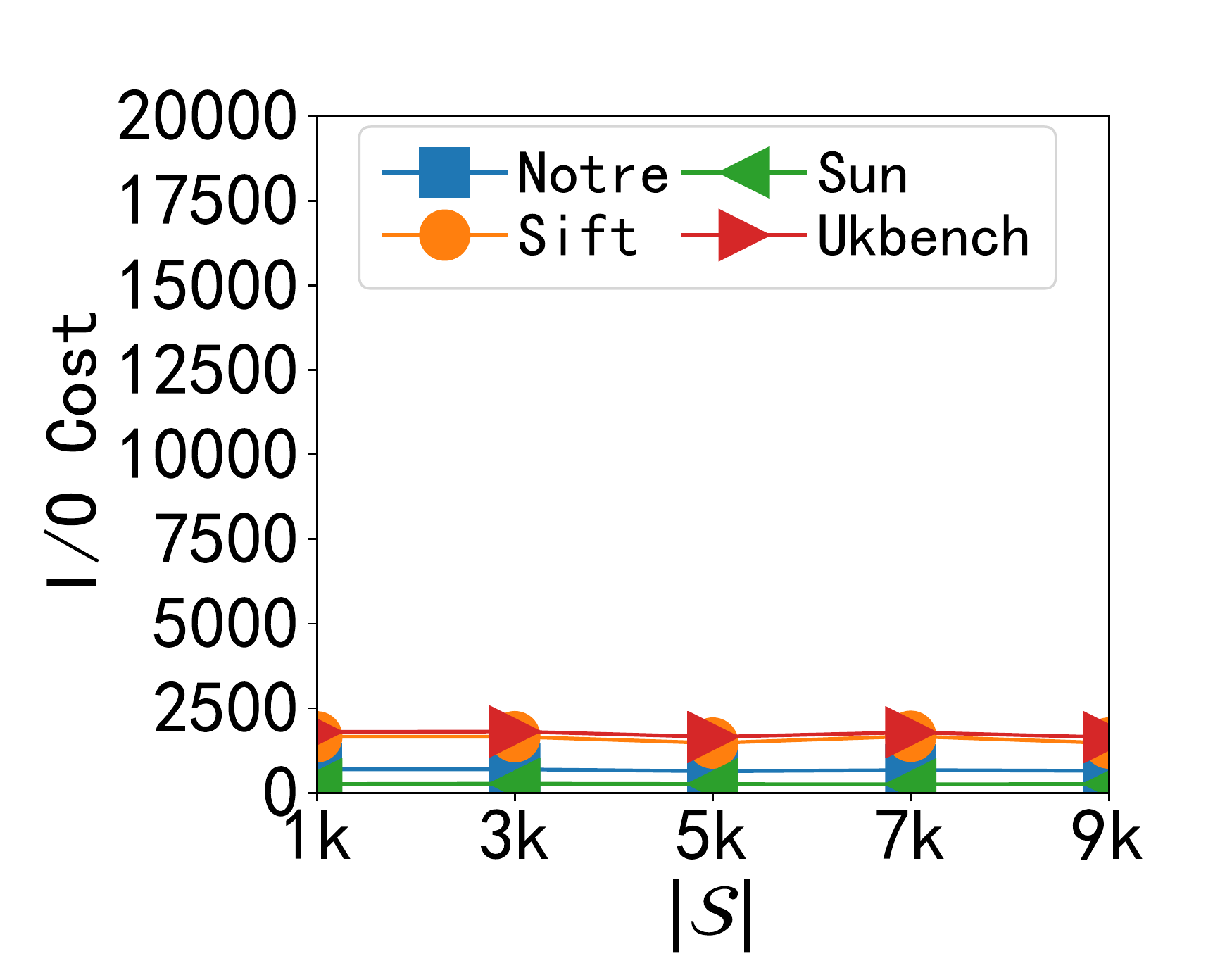}\label{real/S/IO_useCt=1_L2/k=10}}
	\subfigure[Ratio vs. $\left|S\right|$]{\includegraphics[width=0.24\textwidth]{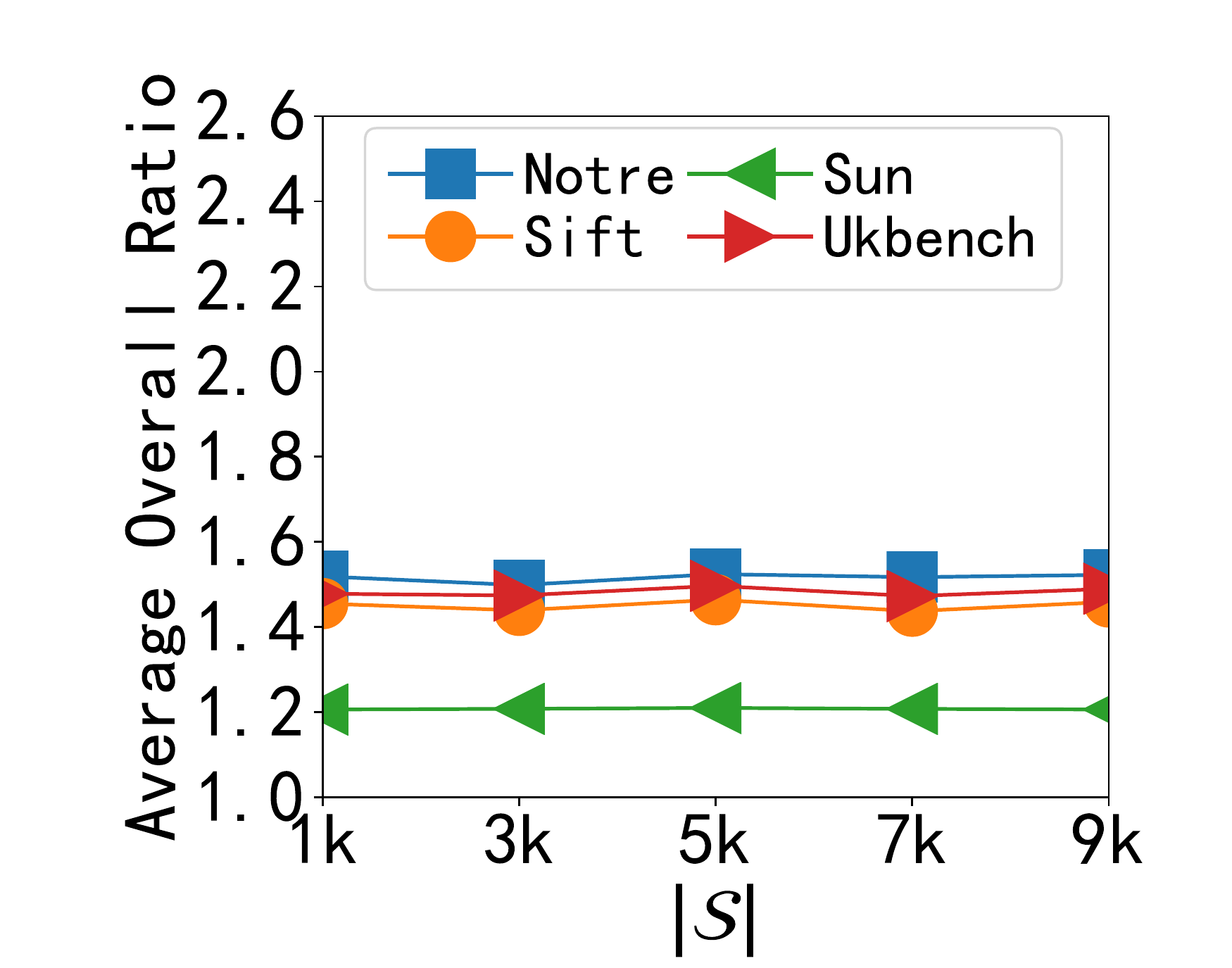}\label{real/S/ratio_useCt=1_L2/k=10}}
	
	\caption{Query efficiency and query accuracy of WLSH on real data when using collision threshold reduction, $l_2$ distance, $k=10$}
	\label{real/efficiency and accuracy/L2/useCt=1/k=10}
\end{figure}

\begin{figure}[t]
	\centering
	\subfigure[I/O cost vs. $c$]{\includegraphics[width=0.24\textwidth]{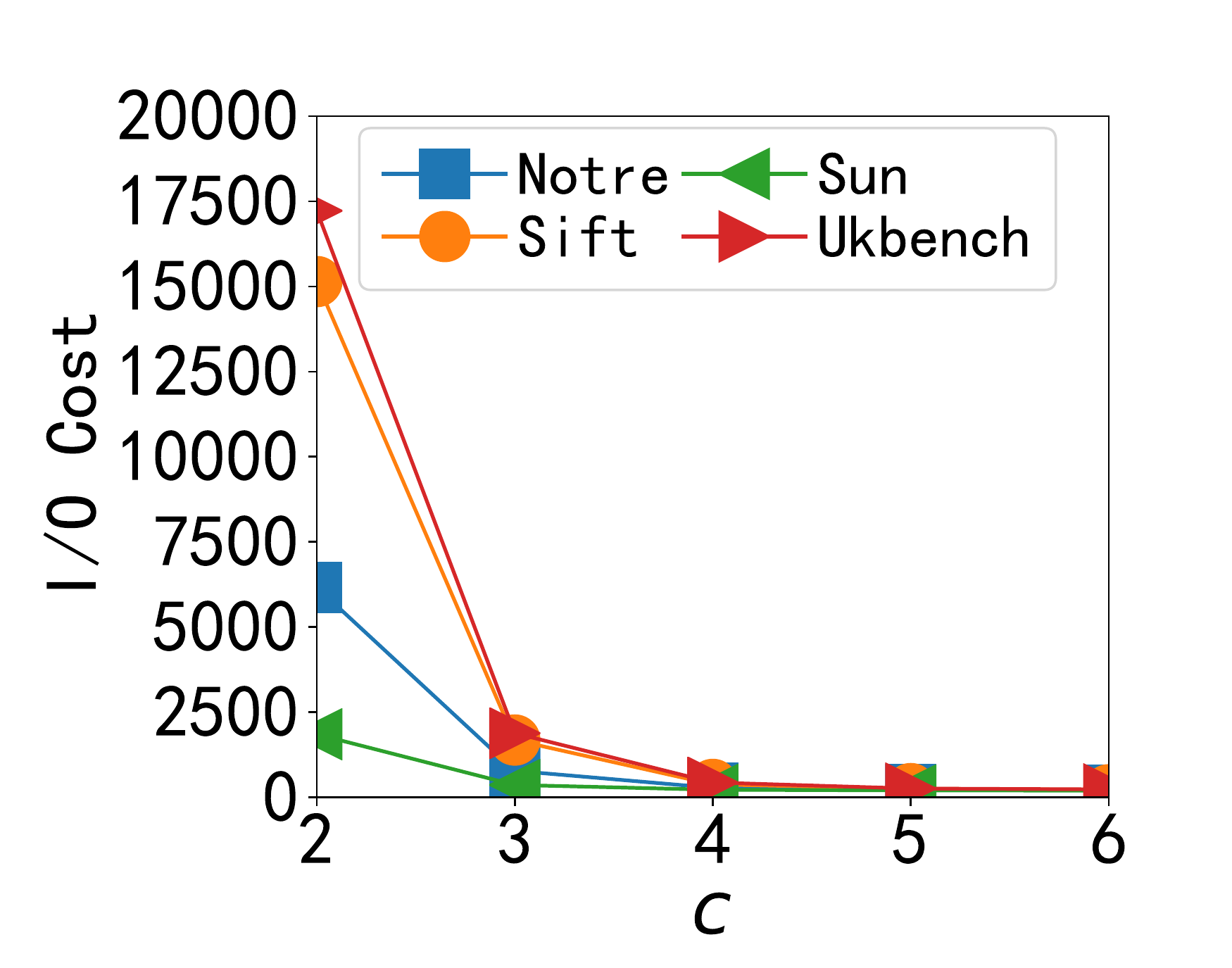}\label{real/c/IO_useCt=1_L2}}
	\subfigure[Ratio vs. $c$]{\includegraphics[width=0.24\textwidth]{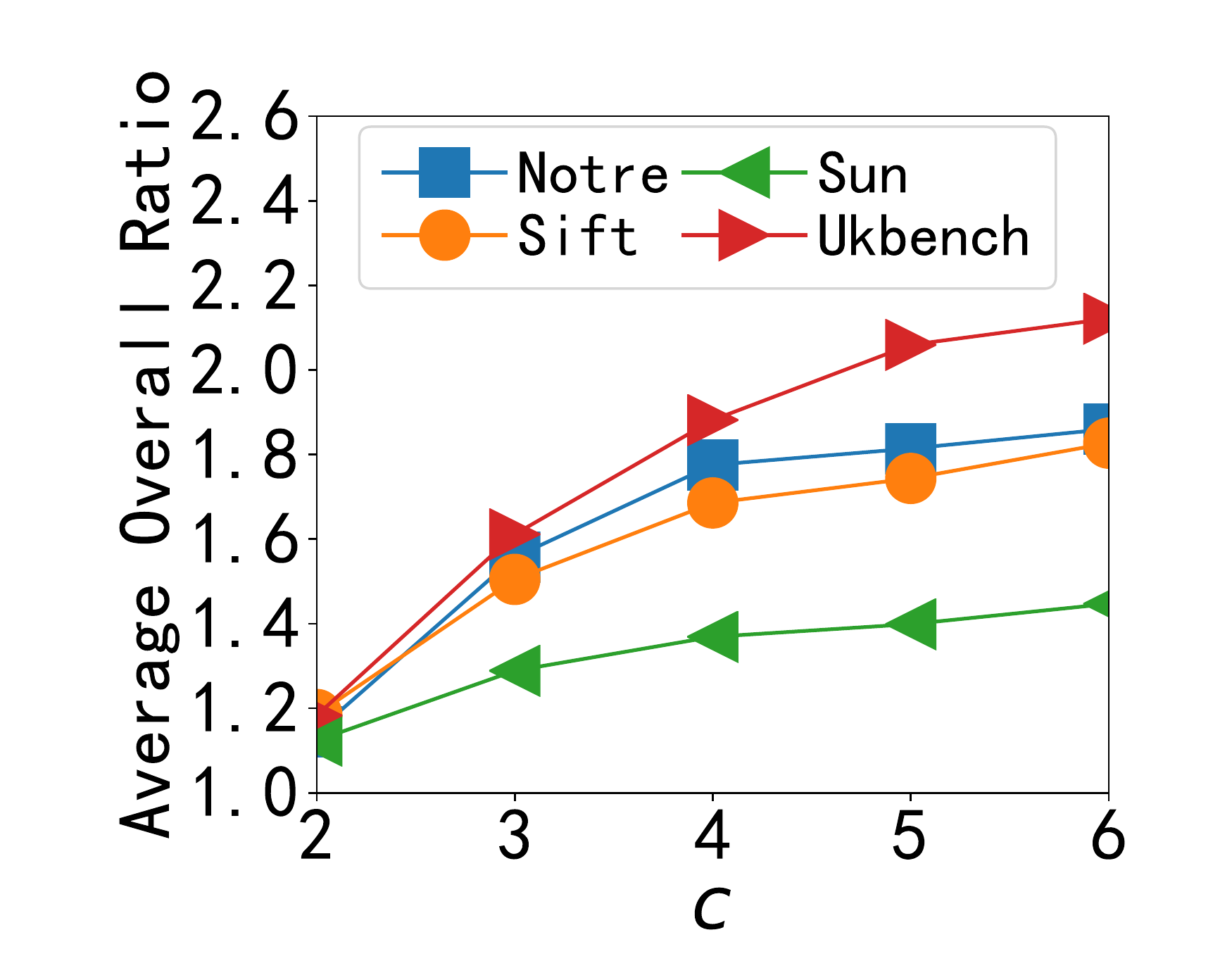}\label{real/c/ratio_useCt=1_L2}}
	
	\subfigure[I/O cost vs. $\#Subrange$]{\includegraphics[width=0.24\textwidth]{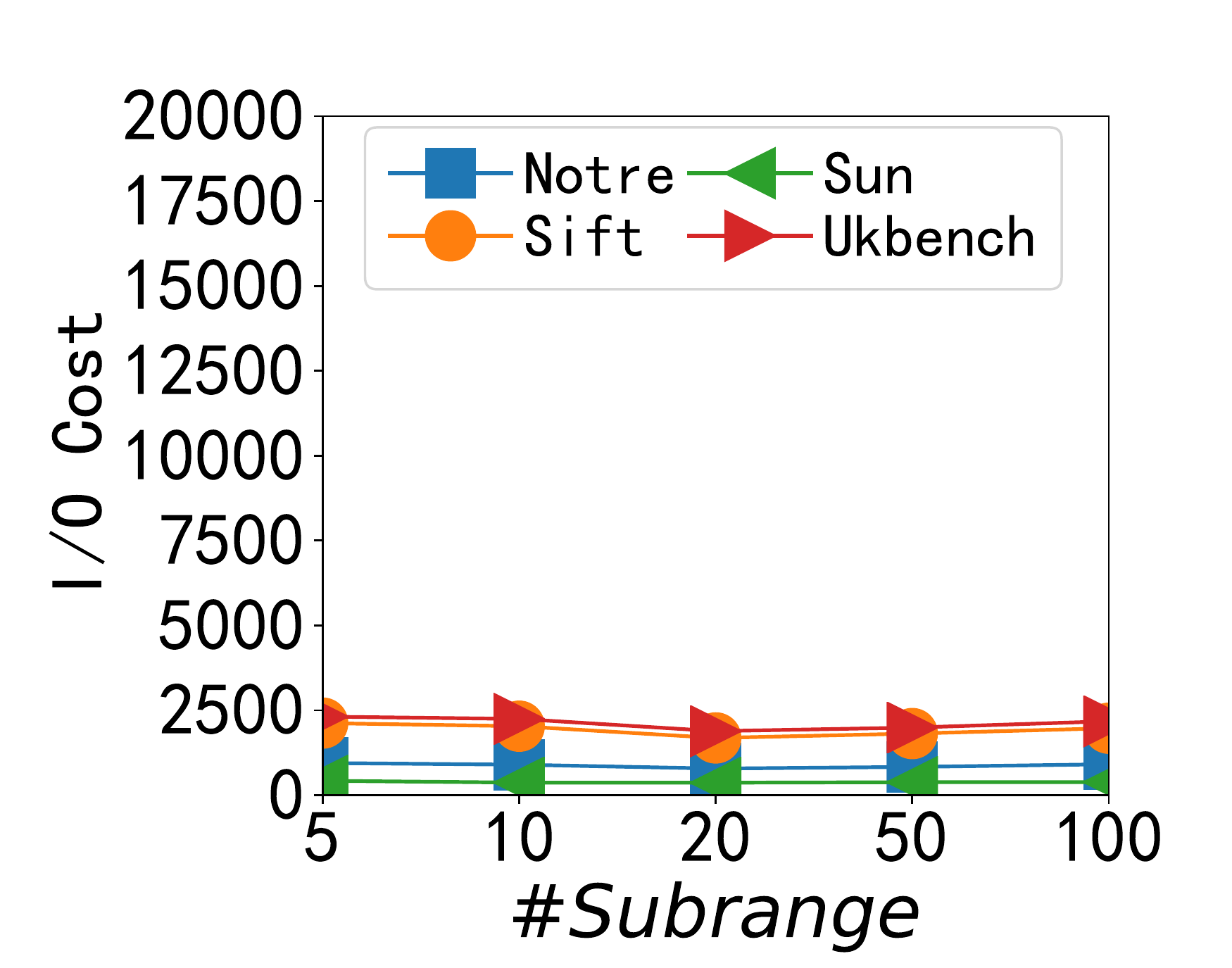}\label{real/subrange/IO_useCt=1_L2}}
	\subfigure[Ratio vs. $\#Subrange$]{\includegraphics[width=0.24\textwidth]{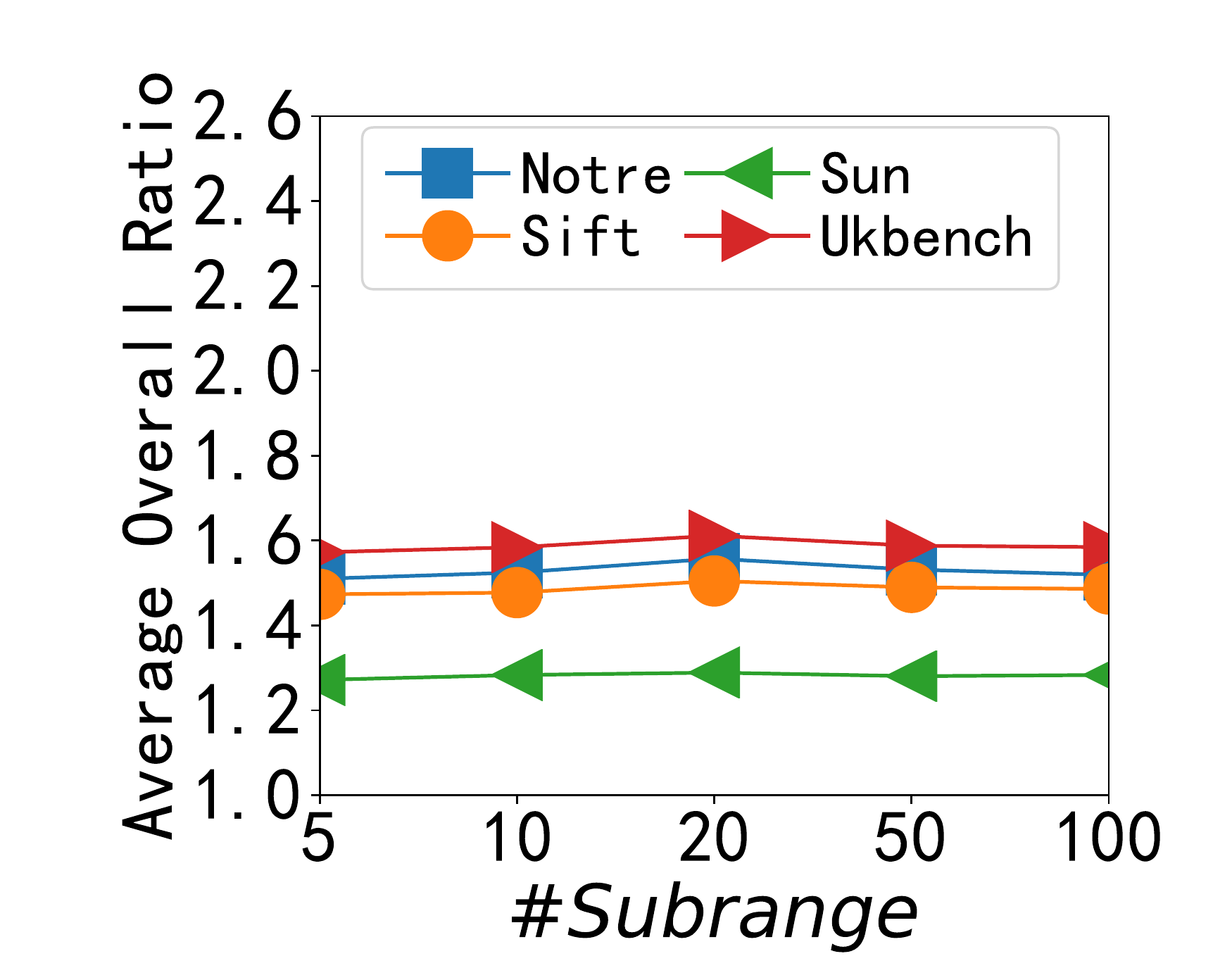}\label{real/subrange/ratio_useCt=1_L2}}
	
	\subfigure[I/O cost vs. $\#Subset$]{\includegraphics[width=0.24\textwidth]{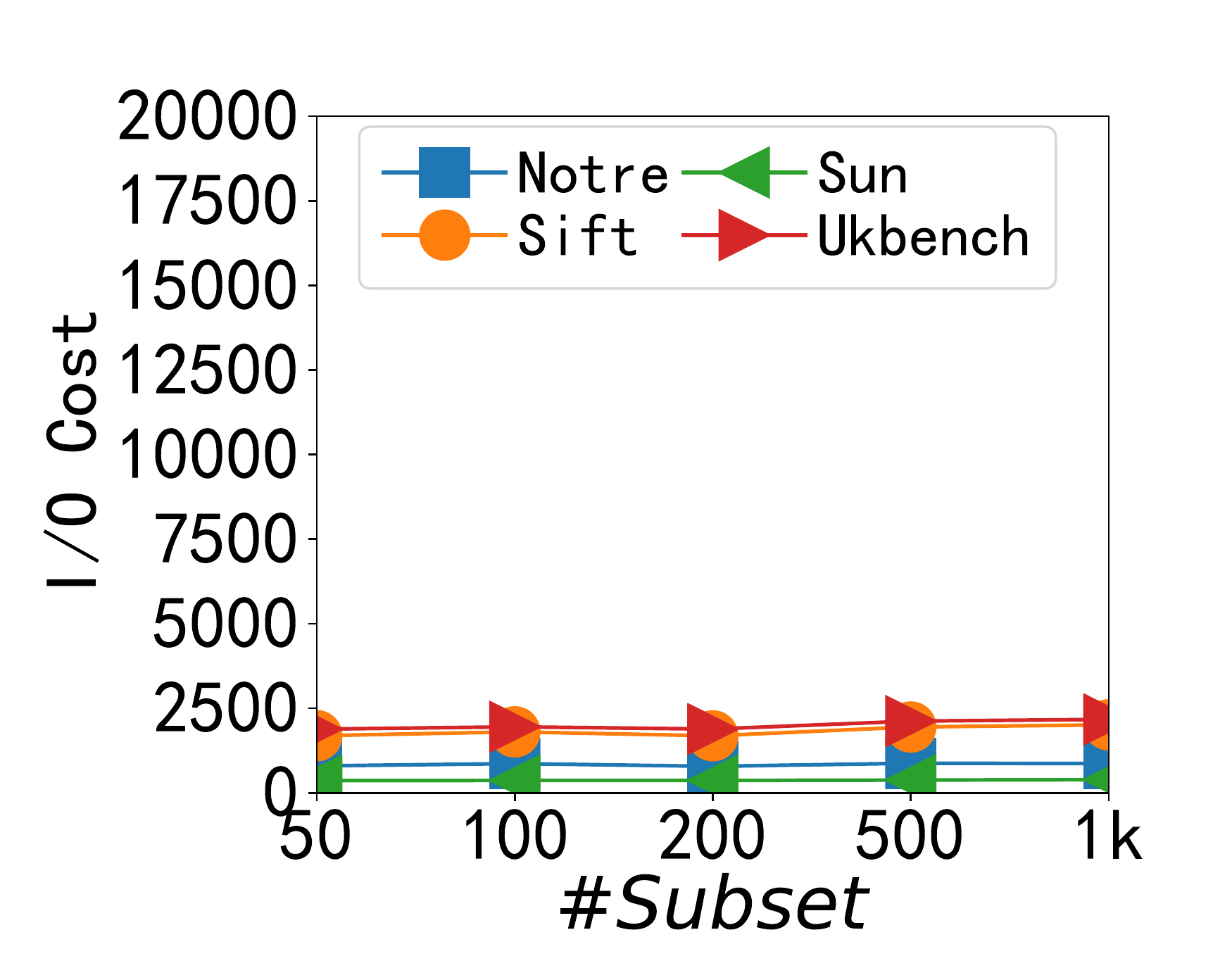}\label{real/subset/IO_useCt=1_L2}}
	\subfigure[Ratio vs. $\#Subset$]{\includegraphics[width=0.24\textwidth]{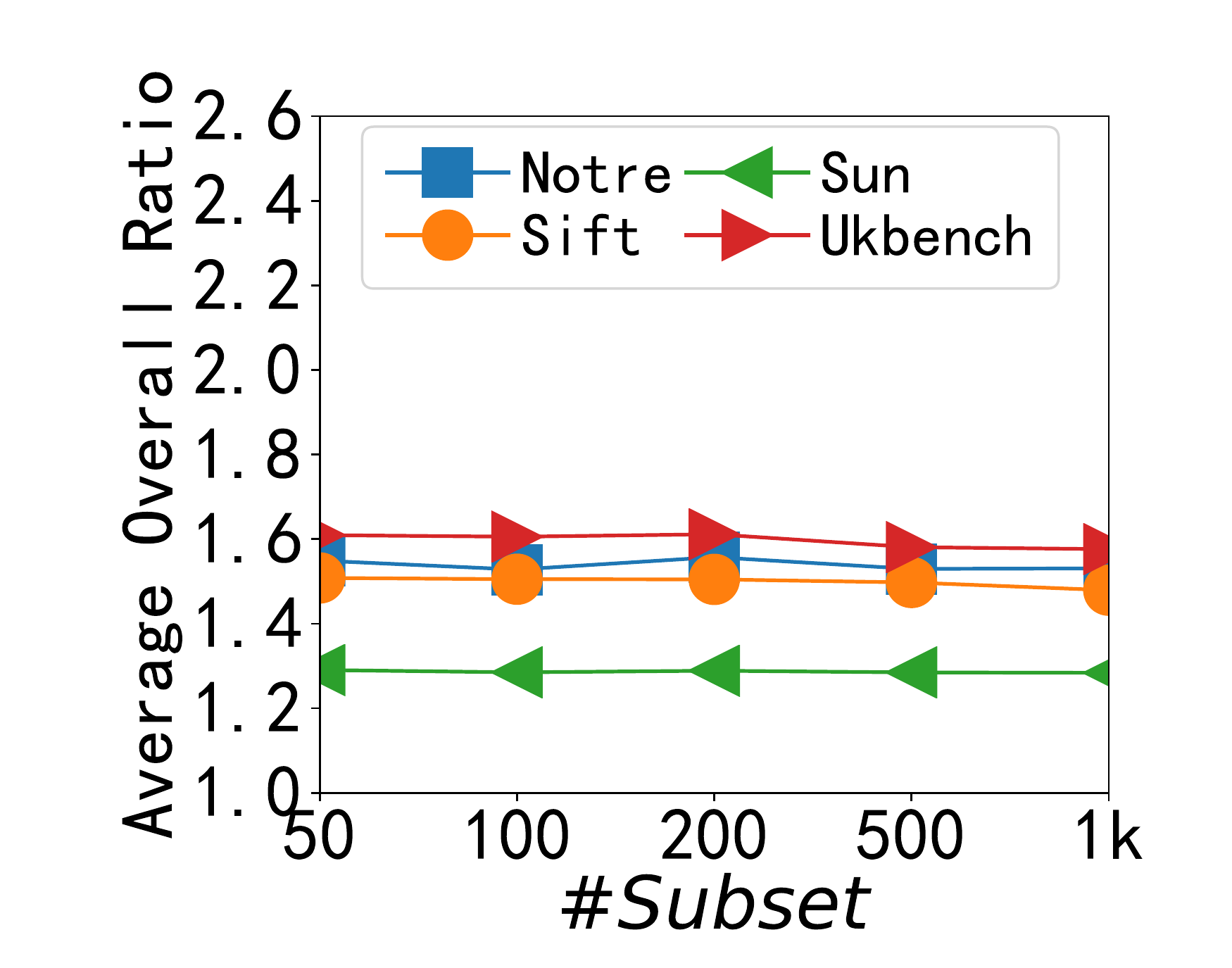}\label{real/subset/ratio_useCt=1_L2}}
	
	\subfigure[I/O cost vs. $\left|S\right|$]{\includegraphics[width=0.24\textwidth]{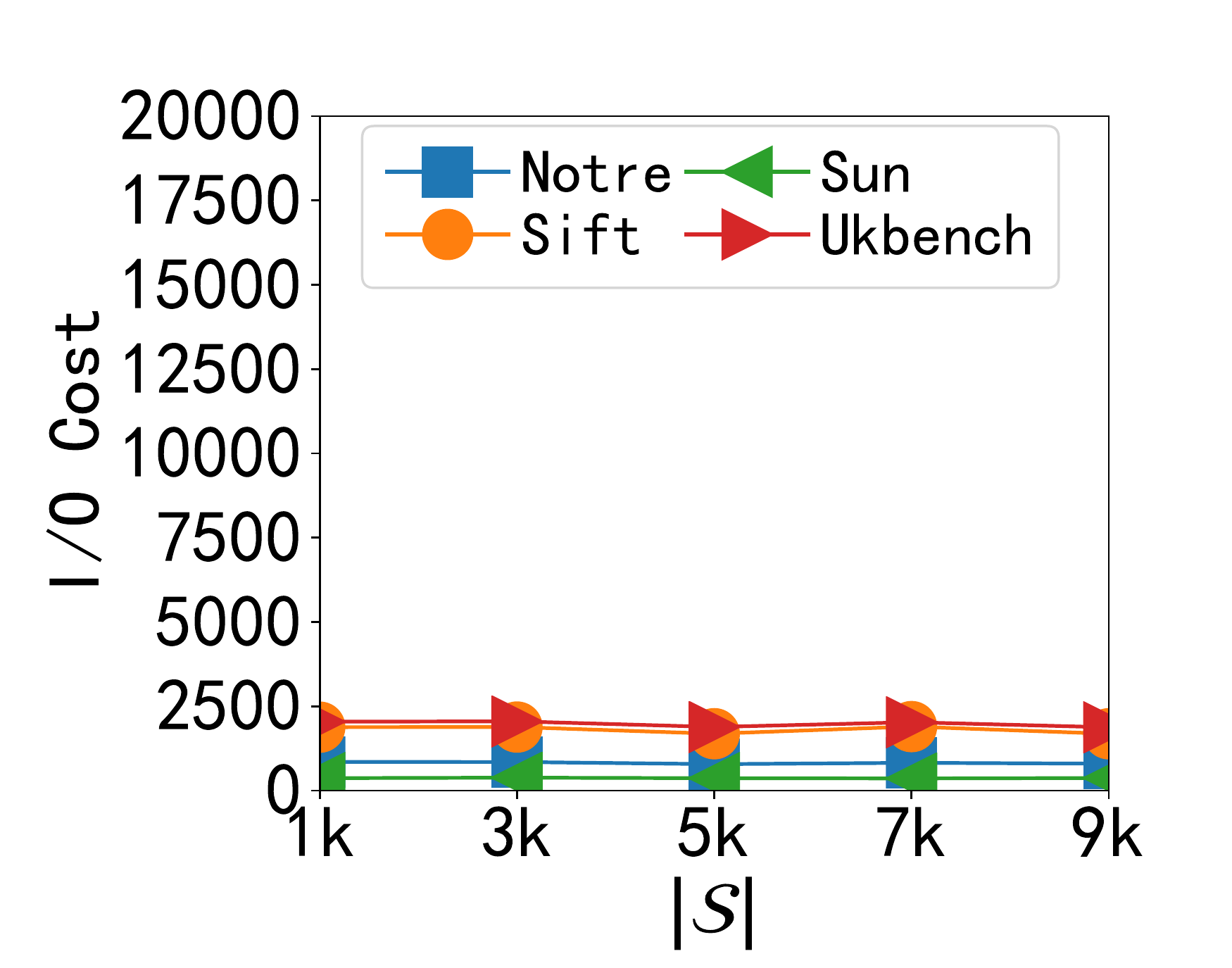}\label{real/S/IO_useCt=1_L2}}
	\subfigure[Ratio vs. $\left|S\right|$]{\includegraphics[width=0.24\textwidth]{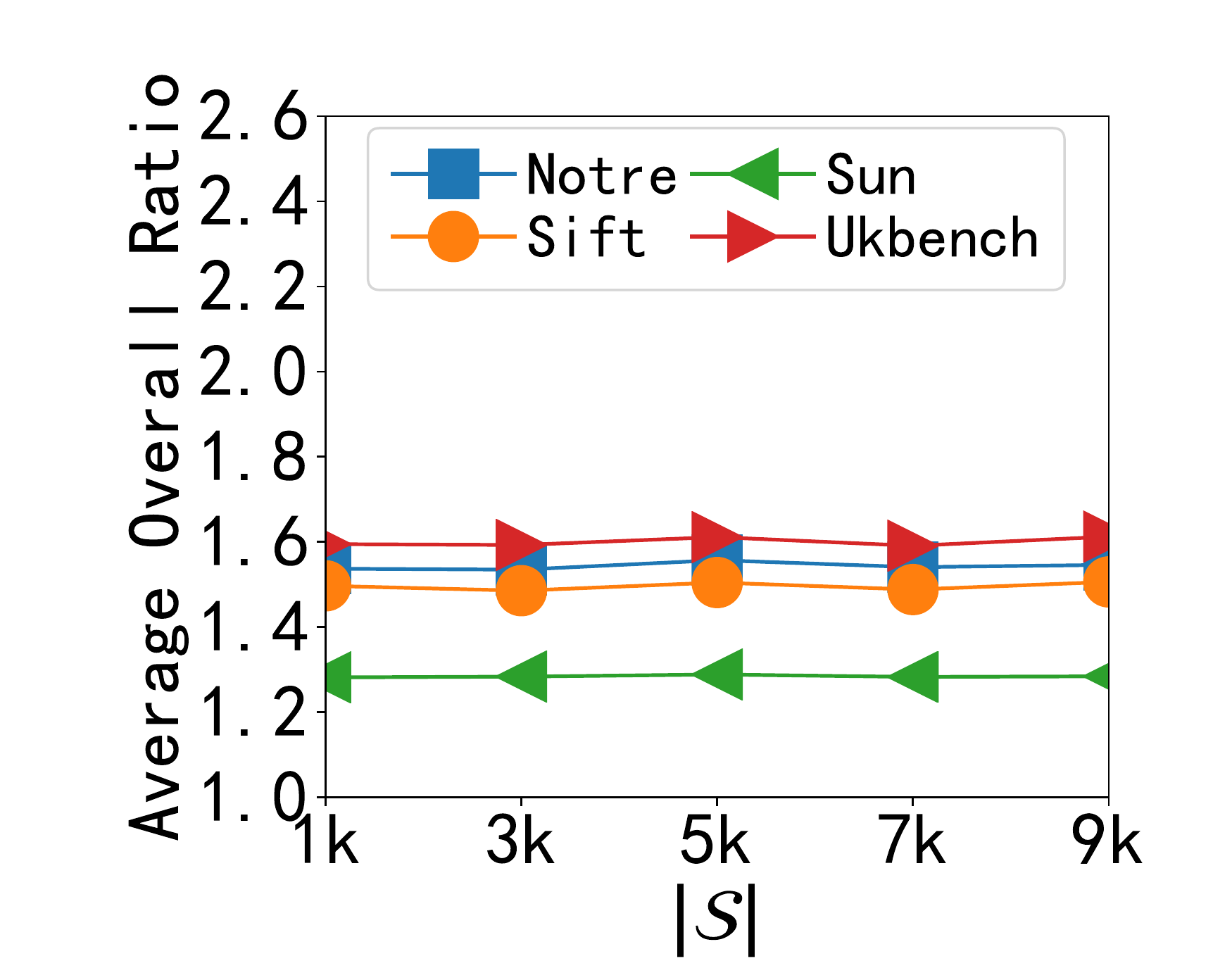}\label{real/S/ratio_useCt=1_L2}}
	
	\caption{Query efficiency and query accuracy of WLSH on real data when using collision threshold reduction, $l_2$ distance, $k=100$}
	\label{real/efficiency and accuracy/L2/useCt=1}
\end{figure}

\begin{figure*}[t]
	\centering
	\subfigure[\textit{Notre}, $c$]{\includegraphics[width=0.246\textwidth]{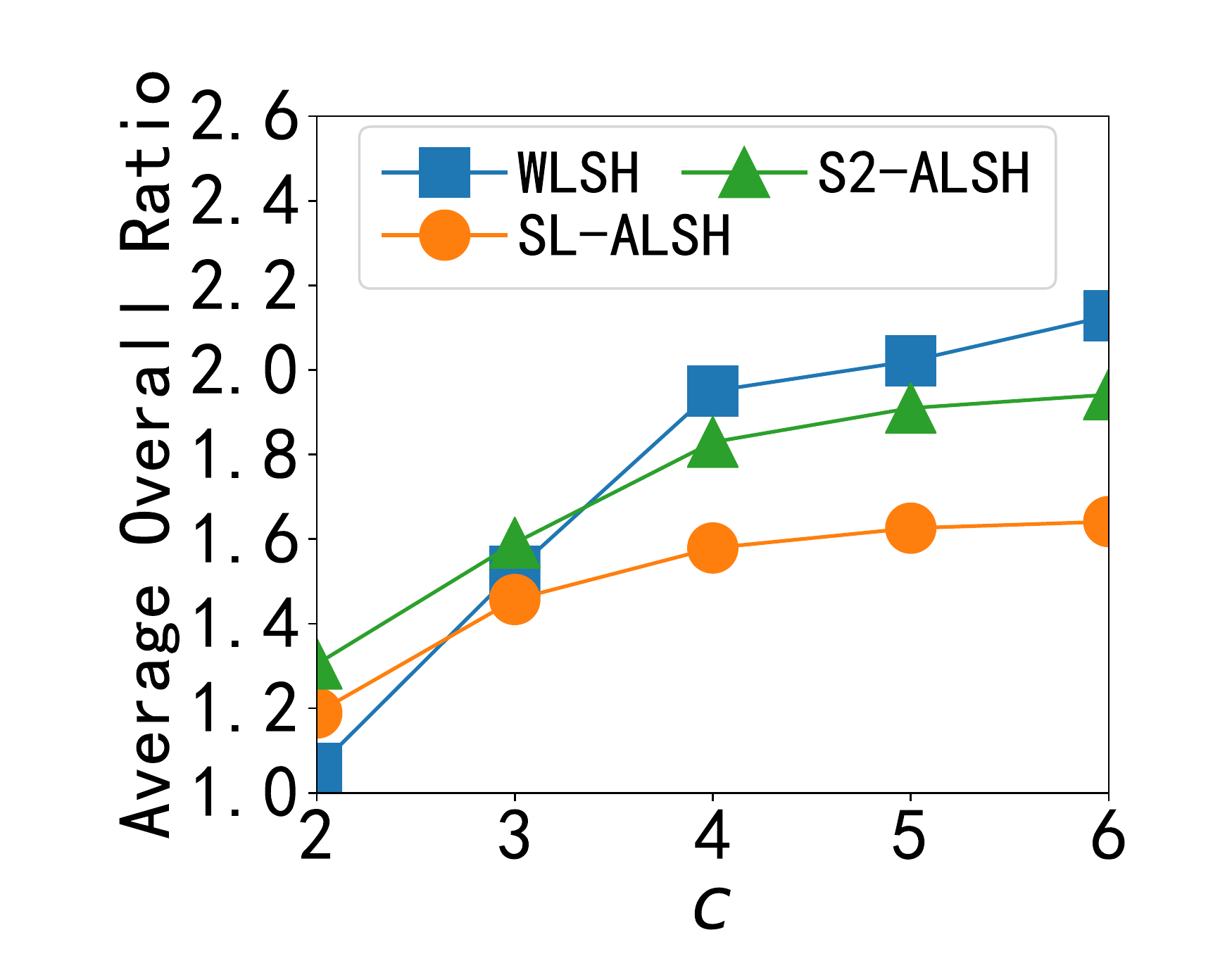}\label{notre/c/ratio_useCt=1_k=10}}
	\subfigure[\textit{Notre}, $\#Subrange$]{\includegraphics[width=0.246\textwidth]{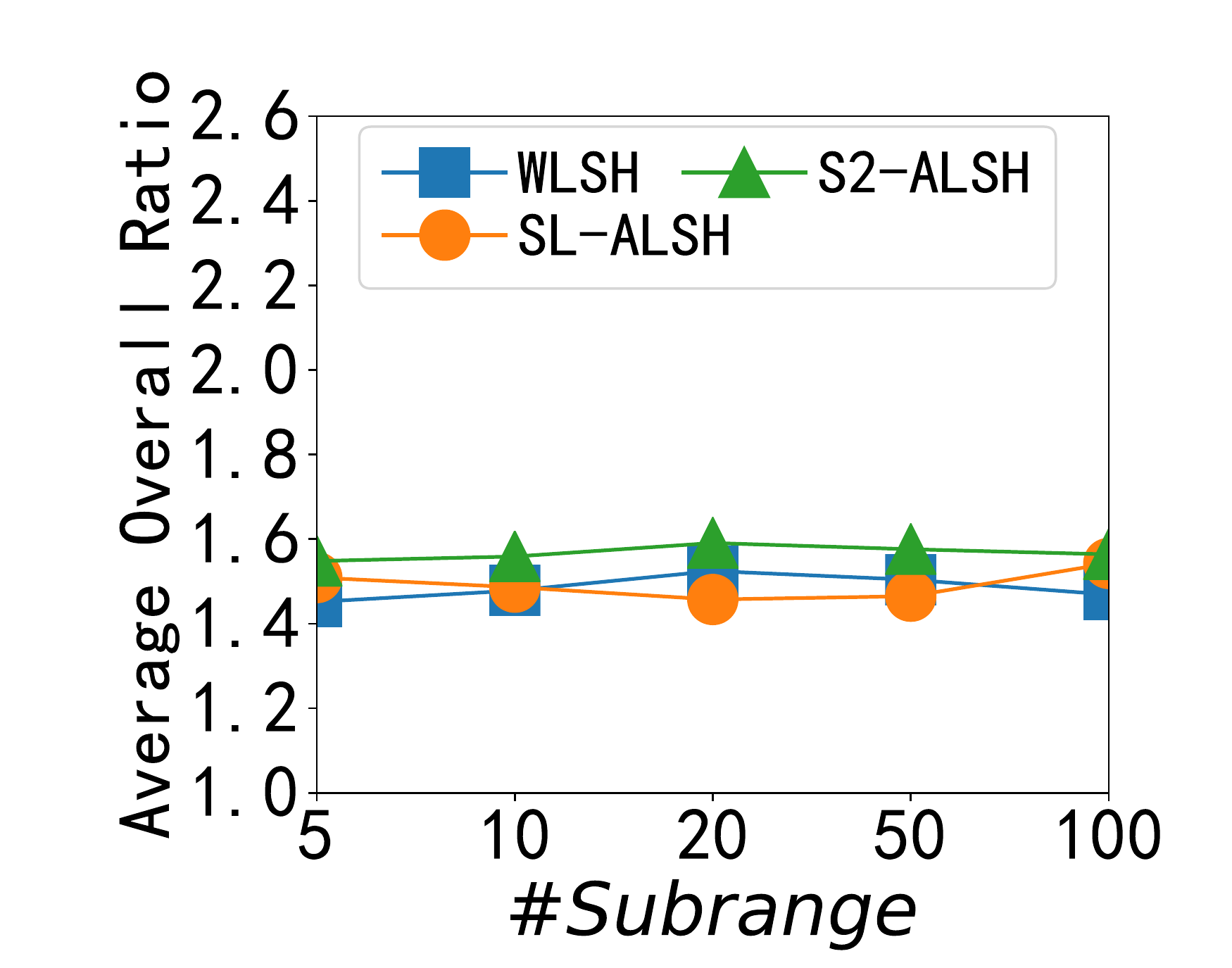}\label{notre/subrange/ratio_useCt=1_k=10}}
	\subfigure[\textit{Notre}, $\#Subset$]{\includegraphics[width=0.246\textwidth]{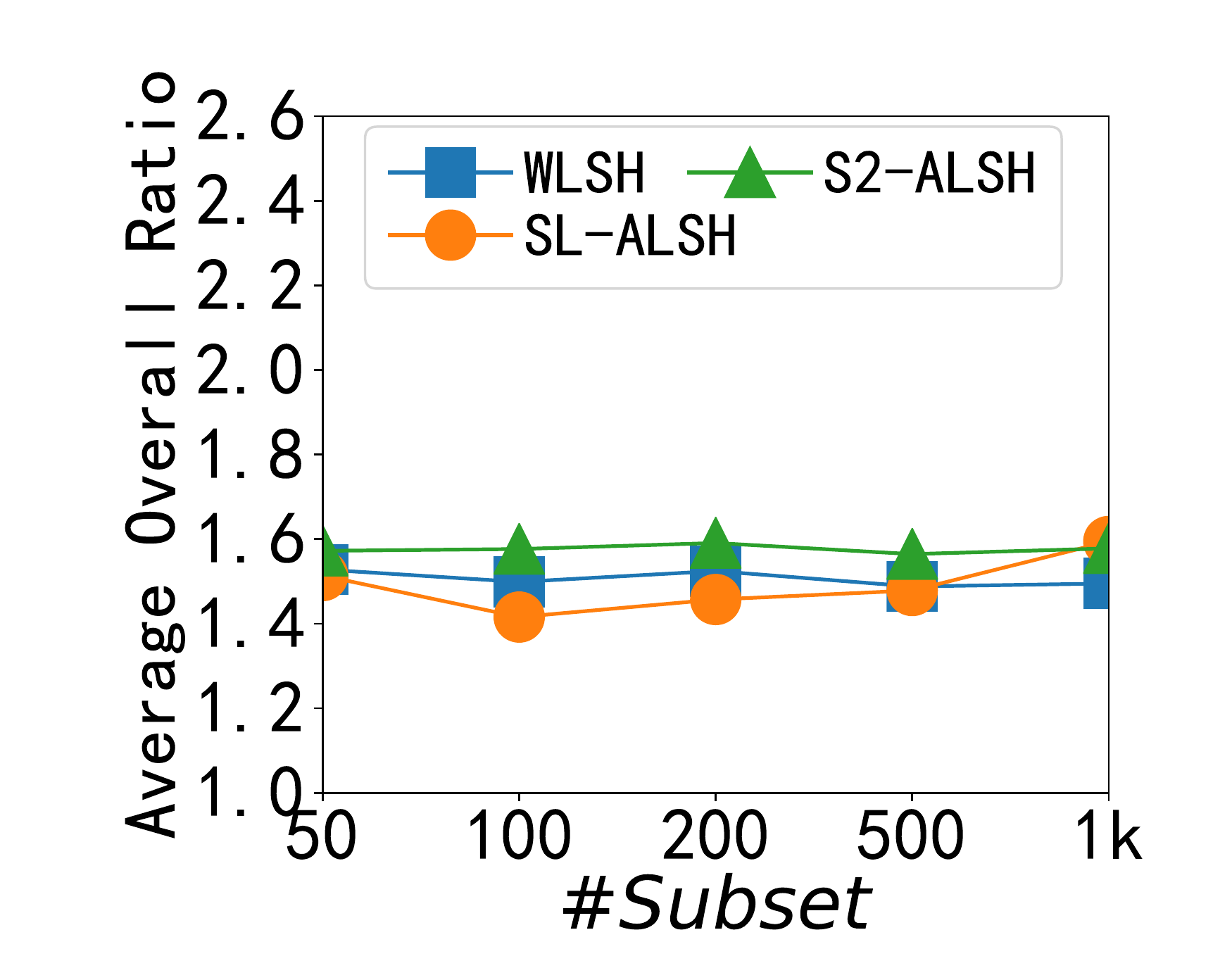}\label{notre/subset/ratio_useCt=1_k=10}}
	\subfigure[\textit{Notre}, $\left|S\right|$]{\includegraphics[width=0.246\textwidth]{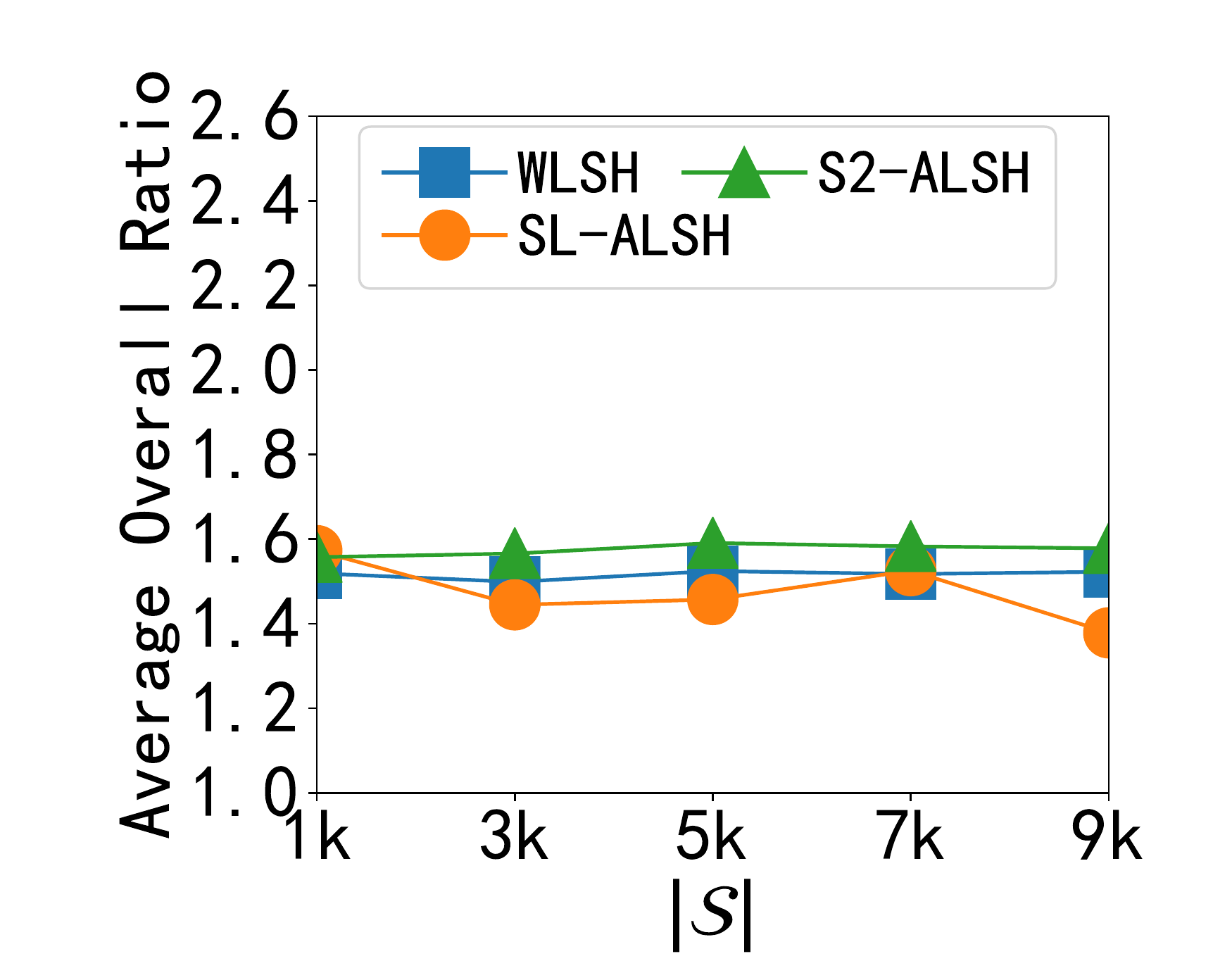}\label{notre/S/ratio_useCt=1_k=10}}
	
	\subfigure[\textit{Ukbench}, $c$]{\includegraphics[width=0.246\textwidth]{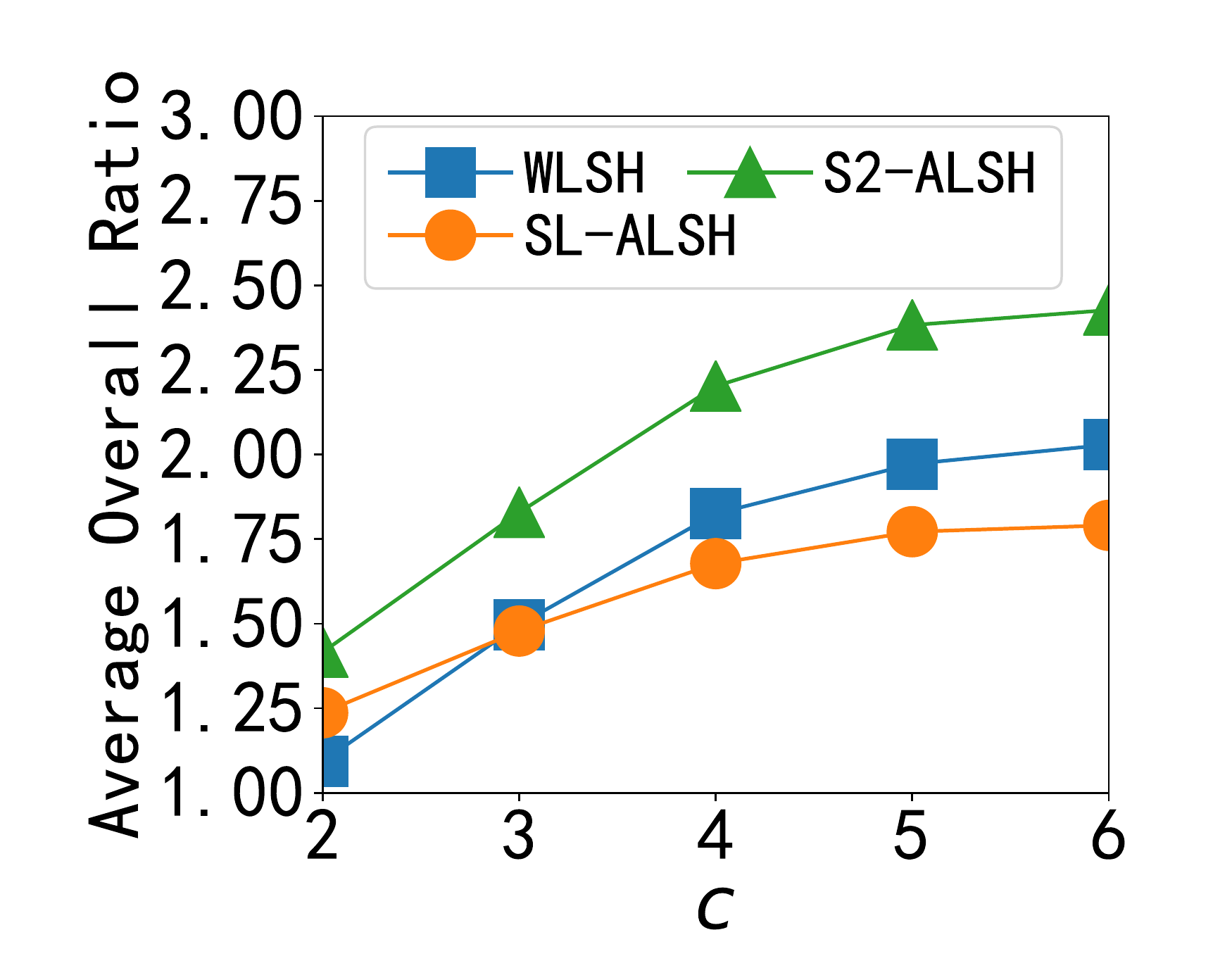}\label{ukbench/c/ratio_useCt=1_k=10}}
	\subfigure[\textit{Ukbench}, $\#Subrange$]{\includegraphics[width=0.246\textwidth]{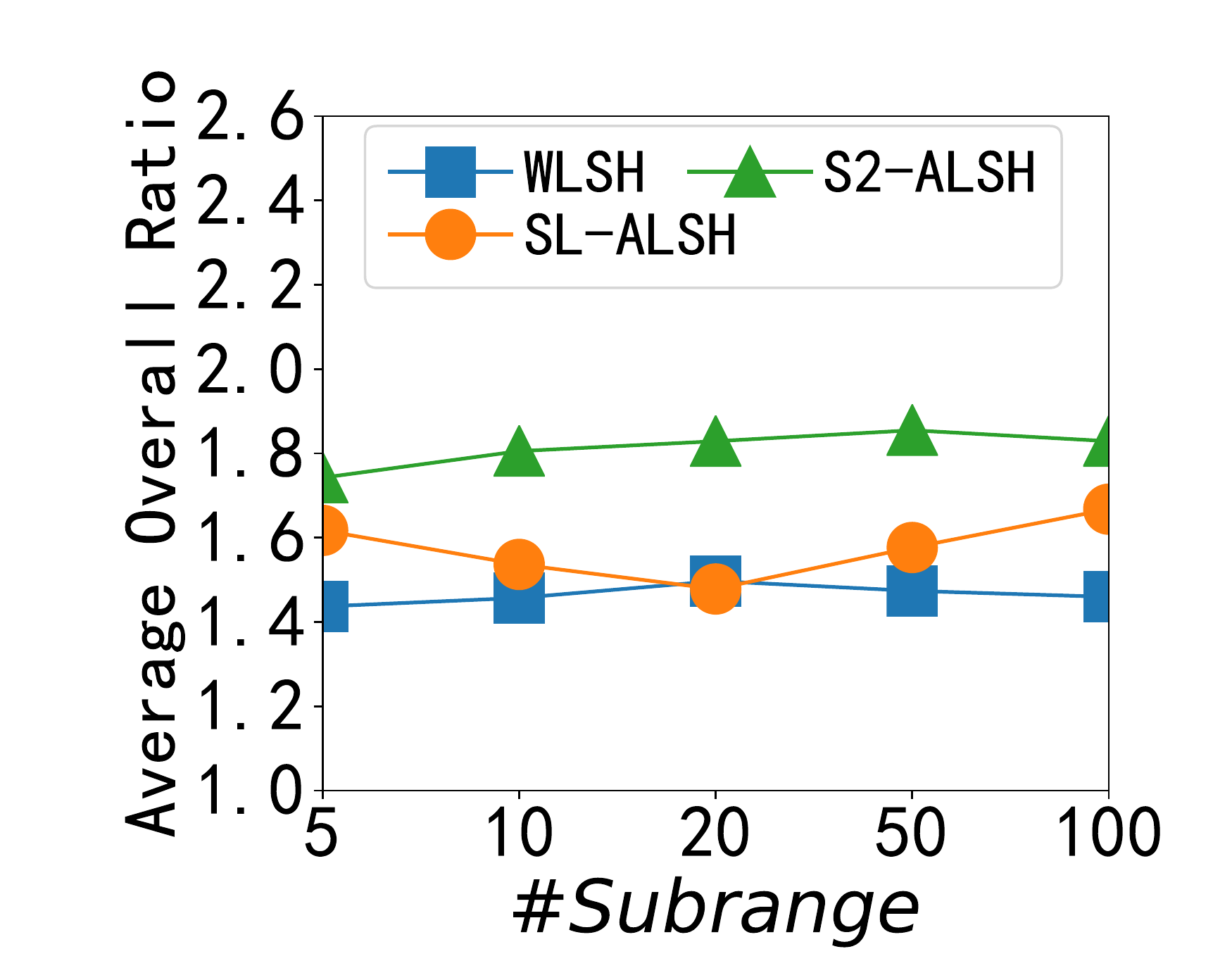}\label{ukbench/subrange/ratio_useCt=1_k=10}}
	\subfigure[\textit{Ukbench}, $\#Subset$]{\includegraphics[width=0.246\textwidth]{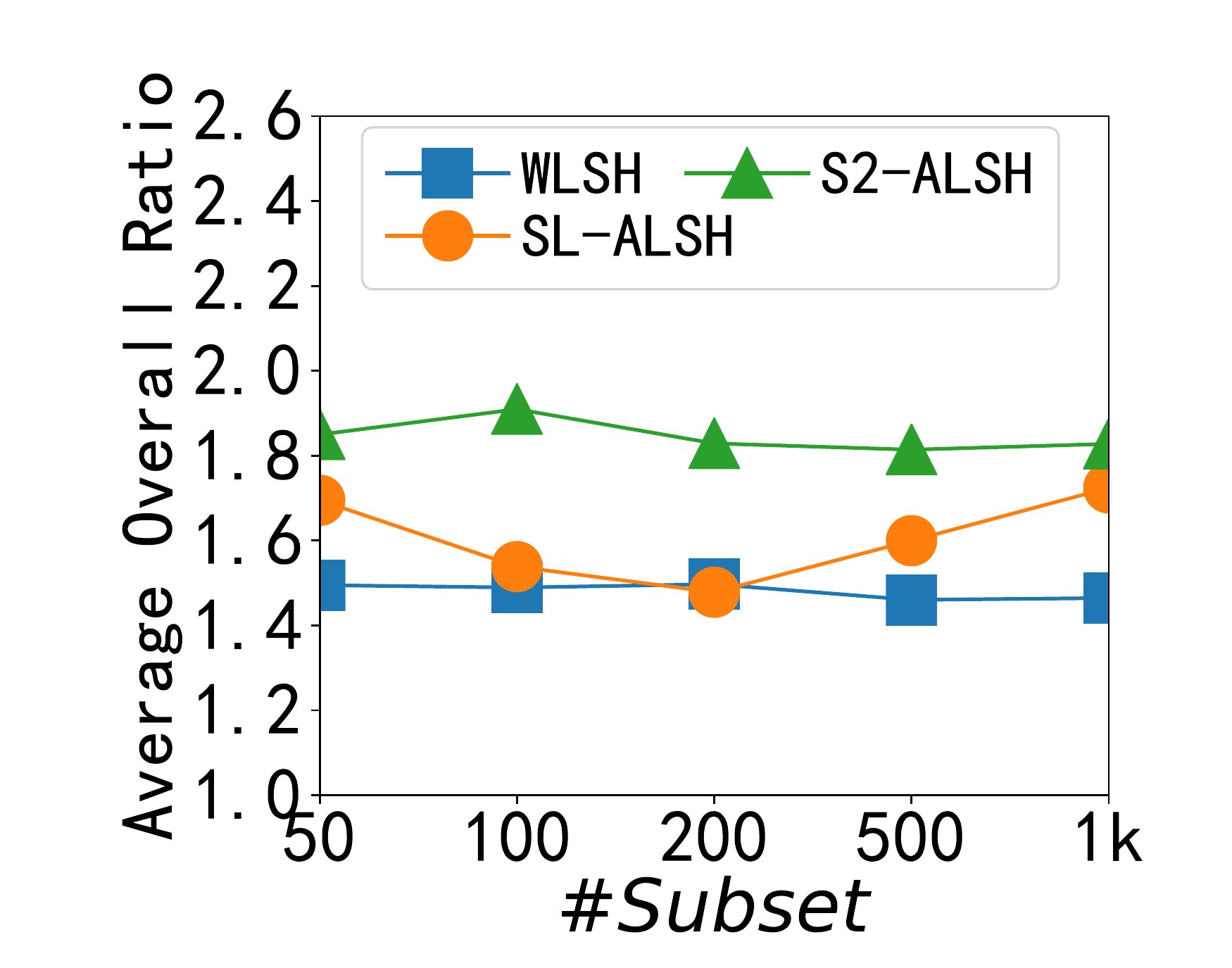}\label{ukbench/subset/ratio_useCt=1_k=10}}
	\subfigure[\textit{Ukbench}, $\left|S\right|$]{\includegraphics[width=0.246\textwidth]{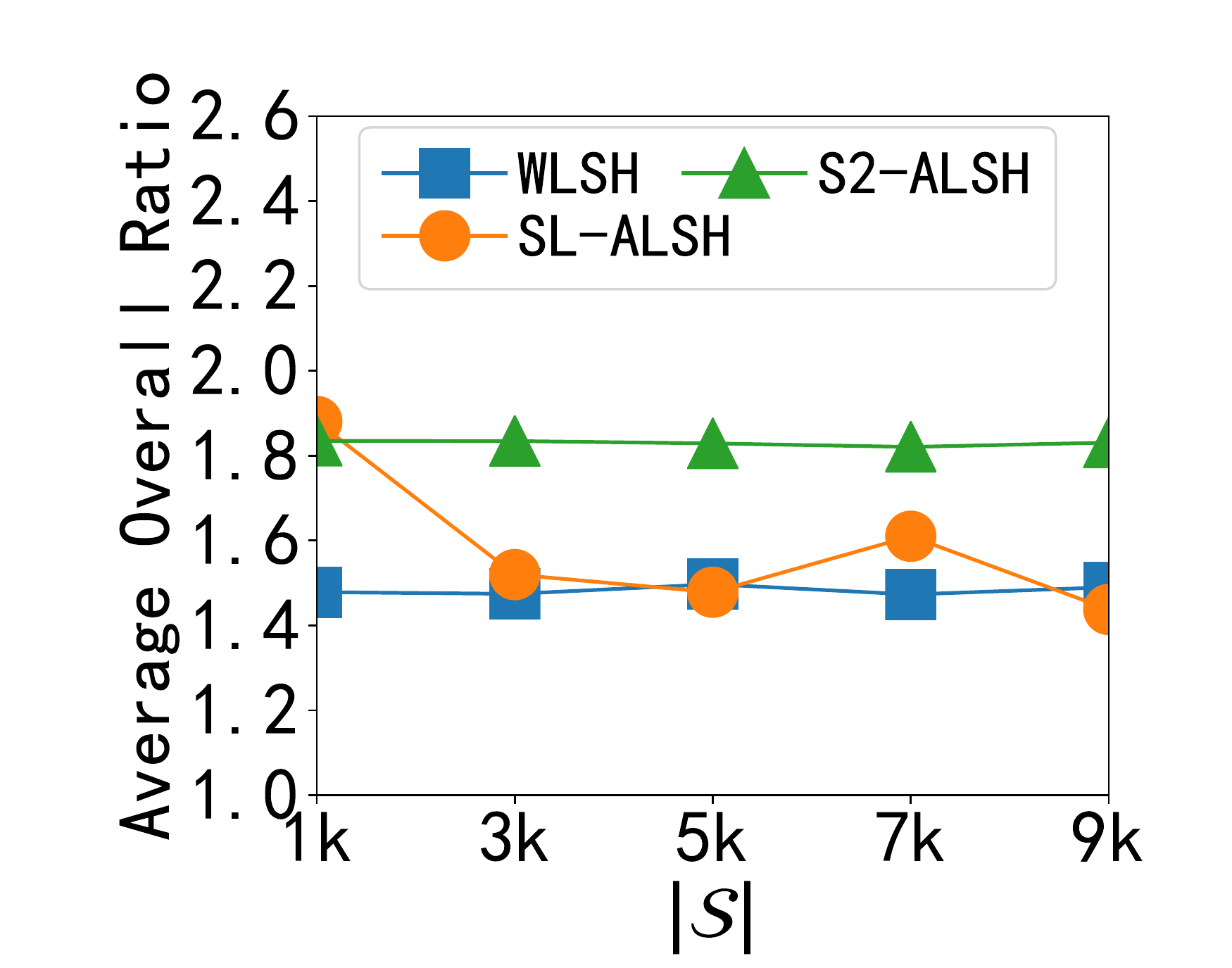}\label{ukbench/S/ratio_useCt=1_k=10}}
	
	\subfigure[\textit{Sift}, $c$]{\includegraphics[width=0.246\textwidth]{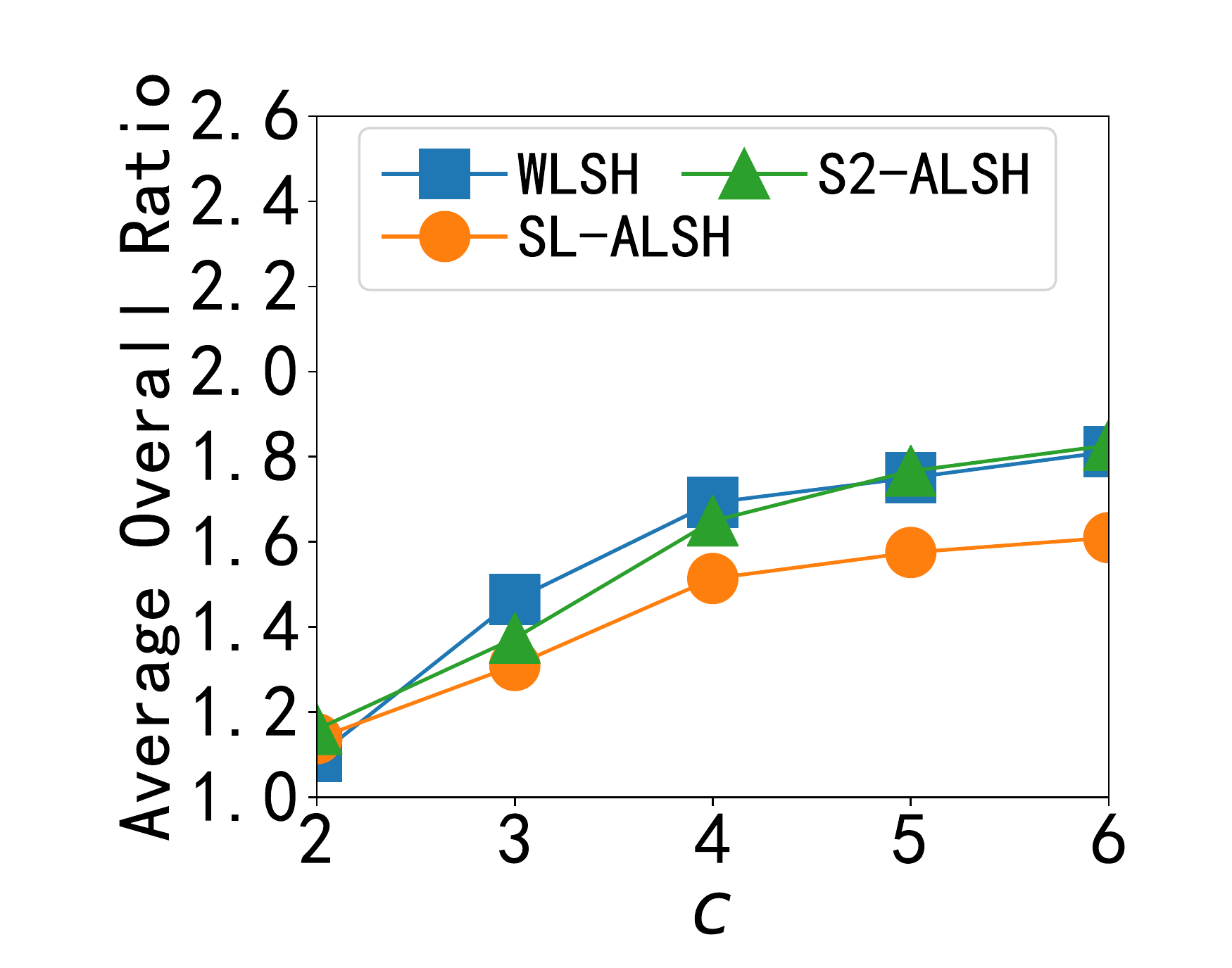}\label{sift/c/ratio_useCt=1_k=10}}
	\subfigure[\textit{Sift}, $\#Subrange$]{\includegraphics[width=0.246\textwidth]{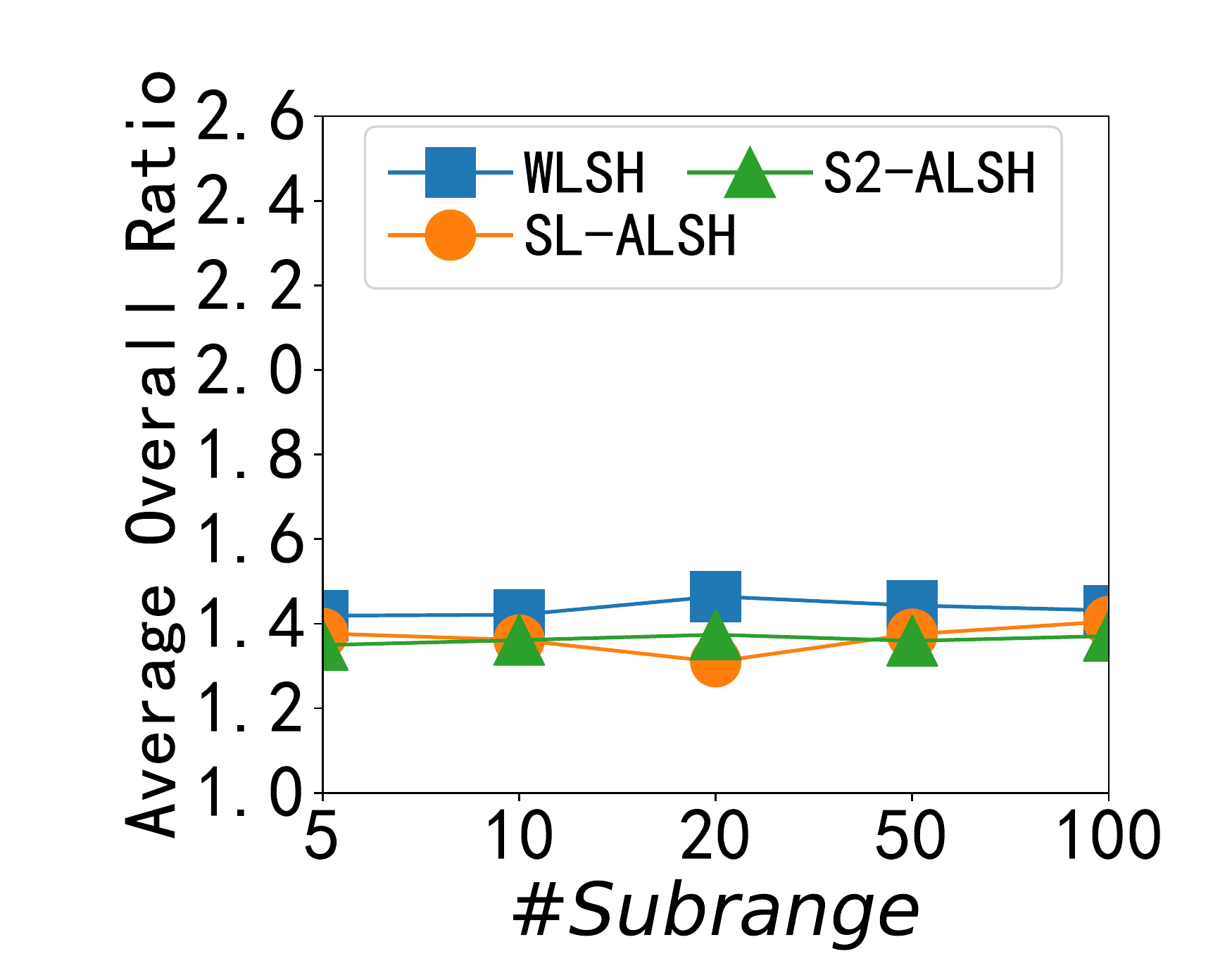}\label{sift/subrange/ratio_useCt=1_k=10}}
	\subfigure[\textit{Sift}, $\#Subset$]{\includegraphics[width=0.246\textwidth]{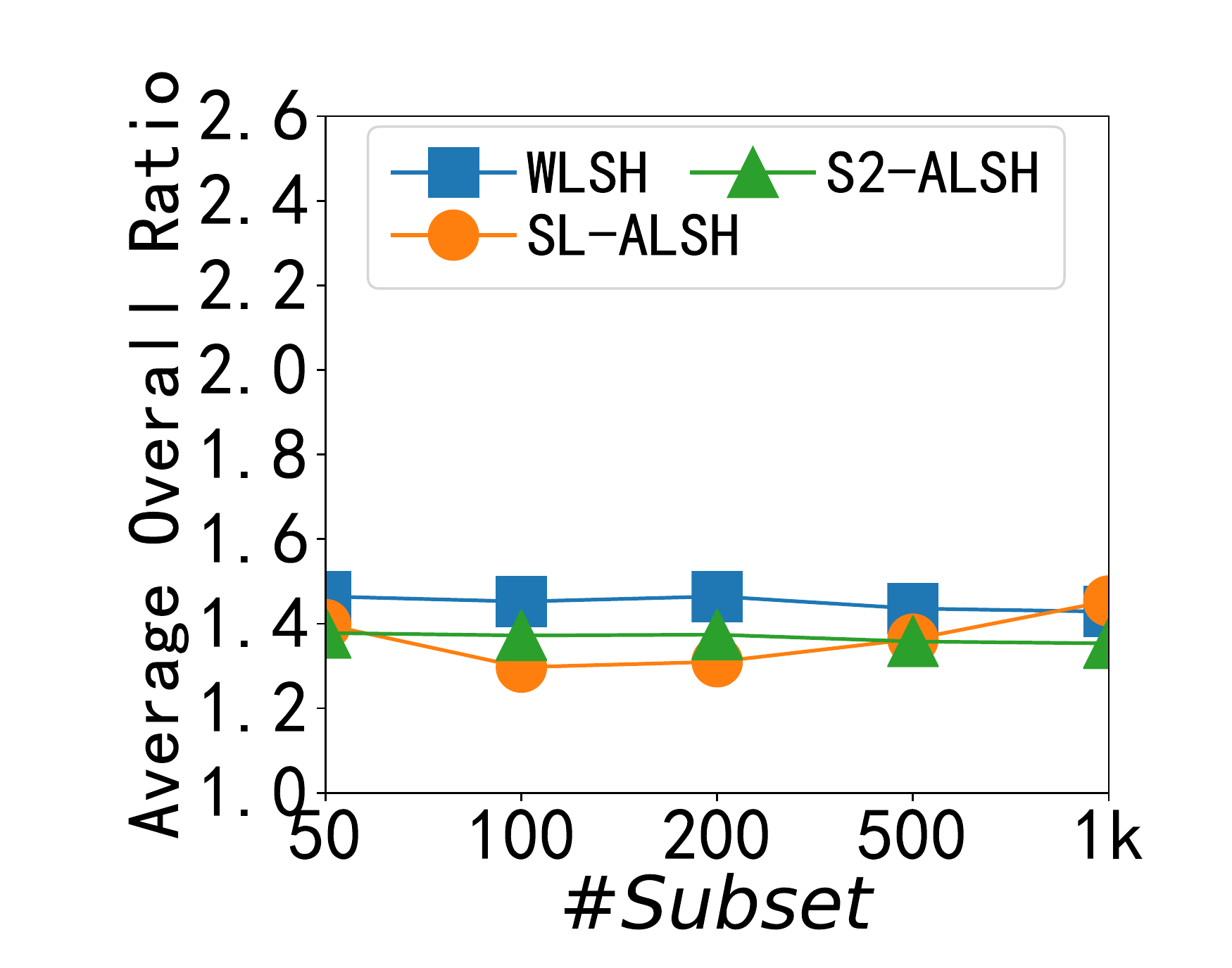}\label{sift/subset/ratio_useCt=1_k=10}}
	\subfigure[\textit{Sift}, $\left|S\right|$]{\includegraphics[width=0.246\textwidth]{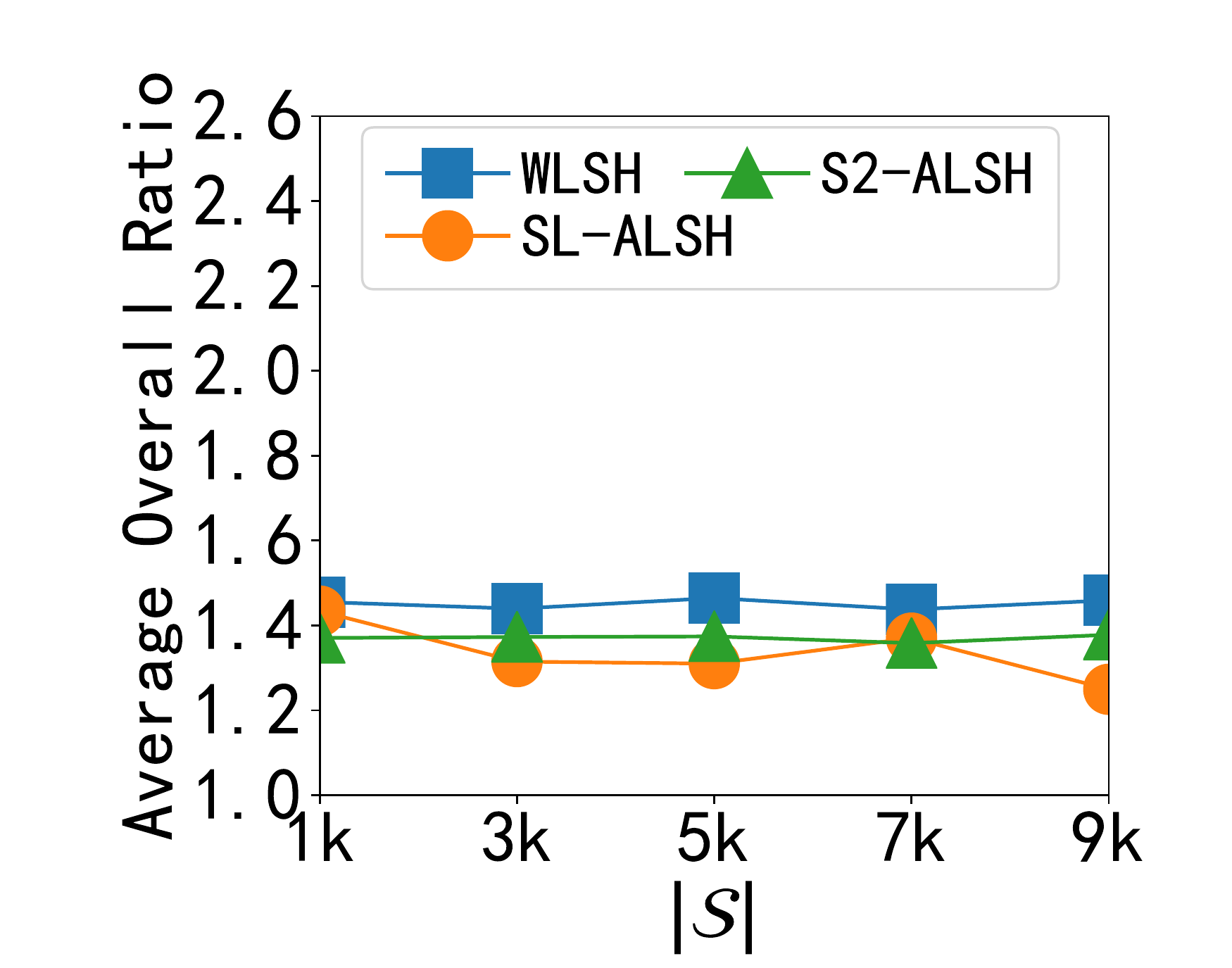}\label{sift/S/ratio_useCt=1_k=10}}
	
	\subfigure[\textit{Sun}, $c$]{\includegraphics[width=0.246\textwidth]{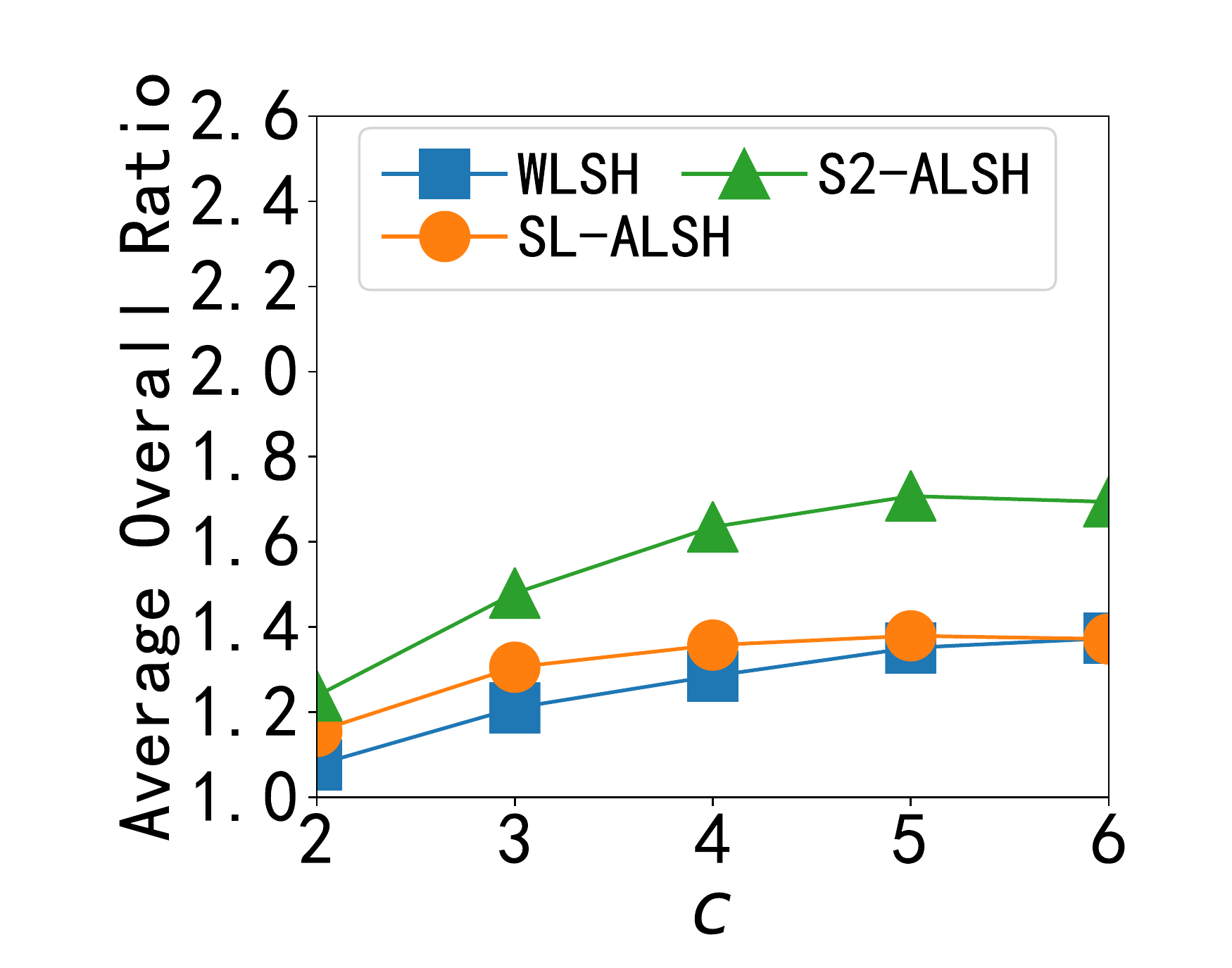}\label{sun/c/ratio_useCt=1_k=10}}
	\subfigure[\textit{Sun}, $\#Subrange$]{\includegraphics[width=0.246\textwidth]{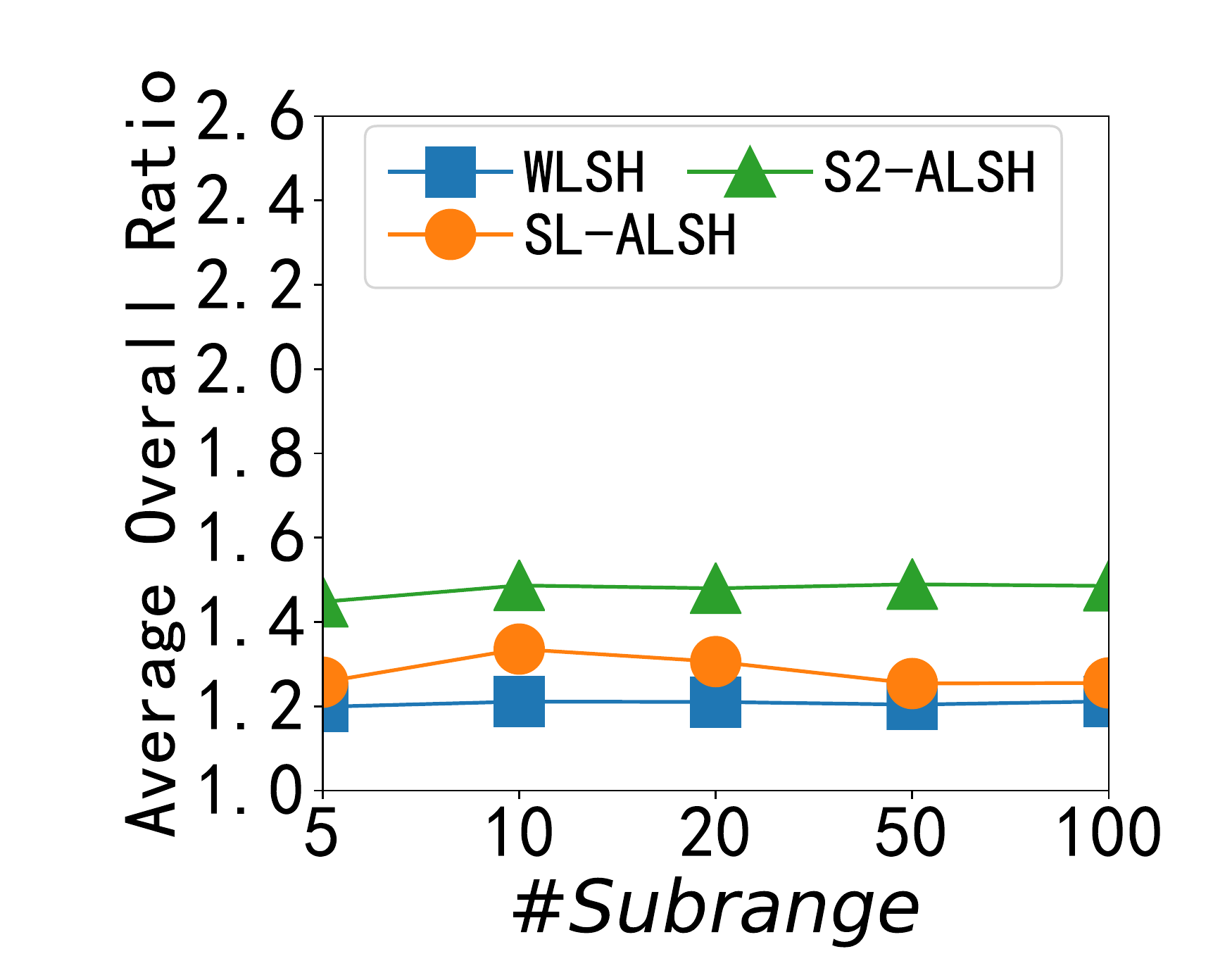}\label{sun/subrange/ratio_useCt=1_k=10}}
	\subfigure[\textit{Sun}, $\#Subset$]{\includegraphics[width=0.246\textwidth]{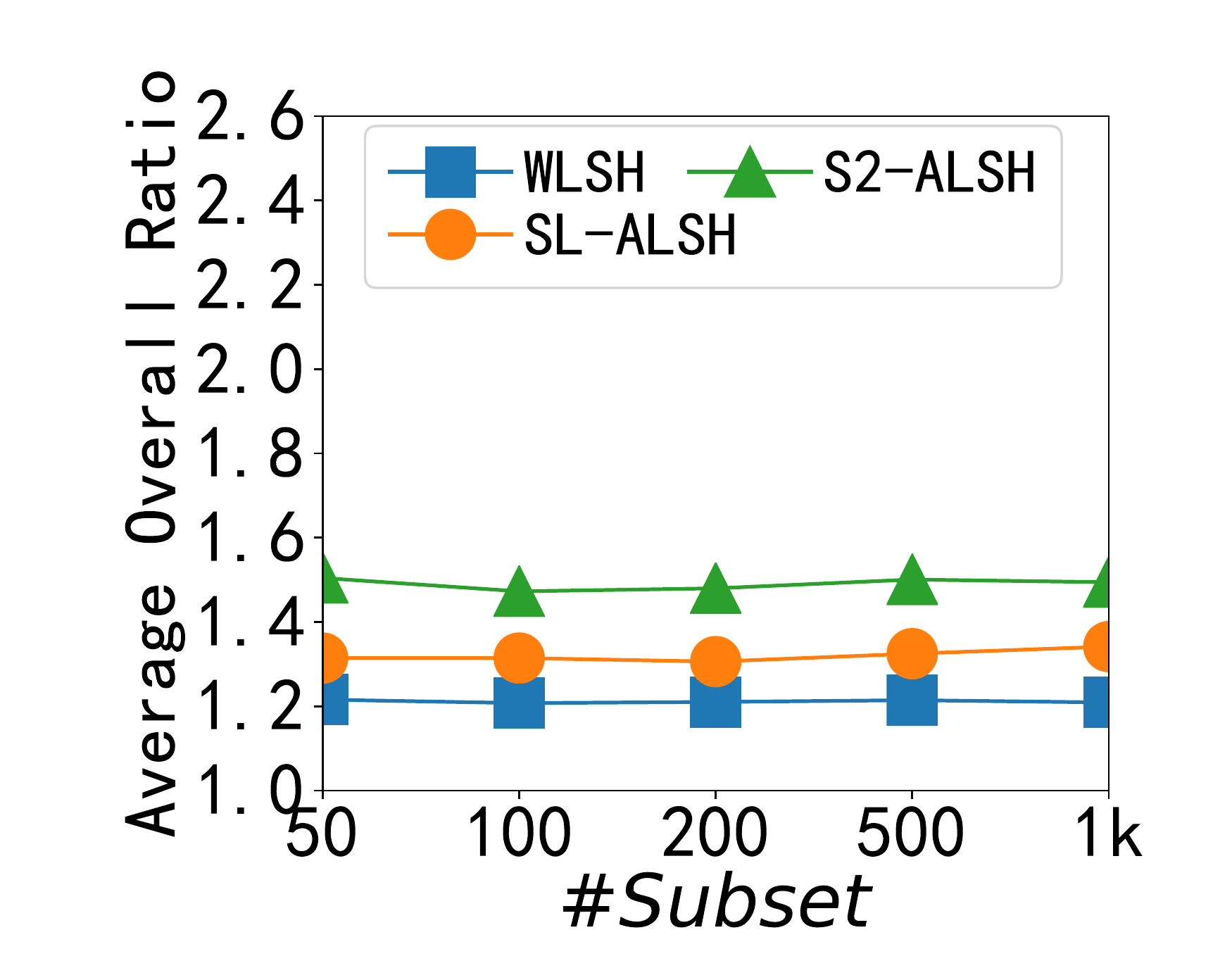}\label{sun/subset/ratio_useCt=1_k=10}}
	\subfigure[\textit{Sun}, $\left|S\right|$]{\includegraphics[width=0.246\textwidth]{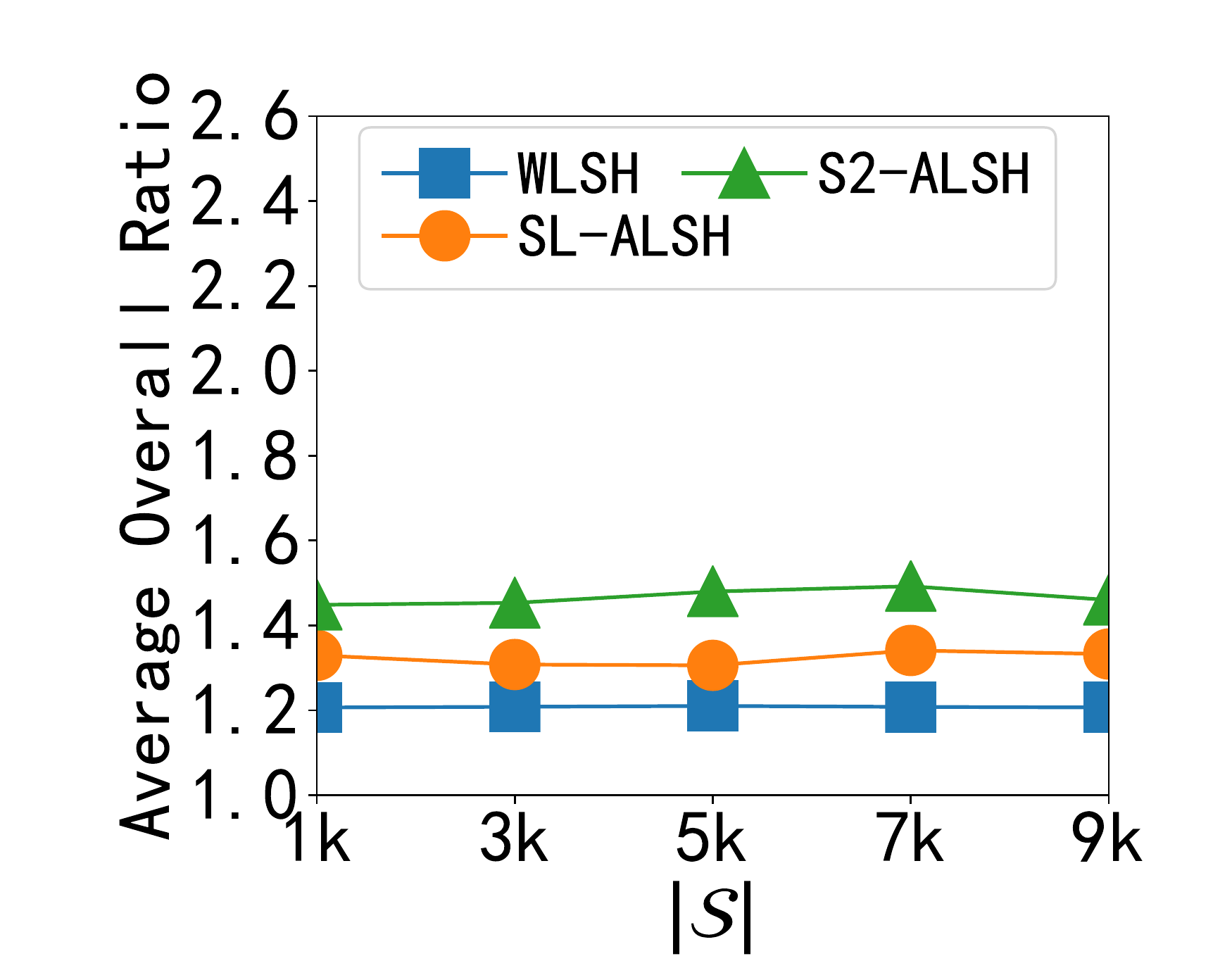}\label{sun/S/ratio_useCt=1_k=10}}
	
	\caption{Comparison of average overall ratios of WLSH, SL-ALSH and S2-ALSH, $k=10$}
	\label{compare/efficiency and accuracy/L2/useCt=1_k=10}
\end{figure*}

\begin{figure*}[t]
	\centering
	\subfigure[\textit{Notre}, $c$]{\includegraphics[width=0.246\textwidth]{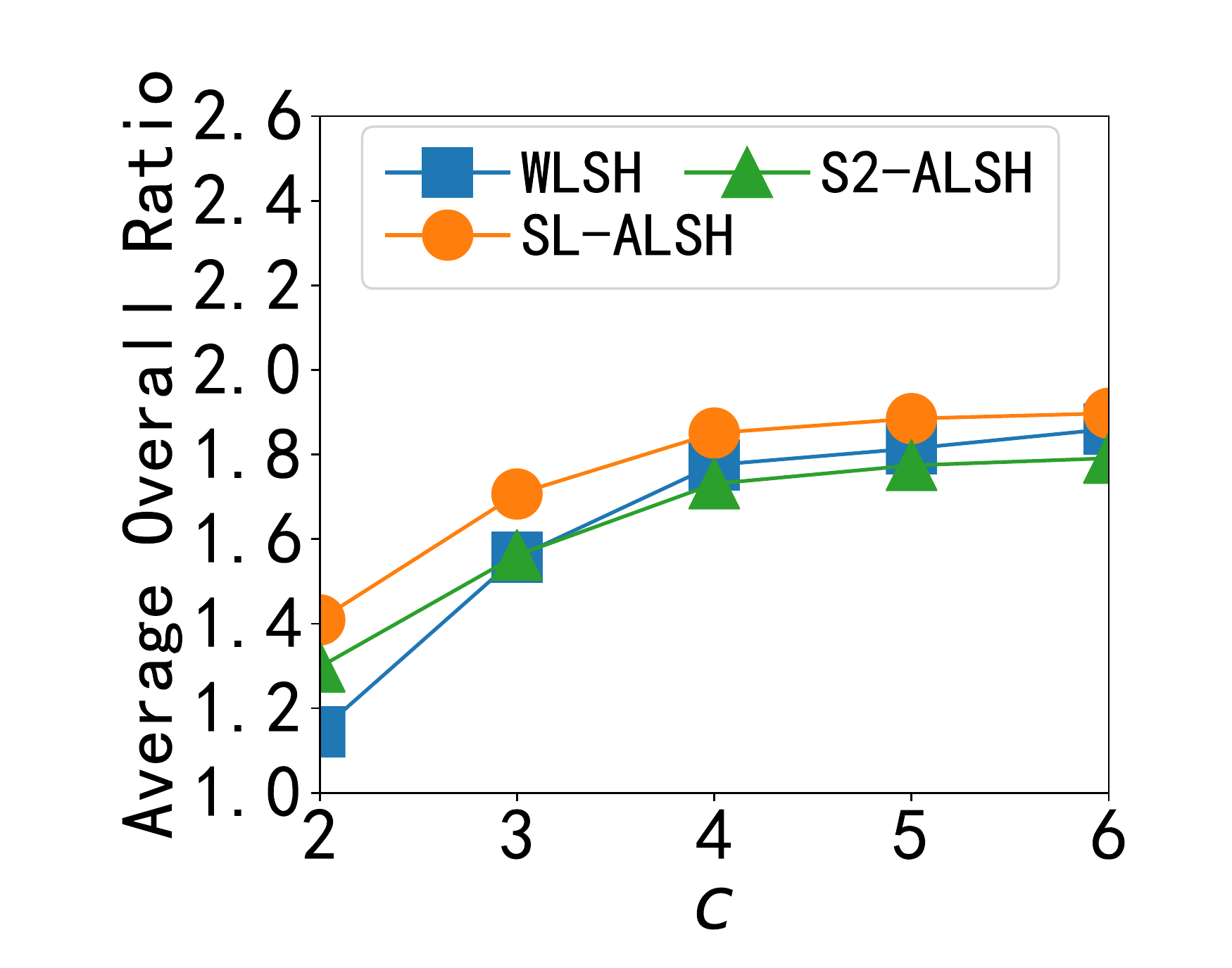}\label{notre/c/ratio_useCt=1_k=100}}
	\subfigure[\textit{Notre}, $\#Subrange$]{\includegraphics[width=0.246\textwidth]{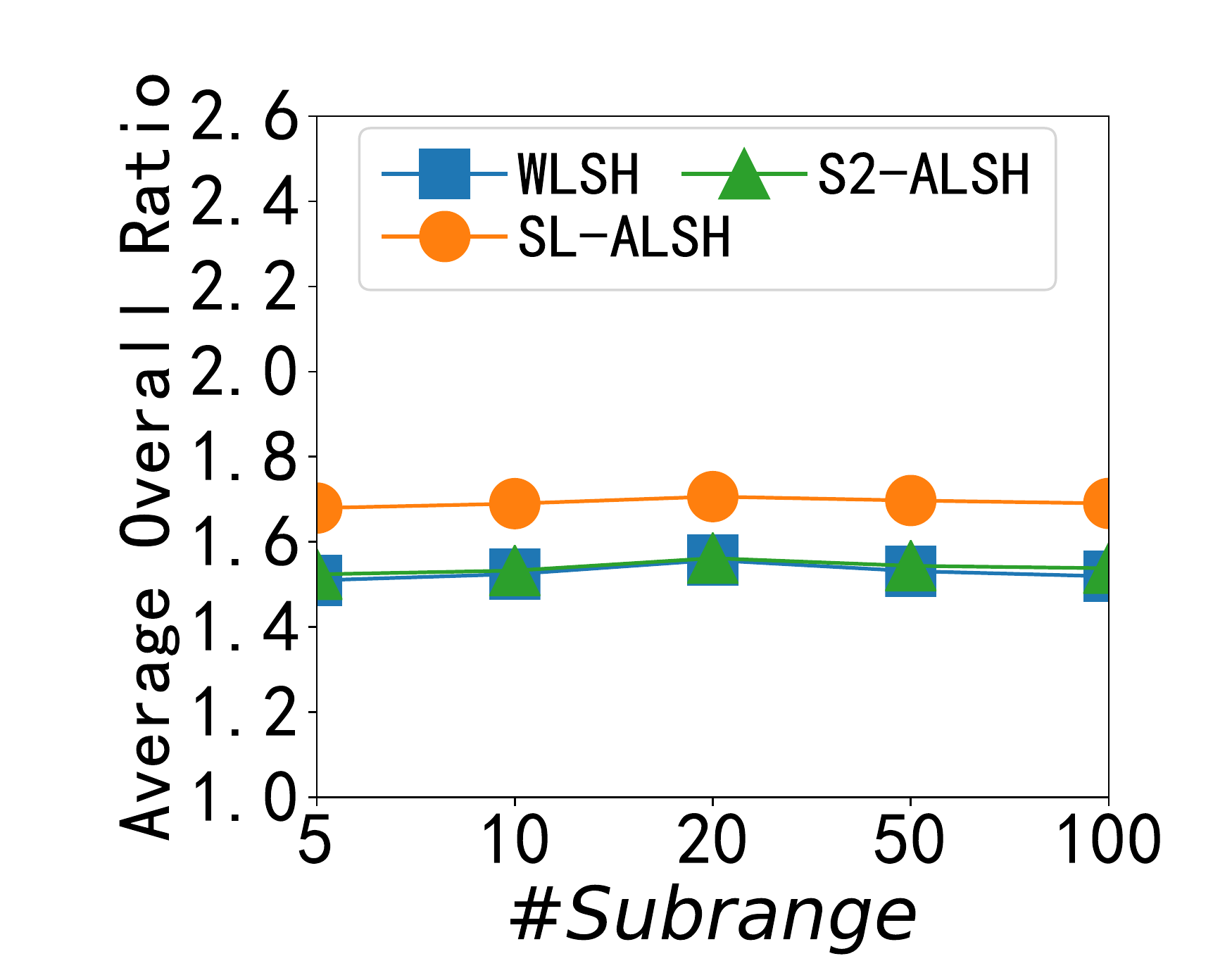}\label{notre/subrange/ratio_useCt=1_k=100}}
	\subfigure[\textit{Notre}, $\#Subset$]{\includegraphics[width=0.246\textwidth]{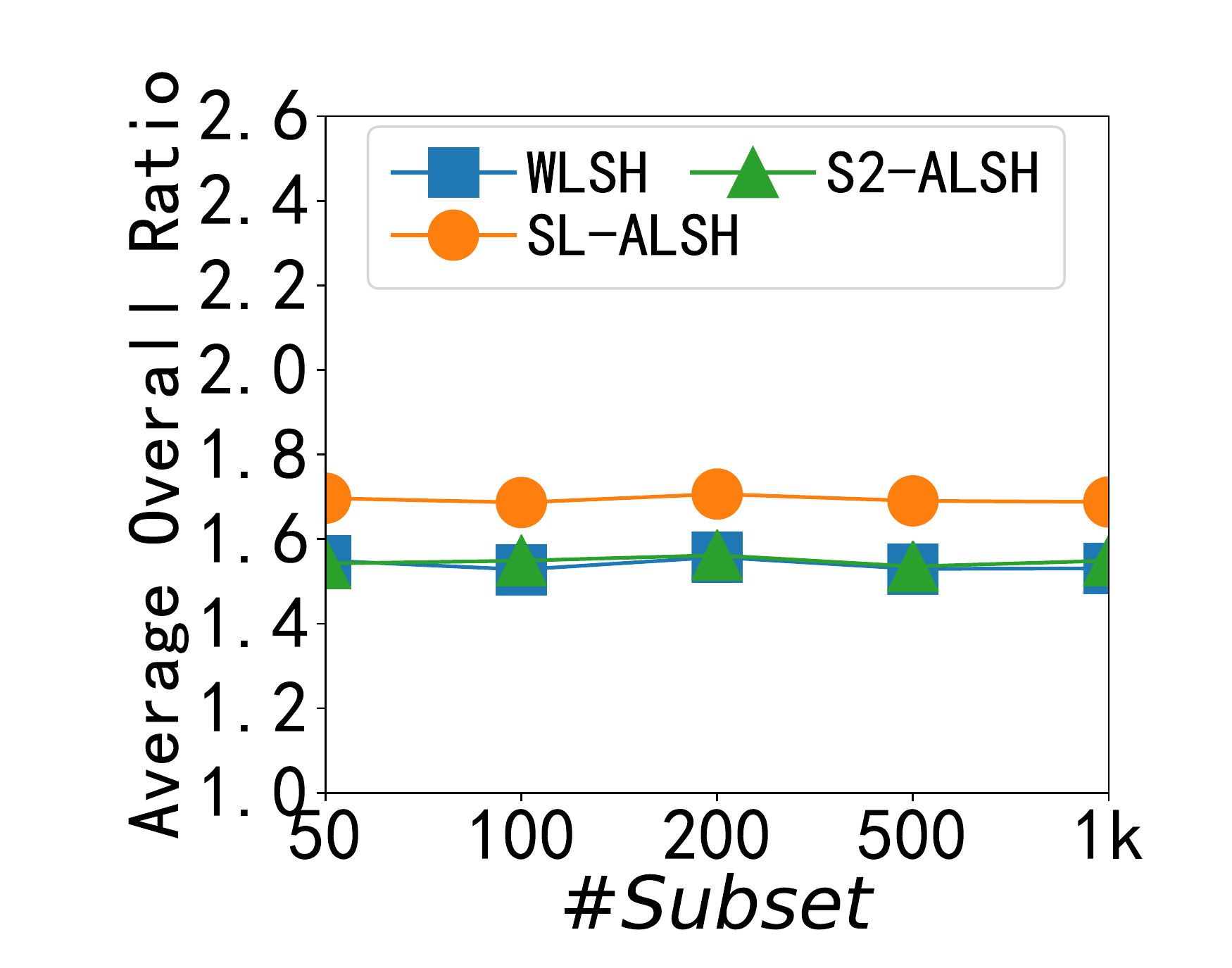}\label{notre/subset/ratio_useCt=1_k=100}}
	\subfigure[\textit{Notre}, $\left|S\right|$]{\includegraphics[width=0.246\textwidth]{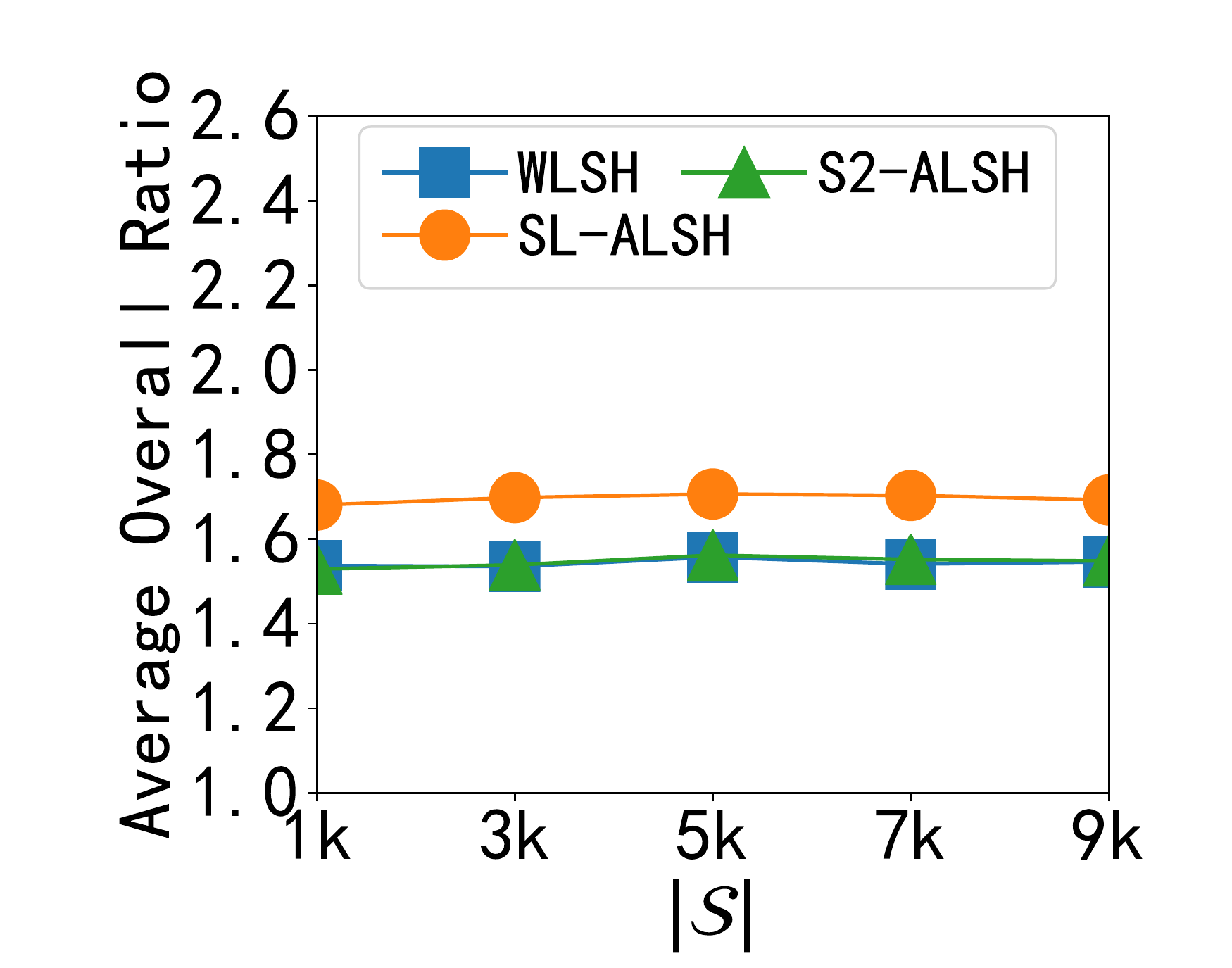}\label{notre/S/ratio_useCt=1_k=100}}
	
	\subfigure[\textit{Ukbench}, $c$]{\includegraphics[width=0.246\textwidth]{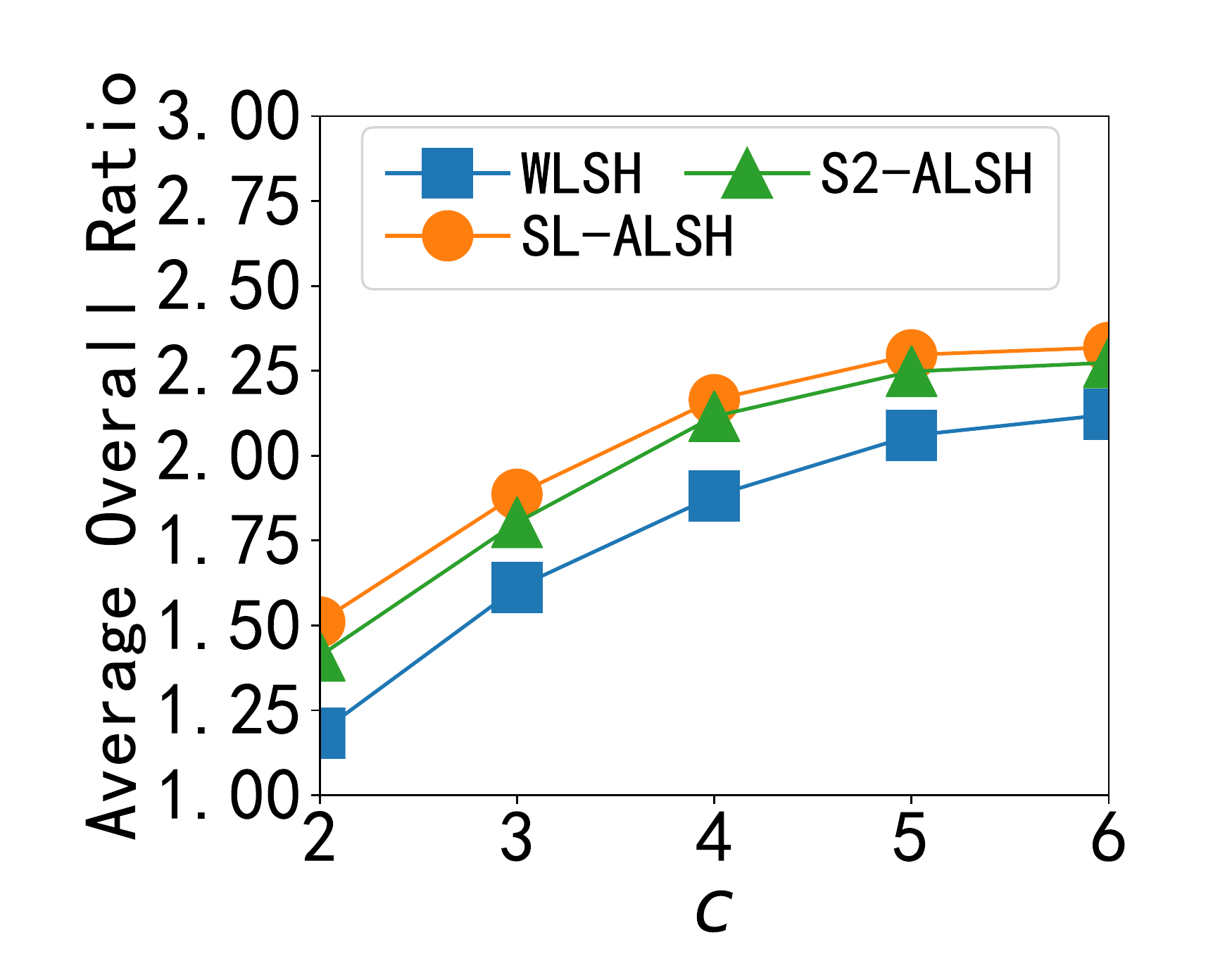}\label{ukbench/c/ratio_useCt=1_k=100}}
	\subfigure[\textit{Ukbench}, $\#Subrange$]{\includegraphics[width=0.246\textwidth]{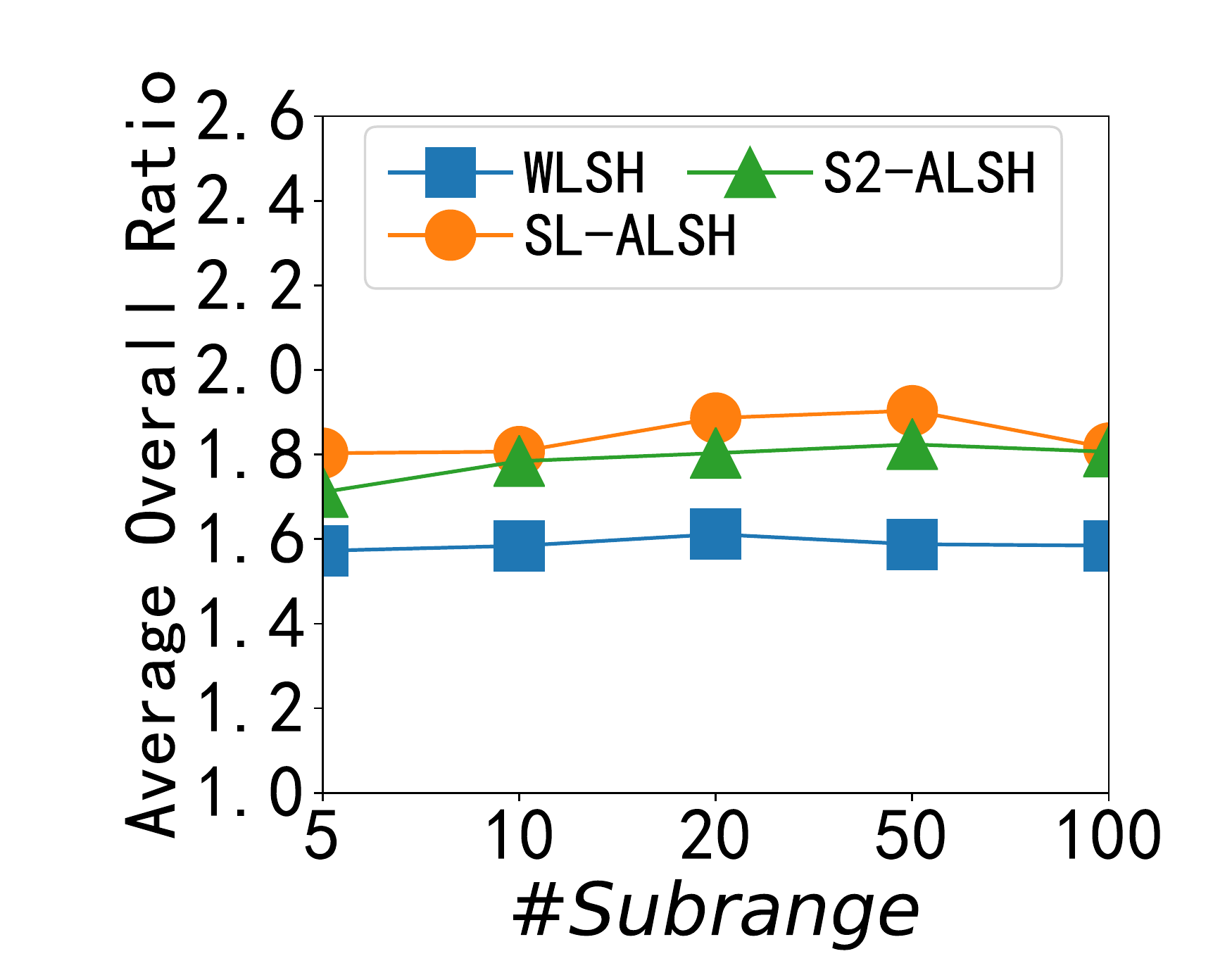}\label{ukbench/subrange/ratio_useCt=1_k=100}}
	\subfigure[\textit{Ukbench}, $\#Subset$]{\includegraphics[width=0.246\textwidth]{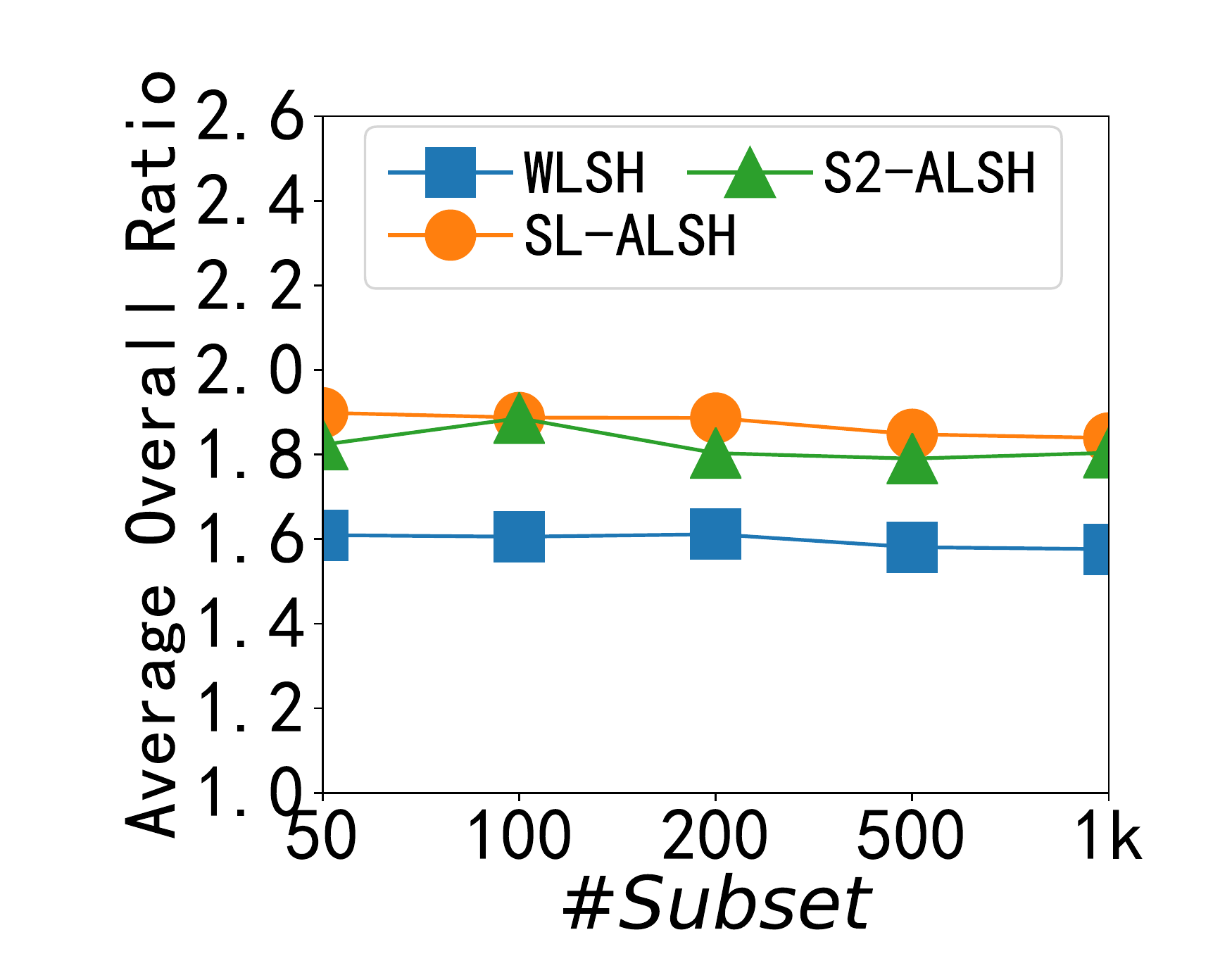}\label{ukbench/subset/ratio_useCt=1_k=100}}
	\subfigure[\textit{Ukbench}, $\left|S\right|$]{\includegraphics[width=0.246\textwidth]{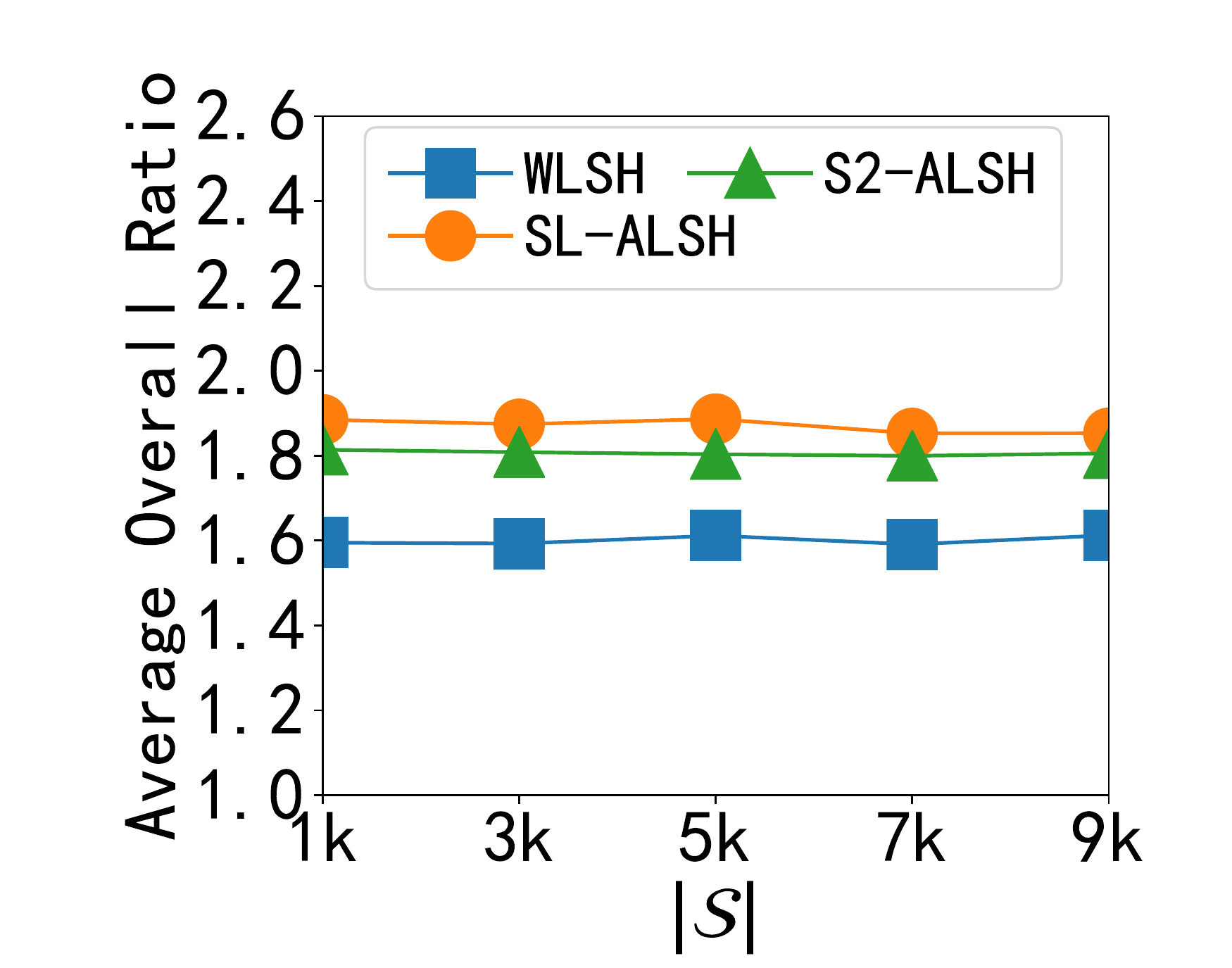}\label{ukbench/S/ratio_useCt=1_k=100}}
	
	\subfigure[\textit{Sift}, $c$]{\includegraphics[width=0.246\textwidth]{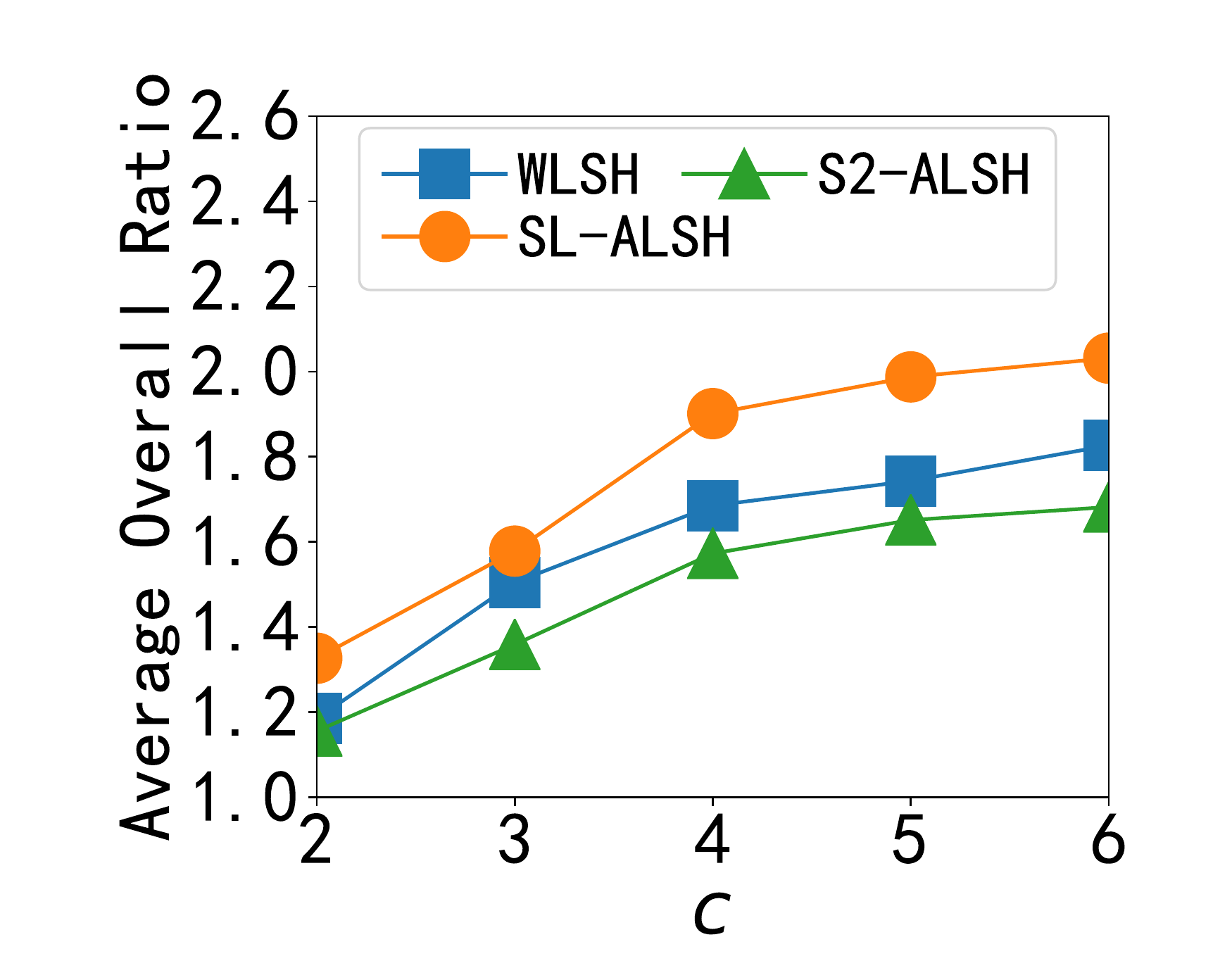}\label{sift/c/ratio_useCt=1_k=100}}
	\subfigure[\textit{Sift}, $\#Subrange$]{\includegraphics[width=0.246\textwidth]{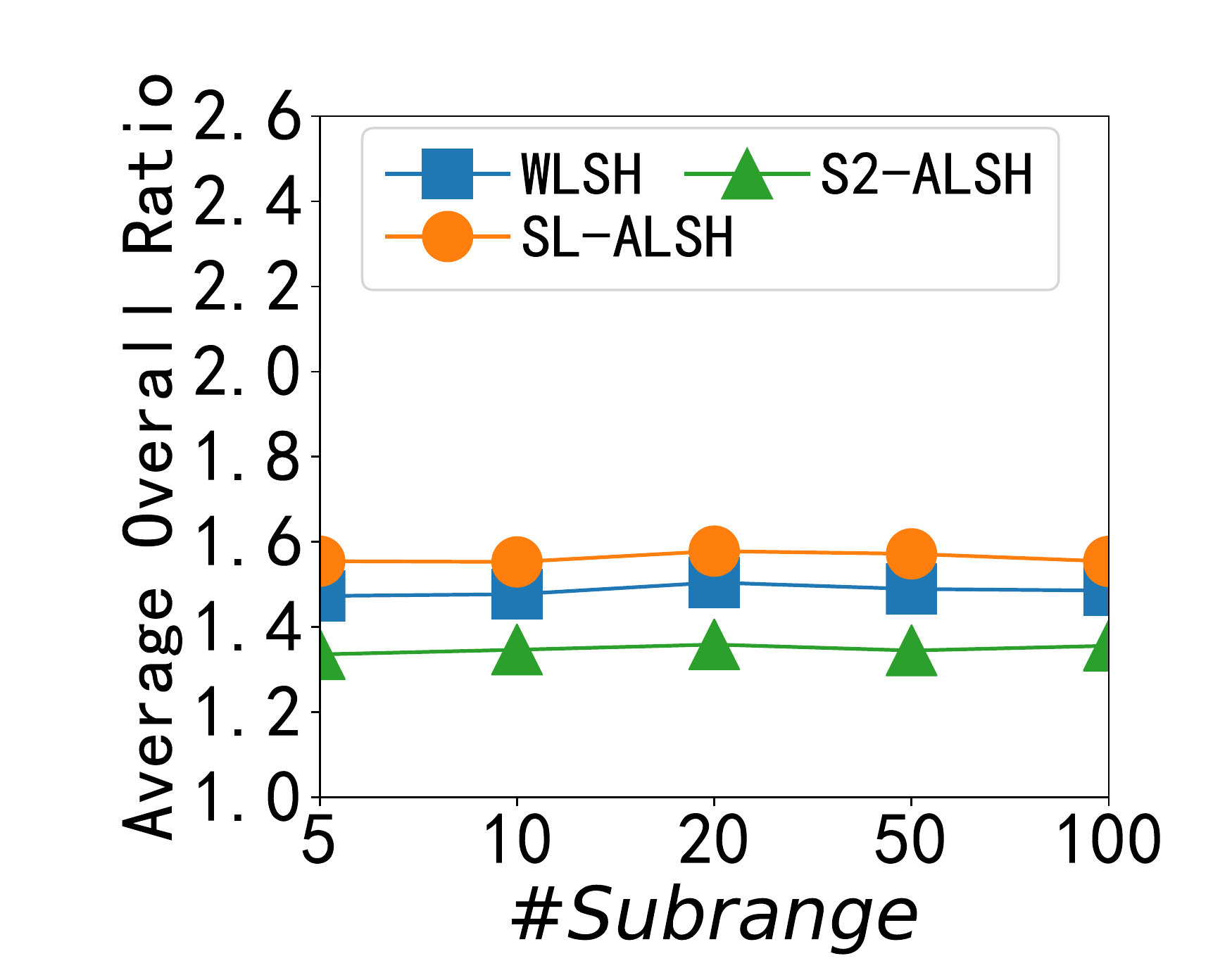}\label{sift/subrange/ratio_useCt=1_k=100}}
	\subfigure[\textit{Sift}, $\#Subset$]{\includegraphics[width=0.246\textwidth]{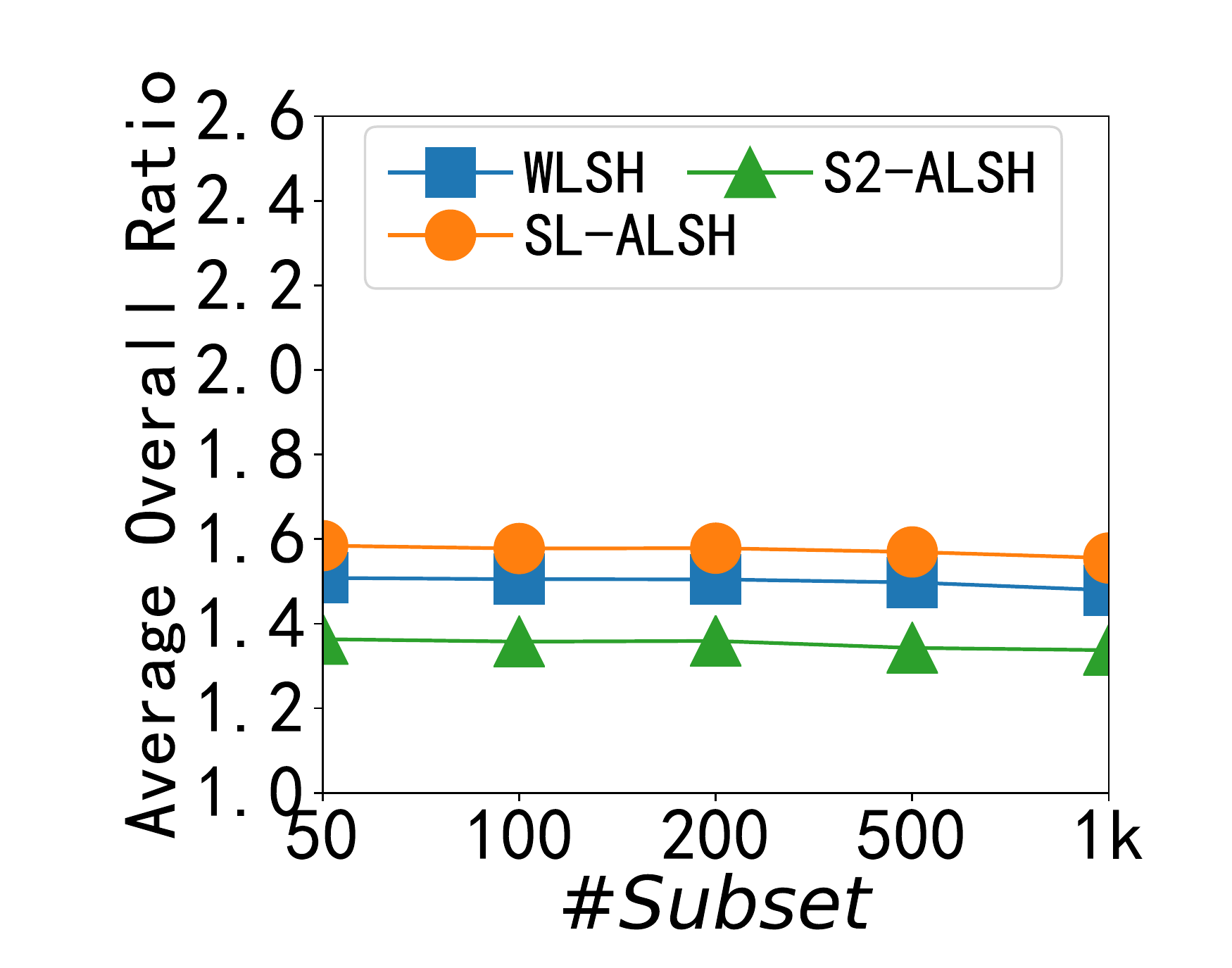}\label{sift/subset/ratio_useCt=1_k=100}}
	\subfigure[\textit{Sift}, $\left|S\right|$]{\includegraphics[width=0.246\textwidth]{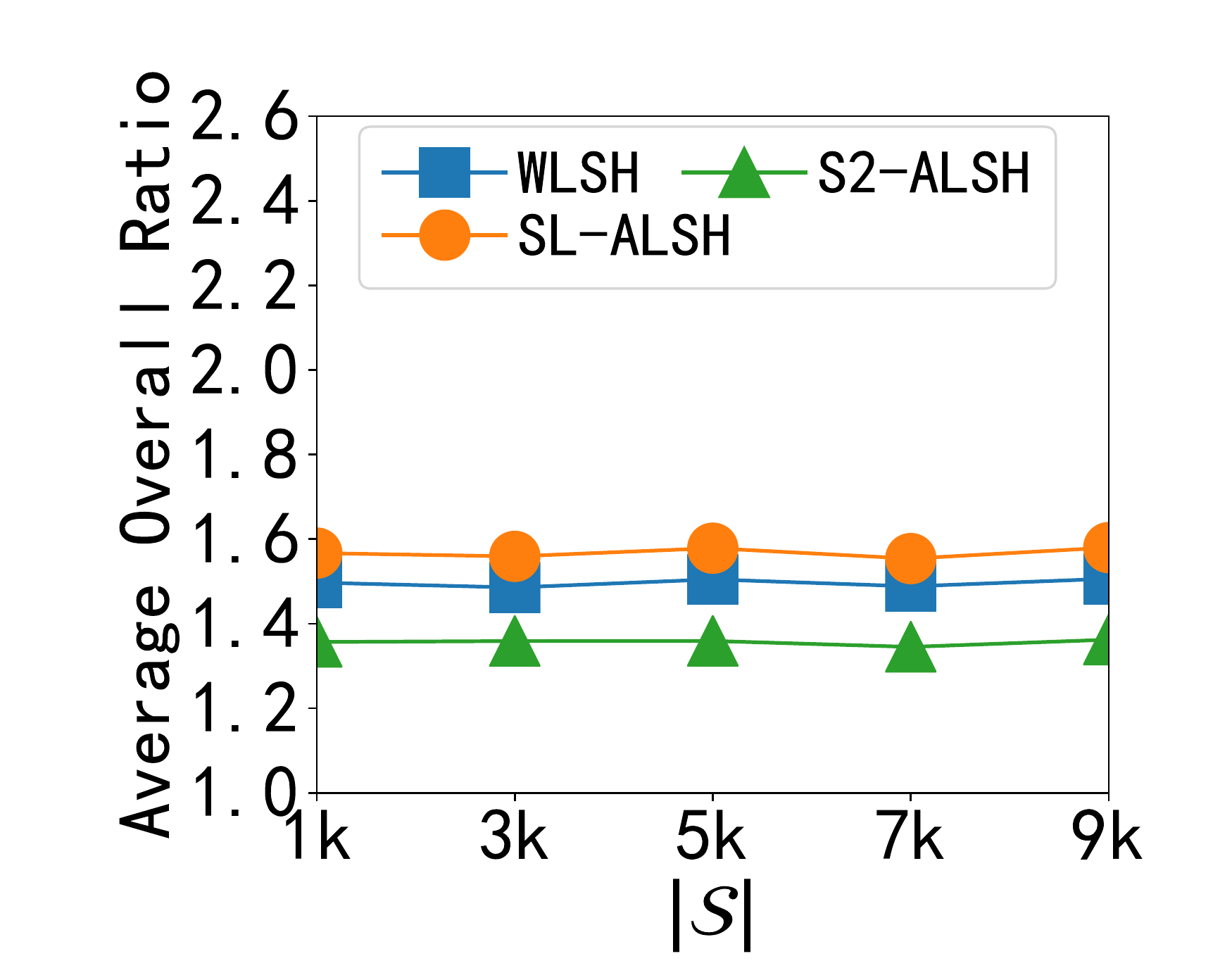}\label{sift/S/ratio_useCt=1_k=100}}
	
	\subfigure[\textit{Sun}, $c$]{\includegraphics[width=0.246\textwidth]{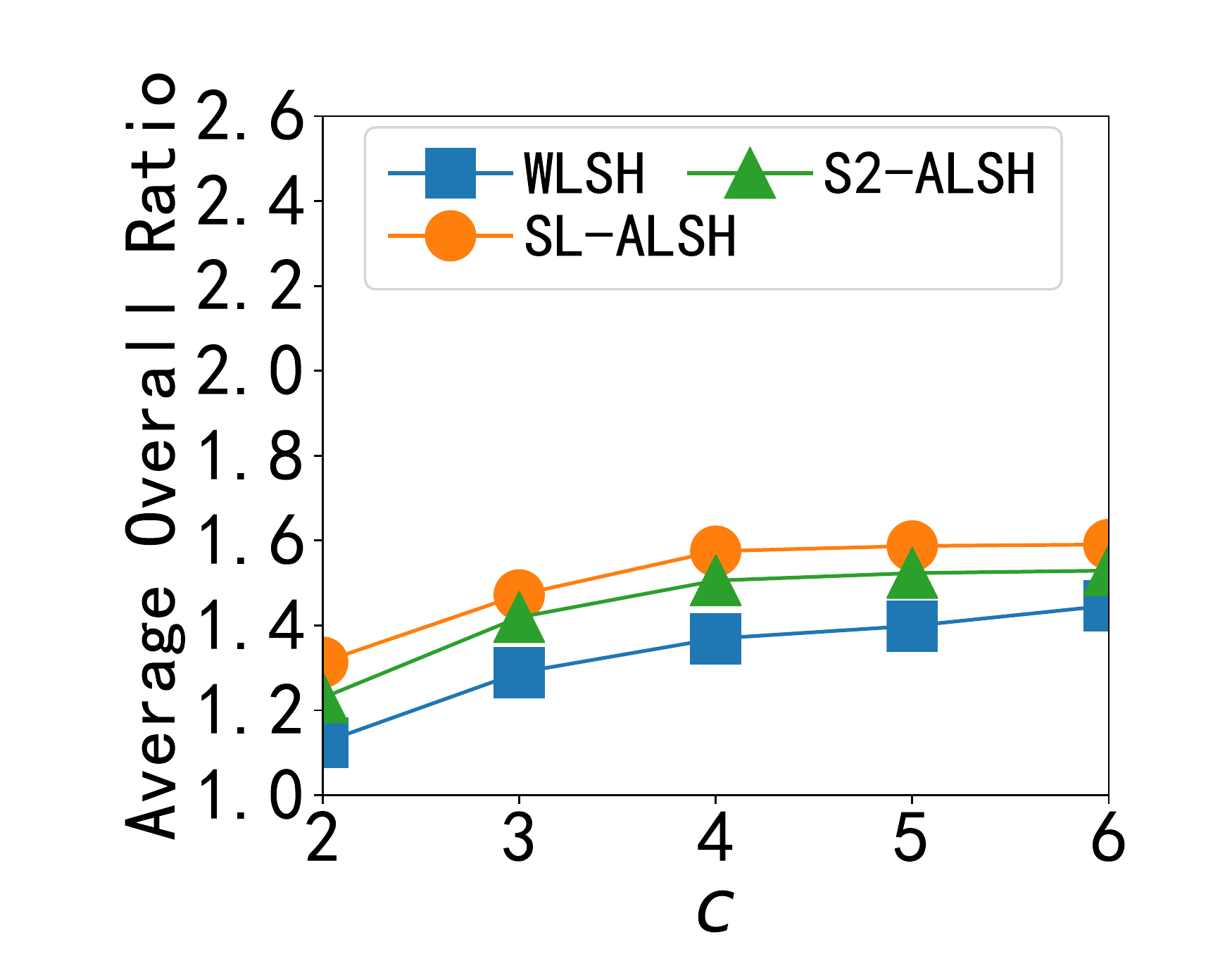}\label{sun/c/ratio_useCt=1_k=100}}
	\subfigure[\textit{Sun}, $\#Subrange$]{\includegraphics[width=0.246\textwidth]{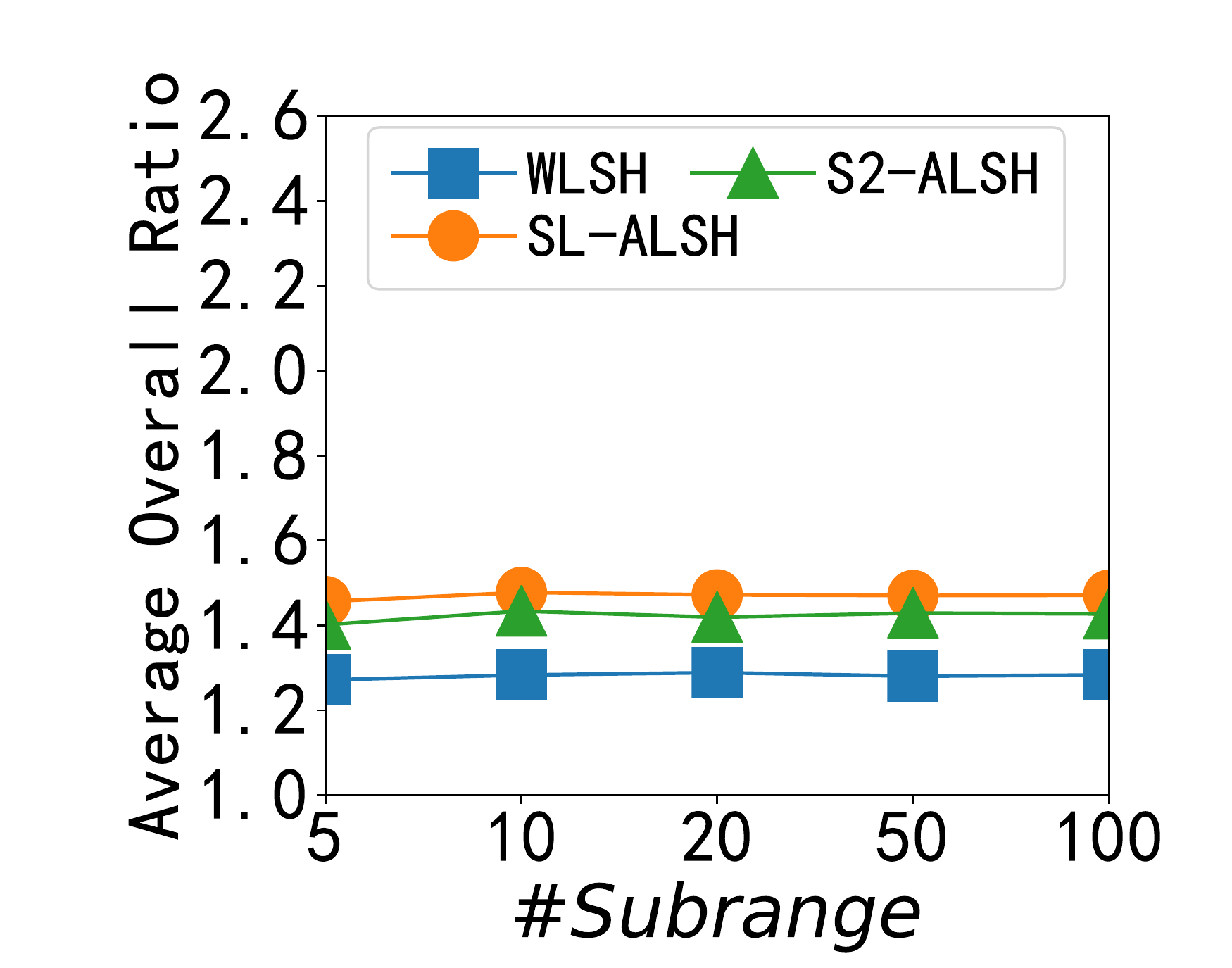}\label{sun/subrange/ratio_useCt=1_k=100}}
	\subfigure[\textit{Sun}, $\#Subset$]{\includegraphics[width=0.246\textwidth]{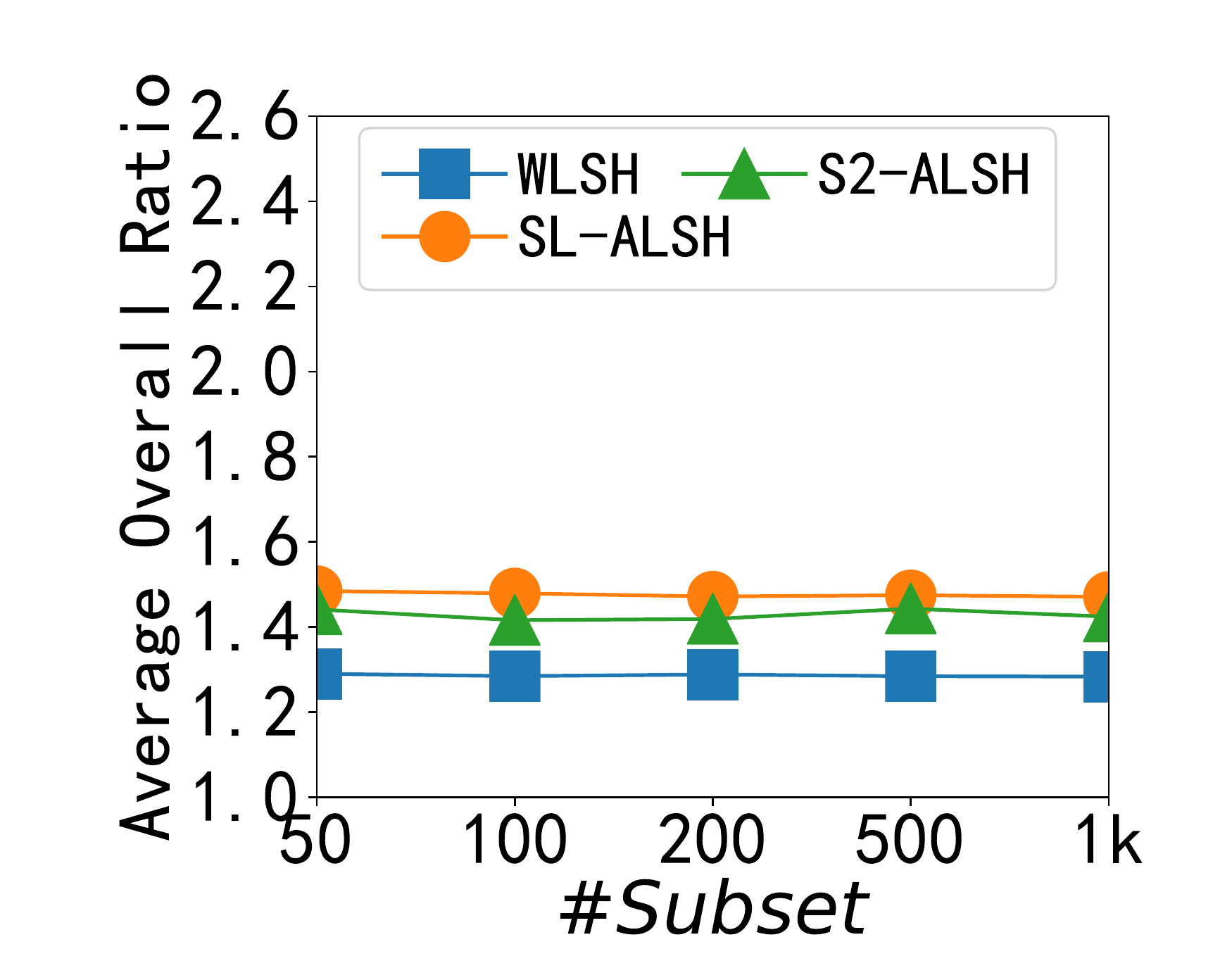}\label{sun/subset/ratio_useCt=1_k=100}}
	\subfigure[\textit{Sun}, $\left|S\right|$]{\includegraphics[width=0.246\textwidth]{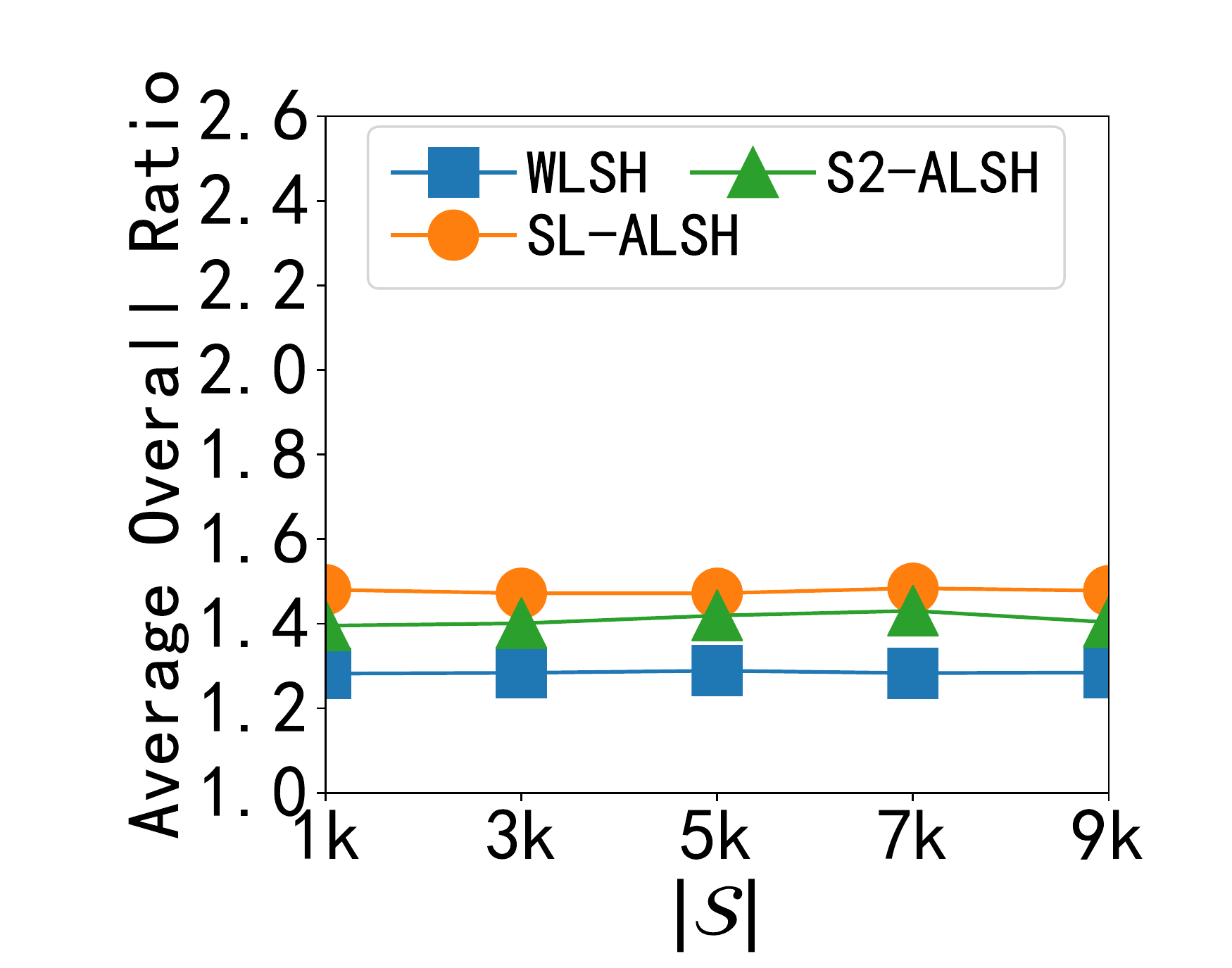}\label{sun/S/ratio_useCt=1_k=100}}
	
	\caption{Comparison of average overall ratios of WLSH, SL-ALSH and S2-ALSH, $k=100$}
	\label{compare/efficiency and accuracy/L2/useCt=1_k=100}
\end{figure*}

Figs. \ref{rand/efficiency and accuracy/L1}-\ref{rand/efficiency and accuracy/L2} show the query efficiency and query accuracy of WLSH on synthetic data sets when not using collision threshold reduction. Figs. \ref{real/efficiency and accuracy/L1/useCt=1/k=10}-\ref{real/efficiency and accuracy/L2/useCt=1} show the query efficiency and query accuracy of WLSH on real data sets when using collision threshold reduction.
Figs. \ref{compare/efficiency and accuracy/L2/useCt=1_k=10}-\ref{compare/efficiency and accuracy/L2/useCt=1_k=100} compare the average overall ratios of WLSH, SL-ALSH and S2-ALSH on real data sets by varying $c$, $\#Subrange$, $\#Subset$ and $\left|\mathcal{S}\right|$.

\begin{table}[h]
	\centering
	\footnotesize
	\caption{Parameter Settings for SL-ALSH and S2-ALSH}
	\begin{tabular}{c|c}
		\hline
		\multicolumn{2}{c}{SL-ALSH}\\
		\hline
		\hline
		\textit{Sift}
		&$m\in\{25, 27, 29, 31, 33\}$, $w=20$, $V=3.14159$\\
		\hline
		\textit{Ukbench}
		&$m\in\{25, 27, 29, 31, 33\}$, $w=20$, $V=3.14159$\\
		\hline
		\textit{Notre}
		&$m\in\{25, 27, 29, 31, 33\}$, $w=20$, $V=3.14159$\\
		\hline
		\textit{Sun}
		&$m\in\{15, 17, 19, 21, 23\}$, $w=20$, $V=3.14159$\\
		\hline
		\hline
		\multicolumn{2}{c}{S2-ALSH}\\
		\hline
		\hline
		\textit{Sift}&$m\in\{25, 27, 29, 31, 33\}$, $V=3.14159$\\
		\hline
		\textit{Ukbench}&$m\in\{25, 27, 29, 31, 33\}$, $V=3.14159$\\
		\hline
		\textit{Notre}&$m\in\{25, 27, 29, 31, 33\}$, $V=3.14159$\\
		\hline
		\textit{Sun}&$m\in\{15, 17, 19, 21, 23\}$, $V=3.14159$\\
		\hline
	\end{tabular}
	\label{parameter settings/SL-ALSH,S2-ALSH}
\end{table}

The parameter settings for SL-ALSH and S2-ALSH in Section \ref{efficiency/comparison} are shown in Table \ref{parameter settings/SL-ALSH,S2-ALSH}, where $w$ and $V$ are the same as in Equations \ref{rho/SL} and \ref{rho/S2}, and $m$ is the number of ALSH functions to form a compound hash function for creating a hash table. We vary $m$ to obtain multiple average overall ratios and take the smallest one as the final result. Please note that it is almost impossible to find the optimal parameter settings for SL-ALSH and S2-ALSH in terms of query efficiency and query accuracy because they vary with the data set and weight vector set.

\ifCLASSOPTIONcompsoc
  \section*{Acknowledgments}
\else
  \section*{Acknowledgment}
\fi

This work is supported by the National Natural Science
Foundation of China under grant NOs 61832003, U1811461,
and 61732003.

\ifCLASSOPTIONcaptionsoff
  \newpage
\fi



\bibliographystyle{IEEEtran}
\bibliography{IEEEabrv,huhuan}
%

%





\end{document}